%% file: dissertation.tex
\documentclass[dissertation]{uathesis}
\usepackage{graphicx}
\usepackage{natbib}			% natbib is available on most systems, and is
\usepackage{deluxetable}		% Allows use of AASTEX deluxe tables
\usepackage{aastex_hack}		% Allows other AASTEX functionality.

\usepackage{amsmath}			% AMS Math (advanced math typesetting)
\usepackage{amssymb} 
\usepackage{lscape}			% Used for making fitting large tables in by putting them landscape
\usepackage[tableposition=t]{caption} % NICK ADDED THIS (makes table captions normal size when font set smaller)

\usepackage{mathtools}
\usepackage{bm}
\usepackage{enumitem}
\usepackage{units} % includes nicefrac

\newcounter{daggerfootnote}

\usepackage[hidelinks]{hyperref}
\usepackage{slashed}

\completetitle{Elementary Particles and Plasma in the First Hour of the Early Universe}
\fullname{Cheng Tao Yang}			% Grad college wants your full name here.
\degreename{Doctor of Philosophy}	% Title of your degree.
\usepackage{titlesec}
 \titleformat*{\section}{\normalfont}
  \titleformat*{\subsection}{\normalfont}
\begin{document}

% Set up the title page
\maketitlepage
{DEPARTMENT OF PHYSICS}	% Title of your department.
{2023}							

% Insert the approval form.  Note that for electronic submission
% of your Ph. D. dissertation, you must bring *two* copies of the
% approval page to your final defense.  These must be signed by
% the committee.  Make two photocopies: one for Pam and the other
% for your records.  Then, bring the two signed originals to the
% graduate college when you submit the final version of the
% dissertation to the University of Arizona.
\approval
{October 20th, 2023}		% Defense Date	
{Johann Rafelski}	% Dissertation Director
{Johann Rafelski}	% 1st committee member
{Shufang Su}		% 2nd committee member
{Sean P Fleming}	% 3rd committee member
{Shufeng Zhang}		% 4th committee member
{Erich W Varnes}		% 5th committee member
%{}		% 6th committee member
% Include the ``Statement by Author'' for Dissertations

%\statementbyauthor
% If this is a Thesis, use the following form, with your thesis director's
% name and title in the square brackets like so (you should also omit the 
% approval form insertion above):
%\statementbyauthor[Jane M. Doe\\Professor of Chemistry]
% Include the ``Acknowledgements''

\incacknowledgements{acknowledgements}

% Include the ``Dedication''
\incdedication{dedication}

% Create a ``Table of Contents''
\tableofcontents

% Create a ``List of Figures''
\listoffigures

% Create a ``List of Tables''
\listoftables
\include{Chapter0}
% Include the ``Abstract''
\incabstract{abstract}

% Include the various chapters
\include{Chapter1}
\include{Chapter2}
\include{Chapter3}

\include{Chapter4}
\include{Chapter5}

\include{Chapter6}

\include{Chapter7}

% Include the various appendices
\appendix
\include{appendix}

% Switch the spacing to single-spaced for the references
\renewcommand{\baselinestretch}{1}		% chaning the value
\small\normalsize						% switch size to make the value take

% Create the References list
\bibliographystyle{uabibnat}
\bibliography{bibliography}

\end{document}

%% file: Chapter0.tex
 %%%%%%%%%%%%%%%%%%%%%%%%%%%%%%%%%%%%%%%
\chapter*{LIST OF PUBLICATIONS AND AUTHOR CONTRIBUTIONS}
%\label{sec:pubs}
\addcontentsline{toc}{chapter}{LIST OF PUBLICATIONS AND AUTHOR CONTRIBUTIONS}
%%%%%%%%%%%%%%%%%%%%%%%%%%%%%%%%%%%%%%%
In the course of satisfying the University of Arizona Department of Physics's requirements for a 
Ph.D. doctoral dissertation, I listed the following publications and point out my contribution to each work which is described under each item.
\begin{itemize}
 \item ''A Short Survey of Matter-Antimatter Evolution in the Primordial 
Universe'' by~[\cite{Rafelski:2023emw}] is a 50 page long review with many novel results 
describing the role of antimatter in the early universe. I collaborate with 
Andrew Steinmetz, Jeremiah Birrell, and Johann Rafelski the document creation, providing the numerical results/figures for the paper to help creating one coherent presentation. I acknowledge the help and consultation of Andrew Steinmetz, Jeremiah Birrell, and Johann Rafelski in research and computation. 

\item ''Matter-antimatter origin of cosmic magnetism'' 
by~[\cite{Steinmetz:2023nsc}] proposes a model of para-magnetization driven by the large 
matter-antimatter (electron-positron) content of the early universe. I collaborate with Andrew Steinmetz to carry out the computation
, contribute key results and five 
technical figures for the paper. I acknowledge the help and consultation of Andrew Steinmetz, Martin Formanek, and Johann Rafelski in research and computation. 

\item ''Electron-positron plasma in BBN: Damped-dynamic screening'' by~[\cite{Grayson:2023flr}] use the linear response theory adapt by C. Grayson to describe the inter nuclear potential in electron/positron plasma. My contribution is the computation of the chemical potential and plasma damping rate which are important properties to  implement into relativistic Boltzmann equation and linear response theory. I acknowledge the help and consultation of Christopher Grayson, Martin Formanek, and Johann Rafelski in research and computation.

\item ''Cosmological strangeness abundance'' by~[\cite{Yang:2021bko}] investigates the strange particle composition of the expanding early Universe and examine their freeze-out temperatures. I performed all computation, writing, and 
figure making for the paper. I acknowledge the help and consultation of Johann Rafelski in 
research, writing and editing.

\item ''The muon abundance in the primordial Universe'' by~[\cite{Rafelski:2021aey}] is a study of the muon abundance and its persistence temperature in early Universe. I performed all 
computation and writing in preparation of the first draft and approved the final draft before 
submission. I acknowledge the help and consultation of Johann Rafelski in research, writing and editing.

\item ''Reactions Governing Strangeness Abundance in Primordial Universe'' by~[\cite{Rafelski:2020ajx}] is a conference proceeding paper for
the 19th International Conference on Strangeness in Quark Matter (SQM 2021). It summarize our earlier work in strangeness reactions. I performed all 
computation and writing in preparation of the first draft and approved the final draft before 
submission. I acknowledge the help and consultation of Johann Rafelski in research, writing and editing.

\item''Possibility of bottom-catalyzed matter genesis near to primordial QGP hadronization'' by~[\cite{Yang:2020nne}] is our fist study of the bottom flavor abundance and show the  nonequilibrium behavior near to QGP hadronization. I performed all computation and writing in preparation of the paper. I acknowledge the help and consultation of Johann Rafelski in research, writing and editing.

\item ''Lepton Number and Expansion of the Universe'' by~[\cite{Yang:2018oqg}] proposes a model of large lepton asymmetry and explore how this large cosmological lepton yield
relates to the effective number of (Dirac) neutrinos. I performed all computation and writing in preparation of the paper. I acknowledge the help and consultation of Johann Rafelski in research, writing and editing.

\item''Temperature Dependence of the Neutron Lifespan'' by~[\cite{Yang:2018qrr}] is a study of neutron lifespan in plasma with Fermi-blocking from electron and neutrino. I performed all computation and writing in preparation of the paper. I acknowledge the help and consultation of Johann Rafelski in research, writing and editing.

\item ''Strong fields and neutral particle magnetic moment dynamics'' by~[\cite{Formanek:2017mbv}] is an overview of our research group's efforts in studying neutral particle dynamics in electromagnetic fields. My contribution is on the neutrino section. I consulted and helped lead author Martin Formanek, and co-authors Stefan Evans, Andrew Steinmetz, and Johann Rafelski in 
editing and revising the manuscript.

\item "Relic Neutrino Freeze-out: Dependence on Natural Constants" by~[\cite{Birrell:2014uka}] is a study of neutrino freeze-out temperature as a function of standard model parameter and its application on the effective number of (Dirac) neutrinos. My contribution is the calculation of all the neutrino-matter weak interaction matrix elements required for the Boltzmann code. I acknowledge the help and consultation of Jeremiah Birrell and Johann Rafelski in research and computation.

\item ''Fugacity and Reheating of Primordial Neutrinos'' by~[\cite{Birrell:2013gpa}] is as study of neutrino fugacity as a funciton of neutrin kinetic freeze-out temperature. My contribution is the calculation of neutrino interaction matrix  elements and helping the evaluation of  neutrino relaxation time. I acknowledge the help and consultation of Jeremiah Birrell, Johann Rafelski, and Pisin Chen in research and computation.

\item''Relic neutrinos: Physically consistent treatment of effective number of neutrinos and neutrino mass'' by~[\cite{Birrell:2012gg}] is a model independent study of the neutrino momentum distribution at freeze-out, treating the freeze-out temperature as a free parameter. I collaborate with Jeremiah Birrell, Johann Rafelski, and Pisin Chen the document creation. I acknowledge the help and consultation of Jeremiah Birrell, Johann Rafelski, and Pisin Chen in research. 

\end{itemize}

%% file: Chapter1.tex
\chapter{Introductory topics: particles and plasma in the Universe}\label{Introduction}
%{Introduction to cosmology and overview}
In this chapter, we will introduce the fundamental concepts in cosmology for us to explore the properties of the Universe during the `first hour'. I will first present the standard cosmological Friedmann-Lemaitre-Robertson-Walker (FLRW) model, then introduce the general Fermi/Bose distribution with and its application in the early Universe. Finally I present an overview of Universe evolution from $300\,\mathrm{MeV}>T>0.02\,\mathrm{MeV}$.
The Natural unit $c=\hbar=k_{B}=1$ is used throughout the thesis for discussion.
%{Introduction\daggerfootnote{This chapter has been published previously as \citet{Gottbrath1999}.}}

%~~~~~~~~~~~~~~~~~~~~~~~~~~~~~~~~~~~~~~~~~~~~~~~~~
%The modern observation in cosmology has demonstrated that the observed universe is highly symmetric in its large-scale structure and 

\section{Textbook review: the standard FLRW-Universe model}
%In this section we will focus on the following:
%\begin{itemize}
%    \item The Robertson –Walker Universe
%    \item The Friedmann equation (Hubble %    \item The composition of the universe
%\end{itemize}
The Friedmann-Lemaitre-Robertson-Walker (FLRW) Universe is a theoretical model used widely to describe the cosmological evolution of the Universe. It is based on the cosmological principles which assumes homogeneity and isotropy of the Universe on large scales. In general, the FLRW metric can be written as
\begin{align}\label{metric}
ds^2=c^2dt^2-a^2(t)\left[ \frac{dr^2}{1-kr^2}+r^2(d\theta^2+\sin^2\theta\,d\phi^2)\right].
\end{align}
The metric is characterized by the scale factor $a(t)$ which measures the size of the Universe as a function of time $t$. The geometric parameter $k$ identifies the Gaussian geometry of the spatial hyper-surfaces defined by co-moving observers. The metrics are qualitatively different depending on the value of $k$. We have $k=1$ which correspond to the closed Universe,  $k=0$ correspond to flat Universe, and $k=-1$ for open geometries of the Universe. Current observation of cosmic microwave background (CMB) anisotropy preferred value $k=0$~[\cite{Planck:2013pxb,Planck:2015fie,Planck:2018vyg}].

The cosmological equations that describe the evolution of the Universe are derived from the Einstein equations. In general, the Einstein equation with cosmological constant $\Lambda$ can be written as:
\begin{equation}\label{Einstein}
  G^{\mu\nu}=R^{\mu\nu}-\left(\frac R 2 +\Lambda\right) g^{\mu\nu}=8\pi G_N T^{\mu\nu}, 
\quad R= g_{\mu\nu}R^{\mu\nu}.  
\end{equation}
where $G_{N}$ is the Newtonian gravitational constant, and $T^{\mu\nu}$ is the stress-energy tensor. Given the homogeneous and isotropic symmetry conditions imply that the matter context of the Universe can be expressed as a perfect fluid. The stress-energy tensor $T^\mu_\nu$ of perfect fluid can be written as
\begin{align}
 T^\mu_\nu =\mathrm{diag}(\rho, -P, -P, -P).
\end{align}
where $\rho$ is energy density and an $P$ is the isotropic pressure.

Substituting the perfect fluid form of the stress-energy tensor into the Einstein equations, one can derive the cosmological equations that describe the evolution of the Universe. We then obtain Friedmann equations as follows:
\begin{align}
\label{Hubble} 
&H^{2}\equiv\left(\frac{\dot a}{a}\right)^2=\frac{8\pi G_{N}}{3}\rho-\frac{k}{a^2}+\frac{\Lambda}{3},\\
\label{q_value}
&qH^2=\frac{4\pi G_{N}}{3}\left(\rho+3P\right)-\frac{\Lambda}{3},\qquad q\equiv -\frac{a\ddot a}{\dot a^2},
\end{align}
where $H$ is the Hubble parameter,  $q$ is the deceleration parameter. These equations relate the dynamics of the scale factor $a(t)$ to the energy density and pressure of the cosmic plasma. On the other hand, considering the divergence freedom of the total stress-energy tensor $\nabla_\nu T^{\mu\nu}=0$. For $\mu=0$ component, we have
\begin{align}\label{energy_eq}
\nabla_\nu T^{0\nu}=\frac{d\rho}{dt}+3H(\rho+P)=0
\end{align}
which provides dynamical evolution equation for $\rho(t)$ and $P(t)$. Solutions of Eq.~(\ref{energy_eq}) describes the time evolution of  energy density and pressures in the Universe. Given the energy density and pressure as a function of time, we can illustrates how the Universe evolves according to the Friedmann equations Eq.~(\ref{Hubble}) and Eq.~(\ref{q_value}). Solving these equations allows us to understand the dynamics and evolution of the Universe such as the Hubble expansion and the behavior of matter and energy over cosmic time.

%~~~~~~~~~~~~~~~~~~~~~~~~~~~~~~~~~~~~~~~~~~~~~~~~~

\section{Approaching abundance equilibrium: Fermi/Bose distribution}

In the early Universe, the reaction rates of particles in the cosmic plasma were much greater than the Universe expansion rate $H$. Therefore, the local thermal equilibrium has been maintained. Assuming the particles are in thermal equilibrium, the dynamical information can be obtained from the single-particle distribution function. The general relativistic covariant Fermi/Bose momentum distribution can be written as
\begin{align}
f_{F/B}(\Upsilon_i,p_i)=\frac{1}{\Upsilon^{-1}_i\exp{\left[(u\cdot p_i-\mu_i)/T\right]}\pm1}
\end{align}
where the plus sign applies for fermions, and the minus sign for bosons. The Lorentz scalar $(u_i\cdot p_i)$ is a scalar product of the particle four momentum $p^\mu_i$ with the local four vector of velocity $u^\mu$. In the absence of local matter flow, the local rest frame is the laboratory frame 
\begin{align}
u^\mu=\left(1,\vec{0}\right),\,\,\,\,\,\,\,\,\, p^\mu_i=\left(E_i,\vec{p}_i\right).
\end{align}  
The parameter $\Upsilon_i$ is the fugacity of a given particle which describes the pair density and it is the same for both particles and antiparticles. For $\Upsilon_i=1$ the distribution maximizes the entropy content at a fixed particle energy. The parameter $\mu_i$ is the chemical potential for a given particle which is associated to the density difference between particles and antiparticles. 

In general there are two types of chemical equilibriums associated with the chemical parameters $\Upsilon$ and $\mu$. We have:
\begin{itemize}
\item Absolute chemical equilibrium:\\
The absolute chemical equilibrium is the level to which energy is shared into accessible degrees of freedom, e.g. the particles can be made as energy is converted into matter.
The absolute equilibrium is reached when the phase space occupancy approaches unity $\Upsilon\to1$. 
 \item Relative chemical equilibrium:\\
 The relative chemical equilibrium is associated with the chemical potential $\mu$ which involves reactions that distribute a certain already existent element/property among different accessible compounds. 
 \end{itemize}
The dynamics of absolute chemical equilibrium, in which energy can be converted to and from particles and antiparticles, is especially important. The consequences for the energy conversion to from particles/antiparticle can be seen in the first law of thermodynamics by introducing the general chemical potential $\mu_N$ for particle and $\mu_{\bar{N}}$ for antiparticle as follows:
\begin{align}
\mu_N\equiv\mu+T\ln\Upsilon,\qquad{\mu_{\bar{N}}}\equiv{-\mu}+T\ln\Upsilon.
\end{align}
Then the first law of thermodynamics can be written as
\begin{align}
dE&=-PdV+TdS+{\mu_N}dN+{\mu_{\bar{N}}}d{\bar{N}}
\\&=-PdV+TdS+{\mu}(dN-d{\bar{N}})+T\ln{\Upsilon}(dN+d{\bar{N}}).
\end{align}
It shows that the chemical potential $\mu$ is the energy required to change the difference between particles and antiparticles, and the $T\ln\Upsilon$ is the energy required to change the total number of particle and antiparticle, and the fugacity $\Upsilon$ is the parameter to adjust the energy.

\subsubsection{Boltzmann equation and particle freeze-out}
The Boltzmann equation describes the evolution of distribution function f in phase space. The Boltzmann equation in the FLRW universe can be written as
\begin{align}
\frac{\partial f}{\partial t}-\frac{\left(E^2-m^2\right)}{E}H\frac{\partial f}{\partial E}=\frac{1}{E}\sum_{q}\mathcal{C}_q[f],
\end{align}
where $H=\dot{a}/a$ is the Hubble parameter. Due to homogeneity and isotropy of the Universe, the distribution function depends on time $t$ and energy $E=\sqrt{p^2+m^2}$ only. The collision term $\sum_qC_q$ represents all elastic and inelastic interactions and $q$ labels the corresponding physical process. In general, the collision term is proportional to the relaxation time for given collision as follows [\cite{ANDERSON1974466}]
\begin{align}
\frac{1}{E}\mathcal{C}_q[f]\propto\frac{1}{\tau_{rel}}
\end{align}
where $\tau_{rel}$ is the relaxation time for the reaction, which is on the order of magnitude of time for the reaction to reach chemical equilibrium.

As the Universe expands, the collision term in the Boltzmann equation competes with the Hubble term. In general, a given particle freeze-out from the cosmic plasma when its interaction rate $\tau_{rel}^{-1}$ becomes smaller than the Hubble expansion rate
\begin{align}
H\geqslant\tau_{rel}^{-1}.
\end{align}
When this happens, the particle's interactions are not rapid enough to maintain thermal distribution, either because the density of particles becomes so low that the chances of any two particles meeting each other becomes negligible, or because the particle energy becomes too low to interact. The freeze-out process can be categorized into three distinct stages based on the type of freeze-out interactions, we have~[\cite{Birrell:2012gg,Rafelski:2023emw}]:

\begin{itemize}
\item Chemical freeze-out :\\
As the Universe expands and the temperature drops, the rate of the inelastic scattering (e.g. production and annihilation reaction) that maintain the equilibrium density becomes smaller than the expansion rate. At this point, the inelastic scattering ceases, and a relic population of particles remain. Prior to the chemical freeze-out temperature, number changing processes are significant and keep the particle in thermal equilibrium, implying that the distribution function has the usual Fermi-Dirac form 
\begin{equation}\label{equilibrium}
f_{th}(t,E)=\frac{1}{\exp[(E-\mu)/T]+1},\qquad \text{ for } T(t)> T_{ch}.
\end{equation}
where $T_{ch}$ represents the chemical freeze-out temperature.

\item Kinetic freeze-out:\\
After chemical freeze-out, particles  still scatter elastically from other particles and keep thermal equilibrium in the primordial plasma. As the temperature drops, the rate of elastic scattering reaction that maintain the thermal equilibrium become smaller than the expansion rate. At that time, elastic scattering processes cease, and the relic particles do not interact with other particles in the primordial plasma anymore. Before the kinetic freeze-out, the distribution function has the form
\begin{equation}\label{kinetic_equilib}
f_k(t,E)=\frac{1}{\Upsilon^{-1}\exp[(E-\mu)/T]+1},\qquad \text{ for }T_f< T(t)< T_{ch},
\end{equation}
where $T_f$ represents the kinetic freeze-out temperature. The generalized fugacity $\Upsilon(t)$ controls the occupancy of phase space and is necessary once $T(t)<T_{ch}$ in order to conserve particle number.

\item{Free streaming:}\\
After kinetic freeze-out, the particles have fully decoupled from the primordial plasma, and thereby ceased influencing the dynamics of the Universe and become free-streaming. The Einstein-Vlasov equation can be solved [\cite{choquet2008general}] and the free-streaming momentum distribution can be written as [\cite{Birrell:2012gg}]
\begin{equation}\label{free_stream_dist}
f_{fs}(t,E)=\frac{1}{\Upsilon^{-1}\exp{\left[\sqrt{\frac{E^2-m^2}{T_{fs}^2}+\frac{m^2}{T^2_f}}-\frac{\mu}{T_f}\right]+1}},\quad T_{fs}(t)=\frac{T_fa(t_k)}{a(t)},
\end{equation}
where the free-streaming effective temperature $T_{fs}$ is obtained by redshifting the temperature at kinetic freeze-out. If a massive particle (e.g. dark matter) freeze-out from cosmic plasma in the nonrelativistic regime, $m\gg T_f$. We can use the
Boltzmann approximation, and the free-streaming distribution for nonrelativistic particle becomes
\begin{align}
&f^B_{fs}(t,p)=\Upsilon\,e^{-(m+\mu)/T_f}\exp\left[-\frac{1}{ T_{eff}}\frac{p^2}{2m}\right],\quad T_{eff}=\left(\frac{a(t_f)}{a(t)}\right)^2T_f,
\end{align}
where we define the effective temperature $T_{eff}$ for massive free-streaming particle. In this scenario, the effective temperature for massive particles decreases faster than the Universe temperature cools. It's worth emphasizing the different temperatures between cold free-streaming particles and hot cosmic plasma would affect the evolution of the early Universe and require more detailed study. 
\end{itemize}

The division of the freeze-out process into these three regimes is a simplification. However, it is a very useful approximation in the study of cosmology~[\cite{Mangano:2005cc,Birrell:2014gea}] . For detailed discussion, see [\cite{Birrell:2012gg,Rafelski:2023emw}].

\section{Thermodynamics of the Early Universe}
In the case of local thermal equilibrium, the laws of thermodynamics can provide a framework for understanding the behavior of particle's energy density, pressure, number density and entropy in the early Universe.

Using the relativistic covariant Fermi/Bose momentum distribution, the corresponding energy density, pressure, and number densities for particle species $i$ are given by
\begin{align}
\rho_i&=g_i\int\!\!\frac{d^3p}{(2\pi)^3}Ef_{F/B}=\frac{g_i}{2\pi^2}\!\int_{m_i}^\infty\!\!\!dE\,\frac{E^2\left(E^2-m_i^2\right)^{1/2}}{\Upsilon_i^{-1}e^{(E-\mu_i)/T}\pm 1},\label{energy_density}\\[0.2cm]
P_i&=g_i\int\!\!\frac{d^3p}{(2\pi)^3}\frac{p^2}{3E}f_{F/B}=\frac{g_i}{6\pi^2}\!\int_{m_i}^\infty\!\!\!dE\,\frac{\left(E^2-m_i^2\right)^{3/2}}{\Upsilon_i^{-1} e^{(E-\mu_i)/T}\pm 1},\label{Pressure_density}\\[0.2cm]
n_i&=g_i\int\!\!\frac{d^3p}{(2\pi)^3}f_{F/B}=\frac{g_i}{2\pi^2}\!\int_{m_i}^\infty\!\!\!dE\,\frac{E(E^2-m_i^2)^{1/2} }{\Upsilon_i^{-1}e^{(E-\mu_i)/T}\pm 1}
\label{number_density}
\end{align}
where $g_i$ is the degeneracy of the particle species. By including the fugacity parameter $\Upsilon_i$ allows us to characterize particle properties in nonchemical equilibrium situations.
On the other hand, the corresponding free-streaming energy density, pressure, and number densities can be written as
\begin{align}
\rho_i&=g_i\int\!\!\frac{d^3p}{(2\pi)^3}Ef_{fs}=\frac{g_i}{2\pi^2}\!\int_{m_i}^\infty\!\!\!dE\,\frac{E^2\left(E^2-m_i^2\right)^{1/2}}{\Upsilon_i^{-1}e^{\sqrt{p^2/T_{fs}^2+m_i^2 /T_f^2}-\mu_i/T_f}\pm 1},\label{free_energy_density}\\[0.2cm]
P_i&=g_i\int\!\!\frac{d^3p}{(2\pi)^3}\frac{p^2}{3E}f_{fs}=\frac{g_i}{6\pi^2}\!\int_{m_i}^\infty\!\!\!dE\,\frac{\left(E^2-m_i^2\right)^{3/2}}{\Upsilon_i^{-1}e^{\sqrt{p^2/T_{fs}^2+m_i^2 /T_f^2}-\mu_i/T_f}\pm1},\label{free_Pressure_density}\\[0.2cm]
n_i&=g_i\int\!\!\frac{d^3p}{(2\pi)^3}f_{fs}=\frac{g_i}{2\pi^2}\!\int_{m_i}^\infty\!\!\!dE\,\frac{E(E^2-m_i^2)^{1/2} }{\Upsilon_i^{-1}e^{\sqrt{p^2/T_{fs}^2+m_i^2 /T_f^2}-\mu_i/T_f}\pm1},
\label{free_number_density}
\end{align} 
which are different from the thermal equilibrium Eq.~(\ref{energy_density}), Eq.~(\ref{Pressure_density}), and Eq.~(\ref{number_density}), by replacing the mass by a time dependant effective mass $m\,T_{fs}(t)/T_f$ in the exponential.

\begin{figure}[t]
%\begin{center}
\centering
\includegraphics[width=\linewidth]
%{./plots/DOF_entropy.jpg}
{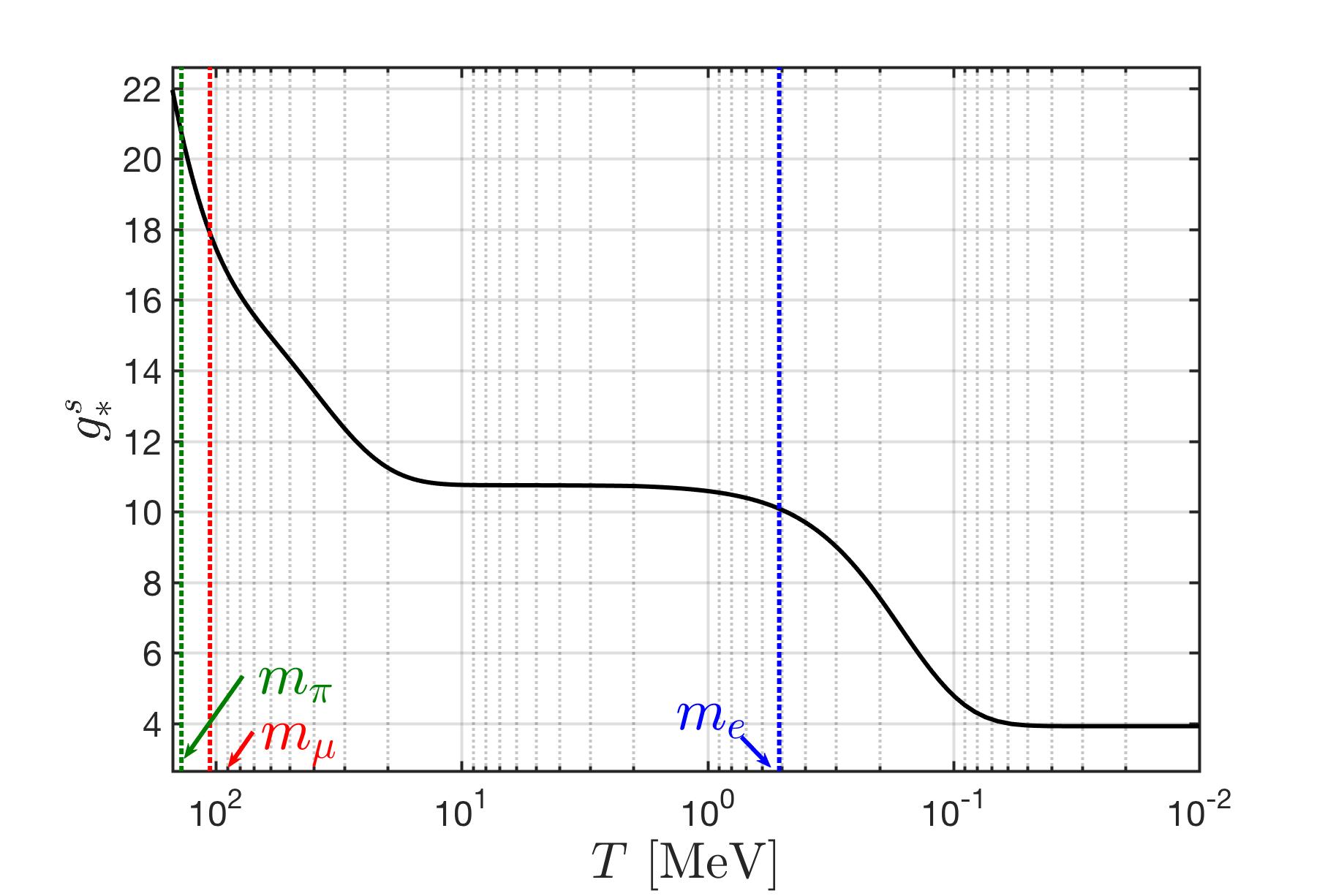}
\caption{The entropy degrees of freedom as a function of $T$ in the early Universe epoch after hadronization $10^{-2}\,\mathrm{MeV} \leqslant  T \leqslant 150 $\,MeV. When particle species becomes nonrelativistic $T\ll m_i$, the contribution to $g^s_\ast$ becomes negligible, as a result creating the dependence $g^s_\ast(T)$.The vertical lines represents the mass of particles: $m_e=0.511$ MeV, $m_\mu=105.6$ MeV, and pion average mass $m_\pi\approx138$ MeV.}
\label{EntropyDOF_Fig}  
\end{figure}

Given the energy density, pressure, and number densities, the entropy density for particle species $i$ can be written as
\begin{align}\label{entropy}
\sigma_i=\frac{S_i}{V}=\left(\frac{\rho_i+P_i}{T}-\frac{\mu_i}{T}\,n_i\right).
\end{align}
In general the chemical potential is associated with the baryon number. Since the net baryon number density relative to the photon number density is of order $10^{-9}$. In this case, we can neglect the small chemical potential when calculating the total entropy density in the Universe. The total entropy density in the early Universe can be written as
\begin{align}
&\sigma=\sum_i\,\sigma_i=\frac{2\pi^2}{45}g^s_\ast\,T^3,\\
&g^s_\ast=\sum_{i=\mathrm{bosons}}g_i\left({\frac{T_i}{T_\gamma}}\right)^3B\left(\frac{m_i}{T_i}\right)+\frac{7}{8}\sum_{i=\mathrm{fermions}}g_i\left({\frac{T_i}{T_\gamma}}\right)^3F\left(\frac{m_i}{T_i}\right),
\end{align}
where $g^s_\ast$ counts the effective number of `entropy' degrees of freedom. The functions $B(m_i/T)$ and $F(m_i/T)$ are defined as
\begin{align}
&B\left(\frac{m_i}{T}\right)=\frac{45}{12\pi^4}\int^\infty_{m_i/T}\,dx\sqrt{x^2-\left(\frac{m_i}{T}\right)^2}\left[4x^2-\left(\frac{m_i}{T}\right)^2\right]\frac{1}{\Upsilon^{-1}_ie^x-1},\\
&F\left(\frac{m_i}{T}\right)=\frac{45}{12\pi^4}\frac{8}{7}\int^\infty_{m_i/T}\,dx\sqrt{x^2-\left(\frac{m_i}{T}\right)^2}\left[4x^2-\left(\frac{m_i}{T}\right)^2\right]\frac{1}{\Upsilon^{-1}_ie^x+1}.
\end{align}
In Fig.\,\ref{EntropyDOF_Fig} we plot the $g^s_\ast$ as a function of temperature, the effect of particle mass threshold~[\cite{Coc:2006rt}] is considered in the calculation for all involved particles. When $T$ decreases below the mass of particle $T\ll m_i$, this particle species becomes nonrelativistic and the contribution to $g^s_\ast$ becomes negligible, creating the dependence on $T$ seen in Fig.\,\ref{EntropyDOF_Fig}.
%~~~~~~~~~~~~~~~~~~~~~~~~~~~~~~~~~~~~~~~~~~~~~~~~~~~~~~~~~~~~~~~~~~~~~~~~~~~~~~~~
%~~~~~~~~~~~~~~~~~~~~~~~~~~~~~~~~~~~~~~~~~~~~~~~~~~~~~~~~~~~~~~~~~~~~~~~~~~~~~~~~

\subsubsection{Relation between time and temperature}
Considering the comoving entropy conservation, we have
\begin{align}
S=\sigma V\propto g^s_\ast T^3a^3=\mathrm{constant},
\end{align}
where $g^s_\ast$ is the entropy degree of freedom and $a$ is the scale factor. Differentiating the entropy with respect to time $t$ we obtain
\begin{align}
\left[\frac{\dot{T}}{g^s_\ast}\frac{dg^s_\ast}{dT}+3\frac{\dot{T}}{T}+3\frac{\dot{a}}{a}\right]g^s_\ast T^3a^3=0,\qquad \dot{T}=\frac{dT}{dt}.
\end{align}
Solving $dT/dt$ and taking the integral, the relation between time and temperature in early universe can be written as
\begin{align}\label{time}
t(T)=t_0-\int^T_{T_0}dT\frac{1}{HT}\left[1+\frac{T}{3g^s_\ast}\frac{dg^s_\ast}{dT}\right],\qquad H^2=\frac{8\pi G_N}{3}\rho_{tot}
\end{align}
where $T_0$ and $t_0$ represent the initial temperature and time respectively, $H$ is the Hubble parameter and $\rho_{tot}$ is the total energy density in early Universe. From Eq. (\ref{time}) we see that the cosmic time depends on the entropy degrees of freedom $g^\ast_s$, which are characterized by the relativistic components in the early Universe. In the temperature range we consider $300\,\mathrm{MeV}>T>0.02\,\mathrm{MeV}$ the Universe is radiation-dominated and $\Lambda$CDM model is not used in this epoch.

%~~~~~~~~~~~~~~~~~~~~~~~~~~~~~~~~~~~~~~~~~~~~~~~~~~~~~~~~~~~~~~~$

\subsubsection{The baryon-per-entropy density ratio}
An important assumption allowing us to explore the early Universe evolution is that following on the era of matter genesis both baryon and entropy content is conserved in the comoving volume. Both baryon and entropy density scale with the third power of the expansion parameter $a(t)$. Therefore the ratio of baryon number density to visible matter entropy density remains constant throughout the evolution of universe. We have
\begin{align}
\frac{n_B-n_{\overline{B}}}{\sigma}= \left.\frac{n_B-n_{\overline{B}}}{ \sigma}\right|_{t_0}=\mathrm{Const.}\;
\end{align}
The subscript $t_0$ denotes the present day condition, and $\sigma$ is the total entropy density.
The observation gives the present baryon-to-photon ratio ~[\cite{ParticleDataGroup:2022pth}] $5.8 \times 10^{-10} \leqslant(n_B-n_{\overline{B}})/n_\gamma\leqslant6.5\times10^{-10}$. This small value quantifies the matter-antimatter asymmetry in the present day Universe, and allows the determination of the present value of baryon per entropy ratio~[\cite{Rafelski:2019twp,Fromerth:2002wb,Fromerth:2012fe}]:
\begin{align}\label{BaryonEntropyRatio}
\left.\frac{n_B-n_{\overline{B}}}{ \sigma}\right|_{t_0}=\eta\left(\frac{n_\gamma}{\sigma_\gamma+\sigma_\nu}\right)_{\!t_0}\!\!\!\!=(8.69\pm0.05)\!\!\times\!\!10^{-11},\qquad \eta=\frac{(n_B-n_{\overline{B}})}{n_\gamma},
\end{align}
where the $\eta=(6.12\pm0.04)\times10^{-10}$~[\cite{ParticleDataGroup:2022pth}] is used in calculation. To obtain the ratio, we consider that the Universe today is dominated by photons and free-streaming massless neutrinos~[\cite{Birrell:2012gg}], and $\sigma_\gamma$ and $\sigma_\nu$ are the entropy densities for photon and neutrino respectively. We have
\begin{align}
    \frac{\sigma_\nu}{\sigma_\gamma}=\frac{7}{8}\,\frac{g_\nu}{g_\gamma}\left(\frac{T_\nu}{T_\gamma}\right)^3\,\qquad\frac{T_\nu}{T_\gamma}=\left(\frac{4}{11}\right)^{1/3}
\end{align}
and the entropy-per-particle for massless bosons and fermions are given by~[\cite{Fromerth:2012fe}]
\begin{align}
s/n|_\mathrm{boson}\approx 3.60\,,\qquad
s/n|_\mathrm{fermion}\approx 4.20\,.
\end{align}
However, from the neutrino oscillation experiment, we know that the the neutrinos are not massless particles. 
The mass differences between neutrino mass eigenstates are~[\cite{ParticleDataGroup:2022pth}]:
\begin{align}
&\Delta{m}_{21}^2=7.39^{+0.21}_{-0.20}\times10^{-5}\,\mathrm{eV}^2,\\
&\Delta{m}_{32}^2=2.45^{+0.03}_{-0.03}\times10^{-3}\,\mathrm{eV}^2.
\end{align}
Neutrino mass eigenvalues can be ordered in the normal mass hierarchy ($m_1\ll m_2<m_3$) or inverted mass hierarchy ($m_3\ll m_1<m_2$). All three mass states remained relativistic until the temperature dropped below their rest mass. These results allow for the possibility that one mass eigenstate or two mass eigenstates of neutrinos may become non-relativistic today, which can affect the baryon-per-entropy ratio.

%~~~~~~~~~~~~~~~~~~~~~~~~~~~~~~~~~~~~~~~~~~~~~~~~~~~

\subsubsection{Nonequilibrium: departure from detailed balance}
Thermal equilibrium implies both chemical equilibrium (particles abundances are balanced) and kinetic equilibrium (energy is evenly distributed). In chemical equilibrium, the rates of the forward and reverse reactions are equal, resulting in a balance between production and annihilation/decay rates, which is called detailed balance. The chemical non-equilibrium can be achieved by breaking this detailed balance and leading to change in particle abundance over time. On the other hand, kinetic equilibrium is usually established much quicker and has less impact on the actual particle abundances.
The chemical nonequilibrium condition is more important than the kinetic equilibrium because it relates to the arrow of time for the particle reactions. 

The chemical nonequilibrium conditions in the early Universe are of general interest: they are understood to be prerequisite for the arrow of time dependent processes to take hold in the Hubble expanding Universe. The arrow of time plays an important role in the evolution of the early Universe, for example:
 1.) The Big Bang Nucleosynthesis (BBN)~[\cite{Pitrou:2018cgg,Kolb:1990vq,Dodelson:2003ft,Mukhanov:2005sc}]  the synthesis of light elements of  e.g. D, $^3$He, $^4$He, and $^7$Li are produced at temperatures around $86\,\mathrm{keV}>T_{BBN}>50\,\mathrm{keV}$. 
 2.) Baryogenesis is believed to occur at or before the Universe underwent electroweak phase transition~[\cite{Kolb:1990vq}] at a temperature $T\simeq 130$\, GeV, which generates the excess of baryon number compared to anti-baryon number in order to create the observed baryon number today.

When Universe expands and temperature cools down, the chemical non-equilibrium can be achieved by breaking the detailed balance between particle production reaction and annihilation/decay as follows:

1.) The particle production rate becomes slower than the rate of Universe expansion and the production reaction freezeout. Once the production reactions freezeout from the cosmic plasma, the corresponding detailed balance is broken and particle abundance decrease via the decay/annihilation reactions.

2.) The non-equilibrium can also be achieved when the production reaction slows down and is not able to keep up with decay/annihilation reaction. In this case, the Hubble expansion rate is much longer than the decay and production rate and is not relevant to the nonequilibrium process. The key factor is competition between production and decay/annihilation  which can result in chemical nonequilibrium in the early Universe.

\noindent We will investigate the nonequilibrium situation for bottom quarks and strang quarks in early universe and their application in Chapter~\ref{Bottom} and Chapter~\ref{Strangeness}, respectively.  

%~~~~~~~~~~~~~~~~~~~~~~~~~~~~~~~~~~~~~~~~~~~~~~~~~~~~~~~~~~~~~~~~~~

%~~~~~~~~~~~~~~~~~~~~~~~~~~~~~~~~~~~~~~~~~~~~~~~~~

\section{Cosmic plasma in early Universe $300\,\mathrm{MeV}>T>0.02\,\mathrm{MeV}$}
%In this section we will focus on the following:
%\begin{itemize}
%    \item Five different plasma epoch from $0.3\mathrm{GeV}>T>20$keV
%\end{itemize}

The primordial hot Universe fireball underwent several practically adiabatic phase changes that dramatically evolved its bulk properties as it expanded and cooled~[\cite{Rafelski:2023emw}]. We present an overview of the Universe evolution as a function of temperature from $300\,\mathrm{MeV}>T>0.02\,\mathrm{MeV}$ and main events constituting the history of the early Universe in Fig.~\ref{Overview_fig}. After the electroweak symmetry breaking epoch and presumably inflation, the comic plasma in the early Universe evolves in the following sequence:

%%%%%%%%%%%%%%%%%%%%%%%%%%%%%%%%%%%%%%%
\begin{figure}[ht]
 \centerline{\includegraphics[width=\textwidth,width=\linewidth]{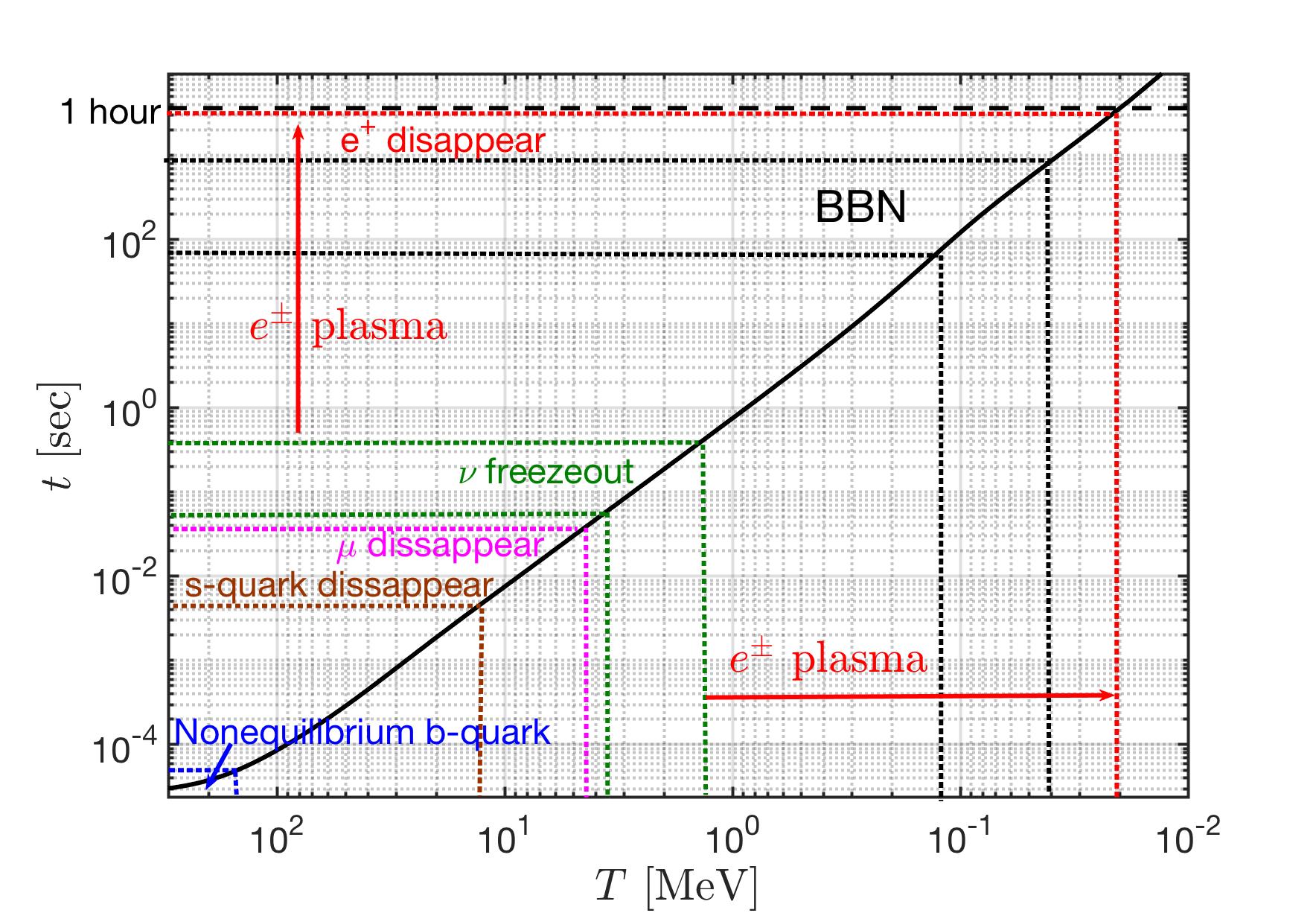}}
 \caption{The time evolution of the early Universe as a function of temperature from $300\,\mathrm{MeV}>T>0.02\,\mathrm{MeV}$ and 
 different sequence of main events are shown with the temperature/time range in the evolution. }
 \label{Overview_fig}
\end{figure}
%%%%%%%%%%%%%%%%%%%%%%%%%%%%%%%%%%%%%%%

\begin{enumerate}
    \item \textbf{Primordial quark-gluon plasma}: 
    At early times when the temperature was between $130\,\mathrm{GeV}>T>0.15\,\mathrm{GeV}$ we have the building blocks of the Universe as we know them today, including the leptons, vector bosons, and all three families of deconfined quarks and gluons which propagated freely in plasma. As all hadrons are dissolved into their constituents during this time, strongly interacting particles $u,d,s,t,b,c,g$ controlled the fate of the Universe. When temperature is near to the QGP phase transition $300\, \mathrm{MeV}>T>150$ MeV, the bottom quark  breaks the detail balance and disappearance from particle inventory provides the arrow in time (see Chapter~\ref{Bottom} for detail).
    
    \item \textbf{Hadronic epoch}: Around the hadronization temperature $T_H\approx150\,\mathrm{MeV}$, a phase transformation occurred, forcing 
    the free quarks and gluons become confined within baryon and mesons [\cite{Letessier:2005qe}]. In the temperature range $ 150\,\mathrm{MeV}>T>20\,\mathrm{MeV}$, the Universe is rich in physics phenomena involving strange mesons and (anti)baryons including (anti)hyperon abundances~[\cite{Fromerth:2012fe,Yang:2021bko}]. The antibaryons disappear from the Universe at temperature $T=38.2$ MeV, and strangeness can be produced by the inverse decay reactions that are in equilibrium via weak, electromagnetic, and strong interactions in the early Universe until $T\approx13$ MeV (see Chapter~\ref{Strangeness} for detailed discussion).

    \item \textbf{Lepton-photon epoch}: For temperature $10\,\mathrm{MeV}>T>2\,\mathrm{MeV}$, the Universe contained relativistic electrons, positrons, photons, and three species of (anti)neutrinos. During this epoch massless leptons and photons controlled the fate of the Universe. Massive $\tau^\pm$ disappear from the plasma at high temperature via decay processes. However $\mu^\pm$ leptons can persist in the early Universe until temperature $T=4.2$ MeV, and positron $e^+$ can persist until the temperature $T=0.02$ MeV (See Chapter~\ref{Electron} for discussion).
    Neutrinos were still coupled to the charged leptons via the weak interaction~[\cite{Birrell:2012gg,Birrell:2014ona}] and freeze-out at temperature range $3\,\mathrm{MeV}>T>2\,\mathrm{MeV}$ which depends on the neutrino's flavors and the magnitude of the Standard Model parameters (See Chapter~\ref{Neutrino} for details). After neutrino freeze-out, they still play a important role in the Universe expansion via the effective number of neutrinos $N_{\nu}^{\mathrm{eff}}$ and affects the Hubble parameter significantly.  
    
    \item \textbf{Electron-positron epoch}: After neutrinos freeze-out at $T=3\sim2\,\mathrm{MeV}$ and become free-streaming in the early Universe, the cosmic plasma was dominated by electrons, positrons, and photons. The $e^\pm$ plasma existed until $T\approx0.02\,\mathrm{MeV}$ such that BBN occurred within a rich electron-positron plasma, and the dense number density of electron/positron also provide the opportunities to investigate magnetization process (See Chapter~\ref{Electron} for detailed discussion). This is the last time the Universe will contain a significant fraction of its content in antimatter.
\end{enumerate}
After $e^\pm$ annihilation, the Universe was still opaque to photons at this point and remained so until the recombination period at $T\approx0.25\,\mathrm{eV}$ starting the era of observational cosmology with the Cosmic Microwave Background. This period has been studied in detail before in~[\cite{Planck:2018vyg}]. Therefore, we focus on the temperature $300\,\mathrm{MeV}>T>0.02\,\mathrm{MeV}$ which corresponds to the first hour of the Universe evolution. We will address the cosmic plasma as follow: In Chapter~\ref{Bottom}, we discuss the heavy quarks (bottom/charm) abundance near to the QGP hadronization and show the nonequilibrium of bottom quark.  In Chapter~\ref{Strangeness} we study the strangeness abundance after hadronization and show the long lasting strangeness in the early Universe. In Chapter~\ref{Neutrino} we focus on the neutrino-matter interactions and the evolution of cosmic neutrino in early universe before/after freeze-out. In Chapter~\ref{Electron} we study the abundance of charged leptons $\mu^\pm$ and $e^\pm$ and show that the present of $e^\pm$ plasma plays an important role in early Universe. In Chapter~\ref{Outlook} we address the ongoing and prospective research projects for future publication.
Finally in Chapter~\ref{Summary} we summarize the important results of our study and conclusion.
%~~~~~~~~~~~~~~~~~~~~~~~~~~~~~~~~~~~~~~~~~~~~~~~~~

%\begin{figure}
%\centering
%\includegraphics[angle=0,width=\columnwidth]{fig1.pdf}
%\caption[]{}
%\label{fig1}
%\end{figure}

%% file: Chapter2.tex
\chapter{Heavy quarks in cosmic plasma}\label{Bottom}

The primordial quark-gluon plasma (QGP) refers to the state of matter that existed in the early Universe, specifically for time $t\approx10\, \mathrm{\mu s}$ after the Big Bang. At that time the Universe was controlled by the strongly interacting particles: quarks and gluons. In this chapter, I study the heavy bottom and charm flavor quarks near to the QGP hadronization temperature $0.3\,\mathrm{GeV}>T>0.15\,\mathrm{GeV}$ and examine the relaxation time for the production and decay of bottom/charm quarks then show that the bottom quark nonequliibrium occur near to QGP –hadronization and create the arrow in time in the early Universe.

%~~~~~~~~~~~~~~~~~~~~~~~~~~~~~~~~~~~~~~~~~~~~~~~~~

\section{Overview of heavy quarks in primordial QGP}
%In this section we will focus on the following:
%\begin{itemize}
%    \item Review of primordial Quark-Gluon Plasma
%    \item Briefly remark the top quark
%\end{itemize}

In the QGP epoch, up and down $(u,d)$ (anti)quarks are effectively massless and remain in equilibrium via quark-gluon fusion. Strange $(s)$ (anti)quarks are in equilibrium via weak, electromagnetic, and strong interactions until $T\approx13$ MeV~[\cite{Yang:2021bko}]. The massive top $(t)$ (anti)quarks decay via the channel $t\to W+b$, with $\Gamma_t=1.4\pm0.2$\;GeV [\cite{ParticleDataGroup:2018ovx}] which implies that no bound state of top quarks have time to form. Given the large value of $\Gamma_t$ we realize that top quarks in hot QGP can be produced by the $ W+b\to t$ fusion process -- given the strength of this process there is no freeze-out of top quarks until $W$ itself freezes out. To address the top quarks in QGP, a dynamic theory for $W$ abundance is needed, a topic we will consider in the future. Finally, the bottom $(b)$ and charm $(c)$ quarks can be produced from strong interactions via quark-gluon pair fusion processes and disappear via weak interaction decays, and their abundance depends on the competition between the strong fusion and weak decay reaction rates.

The properties of QGP can be studied by experimental observations from high-energy heavy-ion collision experiments, such as  the Relativistic Heavy Ion Collider (RHIC) and the Large Hadron Collider (LHC). However, the conditions in the early Universe and those created in relativistic collisions are different. For example, the primordial QGP survives for about $10\mu$s in the cosmological Big Bang. On the other hand, the QGP formed in micro-bangs resulting from ultra-relativistic nuclear collisions has a lifespan of around  $10^{-23}$ s~[\cite{Rafelski:2001hp}]. Due to the considerably slower expansion rate of the Universe compared to quark production reactions and decays, in practicality, quark remained in equilibrium, and the quark fugacity is $\Upsilon=1$ during the QGP epoch.

However near the hadronization temperature, the heavy quarks abundance and deviations from chemical equilibrium have not yet been studied in great detail. In following we will focus on bottom and charm quarks. We will show that the bottom quarks can deviate from chemical equilibrium $\Upsilon\neq1$ by breaking the detailed balance between production and decay reactions of the quarks.

%~~~~~~~~~~~~~~~~~~~~~~~~~~~~~~~~~~~~~~~~~~~~~~~~~

\section{Bottom and Charm quark near QGP hadronization}
%In this section we will focus on the following:
%\begin{itemize}
%    \item Bottom/charm quarks production and decay in primordial QGP 
%    \item Dynamic equation for  bottom abundance (Stationary and non-stationary fugacity)
%    \item Nonstationary departure from detailed balance
%\end{itemize}
In the following we consider the temperature near QGP hadronization $0.3\,\mathrm{GeV}>T>0.15\,\mathrm{GeV}$, and study the bottom and charm abundance by examining the relevant reaction rates of their production and decay.
In thermal equilibrium the number density of light quarks can be evaluated in the massless limit, and we have
\begin{align}\label{FermiN}
n_q=\frac{g_{q}}{2\pi^2}\,T^3 F(\Upsilon_q)\;, \quad F=\int_0^\infty \frac{x^2dx}{1+\Upsilon_q^{-1}e^x}\;,
\end{align}
where $\Upsilon_q$ is the quark fugacity. We have $ F(\Upsilon_q=1)=3\,\zeta(3)/2$ with the Riemann zeta function $\zeta(3)\approx1.202$.
The thermal equilibrium number density of heavy quarks with mass $m\gg T$ can be well described by the Boltzmann expansion of the Fermi distribution function, giving
\begin{align}\label{BoltzN}
n_{q}\!=\!\frac{g_{q}T^3}{2\pi^2}\sum_{n=1}^{\infty}\frac{(-1)^{n+1}\Upsilon_q^n}{n^4}\left(\frac{n\,m_{q}}{T}\right)^{\!2}\!K_2\left(\frac{n\,m_{q}}{T}\right),
\end{align} 
where $K_2$ is the modified Bessel functions of integer order $2$. In the case of interest, when $m\gg T$, it suffices to consider the Boltzmann limit and  keep the first term $n=1$ in the expansion. The first term  $n=1$ also suffices for both charmed $c$-quarks and bottom $b$-quarks, giving
\begin{align}
&n_{b,c}={\Upsilon_{b,c}\,}n^{th}_{b,c},\qquad n^{th}_{b,c}=\frac{g_{b,c}}{2\pi^2}\,T^3\left(\frac{m_{b,c}}{T}\right)^2\,K_2(m_{b,c}/T).
\end{align}
However, for strange $s$ quarks, several terms are needed. 
%~~~~~~~Figure1~~~~~~~~~~~~~~~~~~~~~~~~~~~~~~~~~~~~~~~~~~~~~~~~~~~~~~~~~~~~~~~~~~~~~~~~~~~~~~~~~
\begin{figure}[t]
\begin{center}
\includegraphics[width=0.9\textwidth]{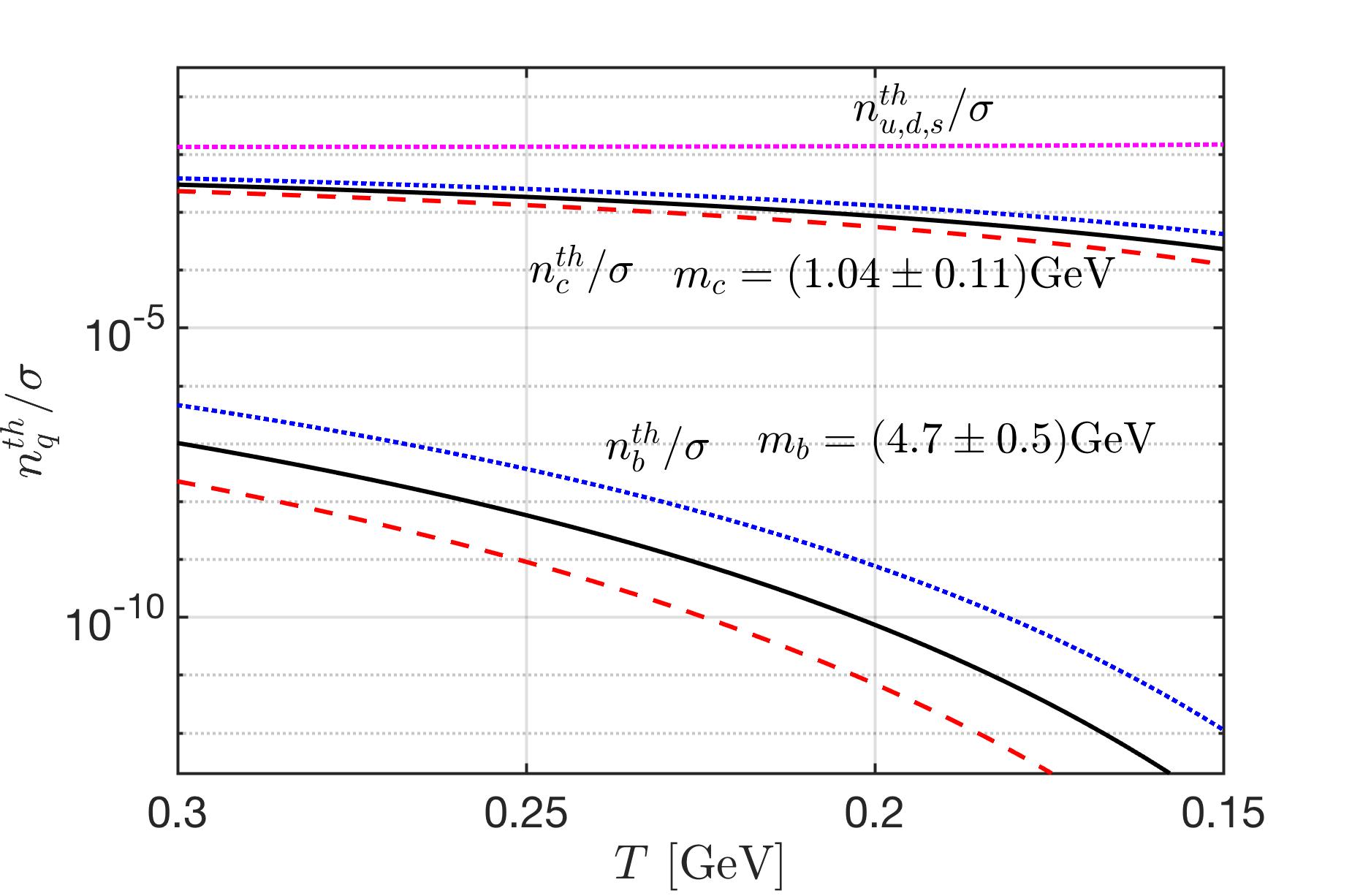}
\caption{
The quark number density normalized by entropy density, as a function of temperature in the early Universe with $\Upsilon=1$. The $b$-quark mass parameters shown are $m_b=4.2\,\mathrm{GeV}$ (blue) dotted line, $m_b=4.7\,\mathrm{GeV}$ (black) solid line, and $m_b=5.2\,\mathrm{GeV}$ (red) dashed line. For  $c$-quark  $m_c=0.93\,\mathrm{GeV}$  (blue) dotted line, $m_c=1.04\,\mathrm{GeV}$ (black) solid line, and $m_c=1.15\,\mathrm{GeV}$ (red) dashed line.}
\label{number_entropy_b002}
\end{center}
\end{figure}
%~~~~~~~~~~~~~~~~~~~~~~~~~~~~~~~~~~~~~~~~~~~~~~~~~~~~~~~~~~~~~~~~~~~~~~~~~~~~~~~~~~~~~~~~~~~~~

In Fig.~\ref{number_entropy_b002} we show the equilibrium ($\Upsilon=1$) number density per entropy density  ratio as a function of temperature $T$ of quarks. The entropy density is given by Eq.~(\ref{entropy}) and only light particles contribute to the entropy density; thus the result we consider is independent of actual abundance of $c$, $b$ and other heavy particles. We evaluated the density-per-entropy ratio for  $m_b=4.2,\;4.7,\;5.2$\,GeV and $m_c=1.04\pm0.11$\,GeV. The $m_b\simeq 5.2\,\mathrm{GeV}$ is  a typical potential model mass used in modeling bound states of bottom, and $m_b=4.2,\,4.7\,\mathrm{GeV}$ is the current quark mass at low and high energy scales. In Fig.~\ref{number_entropy_b002} we see that the charm abundance in the domain of interest $0.3\,\mathrm{GeV}>T>0.15\,\mathrm{GeV}$ is about $10^4\sim\!\!10^{9}$ times greater than the bottom quarks. This implies that the small $b$,$\bar b$ quark abundance is embedded in a large background comprising all lighter $u,d,s,c$ quarks and antiquarks, as well as gluons $g$.

\subsection{Reaction rate for quarks production and decay}
In primordial QGP, the bottom and charm quarks can be produced from strong interactions via quark-gluon pair fusion processes and disappear via weak interaction decays. For production, we have the following processes
\begin{align}
 q+\bar{q}&\longrightarrow b+\bar b,\qquad q+\bar{q}\longrightarrow c+\bar c,\\
 g+g&\longrightarrow b+\bar b,\qquad g+g\longrightarrow c+\bar c,
\end{align}
for bottom and charm we have
\begin{align}
 &b\longrightarrow c+l+\overline{\nu_l}, \qquad b\longrightarrow c+q+\bar{q},\\
&c\longrightarrow s+l+\overline{\nu_l},\qquad c\longrightarrow s+q+\bar{q},
\end{align}
for their decay. In following we will calculate the production rate and decay rate for bottom and charm quarks and compare to the Universe expansion rate. We will show that in the epoch of interest to us the characteristic Universe expansion time $1/H$ is much longer than the lifespan and production time of the bottom/charm quark. In this case, the dilution of bottom/charm quark due to the Universe expansion is slow compare to the the strong interaction production, and the weak interaction decay of the bottom/charm.

%~~~~~~~~~~~~~~~~~~~~~~~~~~~~~~~~~~~~~~~~~~~~~~~~~~~~~~~~~~

\subsubsection{Quark production rate via strong interaction}
For the quark-gluon pair fusion processes
%\begin{align}
% q+q&\longrightarrow b+\bar b,\qquad q+q\longrightarrow c+\bar c,\\
% g+g&\longrightarrow b+\bar b,\qquad g+g\longrightarrow c+\bar c,
%\end{align}
the evaluation of the lowest-order Feynman diagrams yields the cross sections~[\cite{Letessier:2002ony}]:
\begin{align}
&\sigma_{q\bar{q}\rightarrow b\bar{b},c\bar{c}}=\frac{8\pi\alpha_s^2}{27s}\left(1+\frac{2m_{b,c}^2}{s}\right)w(s),\,\qquad w(s)=\sqrt{1-{4m^2_{b,c}}/{s}},\\
&\sigma_{gg\rightarrow b\bar{b},c\bar{c}}=\!\frac{\pi\alpha_s^2}{3s}\bigg[\left(1\!+\!\frac{4m^2_{b,c}}{s}\!+\!\frac{m^4_{b,c}}{s^2}\right)\ln{\left(\frac{1+w(s)}{1-w(s)}\right)}\!-\!\left(\frac{7}{4}\!+\!\frac{31m^2_{b,c}}{4s}\right)w(s)\bigg],
\end{align} 
where $m_{b,c}$ represents the mass of bottom or charm quark, $s$ is the Mandelstam variable, and $\alpha_s$ is the QCD coupling constant. Considering the perturbation expansion of the coupling constant $\alpha_s$ for the two-loop approximation~[\cite{Letessier:2002ony}], we have:
\begin{align}
\alpha_s(\mu^2)=\frac{4\pi}{\beta_0\ln({\mu^2/\Lambda^2})}\bigg[1-\frac{\beta_1}{\beta_0}\frac{\ln(\ln{(\mu^2/\Lambda^2)})}{\ln(\mu^2/\Lambda^2)}\bigg],
\end{align}
where $\mu$ is the renormalization energy scale and $\Lambda^2$ is a parameter that determines the strength of the interaction at a given energy scale in QCD. The energy scale we consider is based on required gluon/quark collisions above $b\bar b$ energy threshold, so we have $\mu=2m_b+T$. For the energy scale $\mu>2m_b$ we have $\Lambda=180\sim230$ MeV( $\Lambda\approx205$ MeV in our calculation), and the parameters $\beta_0=11-2n_f/3$, $\beta_1=102-38n_f/3$ with the number of active fermions $n_f=4$. 

In general the thermal reaction rate per unit time and volume $R$ can be written in terms of the scattering cross section as follows~[\cite{Letessier:2002ony}]:
\begin{align}
R\equiv\sum_i\int_{s_{th}}^\infty\!ds\,\frac{dR_i}{ds}=\sum_i\int_{s_{th}}^\infty\!ds\,\sigma_i(s)\,P_i(s),
\end{align}
where $\sigma_i(s)$ is the cross section of the reaction channel $i$, and $P_i(s)$ is the number of collisions per unit time and volume. Considering the quantum nature of the colliding particles (i.e., Fermi and Bose distribution) with the massless limit and chemical equilibrium condition ($\Upsilon=1$), we obtain~[\cite{Letessier:2002ony}]
\begin{align}
&P_i(s)=\frac{g_1g_2}{32\pi^4}\,\frac{T}{1+I_{12}}\frac{\lambda_2}{\sqrt{s}}\!\sum_{l,n=1}^{\infty}\!(\pm)^{l+n}\frac{K_1(\sqrt{lns}/T)}{\sqrt{ln}},\\
&\lambda_2\equiv\left[s-\left(m_1+m_2\right)^2\right]\,\left[s-\left(m_1-m_2\right)^2\right],
\end{align}
where $+$ is for boson and $-$ is for fermions, and the factor $1/(1+I_{12})$ is introduced to avoid double counting of indistinguishable pairs of particles. $I_{12}=1$ for identical pair of particles, otherwise $I_{12}=0$. Hence the total thermal reaction rate per volume for bottom quark production can be written as
\begin{align}
\label{Bquark_Source}
R^{\mathrm{Source}}_{b,c}=\int^\infty_{s_{th}}ds\,\bigg[\sigma_{q\bar{q}\rightarrow b\bar{b},c\bar{c}}\,P_q+\sigma_{gg\rightarrow b\bar{b},c\bar{c}}\,P_g\bigg]%=R_{q\bar{q}\rightarrow b\bar{b},c\bar{c}}+R_{gg\rightarrow b\bar{b},c\bar{c}}.
\end{align}
We introduce the bottom/charm quark relaxation time for the quark-gluon pair fusion as follows:
\begin{align}
\label{relaxation_time}
&{\tau_{b,c}^{\mathrm{Source}}}\equiv\frac{dn_{b,c}/d\Upsilon_{b,c}}{R^{\mathrm{Source}}_{b,c}}\;,\quad
\end{align}
where $dn_{b,c}/d\Upsilon_{b,c}=n^{th}_{b,c}$ in the Boltzmann approximation. The relaxation time is on the order of magnitude of time needed to reach chemical equilibrium. In Fig.~\ref{BCreaction_fig} we show the characteristic time for bottom/charm  quark strong interaction production in the domain of interest, $ 0.3\,\mathrm{GeV}>T> 0.15\,\mathrm{GeV}$. 

\subsubsection{Quark decay rate via weak interaction}
\begin{figure} %[ht]
    \centering
    \includegraphics[width=0.85\textwidth]{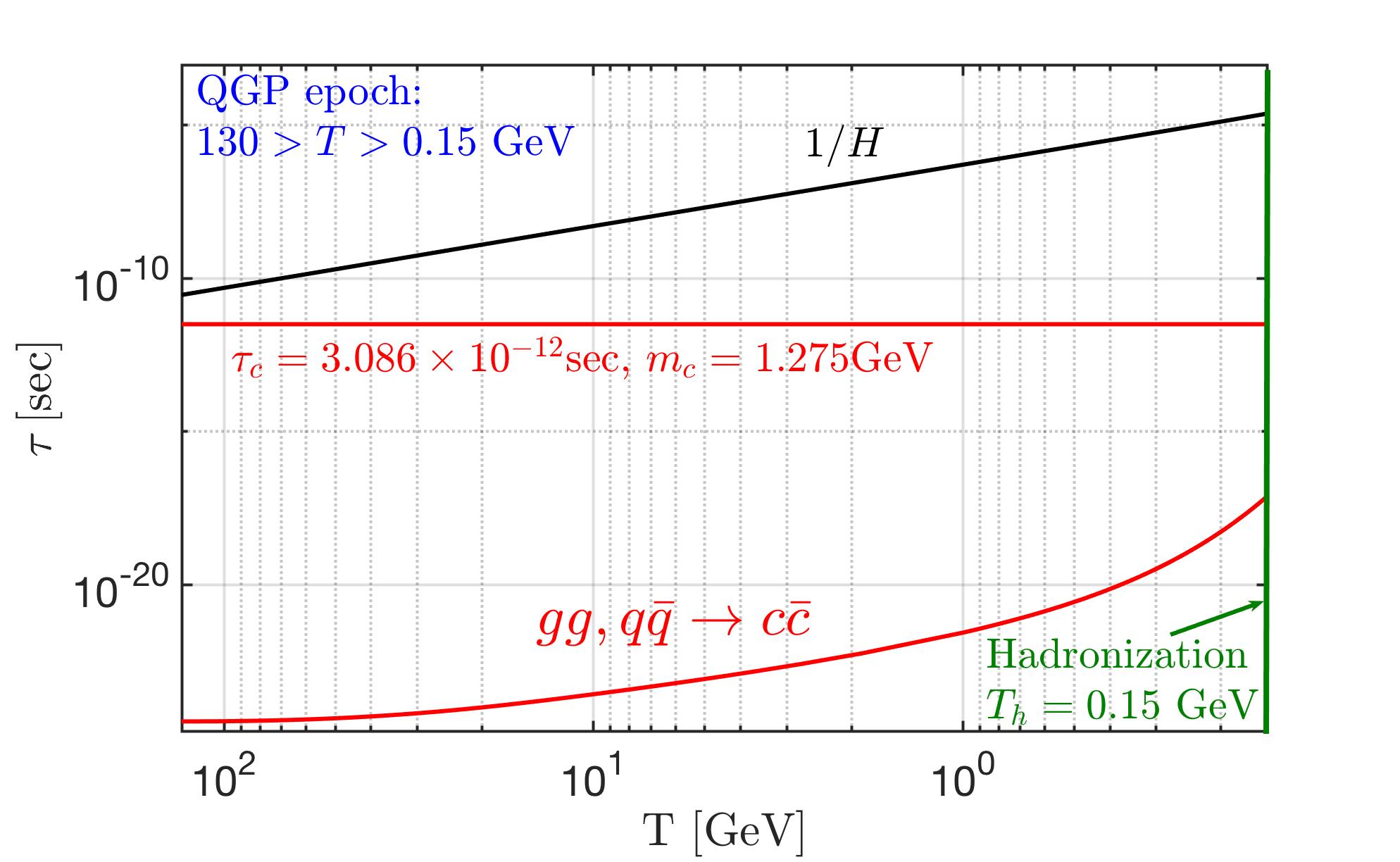}
    \includegraphics[width=0.85\textwidth]{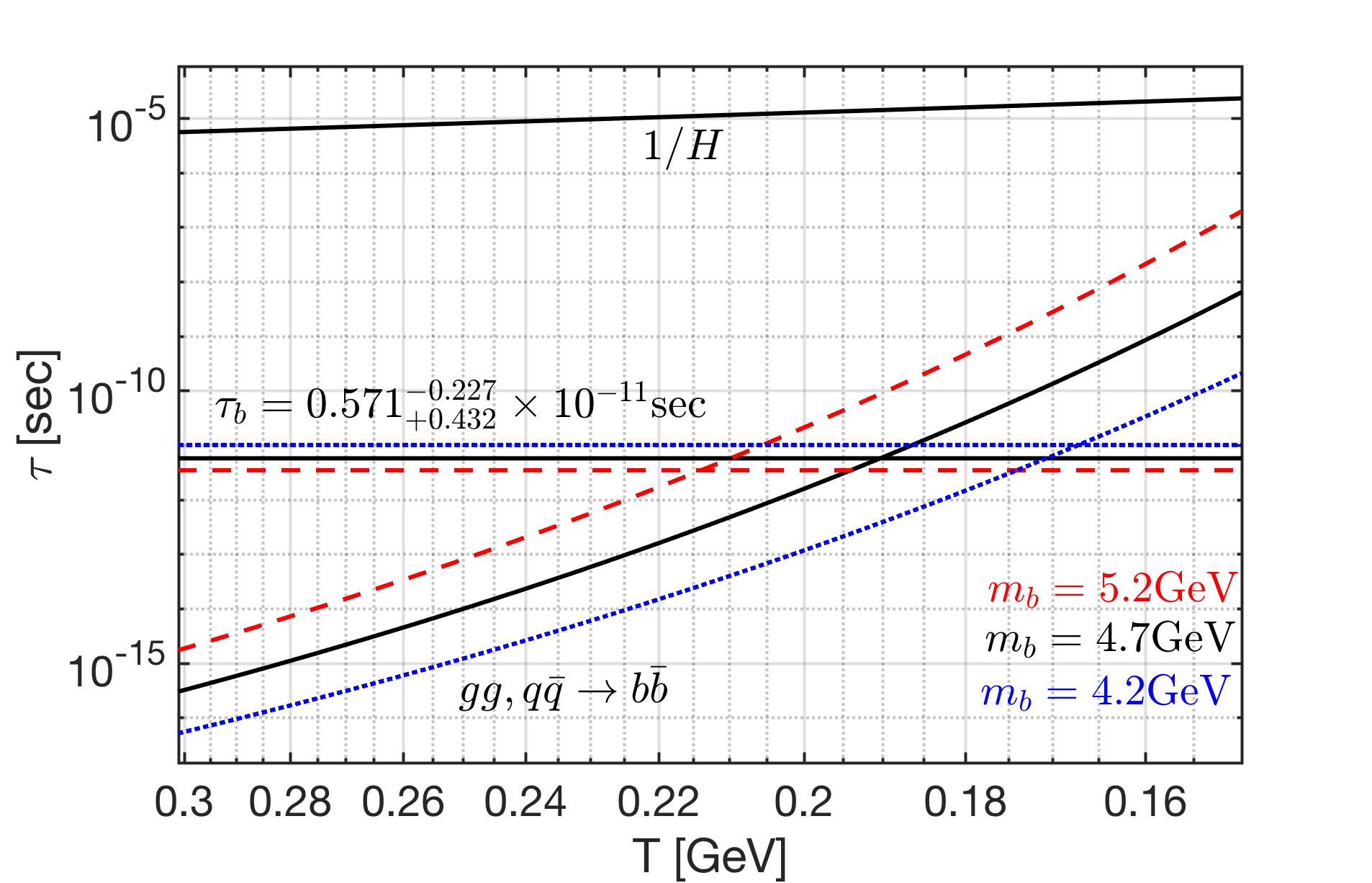}
    \caption{Comparison of Hubble time $1/H$, quark lifespan $\tau_{q}$, and characteristic time for production via quark-gluon pair fusion for (top figure) charm and (bottom figure) bottom quarks as a function of temperature. Both figures end at approximately the hadronization temperature of $T_{H}\approx150$ MeV. Three different masses $m_{b}=4.2$ GeV (blue short dashes), $4.7$ GeV, (solid black), $5.2$ GeV (red long dashes) for bottom quarks are plotted to account for its decay width.}
\label{BCreaction_fig}
\end{figure}

The bottom/charm quark decay via the weak interaction. 
%\begin{align}
% &b\longrightarrow c+l+\nu_l, \qquad %b\longrightarrow c+q+\bar{q},\\
%&c\longrightarrow s+l+\nu_l,\qquad c\longrightarrow s+q+\bar{q},
%\end{align}
The vacuum decay rate for $1\to2+3+4$ in vacuum can be evaluated via the weak interaction:
\begin{align}
\frac{1}{\tau_1}=&\frac{64G^2_F\,V^2_{12}\,V^2_{34}}{(4\pi)^3g_1}\,m^5_1\times\left[\frac{1}{2}{\left(1-\frac{m^2_2}{m^2_1}-\frac{m^2_3}{m^2_1}+\frac{m^2_4}{m^2_1}\right)}\mathcal{J}_1-\frac{2}{3}\mathcal{J}_2\right],
\end{align}
where the Fermi constant is $G_F=1.166\times10^{-5}\,\mathrm{GeV}^{-2}$, $V_{ij}$ is the element of the Cabibbo-Kobayashi-Maskawa (CKM) matrix~[\cite{Czarnecki:2004cw}] for quark channel and $V_{l\nu_l}=1$ for lepton channel. The functions $\mathcal{J}_1$ and $\mathcal{J}_2$ are given by
\begin{align}
&\mathcal{J}_1\!=\!\!\!\int_0^{(1-m^2_2/m^2_1)/2}\!\!\!\!\!\!\!\!dx\left(1\!-\!2x\!-\!\frac{m^2_2}{m_1^2}\right)^{\!\!2}\left[\frac{1}{(1-2x)^2}-1\right]\\
&\mathcal{J}_2\!=\!\!\!\int_0^{(1-m^2_2/m^2_1)/2}\!\!\!\!\!\!\!\!dx\left(1\!-\!2x\!-\!\frac{m^2_2}{m_1^2}\right)^{\!\!3}\left[\frac{1}{(1-2x)^3}-1\right]
\end{align}
The modification due to the heat bath(plasma) is small because the bottom and charm  mass $m_{b,c}\gg T$~[\cite{Kuznetsova:2008jt}]. In the temperature range we are interested in, the decay rate in the vacuum is a good approximation for our calculation. We show the lifespan for bottom/charm quark in Fig.~\ref{BCreaction_fig}. 

%there is no modification due to the phase space blocking because the bottom/charm mass is too heavier $m_{b,c}\gg T$~[\cite{Kuznetsova:2008jt}]. We did not include above either base enhancement nor fermi blocking factors since process of bottom decay involve energies beyond those available for $ 150< T< 300\,\mathrm{MeV}$.

%%%%%%%%%%%%%%%%%%%%%%%%%%%%%%%%%%%%%%%
%%%%%%%%%%%%%%%%%%%%%%%%%%%%%%%%%%%%%%%

%~~~~~~~~~~~~~~~~~~~~~~~~~~~~~~~~~~~~~~~~~~~~~~~~~~~~~~~~~~~~~~~~~~~~~~~~~~~~
\subsubsection{Hubble expansion rate}
In the early Universe, within a temperature range $130\, \mathrm{GeV}>T>0.15\,\mathrm{GeV}$,  we have the following particles:  photons, $8$ color charge gluons, $W^\pm$, $Z^0$, three generations of $3$ color charge quarks and leptons in the primordial QGP.  The Hubble parameter can be written in terms of particle energy density $\rho_i$
\begin{align}
H^2=\frac{8\pi G_N}{3}\left(\rho_\gamma+\rho_{\mathrm{lepton}}+\rho_{\mathrm{quark}}+\rho_{g,{W^\pm},{Z^0}}\right),
\end{align}
where $G_N$ is the Newtonian constant of gravitation. The effectively massless particles and radiation dominate the speed of expansion of the Universe. The characteristic Universe expansion time constant $1/H$ is seen in Fig.~\ref{BCreaction_fig}. In the epoch of interest to us $0.3\,\mathrm{GeV}>T>0.15\,\mathrm{GeV}$, the Hubble time $1/H\approx10^{-5}$ sec which is much longer than the lifespan and production time of the bottom and charm quarks. 
%~~~~~~~~~~~~~~~~~~~~~~~~~~~~~~~~~~~~~~~~~~~~~~~~~~~~~~~~~~~~~~~~~~~~~~~~~~~~~~~~~~~~~~~

\subsubsection{Rate Comparison: Strong fusion, Weak decay, and Hubble expansion}
In Fig.~\ref{BCreaction_fig} (top), we plot the relaxation time of the production/decay for charm quarks and Hubble time $1/H$ as a function of temperature. Throughout the entire duration of QGP, the Hubble time is larger than the lifespan and production times of the charm quark. %Therefore, the heavy charm quark remains in equilibrium as its processes occur faster than the expansion of the Universe. 
Additionally, the charm quark production time is faster than the decay. The faster quark-gluon pair fusion keeps the charm in chemical equilibrium up until hadronization. After hadronization, charm quarks form heavy mesons that decay into multi-particles quickly in plasma. The daughter particles from charm meson decay can interact and reequilibrate with
the plasma quickly. In this case the energy required for the inverse decay reaction to produce
charm meson is difficult to overcome and causing the charm quark to vanish from the inventory of particles via decay in the Universe.

In Fig.~\ref{BCreaction_fig} (bottom) we present the relaxation time for production and decay of the bottom quark with different masses as a function of temperature. It shows that both production and decay are faster than the Hubble time $1/H$ for the duration of QGP. However, unlike charm quarks, the relaxation time for bottom quark production intersects with bottom quark decay at different temperatures which depends on the mass of the bottom. The intersection implies that the bottom quark freeze-out from the primordial plasma before hadronization as the production process slows down at low temperatures and the subsequent weak interaction decay leads to a dilution of the bottom quark content within the QGP plasma. All of this occurs with rates faster than Hubble expansion and thus as the Universe expands, the system departs from a detailed chemical balance because of the competition between decay and production reactions in QGP. In this scenario, the dynamic equation on bottom abundance is required and causes the distribution to deviate from equilibrium with $\Upsilon\neq1$ in the temperature range below the crossing point but before the hadronization.

\subsection{Bottom quark abundance nonequilibrium}

\begin{figure}[ht]
\begin{center}
\includegraphics[width=0.85\textwidth]{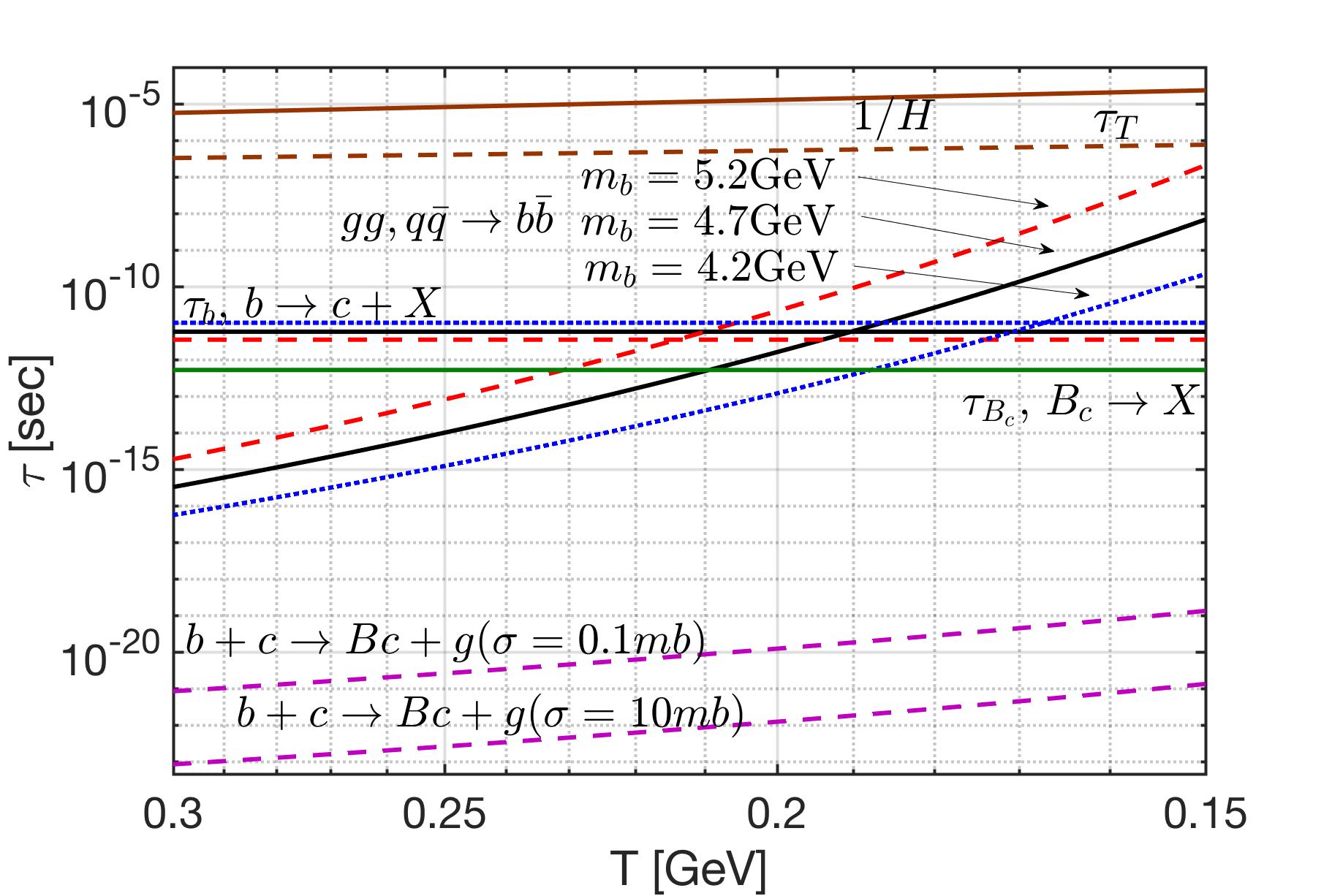}
\caption{Production and decay characteristic times of bottom quark and the Hubble time $1/H$ within the temperature range of interest  $ 0.3\,\mathrm{GeV}>T> 0.15\,\mathrm{MeV}$. Near the top of figure  $1/H$ (brown solid line) and $\tau_T$ (brown dashed line); other horizontal lines are bottom-quark (in QGP) weak interaction lifetimes $\tau_b$ for the three different masses: $m_b=4.2\,\mathrm{GeV}$ (blue dotted line), $m_b=4.7\,\mathrm{GeV}$ (black solid  line), $m_b=5.2\,\mathrm{GeV}$ (red dashed line), and the vacuum lifespan $\tau_B$ of the  B$_c$ meson (green solid  line). The relaxation time for strong interaction bottom production $g+g, q+\bar q\rightarrow b+\bar{b}$ is shown with three different bottom masses and same type-color coding as weak interaction decay rate. At bottom of figure the in plasma formation process (dashed lines, purple) $b+c\rightarrow \mathrm{B}_c+g$ with cross section range $\sigma=0.1,10\,\mathrm{mb}$.}
\label{ReactionTime}
\end{center}
\end{figure}

The competition between weak interaction decay and strong interaction production rates leads the dynamic bottom abundance in QGP. The dynamic equation for bottom quark abundance in QGP can be written as 
\begin{align}
\label{Bquark_eq}
\frac{1}{V}\frac{dN_b}{dt}=\big(\,1-\Upsilon^2_{b}\,\big)\,R^{\mathrm{Source}}_{b}-\Upsilon_b\,R^{\mathrm{Decay}}_{b}\;,
\end{align}
where $R^{\mathrm{Source}}_{b}$ and $R^{\mathrm{Decay}}_{b}$ are the thermal reaction rates per volume of production and decay of bottom quark, respectively. The bottom source rates are the gluon and quark fusion rates Eq.~(\ref{Bquark_Source}). The decay rate depends on whether the bottom quarks are freely present in the plasma or are bounded within mesons. We consider two extreme scenarios for the bottom quark population: 1.) all bottom flavor is free, and 2.) all bottom flavor is bounded into mesons in QGP. In Fig.~\ref{ReactionTime}  we show the characteristic interaction times relevant to the abundance of bottom quarks, as well as the Hubble time $1/H$ for the temperature range of interest, $0.3\,\mathrm{GeV}> T> 0.15\,\mathrm{GeV}$.

%~~~~~~~~~~~~~~~~~~~~~~~~~~~~~~~~~~~~~~~~~~~~~~~~~~~~~~~~~~~~~
%~~~~~~~~~~~~~~~~~~~~~~~~~~~~~~~~~~~~~~~~~~~~~~~~~~~~~~~~~~~~~
Considering all bottom flavor is free in QGP, the bottom decay rate per volume is the bottom lifespan weighted with density of particles Eq.~(\ref{BoltzN})~~[\cite{Kuznetsova:2008jt}]. We have
\begin{align}\hspace{0.5cm}
R^{\mathrm{Decay}}_b=\frac{dn_b/d\Upsilon_b}{\tau_b},\,\,\,\,\, \tau_b\approx0.57\times10^{-11} \mathrm{sec}.
\end{align}
On the other hand, $b$,$\bar b$ quark abundance is embedded in a large background comprising all lighter quarks and antiquarks (see Fig.~\ref{number_entropy_b002}). After formation the heavy $b, \bar b$ quark can bind with any of the available lighter quarks, with the most likely outcome being a chain of reactions 
\begin{align}
&b+q\longrightarrow\mathrm{B}+g\;,\\
&\mathrm{B}+s\longrightarrow\mathrm{B}_s+q\;,\\
&\mathrm{B}_s+c\longrightarrow\mathrm{B}_c+s\;,
\end{align}
with each step providing a gain in binding energy and reduced speed due to the diminishing abundance of heavier quarks $s, c$. To capture the lower limit of the rate of $\mathrm{B}_c$ production we show in Fig.~\ref{ReactionTime} the expected formation rate by considering the direct process $b+\overline c\rightarrow \mathrm{B}_c+g$, considering the range of cross section $\sigma=0.1\sim10\,\mathrm{mb}$ ~[\cite{Schroedter:2000ek}]. The rapid formation rate of B$_c(b\bar c)$ states in primordial plasma is shown by purple dashed lines at bottom in Fig.~\ref{ReactionTime}, we have
\begin{align}
\tau (b+\overline c\rightarrow \mathrm{B}_c+g)\approx(10^{-16}\sim10^{-14})\times\frac{1}{H} \;.
\end{align}

Despite the low abundance of charm, the rate of $\mathrm{B}_c$ formation is relatively fast, and that of lighter flavored B-mesons is substantially higher. Note that as long as we have bottom quarks made in gluon/quark fusion bound practically immediately with any quarks $u, d, s$ into B-mesons, we can use the production rate of $b, \bar b$ pairs as the rate of B-meson formation in the primordial-QGP, which all decay with lifespan of pico-seconds. We believe that this process is fast enough to allow consideration of bottom decay from the B$_c(b\bar c)$, $\overline{\mathrm{B}}_c(\bar b c)$ states~[\cite{Yang:2020nne}].  
 Based on the hypothesis that all bottom flavor is bound rapidly into $\mathrm{B}_c^\pm$ mesons, we have 
\begin{align}\label{Bc_source}
g+g, q+q \longleftrightarrow &b+\bar b\;[b(\bar{b})+\bar{c}(c)]\longrightarrow \mathrm{B}_c^\pm\longrightarrow\mathrm{anything}.
\end{align}
In this case, the decay rate per volume can be written as
\begin{align}\hspace{0.5cm}
 R^{\mathrm{Decay}}_b=\frac{dn_b/d\Upsilon_b}{\tau_{\mathrm{B}_c}},\,\,\,\,\, \tau_{\mathrm{B}_c}\approx0.51\times10^{-12} \mathrm{sec}.
 \end{align}

To investigate the nonequilibrium phenomena of bottom quarks, we aim to replace the variation of particle abundance seen on LHS in Eq.~(\ref{Bquark_eq}) by the time variation of abundance fugacity $\Upsilon$.
This substitution allows us to derive the dynamic equation for the fugacity parameter and enables us to study the fugacity as a function of time . Considering the expansion of the Universe we have
\begin{align}\label{number_dilution}
\frac{1}{V}\frac{dN_b}{dt}=\frac{dn_b}{d\Upsilon_b}\frac{d\Upsilon_b}{dt}+\frac{dn_b}{dT}\frac{dT}{dt}+3Hn_b,\;
\end{align}
where we use $d\ln(V)/dt=3H$ for the Universe expansion. Substituting Eq.~(\ref{number_dilution}) into Eq.~(\ref{Bquark_eq}) and dividing both sides of equation by $dn_b/{d\Upsilon_b}=n^{th}_b$, the fugacity equation becomes
\begin{align}
\frac{d\Upsilon_b}{dt}+&3H\Upsilon_b+\Upsilon_b\frac{dn^{th}_b/dT}{n^{th}_b}\frac{dT}{dt}=\left(1-\Upsilon_b^2\right)\frac{1}{\tau_{b}^{\mathrm{Source}}}-\Upsilon_b\frac{1}{\tau^{\mathrm{Decay}}_b}\;,
\end{align}
where relaxation time for bottom production is obtained using Eq.~(\ref{relaxation_time}). It is convenient to introduce the relaxation time $1/\tau_T$ as follows,
\begin{align}
\frac{1}{\tau_T}\equiv-\frac{dn^{th}_b/dT}{n^{th}_b}\frac{dT}{dt},
\end{align}
where we put '$-$' sign in the definition to have $\tau_T>0$. The relaxation time $\tau_T$ represents how the bottom density changes due to the Universe temperature cooling. In this case, the fugacity equation can be written as
\begin{align}\label{Fugacity_Eq0}
\frac{d\Upsilon_b}{dt}\!\!=&(1-\Upsilon_{b}^2)\frac{1}{\tau_{b}^{\mathrm{Source}}}
\!-\!\Upsilon_{b}\left(\frac{1}{\tau^{\mathrm{Decay}}_b}+3H\!-\!\frac{1}{\tau_T}\right).
\end{align}
In following sections we will solve the fugacity differential equation in two different scenarios: stationary and nonstationary Universe.

\subsubsection{First solution: stationary Universe}
In Fig.~\ref{BCreaction_fig} (bottom) we show that the relaxation time for both production and decay are faster than the Hubble time $1/H$ for the duration of QGP, which implies that $H,1/\tau_T\ll1/\tau_{b}^{\mathrm{Source}},1/\tau^{\mathrm{Decay}}_b$. In this scenario, we can solve the fugacity equation by considering the stationary Universe first, i.e., the Universe is not expanding and we have
\begin{align}\label{stationary}
H=0,\qquad1/\tau_T=0.
\end{align} 
In the stationary Universe at each given temperature we consider the dynamic equilibrium condition (detailed balance) between production and decay reactions that keep
\begin{align}
\frac{d\Upsilon_b}{dt}=0.
\end{align}
Neglecting the time dependence of the fugacity $d\Upsilon_b/dt$ and substituting the condition Eq.~(\ref{stationary}) into the fugacity equation Eq.~(\ref{Fugacity_Eq0}), then we can solve the quadratic equation to obtain the stationary fugacity as follows:
\begin{align}
\label{Fugacity_Sol}
\Upsilon_{\mathrm{st}}&=\sqrt{1+\left(\frac{\tau_{source}}{2\tau_{decay}}\right)^2}-\left(\frac{\tau_{source}}{2\tau_{decay}}\right).
\end{align}
%~~~~~~~Figure~~~~~~~~~~~~~~~~~~~~~~~~~~~~~~~~~~~~~~~~~~~~~~~~~~~~~~~~~~~~~~~~~~~~~~~~~~~~~~~~~
\begin{figure}[ht]
\begin{center}
\includegraphics[width=\textwidth]{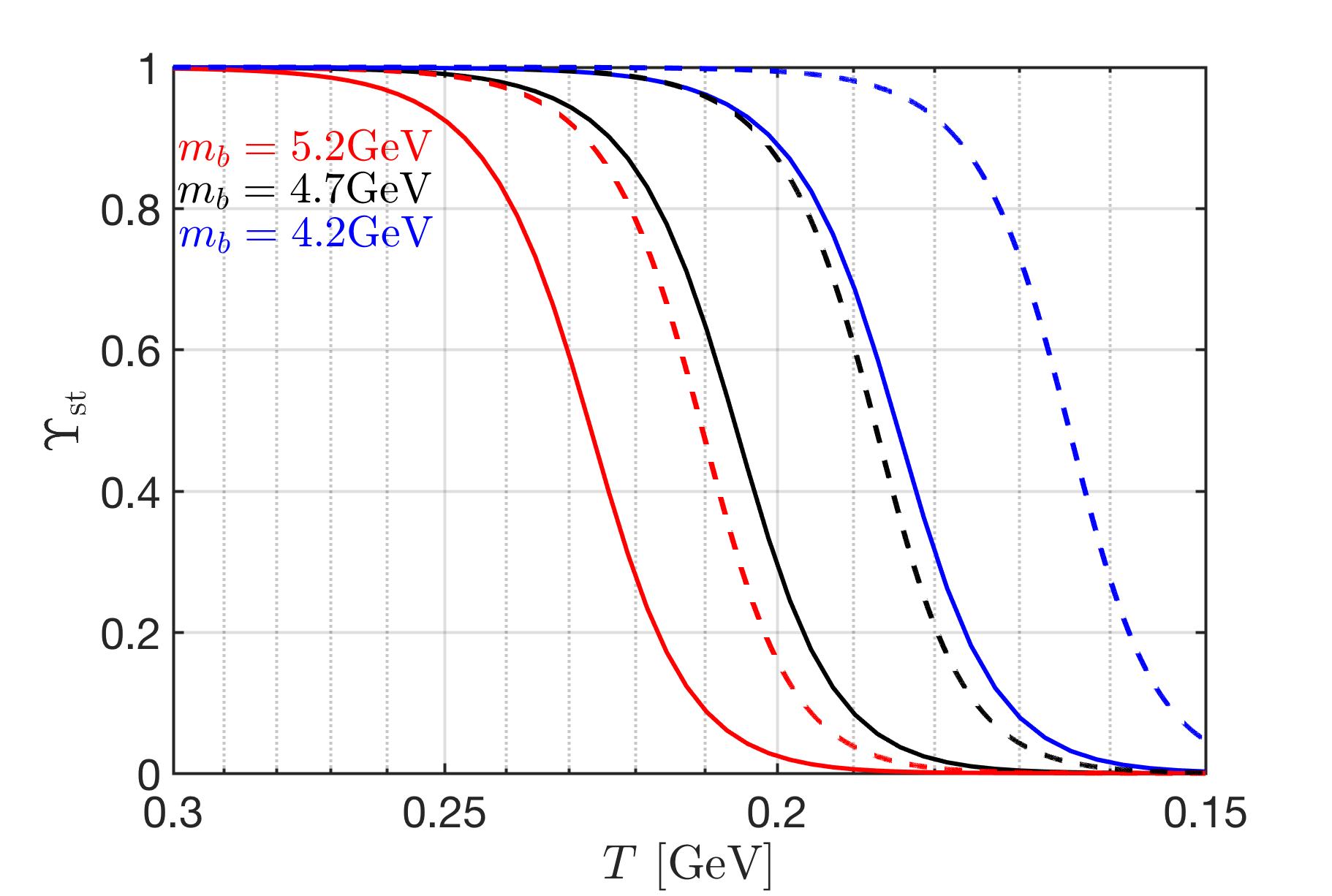}
\caption{The fugacity of free/ bounded bottom quark as a function of temperature in the early Universe for $m_b=4.2\,\mathrm{GeV}$ (blue), $m_b=4.7\,\mathrm{GeV}$ (black), and $m_b=5.2\,\mathrm{GeV}$ (red). The solid lines represent the case bottom quark bound into $B_c$ mesons, and the dashed lines label the case of  free bottom quark.}
\label{fugacity_bc}
\end{center}
\end{figure}
%~~~~~~~~~~~~~~~~~~~~~~~~~~~~~~~~~~~~~~~~~~~~~~~~~~~~~~~~~~~~~~~~~~~~~~~~~~~~~~~~~~~~~~~~~~~~~~~
\,
In Fig.~\ref{fugacity_bc} the fugacity of bottom quark $\Upsilon_{\mathrm{st}}$ as a function of temperature, Eq.~(\ref{Fugacity_Sol}) is shown around the temperature $T=0.3\,\mathrm{GeV}>T>0.15\,\mathrm{GeV}$ for different masses of bottom quarks. In all cases we see prolonged non-equilibrium, this happens since the decay and reformation rates of bottom quarks are comparable to each other as we have noted in Fig.~\ref{ReactionTime} where both lines cross. One of the key results shown in Fig.~\ref{fugacity_bc} is that the smaller mass of bottom quark slows the strong interaction formation rate to the value of weak interaction decays just near the phase transformation of QGP to HG phase. Finally, the stationary fugacity corresponds to the reversible reactions in the stationary Universe. In this case, there is no arrow in time for bottom quark because of the detailed balance.

\subsubsection{Non-stationary correction in expanding Universe}

The Universe is expanding and temperature is a function of time. In this section we now consider the fugacity as a function of time and study the correction in fugacity due to the expanding Universe. In general, the fugacity of bottom quark can be written as 
\begin{align}\label{Nonstationary_sol}
&\Upsilon_b=\Upsilon_{\mathrm{st}}+\Upsilon^{\mathrm{non}}_{\mathrm{st}}=\Upsilon_\mathrm{st}\left(1+x\right),\quad x\equiv{\Upsilon_\mathrm{st}^{\mathrm{non}}}/{\Upsilon_\mathrm{st}},
\end{align}
where the variable $x$ corresponds to the correction due to non-stationary Universe. Substituting the general solution Eq.(\ref{Nonstationary_sol}) into differential equation Eq.(\ref{Fugacity_Eq0}), we obtain
\begin{align}\label{Nonstationary_eq}
\frac{dx}{dt}=-x^2\frac{\Upsilon_\mathrm{st}}{\tau_{source}}&-x\left[\frac{1}{\tau_{eff}}+3H-\frac{1}{\tau_T}\right]-\left[\frac{d\ln\Upsilon_\mathrm{st}}{dt}+3H-\frac{1}{\tau_T}\right],
\end{align}
where the effective relaxation time $1/\tau_{eff}$ is defined as
\begin{align}
\frac{1}{\tau_{eff}}\equiv\left[\frac{2\Upsilon_\mathrm{st}}{\tau_{source}}+\frac{1}{\tau_{decay}}+\frac{d\ln\Upsilon_\mathrm{st}}{dt}\right].
\end{align}
%~~~~~~~Figure~~~~~~~~~~~~~~~~~~~~~~~~~~~~~~~~~~~~~~~~~~~~~~~~~~~~~~~~~~~~~~~~~~~~~~~~~~~~~~~~~
\begin{figure}[t]
\begin{center}
\includegraphics[width=\textwidth]{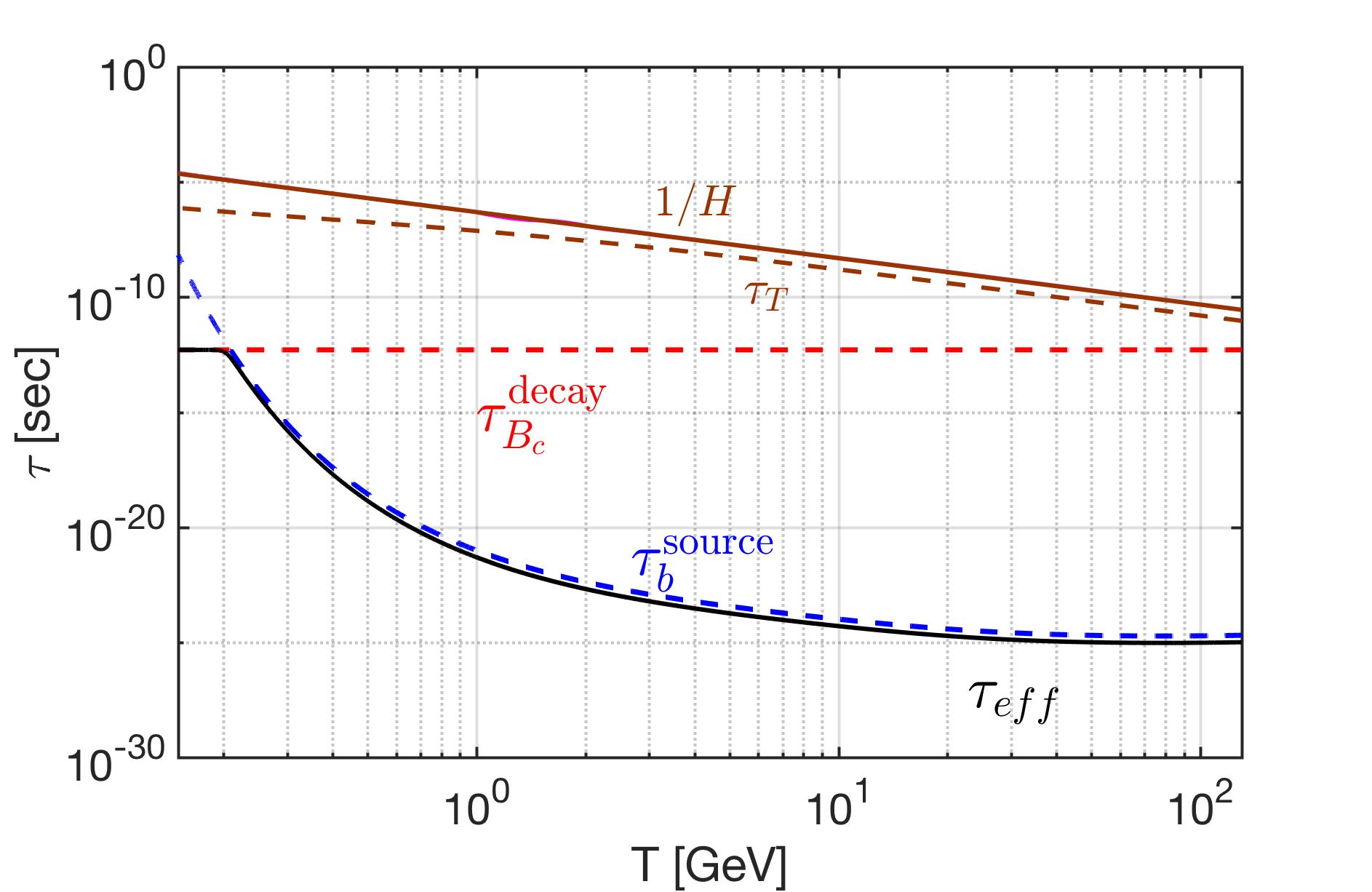}
\caption{The effective relaxation time $\tau_{eff}$ as a function of temperature in the early Universe for bottom mass $m_b=4.7$GeV.  For comparison, we also plot the vacuum lifespan of $B_c$ meson $\tau_{B_c}^{decay}$(red dashed-line), the relaxation time for bottom production $\tau^b_{source}$ (blue dashed-line), Hubble expansion time $1/H$(brown solid line) and relaxation time for temperature cooling $\tau_T$(brown dashed-line).  }
\label{RelaxationTime_eff}
\end{center}
\end{figure}
%~~~~~~~~~~~~~~~~~~~~~~~~~~~~~~~~~~~~~~~~~~~~~~~~~~~~~~~~~~~~~~~~~~~~~~~~~~~~~~~~~~~~~~~~~~~~~~~
In Fig.~\ref{RelaxationTime_eff} we see that when temperature is near to $T=0.2$ GeV, we have $1/\tau_{eff}\approx10^{7}H$, and $1/\tau_{eff}\approx10^5/\tau_T$. In this case, the last two terms in Eq.~(\ref{Nonstationary_eq}) compare to $1/\tau_{eff}$ can be neglected, and the differential equation becomes
\begin{align}\label{nonstationary_eq}
\frac{dx}{dt}=-\frac{x^2\,\Upsilon_\mathrm{st}}{\tau_{source}}&-\frac{x}{\tau_{eff}}-\left[\frac{d\ln\Upsilon_\mathrm{st}}{dt}+3H-\frac{1}{\tau_T}\right],
\end{align}

To solve the variable $x$ we consider the case $dx/dt,x^2\ll1$ first, we neglect the terms $dx/dt$ and $x^2$ in Eq.~(\ref{nonstationary_eq})  then solve the linear fugacity equation.  We will establish that these approximations are justified by checking the magnitude of the solution. Neglecting terms $dx/dt$ and $x^2$ in Eq.~(\ref{nonstationary_eq}) we obtain
\begin{align}
x\approx\tau_{eff}\left[\frac{d\ln\Upsilon_\mathrm{st}}{dt}+3H-\frac{1}{\tau_T}\right].
\end{align}
It is convenient to change the variable from time to temperature. For an isentropically-expanding universe, we have
\begin{align}\label{tau_H}
\frac{dt}{dT}=-\frac{\tau^\ast_H}{T},\qquad \tau^\ast_H=\frac{1}{H}\left(1+\frac{T}{3g^s_\ast}\frac{dg^s_\ast}{dT}\right).
\end{align}
In this case, we have
\begin{align}
x=\tau_{eff}\left[\frac{1}{\Upsilon_\mathrm{st}}\frac{d\Upsilon_\mathrm{st}}{dT}\frac{T}{\tau^\ast_H}+3H-\frac{1}{\tau_T}\right].
\end{align}
Finally, we can obtain the nonstationary fugacity by multiplying the fugacity ratio $x$ with $\Upsilon_\mathrm{st}$, giving
\begin{align}
\Upsilon_{\mathrm{st}}^{\mathrm{non}}
&\approx\left(\frac{\tau_{eff}}{\tau^\ast_H}\right)\left[\frac{d\Upsilon_\mathrm{st}}{dT}T-\Upsilon_{\mathrm{st}}\left(3H\tau^\ast_H-\frac{\tau^\ast_H}{\tau_T}\right)\right].
\end{align}

\begin{figure}[t]
\begin{center}
\includegraphics[width=\textwidth]{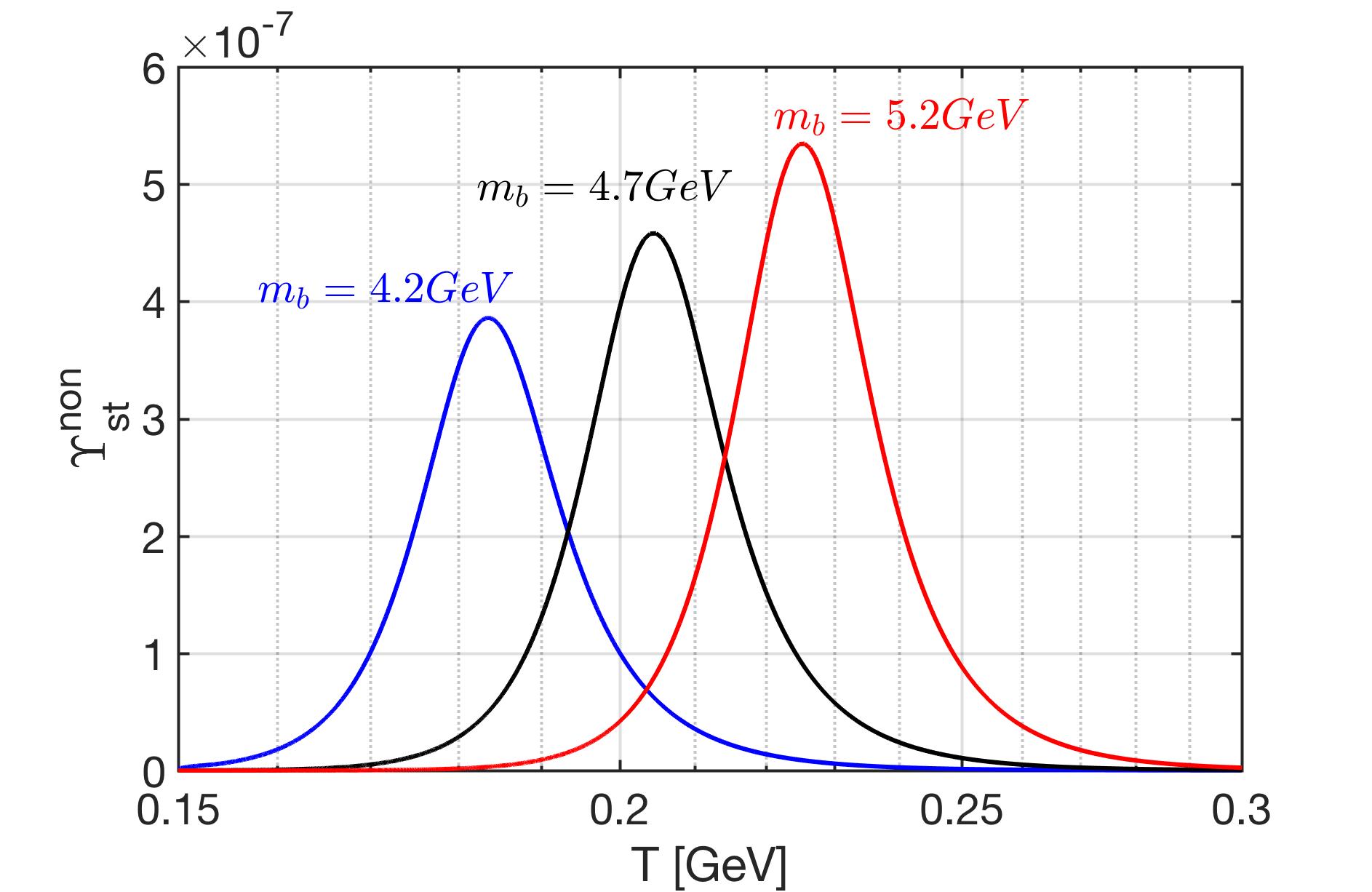}
\caption{The non-stationary fugacity $\Upsilon_\mathrm{st}^{\mathrm{non}}$ as a function of temperature in the Universe for different bottom mass $m_b=4.2\,\mathrm{GeV}$ (blue), $m_b=4.7\,\mathrm{GeV}$ (black), and $m_b=5.2\,\mathrm{GeV}$ (red) for the case bottom  quarks bound into $B_c$ mesons.}
\label{NonFugacity}
\end{center}
\end{figure}

In Fig.~\ref{NonFugacity} we plot the nonstationary $\Upsilon^{\mathrm{non}}_\mathrm{st}$ as a function of temperature. The nonstationary fugacity $\Upsilon^{\mathrm{non}}_\mathrm{st}$ follows the behavior of $d\Upsilon_{\mathrm{st}}/dT$, which corresponds to the irreversible process in expanding Universe. In this case, the irreversible nonequilibrium process creates the arrow in time for bottom quark in the Universe. The large value of Hubble time compares to the effective relaxation time suppressing the value of nonstationary fugacity to $\mathcal{O}\sim10^{-7}$, which shows that the neglecting $dx/dt,x^2\ll1$ is a good approximation for solving the non-stationary fugacity in the early Universe.
%~~~~~~~Figure7~~~~~~~~~~~~~~~~~~~~~~~~~~~~~~~~~~~~~~~~~~~~~~~~~~~~~~~~~~~~~~~~~~~~~~~~~~~~~~~~~
%~~~~~~~~~~~~~~~~~~~~~~~~~~~~~~~~~~~~~~~~~~~~~~~~~~~~~~~~~~~~~~~~~~~~~~~~~~~~~~~~~~~~~~~~~~~~~~~

To conclude this chapter, we have demonstrated that the bottom quark nonequliibrium occurs near the QGP phase transition around the temperature $T=0.3\sim0.15$ GeV in Fig.~\ref{fugacity_bc} and Fig.~\ref{NonFugacity}. We show the competition between weak interaction decay and the strong interaction $g+g\to b+\bar b$, $q+\bar q \to b+\bar b$ fusion processes drive the bottom quark departure from the equilibrium and create the arrow of time in the early Universe at relatively low QGP temperature. The results provide a strong motive for exploring the physics of baryon nonconservation involving the bottomnium mesons or/and bottom quarks in a thermal environment.

%~~~~~~~~~~~~~~~~~~~~~~~~~~~~~~~~~~~~~~~~~~~~~~~~~
%~~~~~~~~~~~~~~~~~~~~~~~~~~~~~~~~~~~~~~~~~~~~~~~~~
%{Introduction\daggerfootnote{This chapter has been published previously as \citet{Gottbrath1999}.}}

%\begin{figure}
%\centering
%\includegraphics[angle=0,width=\columnwidth]{fig1.pdf}
%\caption[]{}
%\label{fig1}
%\end{figure}

%% file: Chapter3.tex
\chapter
{Strangeness abundance in cosmic plasma}
\label{Strangeness}
%{Heavy-quark in primordial QGP: after hadronization}
As the Universe expanded and cooled down to the hadronization temperature $T_H\approx150$ MeV, the primordial QGP underwent a phase transformation called hadronization. This transition resulted in the confinement of the strong force, causing quarks and gluons to combine and form matter 
and antimatter. After hadronization, one may think the relatively short lived massive hadrons decay rapidly and disappear from the Universe. However, the most abundant hadrons, pions $\pi(q\bar q)$, can be produced via their inverse decay process $\gamma\gamma\rightarrow\pi^0$ and retain their chemical equilibrium until temperature $T=3\sim5$ MeV~[\cite{Kuznetsova:2008jt}]. 

Following the idea and the framework presented by~[\cite{Kuznetsova:2008jt}], we investigate the strange particle composition of the expanding early Universe in the epoch $150\,\mathrm{MeV}\ge T\ge 10$\,MeV, and examine the freeze-out temperature for strangeness-producing  by comparing the relevant reaction rates to the Hubble expansion rate. We show that strangeness is kept in equilibrium via weak, electromagnetic, and strong interactions in the early Universe until $T\approx13$ MeV.

%~~~~~~~~~~~~~~~~~~~~~~~~~~~~~~~~~~~~~~~~~~~~~~~~~

%~~~~~~~~~~~~~~~~~~~~~~~~~~~~~~~~~~~~~~~~~~~~~~~~~

\section{Chemical equilibrium in the hadronic Universe}
%In this section we will focus on the following:
%\begin{itemize}
%    \item Chemical potentials of $\mu_B$ and $\mu_s$
%    \item Composition of universe (strangeness abundance)
%\end{itemize}

In this section, we explore the Universe composition assuming both kinetic and particle abundance equilibrium (chemical equilibrium) by considering the charge neutrality and prescribed conserved baryon-per-entropy-ratio ${(n_B-n_{\overline{B}})}/{\sigma}$ to determine the baryon chemical potential $\mu_B$~[\cite{Fromerth:2012fe,Rafelski:2013yka}]. With the chemical potential as a function of temperature, we can obtain the particle number densities for different species and study their composition in the early Universe.

We improve the prior work~[\cite{Fromerth:2012fe}] by considering the conserved entropy per baryon ratio with conservation of strangeness in the early Universe. To study the baryon and strange quark chemical potential, it is convenient to introduce the chemical fugacity for strangeness $\lambda_s$ and quark $\lambda_q$ as follows:
\begin{align}
\lambda_s=\exp(\mu_s/T)\,\quad \lambda_q=\exp(\mu_B/3T),
\end{align}
where $\mu_s$ and $\mu_B$ are the chemical potential of strangeness and baryon, respectively. For the quark fagucity $\lambda_q$, we divide the chemical potential of baryons by 3 as an approximation for quark chemical potential. Imposing the conservation of strangeness  
$\langle s-\bar s \rangle=0$, we have, when the baryon chemical potential does not vanish the chemical potential of strangeness in the early Universe satisfying (see Section 11.5 in \,[\cite{Letessier:2002ony}])
\begin{align}\label{museq}
\lambda_s=\lambda_q\sqrt{\frac{F_K+\lambda^{-3}_q\,F_Y}{F_K+\lambda^3_q\,F_Y}}.
\end{align}
where we employ the phase-space function $F_i$ for sets of nucleon $N$, kaon $K$, and hyperon $Y$ particles defined as (see [\cite{Letessier:2002ony}], Section 11.4):
\begin{align}
&F_N=\sum_{N_i}\,g_{N_i}W(m_{N_i}/T)\;, \quad N_i=n, p, \Delta(1232),\\
&F_K=\sum_{K_i}\,g_{K_i}W(m_{K_i}/T)\;, \quad K_i=K^0, \overline{K^0}, K^\pm, K^\ast(892),\\
&F_Y=\sum_{Y_i}\,g_{Y_i}W(m_{Y_i}/T)\;, \quad Y_i=\Lambda, \Sigma^0,\Sigma^\pm, \Sigma(1385),
\end{align}
where $g_{N_i,K_i,Y_i}$ are the degenerate factors, $W(x)=x^2K_2(x)$ with $K_2$ is the modified Bessel functions of integer order "$2$".  

Considering the Boltzmann approximation for the massive particle number density we have
\begin{align}
\label{Density_N}
&n_N=\frac{T^3}{2\pi^2}\lambda_q^3F_N,\quad\qquad\qquad n_{\overline N}=\frac{T^3}{2\pi^2}\lambda^{-3}_qF_N,\\
\label{Density_K}
&n_K=\frac{T^3}{2\pi^2}\left(\lambda_s\lambda_q^{-1}\right)F_K,\,\qquad n_{\overline{K}}=\frac{T^3}{2\pi^2}\left(\lambda_s^{-1}\lambda_q\right)F_K,\\
\label{Density_Y}
&n_Y=\frac{T^3}{2\pi^2}\left(\lambda_q^2\lambda_s\right)F_Y,\quad\qquad n_{\overline Y}=\frac{T^3}{2\pi^2}\left(\lambda^{-2}_q\lambda_s^{-1}\right)F_Y.
\end{align}
In this case, the net baryon density in the early Universe with temperature range $150\,\mathrm{MeV}> T>10$\,MeV can be written as 
\begin{align}
\frac{\left(n_B-n_{\overline{B}}\right)}{\sigma}&=\frac{1}{\sigma}\left[\left(n_p-n_{\overline{p}}\right)+\left(n_n-n_{\overline{n}}\right)+\left(n_Y-n_{\overline{Y}}\right)\right]\notag\\
&=\frac{T^3}{2\pi^2\,\sigma}\left[\left(\lambda_q^3-\lambda^{-3}_q\right)F_N+\left(\lambda_q^2\lambda_s-\lambda^{-2}_q\lambda_s^{-1}\right)F_Y\right]\notag\\
&=\frac{T^3}{2\pi^2\sigma}\left(\lambda_q^3-\lambda_q^{-3}\right)F_N\left[1+\frac{\lambda_s}{\lambda_q}\left(\frac{\lambda_q^3-\lambda^{-1}_q\lambda_s^{-2}}{\lambda^3_q-\lambda^{-3}_q}\right)\,\frac{F_Y}{F_N}\right]\notag\\
&\approx\frac{T^3}{2\pi^2\sigma}\left(\lambda_q^3-\lambda_q^{-3}\right)F_N\left[1+\frac{\lambda_s}{\lambda_q}\,\frac{F_Y}{F_N}\right],
\end{align}
where we can neglect the term $F_Y/F_K$ in the expansion of Eq.(\ref{museq}) in our temperature range. Introducing the strangeness $\langle s-\bar s\rangle=0$ constraint and using the entropy density in early universe, the explicit relation for baryon to entropy ratio becomes
\begin{align}\label{muBeq}
\frac{n_B-n_{\overline{B}}}{\sigma}&=\frac{45}{2\pi^4g^s_\ast}\sinh\left[\frac{\mu_B}{T}\right]F_N\times\left[1+\frac{F_Y}{F_N}\sqrt{\frac{1+e^{-\mu_B/T}\,F_Y/F_K}{1+e^{\mu_B/T}\,F_Y/F_K}}\right].
\end{align}
Governing Eq.\,(\ref{muBeq}) is the present-day baryon-per-entropy-ratio, and we obtain the value 
\begin{align}\label{BdS}
\frac{n_B-n_{\overline{B}}}{\sigma}= \left.\frac{n_B-n_{\overline{B}}}{ \sigma}\right|_{t_0}=(0.865\pm0.008)\times10^{-10} \;.
\end{align}
For a detailed evaluation method we refer to this earlier work now using a baryon-to-photon ratio~[\cite{ParticleDataGroup:2018ovx}]: $\left(n_B-n_{\overline{B}}\right)/n_\gamma= (0.609\pm0.06)\times10^{-9}$, as well as the entropy per particle for a massless boson $\sigma/n|_\mathrm{boson}\approx 3.60$ and a massless fermion $\sigma/n|_\mathrm{fermion}\approx 4.20$. 

%~~~~~~~~~~~~~~~~~~~~~~~~~~~~~~~~~~~~~~~~~~~~~~~~~~~~~~~~~~~~~~~~~~~~~~~~~~~~~~~~
\begin{figure}[t]
%\begin{center}
\centering
\includegraphics[width=0.8\linewidth]{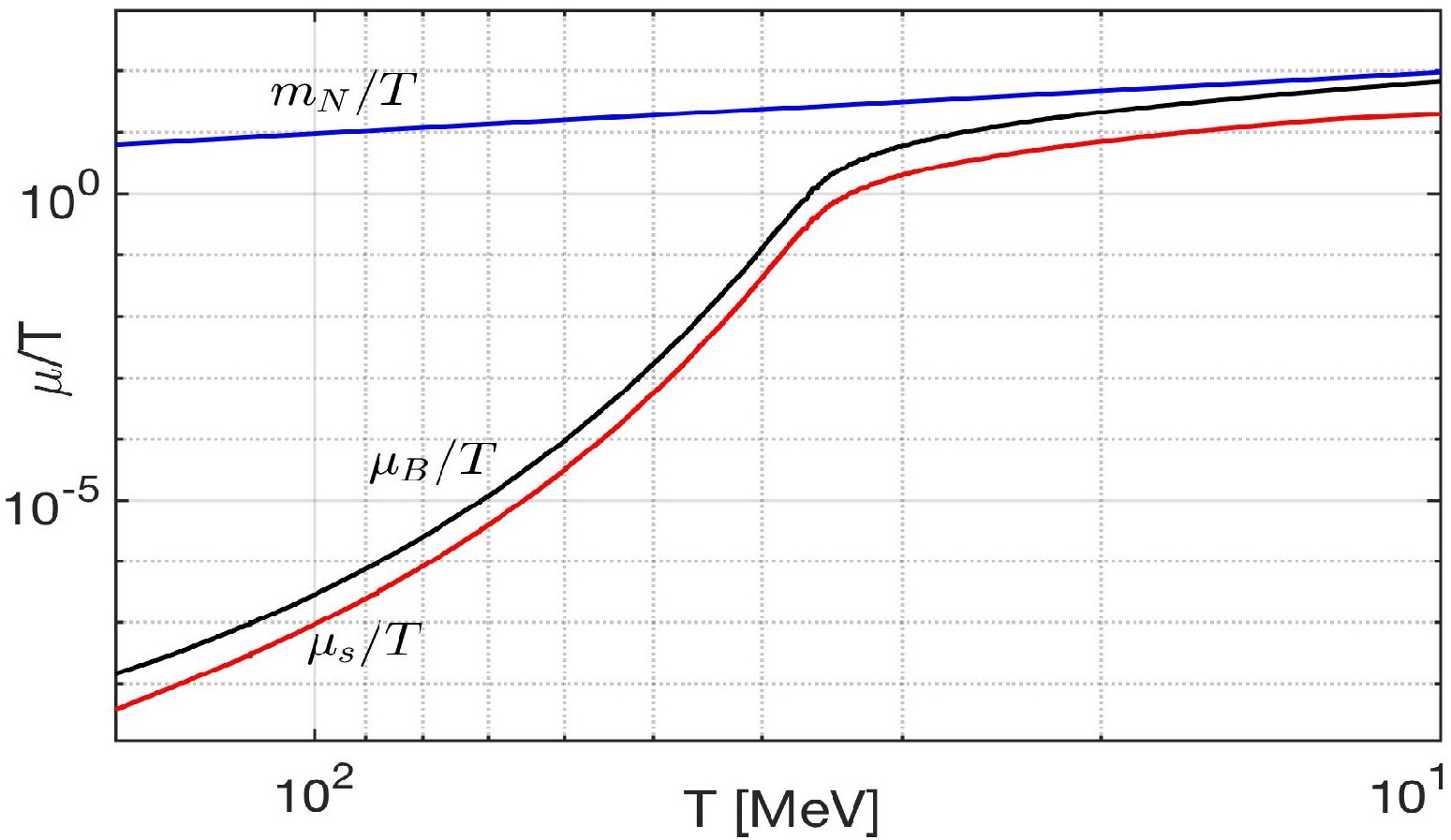}
\caption{The chemical potential of baryon $\mu_B/T$ and strangeness $\mu_s/T$ as a function of temperature $150\,\mathrm{MeV}> T>10\,\mathrm{MeV}$ in the early Universe; for comparison we show $m_N/T $ with $m_N=938.92$\,MeV, the average nucleon mass.}
\label{ChemPotFig}
%\end{center}
\end{figure}
%~~~~~~~~~~~~~~~~~~~~~~~~~~~~~~~~~~~~~~~~~~~~~~~~~~~~~~~~~~~~~~~~~~~~~~~~~~~~~~
%%%%%%%%%%%%%%%%%%%%%%%%%%%%%%%%%%%%%%%
\begin{figure}[h]
\centering
\includegraphics[width=\textwidth]{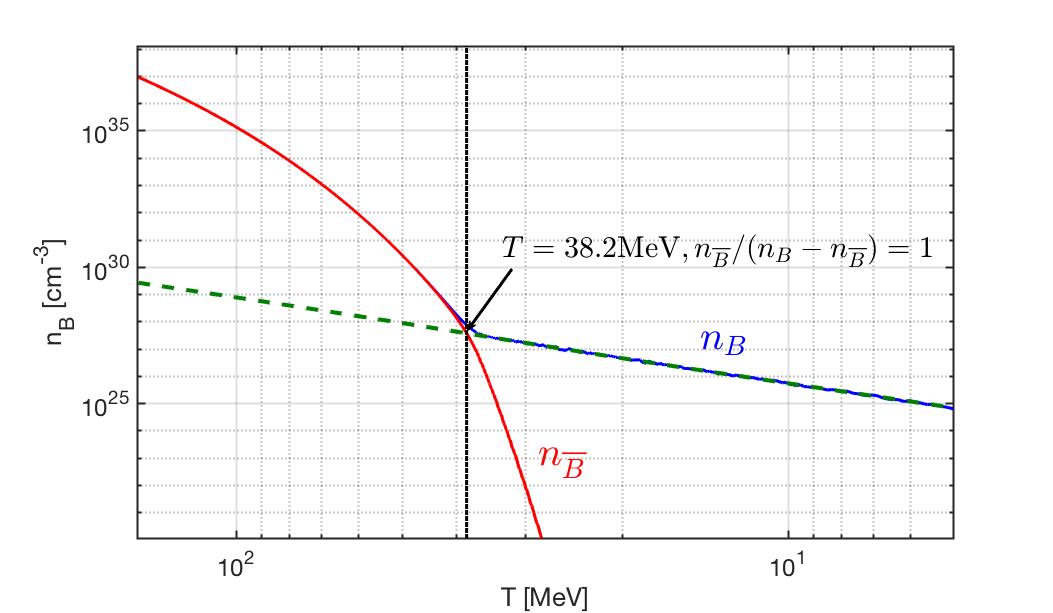}
\caption{The baryon (blue solid line) and antibaryon (red solid line) number density as a function of temperature in the range $150\,\mathrm{MeV}>T>5\,\mathrm{MeV}$. The green dashed line is the extrapolated value for baryon density. The temperature $T=38.2\,\mathrm{MeV}$ (black dashed vertical line) is denoted when the ratio $n_{\overline B}/(n_B-n_{\overline B})=1$ which defines the condition where antibaryons disappear from the Universe.}
\label{Baryon_fig}
\end{figure}
%%%%%%%%%%%%%%%%%%%%%%%%%%%%%%%%%%%%%%%

We solve Eq.~(\,\ref{museq}) and Eq~(\ref{muBeq}) numerically to obtain baryon and strangeness chemical potentials as a function of temperature in Fig.~\ref{ChemPotFig}. The chemical potential changes dramatically in the temperature window $50\,\mathrm{MeV}\le T\le 30$\,MeV, its behavior describing the process of antibaryon disappearance. Substituting the chemical potential $\lambda_q$ and $\lambda_s$ into particle density Eq.~(\ref{Density_N}), Eq.~(\ref{Density_K}), and Eq.~(\ref{Density_Y}), we can obtain the particle number densities for different species as a function of temperature.

In Fig.~\ref{Baryon_fig} we plot the number density of baryon and antibaryon as a function of temperature. We consider that when the  $n_{\overline B}\ll(n_B-n_{\overline B})$ the anitbaryons density is sufficient low and disappear from the Universe inventory quickly. To determine the temperature where antibaryons is sufficient law in the Universe inventory we defined the condition when the ratio $n_{\overline B}/(n_B-n_{\overline B})=1$. This condition is reached in an expanding Universe at $T=38.2$\,MeV, which is in agreement with the qualitative result in [\cite{kolb1990early}]. After this temperature, the net baryon density dilutes with a residual co-moving conserved quantity determined by the observed baryon asymmetry.

%~~~~~~~~~~~~~~~~~~~~~~~~~~~~~~~~~~~~~~~~~~~~~~~~~~~~~~~~~~~~~~~~~~~~~~~~~~~~~~~~
\begin{figure}[bt]
%\begin{center}
\centering
\includegraphics[width=0.85\linewidth]{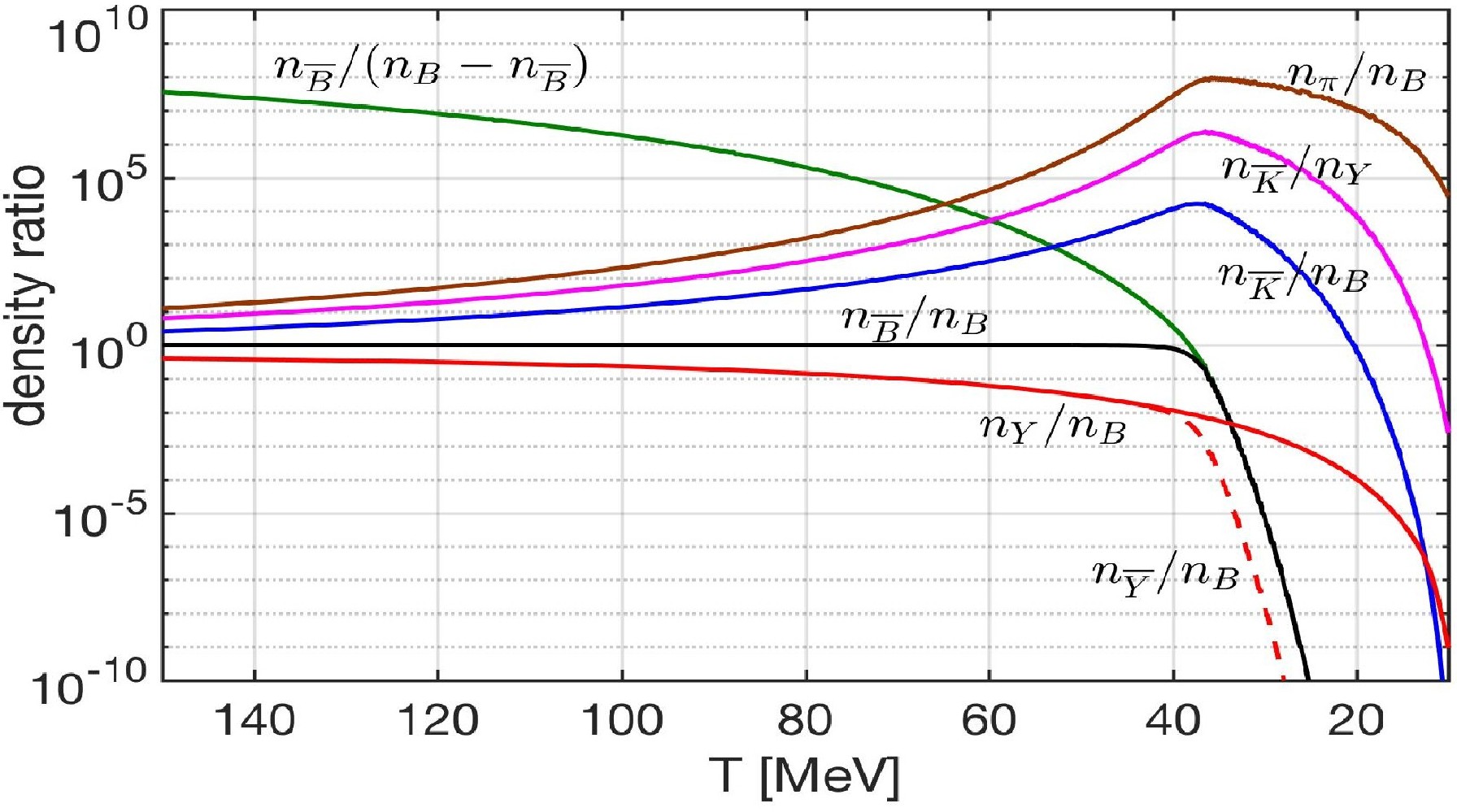}
\caption{Ratios of hadronic particle number densities as a function of temperature $150\,\mathrm{MeV}> T>10\,\mathrm{MeV}$ in the early Universe, with baryon $B$ yields: pions $\pi$ (brown line), kaons $K( q\bar s)$ (blue), antibaryon $\overline B$ (black), hyperon $Y$ (red) and anti-hyperons $\overline Y$ (dashed red). Also shown $\overline K/Y$(purple).}
\label{EquilibPartRatiosFig}
%\end{center}
\end{figure}
%~~~~~~~~~~~~~~~~~~~~~~~~~~~~~~~~~~~~~~~~~~~~~~~~~~~~~~~~~~~~~~~~~~~~~~~~~~~~~~

In Fig.~\ref{EquilibPartRatiosFig} we show examples of particle abundance ratios of interest. %Considering $n_Y/n_B$ we see that hyperons $Y(sqq)$ remain a noticeable 1\% component in baryon yield through this domain of antibaryon decoupling.
Pions $\pi(q\bar q)$ are the most abundant hadrons $n_\pi/n_B\gg1$, because of their low mass and the reaction $\gamma\gamma\rightarrow\pi^0$, which assures chemical yield equilibrium~[\cite{Kuznetsova:2008jt}]. For $150\,\mathrm{MeV}>T>20.8\,\mathrm{MeV}$, we see the ratio $n_{{\overline K}(\bar q s)}/n_B\gg1$, which implies pair abundance of strangeness is more abundant than baryons, and is dominantly present in mesons, since $n_{\overline K}/n_Y\gg1$. 
For $20.8\,\mathrm{MeV}>T$, the baryon becomes dominant $n_{\overline K}/n_B<1$, which implies that the strange meson is embedded in a large background of baryons, and the exchange reaction $\overline{K}+N\rightarrow \Lambda+\pi$ can re-equilibrate kaons and hyperons in the temperature range; therefore strangeness symmetry $s=\bar s$ is maintained. For $12.9\,\mathrm{MeV}>T$ we have $n_Y/n_B>n_{\overline K}/n_B$, now the still existent tiny abundance of strangeness is found predominantly in hyperons.

%~~~~~~~~~~~~~~~~~~~~~~~~~~~~~~~~~~~~~~~~~~~~~~~~~

\section{Seeking strangeness freeze-out chemical nonequilibrium
}
%In this section we will focus on the following:
%\begin{itemize}
%    \item Relevant strangeness reactions
%    \item Strangeness creation/annihilation in mesons
%    \item Strangeness production/ exchange in hyperons
%    \item Strangeness epochs in  the Universe 
%\end{itemize}

%In this section, we focus on investigating the strangeness abundance and decoupling following the formation of normal matter during the hadronization process of the quark-gluon plasma (QGP). Nonequilibrium conditions in the early Universe are of general interest: they are understood to be prerequisite for the arrow of time dependent processes to take hold in the Hubble expanding Universe.
%~~~~~~~~~~~~~~~~~~~~~~~~~~~~~~~~~~~~~~~~~~~~~~~~~~~~~~~~~~~~

%\subsection{Relevant strangeness reactions}
This section considers an unstable strange particle $S$ decaying into two particles $1$ and $2$, which themselves have no strangeness content. In a dense and high-temperature plasma with particles $1$ and $2$ in thermal equilibrium, the inverse reaction populates the system with particle $S$. This is written schematically as
\begin{align}
 S\Longleftrightarrow1+2,\qquad \mathrm{Example}: K^0\Longleftrightarrow\pi+\pi\,.
\end{align}
The natural decay of the daughter particles provides the intrinsic strength of the inverse strangeness production reaction rate. As long as both decay and production reactions are possible, particle $S$ abundance remains in thermal equilibrium. This balance between production and decay rates is called a detailed balance.

Once the primordial Universe expansion rate $1/H$ overwhelms the strongly temperature dependent back-reaction and the back reaction freeze-out, then the decay $S\rightarrow 1+2$ occurs out of balance and particle $S$ disappears from the inventory. The two-on-two strangeness producing reactions have a significantly higher strangeness production reaction threshold, thus especially near to strangeness decoupling their influence is negligible. Such reactions are more important near the QGP hadronization temperature $T_H\simeq 150$\,MeV, and they characterize strangeness exchange reactions such as $\mathrm{K}+N\leftrightarrow \Lambda+\pi$, (see Chapter 18 in [\cite{Letessier:2002ony}]).

\begin{figure} %[h]
%\begin{center}
\centering
\includegraphics[width=0.75\linewidth]{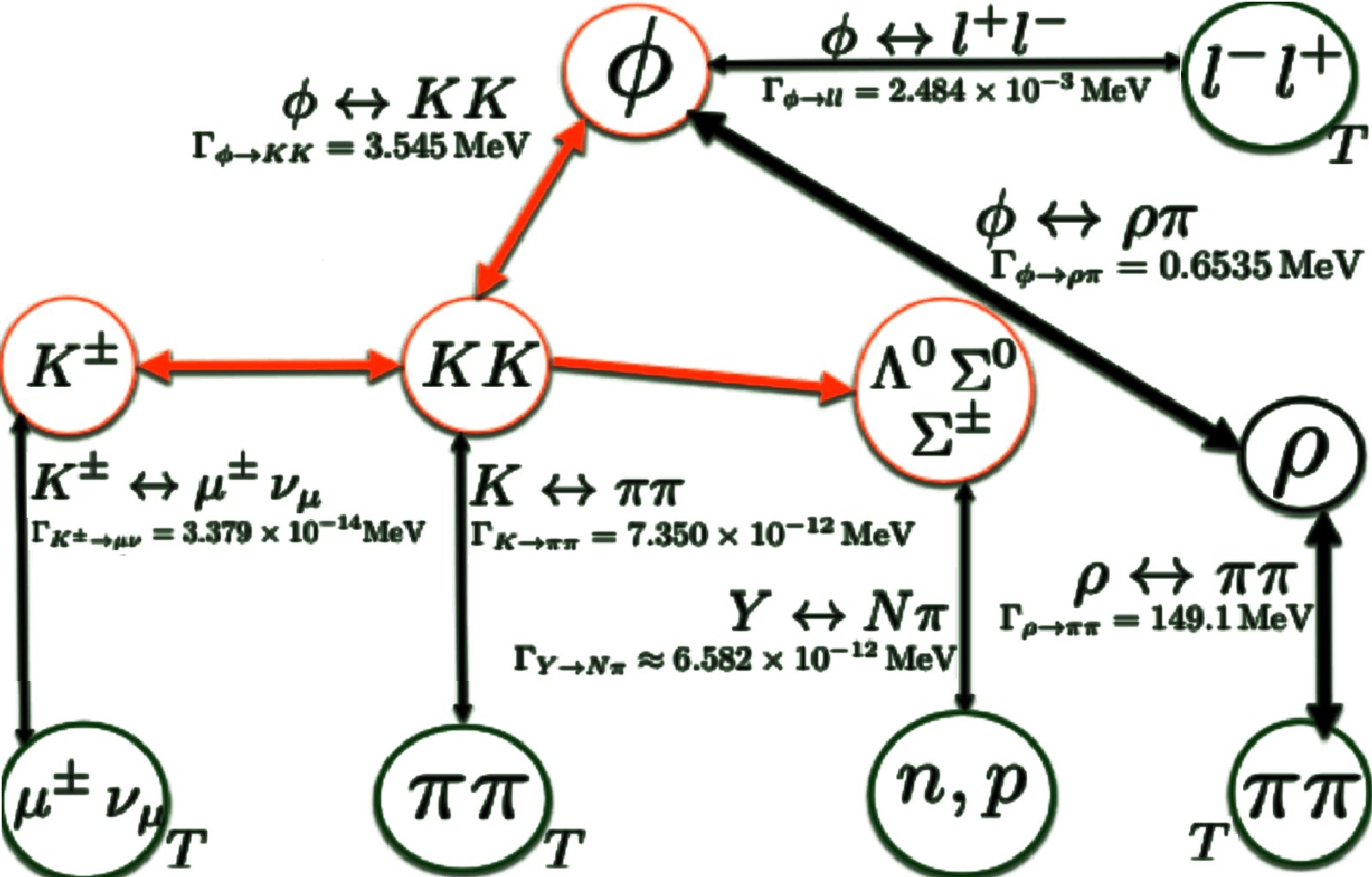}
\caption{
The strangeness abundance changing reactions in the primordial Universe. The red circles show strangeness carrying hadronic particles; red thick lines denote effectively instantaneous reactions. Black thick lines show relatively strong hadronic reactions. The reaction rates required to describe  strangeness time evolution are shown in [\cite{Rafelski:2020ajx}].
}
\label{Strangeness_map2}
%\end{center}
\end{figure}

In Fig.~\ref{Strangeness_map2} we show reactions relevant to strangeness evolution in the considered Universe evolution epoch $150\,\mathrm{MeV}\ge T\ge 10$\,MeV  and their pertinent reaction strength. As shown:
\begin{itemize}
\item
We study strange quark abundance in baryons and mesons, considering both open and hidden strangeness (hidden: $s\bar s$-content). Important source reactions are $l^-+l^+\rightarrow\phi$, $\rho+\pi\rightarrow\phi$, $\pi+\pi\rightarrow K_\mathrm{S}$, $\Lambda \leftrightarrow \pi+ N$, and $\mu^\pm+\nu\rightarrow K^\pm$. 
\item
Muons and pions are coupled through electromagnetic reactions $\mu^++\mu^-\leftrightarrow\gamma+\gamma$ and $\pi\leftrightarrow\gamma+\gamma$ to the photon background and retain their chemical equilibrium until the temperature $T =4$\, MeV and $T=5$\,MeV, respectively~[\cite{Rafelski:2021aey,Kuznetsova:2008jt}]. The large $\phi\leftrightarrow K+K$ rate assures $\phi$ and $K$ are in relative chemical equilibrium.
\end{itemize}
In order to determine where exactly strangeness disappears from the Universe inventory, we explore the magnitudes of different rates of production and decay processes in mesons and hyperons.
%~~~~~~~Figure~~~~~~~~~~~~~~~~~~~~~~~~~~~~~~~~~~~~~~~~~~~~~~~~~~~~~~~~~~~~~~~~~~~~
%~~~~~~~~~~~~~~~~~~~~~~~~~~~~~~~~~~~~~~~~~~~~~~~~~~~~~~~~~~~~~~~~~~~~~~~~~~

\subsection{Strangeness creation/annihilation rate in mesons}
From Fig.~\ref{Strangeness_map2} in the meson domain, the relevant interaction rates competing with Hubble time are the reactions
\begin{align}
 &\pi+\pi\leftrightarrow K\,,\quad\mu^\pm+\nu\leftrightarrow K^\pm\,,\quad l^++l^-\leftrightarrow\phi\,,\\
 &\rho+\pi\leftrightarrow\phi\,,\quad \pi+\pi\leftrightarrow\rho\,.
\end{align}
The thermal reaction rate per time and volume for two body-to-one particle reactions $1+2\rightarrow 3$ has been presented before~[\cite{Koch:1986ud,Kuznetsova:2008jt,Kuznetsova:2010pi}]. In full kinetic and chemical equilibrium, the reaction rate per time per volume can be written as~[\cite{Kuznetsova:2010pi}] :
\begin{align}
&R_{12\to 3}=\frac{g_3}{(2\pi)^2}\,\frac{m_3}{\tau^0_3}\,\int^\infty_0\frac{p^2_3dp_3}{E_3}\frac{e^{E_3/T}}{e^{E_3/T}\pm1}\Phi(p_3)\;,
\end{align}
where $\tau^0_3$ is the vacuum lifetime of particle $3$. The positive sign $``+"$ is for the case when particle $3$ is a boson, and negative sign $``-"$ for fermion. The function $\Phi(p_3)$ for the non-relativistic limit $m_3\gg p_3,T$ can be written as 
\begin{align}
\Phi(p_3\to0)=2\frac{1}{(e^{E_1/T}\pm1)(e^{E_2/T}\pm1)}.
\end{align}

Considering the Boltzmann limit, the thermal reaction rate per unit time and volume becomes
\begin{align}
\label{Thermal_Rate}
R_{12\rightarrow3}=\frac{g_3}{2\pi^2}\left(\frac{T^3}{\tau^0_3}\right)\left(\frac{m_3}{T}\right)^2\,K_1(m_3/T),
\end{align}
where $K_1$ is the modified Bessel functions of integer order "$1$". In order to compare the reaction time with Hubble time $1/H$, it is convenient to define the relaxation time for the process $1+2\rightarrow 3$ as follows:
\begin{align}
\label{Reaction_Time}
\tau_{12\rightarrow 3}\equiv\frac{n^{eq}_{1}}{R_{12\rightarrow n}}\,,\quad
n^{eq}_1=\frac{g_1}{2\pi^2}\int_{m_1}^\infty\!\!\!\!dE\,\frac{E\,\sqrt{E^2-m_1^2}}{\exp{\left(E/T\right)}\pm1}\;, 
\end{align}
where $n^{eq}_1$\,is the thermal equilibrium number density of particle\,$1$ with the `heavy' mass $m_1>T$.  Combining Eq.\,(\ref{Thermal_Rate}) with  Eq.\,(\ref{Reaction_Time}) we obtain
\begin{align}\label{RelaxationTime}
&\frac{\tau_{12\rightarrow3}}{ \tau^0_3}=  
\frac{2\pi^2 n^{eq}_1/T^3}{g_3(m_3/T)^2\,K_1(m_3/T)}\,, \quad 
n^{eq}_1\simeq g_1\left(\frac{m_1 T}{2\pi}\right)^{3/2}e^{-m_1/T},
\end{align}
where, conveniently, the relaxation time does not depend on the abundant and often relativistic heat bath component $2$, {\it e.g.\/} $l^\pm,\pi,\nu,\gamma$. The density of heavy particles\,$1$\,and\,$3$ can in general be well approximated using the leading and usually nonrelativistic Boltzmann term as shown above.

In general, the reaction rates for inelastic collision process capable of changing particle number, for example $\pi\pi\to K^0$, is suppressed by the factor $\exp{(-m_{K^0}/T)}$. On the other hand, there is no suppression for the elastic momentum and energy exchanging particle collisions in plasma. We conclude that for the case $m\gg T$, the dominant collision term in the relativistic Boltzmann equation is the elastic collision term, keeping all heavy particles in kinetic energy equilibrium with the plasma. This allows us to study the particle abundance in plasma presuming the energy-momentum statistical distribution equilibrium exists. This insight was discussed in detail in the preparatory phase of laboratory exploration of hot hadron and quark matter, see~[\cite{Koch:1986ud}]. In order to study the particle abundance in the Universe when $m\gg T$, instead of solving the exact Boltzmann equation, we can separate the fast energy-momentum equilibrating collisions from the slow particle number changing inelastic collisions. In the following we explore the rates of inelastic collision and compare the relaxation times of particle production in all relevant reactions with the Universe expansion rate.

It is common to refer to particle freeze-out as the epoch where a given type of particle ceases to interact with other particles. In this situation the particle abundance decouples from the cosmic plasma, a chemical nonequilibrium and even complete abundance disappearance of this particle can happen; the condition for the given reaction $1+2\rightarrow 3$ to decouple is
\begin{align}
\tau_{12\rightarrow 3}(T_f)=1/H(T_f),
\end{align}
where $T_f$ is the freeze-out temperature.
In the epoch of interest, $150\,\mathrm{MeV}>T>10\,\mathrm{MeV}$, the Universe is dominated by radiation and effectively massless matter behaving like radiation. The Hubble parameter can be written as~[\cite{Kolb:1990vq}]
\begin{align}\label{H2g}
H^2=H^2_{rad}\left(1+\frac{\rho_{\pi,\,\mu,\,\rho}}{\rho_\mathrm{rad}}+\frac{\rho_\mathrm{strange}}{\rho_\mathrm{rad}}\right)=\frac{8\pi^3G_\mathrm{N}}{90}g^e_\ast T^4,\qquad H^2_\mathrm{rad}=\frac{8\pi G_\mathrm{N}\,\rho_\mathrm{rad}}{3},
\end{align}
where: $g^e_\ast$ is the total number of effective relativistic `energy' degrees of freedom; $G_\mathrm{N}$ is the Newtonian constant of gravitation; the `radiation' energy density includes $\rho_\mathrm{rad}=\rho_\gamma+\rho_\nu+\rho_{e^\pm}$ for photons, neutrinos, and massless electrons(positrons). The massive-particle correction is $\rho_{\pi,\,\mu,\,\rho}=\rho_\pi+\rho_\mu+\rho_\rho$; and at highest $T$ of interest, also of (minor) relevance, $\rho_\mathrm{strange}=\rho_{K^0}+\rho_{K^\pm}+\rho_{K^\ast}+\rho_{\eta}+\rho_{\eta^\prime}$.
%~~~~~~~Figure~~~~~~~~~~~~~~~~~~~~~~~~~~~~~~~~~~~~~~~~~~~~~~~~~~~~~~~~~~~~~~~~~~~~~~~~~~
\begin{figure}[ht]
%\begin{center}
\centering
\includegraphics[width=0.95\linewidth]{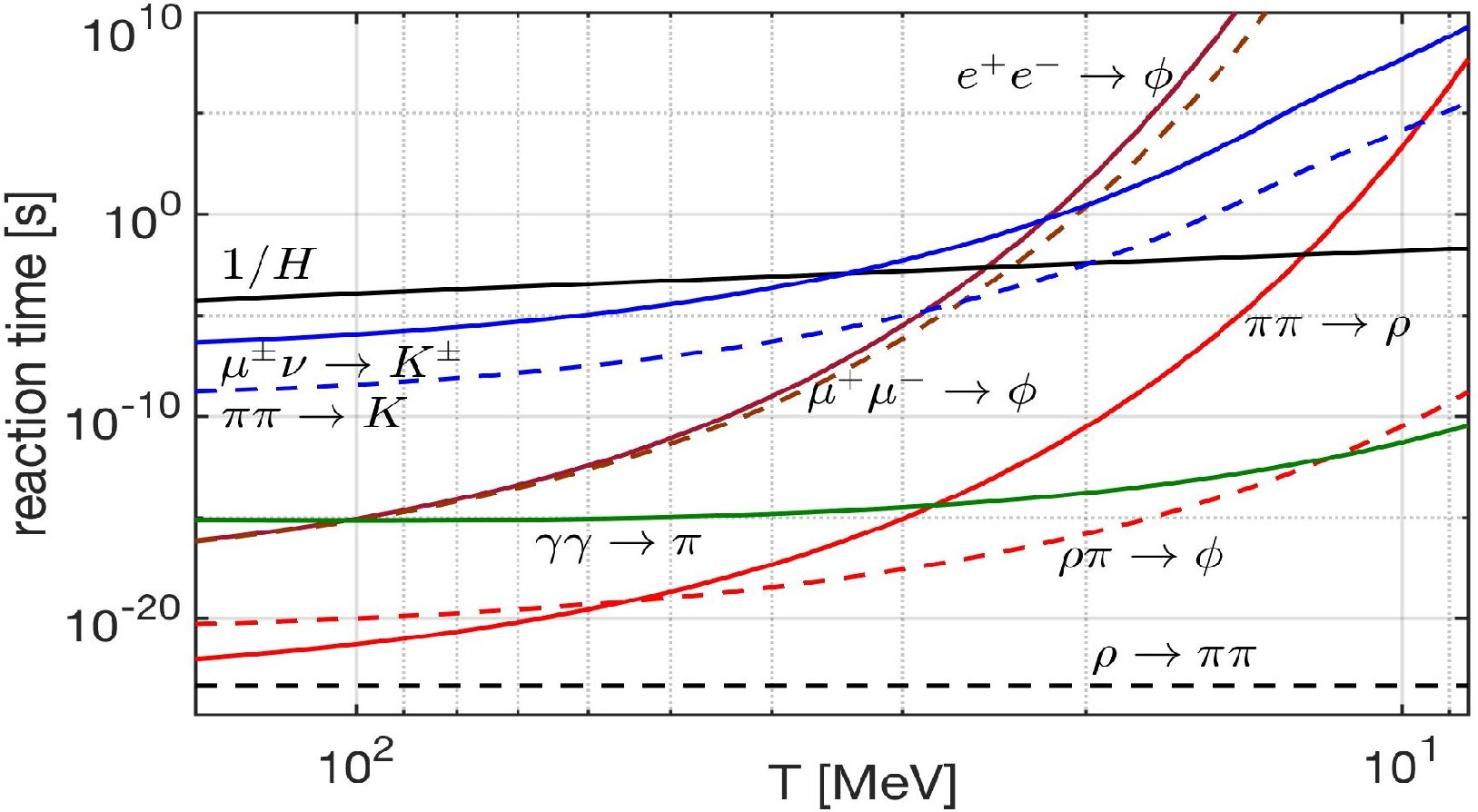}
\caption{Hadronic relaxation reaction times, see Eq.\,(\ref{Reaction_Time}), as a function of temperature $T$, are compared to Hubble time $1/H$ (black solid line). At bottom the horizontal black-dashed line is the natural (vacuum) lifespan of $\rho$.}
\label{reaction_time_tot}
%\end{center}
\end{figure}
%~~~~~~~~~~~~~~~~~~~~~~~~~~~~~~~~~~~~~~~~~~~~~~~~~~~~~~~~~~~~~~~~~~~~~~~~~~~~~~~~~~~~~

When presenting the reaction rates and quoting decoupling as a function of temperature $T$, we must remember that for a temperature range $50\,\mathrm{MeV}>T>5$\,MeV, we have $10^{-1}<dT/dt<10^{-4}$\,MeV/$\mu$s. We estimate the width of freezeout temperature interval $\Delta T_f$  as follows:
\begin{align}
\frac{1}{\Delta T_f}\equiv \left[\frac{1}{(\Gamma_{12\to3}/H)}\frac{d(\Gamma_{12\to3}/H)}{dT}\right]_{T_f},\quad \Gamma_{12\to3}\equiv\frac{1}{\tau_{12\to3}}.
\end{align}
Using Eq.(\ref{H2g}) and Eq.(\ref{RelaxationTime}) and considering the temperature range $50\,\mathrm{MeV}>T>5$\,MeV with $g^e_\ast\approx\mathrm{constant}$ we obtain using the Boltzmann approximation to describe the  massive particles\,$1$\,and\,$3$
\begin{align}\label{DeltaFreezeout}
 \frac{\Delta T_f}{ T_f} \approx\frac{T_f  }{ m_3 - m_1 -2T_f}\,,\quad m_3 - m_1>> T_f\,.
\end{align}
The width of freeze-out is shown in the right column in Table~\ref{FreezeoutTemperature_table}. We see a range of $2$-$10\%$. Therefore it is justified to consider as a decoupling condition in time the value of temperature at which the pertinent rate cross the Hubble expansion rate, see Fig.~\ref{reaction_time_tot}.
 
In Fig.~\ref{reaction_time_tot} we plot the  hadronic reaction relaxation times $\tau_{i}$ in the meson sector as a function of temperature compared to Hubble time $1/H$.
It shows that the weak interaction reaction $\mu^\pm+\nu_{\mu}\rightarrow K^\pm$ becomes slower compared to the Universe expansion near temperature $T_f^{K^\pm}=33.8\,\mathrm{MeV}$, signaling the onset of abundance nonequilibrium for $K^\pm$. For $T<T_f^{K^\pm}$, the reactions $\mu^\pm+\nu_{\mu}\rightarrow K^\pm$ decouples from the cosmic plasma; the corresponding detailed balance can be broken and the decay reactions $K^\pm\rightarrow\mu^\pm+\nu_{\mu}$ are acting like a (small) ``hole'' in the strangeness abundance ``pot''. If other strangeness production reactions did not exist, strangeness would disappear as the Universe cools below $T_f^{K^\pm}$. However, we have other reactions: $l^++l^-\leftrightarrow\phi$, $\pi+\pi\leftrightarrow K$, and $\rho+\pi\leftrightarrow\phi$ can still produce the strangeness in cosmic plasma and the rate is very large compared to the weak interaction decay.
%~~~~~~~~~~~~~~~~~~~~~~~~~~~~~~~~~~~~~~~~~~~~~~~~~~~~~~~~~~~~~~~~~~~~~~~~~~~~~~~~~~~~~~~~~
\begin{table}%[h]
\centering
\begin{tabular}{c| c| c}
\hline\hline
Reactions &Freeze-out Temperature (MeV) & {$\Delta T_f$\,(MeV)} \\
\hline
$\mu^\pm\nu\rightarrow K^\pm$ & $T_f=33.8$\,MeV & {$3.5$ \,MeV}\\ 
\hline
$e^+e^-\rightarrow \phi$ & $T_f=24.9$\,MeV &{$0.6$\,MeV}\\
$\mu^+\mu^-\rightarrow\phi$ & $T_f=23.5$\,MeV &{$0.6$\,MeV}\\
\hline
 $\pi\pi\rightarrow K$ & $T_f=19.8$\,MeV&{$1.2$\,MeV}\\
\hline
$\pi\pi\rightarrow\rho$ & $T_f=12.3$\,MeV&{$0.2$\,MeV}\\
\hline\hline
\end{tabular}
\caption{The characteristic strangeness reaction and their freeze-out temperature and temperature width in early Universe.}
\label{FreezeoutTemperature_table} 
\end{table}

%~~~~~~~~~~~~~~~~~~~~~~~~~~~~~~~~~~~~~~~~~~~~~~~~~~~~~~~~~~~~~~~~~~~~~~~~~ 

In Table~\ref{FreezeoutTemperature_table} we show the characteristic strangeness reactions and their freeze-out temperatures in the early Universe. The intersection of strangeness reaction times with $1/H$ occurs for $l^-+l^+\rightarrow\phi$ at $T_f^\phi=25\sim23\,\mathrm{MeV}$, and for $\pi+\pi\rightarrow K$ at $T_f^K=19.8\,\mathrm{MeV}$, for $\pi+\pi\rightarrow\rho$ at $T_f^\rho=12.3\,\mathrm{MeV}$. The reactions $\gamma+\gamma\rightarrow\pi$ and $\rho+\pi\leftrightarrow\phi$ are faster compared to $1/H$. However, the $\rho\to\pi+\pi$ lifetime (black dashed line in Fig.~\ref{reaction_time_tot}) is smaller than the reaction $\rho+\pi\leftrightarrow\phi$; in this case, most of $\rho$-meson decays faster, thus are absent and cannot contribute to the strangeness creation in the meson sector. Below the temperature $T<20$\,MeV, all the detail balances in the strange meson reactions are broken and the strangeness in the meson sector should disappear rapidly, were it not for the small number of baryons present in the Universe.

%~~~~~~~~~~~~~~~~~~~~~~~~~~~~~~~~~~~~~~~~~~~~~~~~~~~~~~~~

\subsection{Strangeness production/ exchange rate in hyperons}
In order to understand strangeness in hyperons in the baryonic domain, we now consider the strangeness production reaction $\pi +N\rightarrow K+\Lambda$, the strangeness exchange reaction $\overline{K}+N\rightarrow \Lambda+\pi$; and the strangeness decay $\Lambda\rightarrow N+\pi$. The competition between different strangeness reactions allows strange hyperons and antihyperons to influence the dynamic nonequilibrium condition, including development of $\langle s-\bar s\rangle \ne 0$. %The cross sections $\sigma_{\overline{K}N\rightarrow \Lambda\pi}$ and $\sigma_{\pi N\rightarrow K\Lambda}$ are obtained from experiment.

To evaluate the reaction rate in two-body reaction $1+2\rightarrow3+4$ in the Boltzmann approximation we can use the reaction cross section $\sigma(s)$ and the relation~[\cite{Letessier:2002ony}]:
\begin{align}
R_{12\rightarrow34}=\frac{g_1g_2}{32\pi^4}\frac{T}{1+I_{12}}\!\!\int^\infty_{s_{th}}\!\!\!\!ds\,\sigma(s)\frac{\lambda_2(s)}{\sqrt{s}}\!K_1\!\!\left({\sqrt{s}}/{T}\right),
\end{align}
where $K_1$ is the Bessel function of order $1$ and the function $\lambda_2(s)$ is defined as
\begin{align}
\lambda_2(s)=\left[s-(m_1+m_2)^2\right]\left[s-(m_1-m_2)^2\right],
\end{align}
with $m_1$ and $m_2$, $g_1$ and $g_2$ as the masses and degeneracy of the initial interacting particle. The factor $1/(1+I_{12})$ is introduced to avoid double counting of indistinguishable pairs of particles; we have $I_{12}=1$ for identical particles and $I_{12}=0$ for others. 

The thermal averaged cross sections for the strangeness production and exchange processes are about $\sigma_{\pi N\rightarrow K\Lambda}\sim0.1\,\mathrm{mb}$ and $\sigma_{\overline{K}N\rightarrow \Lambda\pi}=1\sim3\,\mathrm{mb}$ in the energy range in which we are interested~[\cite{Koch:1986ud}]. The cross section can be parameterized as follows:\\
1) For the cross section $\sigma_{\overline{K}N\rightarrow \Lambda\pi}$ we use~[\cite{Koch:1986ud}]
 \begin{align}
 \sigma_{\overline{K}N\rightarrow \Lambda\pi}=\frac{1}{2}\left(\sigma_{K^-p\rightarrow \Lambda\pi^0}+\sigma_{K^-n\rightarrow \Lambda\pi^-}\right)\,.
\end{align}
Here the experimental cross sections can be parameterized as 
\begin{align}
&\sigma_{K^-p\rightarrow \Lambda\pi^0}\!\!=\!\!\left(\begin{array}{l}\!\!1479.53\mathrm{mb}\!\cdot\!\exp{\left(\frac{-3.377\sqrt{s}}{\mathrm{GeV}}\right)},\; \mathrm{for}\,\sqrt{s_m}\!\!<\!\!\sqrt{s}\!<\!3.2\mathrm{GeV} \\ \\0.3\mathrm{mb}\!\cdot\!\exp{\left(\frac{-0.72\sqrt{s}}{\mathrm{GeV}}\right)},\; \mathrm{for}\sqrt{s}>3.2\mathrm{GeV}\end{array}\right.\\
&\sigma_{K^-n\rightarrow \Lambda\pi^-}\!\!=\!\!1132.27\mathrm{mb}\!\cdot\!\exp{\left(\frac{-3.063\sqrt{s}}{\mathrm{GeV}}\right)},\; \mathrm{for}\sqrt{s}>1.699\mathrm{GeV},
\end{align}
where $\sqrt{s_m}=1.473$ GeV.\\
2) For the cross section $\sigma_{\pi N\rightarrow K\Lambda}$ we use~[\cite{Cugnon:1984pm}]
\begin{align}
&\sigma_{\pi N\rightarrow K\Lambda}=\frac{1}{4}\times\sigma_{\pi p\rightarrow K^0\Lambda}\,.
\end{align}
The experimental $\sigma_{\pi p\rightarrow K^0\Lambda}$  can be approximated as follows
\begin{align}
\sigma_{\pi p\rightarrow K^0\Lambda}=\left(\begin{array}{l}\frac{0.9\mathrm{mb}\cdot\left(\sqrt{s}-\sqrt{s_0}\right)}{0.091\mathrm{GeV}},\; \mathrm{for} \sqrt{s_0}<\sqrt{s}<1.7\mathrm{GeV} \\ \\ \frac{90\mathrm{MeV\cdot mb}}{\sqrt{s}-1.6\mathrm{GeV}},\; \mathrm{for}\sqrt{s}>1.7\mathrm{GeV},\end{array}\right.
 \end{align}
 with $ \sqrt{s_0}=m_\Lambda+m_K$.

\begin{figure}[ht]
%\begin{center}
\centering
\includegraphics[width=0.9\linewidth]{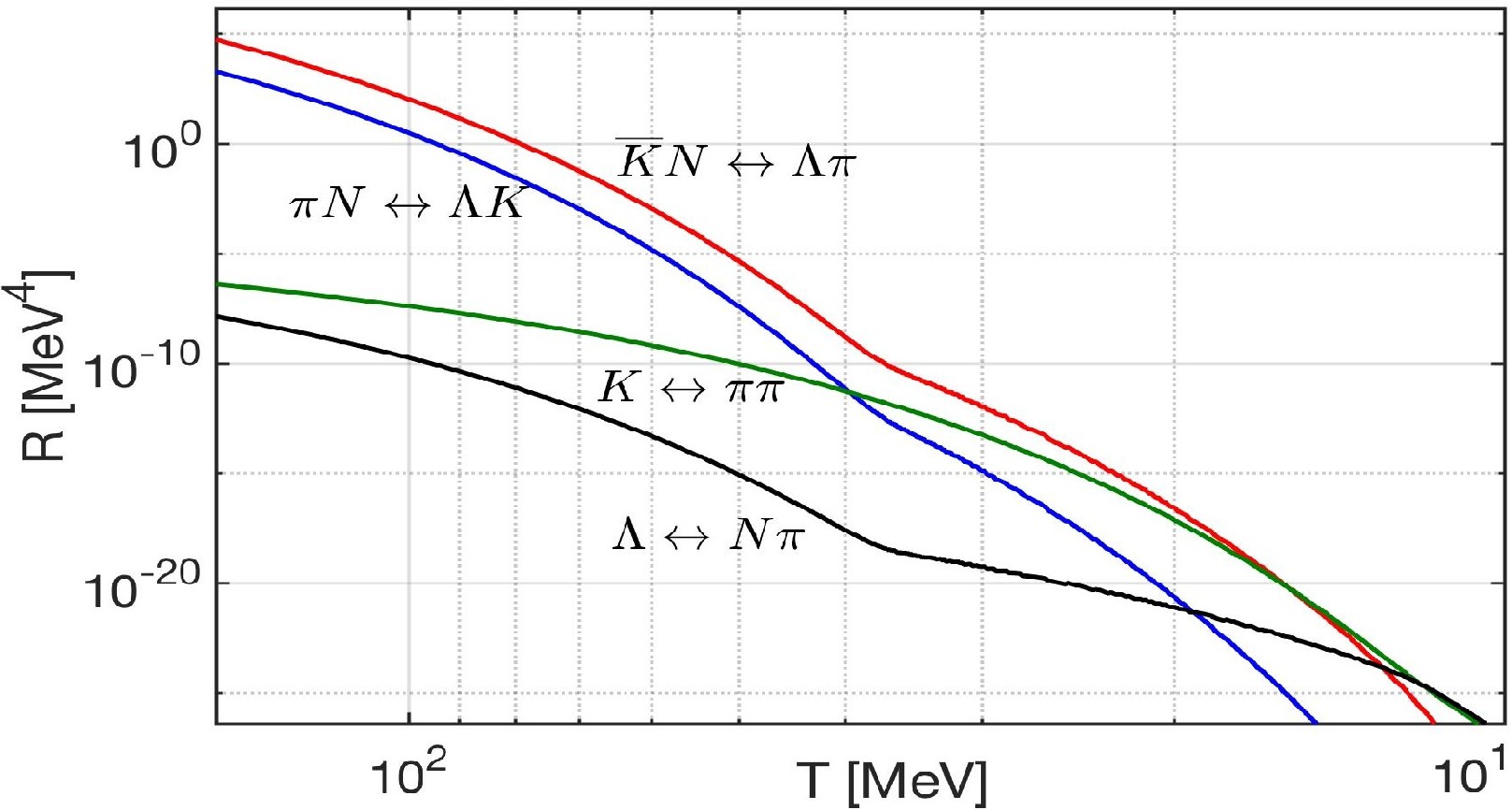}
\caption{Thermal reaction rate $R$ per volume and time for important hadronic strangeness production and exchange processes as a function of temperature $150\,\mathrm{MeV}> T>10\,\mathrm{MeV}$ in the early Universe.}
\label{Lambda_Rate_volume.fig}
%\end{center}
\end{figure}

Given the cross sections, we obtain the thermal reaction rate per volume for strangeness exchange reaction seen in Fig.~\ref{Lambda_Rate_volume.fig}. We see that around $T=20$\,MeV, the dominant reactions for the hyperon $\Lambda$ production is $\overline{K}+N\leftrightarrow\Lambda+\pi$. At the same time, the $\pi+\pi\to K$ reaction becomes slower than Hubble time and kaon $K$ decay rapidly in the early Universe. However, the anti-kaons $\overline K$ produce the hyperon $\Lambda$ because of the strangeness exchange reaction $\overline{K}+N\rightarrow\Lambda+\pi$ in the baryon-dominated Universe. We have strangeness in $\Lambda$ and it disappears from the Universe via the decay $\Lambda\rightarrow N+\pi$. Both strangeness and anti-strangeness disappear because of the $K\rightarrow\pi+\pi$ and $\Lambda\rightarrow N+\pi$, while the strangeness abundance $s = \bar{s}$ in the early Universe remains.

%~~~~~~~~~~~~~~~~~~~~~~~~~~~~~~~~~~~~~~~~~~~~~~~~~~~~~~~~~~~~~~~~~~~~~~~~~~~~~~~~
%~~~~~~~~~~~~~~~~~~~~~~~~~~~~~~~~~~~~~~~~~~~~~~~~~~~~~~~~~~~~~~~~~~~~~~~~~~~~~~

Around $T=12.9$\,MeV, the reaction $\Lambda+\pi\rightarrow\overline{K}+N$ becomes slower than the strangeness decay $\Lambda\leftrightarrow N+\pi$ and shows that at the low temperature the $\Lambda$ particles are still in equilibrium via the reaction $\Lambda\leftrightarrow N+\pi$ and little strangeness remains in the $\Lambda$. Then strangeness abundance becomes asymmetric $s\gg \bar{s}$, which implies that the assumption for strangeness conservation can only be valid until the temperature $T\sim13$\,MeV. Below this temperature a new regime opens up in which the tiny residual strangeness abundance is governed by weak decays with no re-equilibration with mesons. Also, in view of baron asymmetry, $\langle s-\bar s\rangle \ne 0$.

The primary conclusion of this first study of strangeness production and content in the early Universe, following on QGP hadronization, is that the relevant temperature domains indicate a complex interplay between baryon and meson (strange and non-strange) abundances and non-trivial decoupling from equilibrium for strange and non-strange mesons. We believe that this work contributes to the opening of a new and rich domain in the study of the Universe evolution in the future. 

%~~~~~~~~~~~~~~~~~~~~~~~~~~~~~~~~~~~~~~~~~~~~~~~~~

%{Introduction\daggerfootnote{This chapter has been published previously as \citet{Gottbrath1999}.}}

%% file: Chapter4.tex
\chapter{Neutrinos in cosmic plasma}\label{Neutrino}
Neutrinos are fundamental particles and play an important role in the evolution of the Universe. In the early Universe the neutrinos are kept in equilibrium with cosmic plasma via the weak interaction. The neutrino-matter interactions play a crucial role in  understanding of neutrinos evolution in the early Universe (such as the neutrino freezeout) and the later Universe (the property of today's neutrino background). In this chapter, I will examine the neutrino coherent and incoherent scattering with matter and their application in cosmology. The investigation of the relation between the effective number of neutrinos $N^{\mathrm{eff}}_\nu$ and lepton asymmetry $L$ after neutrino freezeout and its impact on Universe expansion is also discussed in this chapter.

\section{Matrix element for neutrino coherent/ incoherent scattering}
 According to the standard model, neutrinos interact with other particles via the Charged-Current(CC) and Neutral-Current(NC) interactions. Their Lagrangian can be written as~[\cite{Giunti:2007ry}]
\begin{align}
&\mathcal{L}^{CC}=\frac{g}{2\sqrt{2}}\left(j^\mu_W\,W_\mu+{j^\mu_W}^\dagger\,W^\dagger_\mu\right),\qquad\mathcal{L}^{NC}=-\frac{g}{2\cos{\theta_w}}\,j^\mu_Z\,Z_\mu,
\end{align}
where $g=e\sin\theta_w$, $W^\mu$ and $Z^\mu$ are W and Z boson gauge fields, and $j^\mu_W$ and $j^\mu_Z$ are the charged-current and neutral-current separately. In the limit of energies lower than the $W(m_w=80\,\mathrm{GeV})$ and $Z(m_z=91\,\mathrm{GeV})$ gauge bosons, the effective Lagrangians are given by
\begin{align}\label{L_low}
\mathcal{L}^{CC}_{eff}=-\frac{G_F}{\sqrt{2}}\,j^\dagger_{W\,\mu}\,j^\mu_W,\qquad
\mathcal{L}^{NC}_{eff}=-\frac{G_F}{\sqrt{2}}\,j^\dagger_{Z\,\mu}\,j^\mu_Z,\qquad \frac{G_F}{\sqrt{2}}=\frac{g^2}{8m^2_W},
\end{align}
where $G_F=1.1664\times10^{-5}\,\mathrm{GeV}^{-2}$ is the Fermi constant, which is one of the important parameters that determine the strength of the weak interaction rate. When neutrinos interact with matter, based on the neutrino's wavelength, they can undergo two types of scattering processes: coherent scattering and incoherent scattering with the particles in the medium. 

With coherent scattering, neutrinos interact with the entire  composite system rather than individual particles within the system. The coherent scattering is particularly relevant for low-energy neutrinos when the wavelength of neutrino is much larger than the size of system. In $1978$, Lincoln Wolfenstein pointed out that the coherent forward scattering of neutrinos off matter could be very important in studying the behavior of neutrino flavor oscillation in a dense medium~[\cite{PhysRevD.17.2369}]. The fact that neutrinos propagating in matter may interact with the background particles can be described by the picture of free neutrinos traveling in an effective potential.

For incoherent scattering, neutrinos interact with particles in the medium individually. Incoherent scattering is typically more prominent for high-energy neutrinos, where the wavelength of neutrino is smaller compared to the spacing between particles. Study of incoherent scattering of high-energy neutrinos is important for understanding the physics in various astrophysical systems (e.g. supernova, stellar formation) and the evolution of the early Universe.

In this section, we discuss the coherent scattering between long wavelength neutrinos and atoms, and study the effective potential for neutrino coherent interaction. Then we present the matrix elements that describe the incoherent interaction between high energy neutrinos and other fundamental particles in the early Universe. Understanding these matrix elements is crucial for comprehending the process of neutrino freeze-out in the early Universe.

%~~~~~~~~~~~~~~~~~~~~~~~~~~~~~~~~~~~~~~~~~~~~~~~~~~~~~~~~~~~~~~~~~~~~~~~

\subsection{Long wavelength limit of neutrino-atom coherent scattering}\label{LongWavelength}
According to the standard cosmological model, the Universe today is filled with the cosmic neutrinos with temperature $T_{\nu}^0=1.9 \,\mathrm{K}=1.7\times10^{-4}\,\mathrm{eV}$.
The average momentum of present-day relic neutrinos is given by $\langle p_\nu^0\rangle\approx3.15\,T_\nu^0$ and the typical wavelength $\lambda_\nu^0={2\pi}/{\langle p_\nu^0\rangle}\approx2.3\times10^5\,\mathrm{\AA}$, which is much larger than the radius at the atomic scale, such as the Bohr radius $R_{\mathrm{atom}}=0.529\,\mathrm{\AA}$. In this case we have the long wavelength condition $\lambda_\nu\gg\,R_{\mathrm{atom}}$ for cosmic neutrino background today.  

Under the condition $\lambda_\nu\gg\,R_{\mathrm{atom}}$, when the neutrino is scattering off an atom, the interaction can be coherent scattering [\cite{PhysRevD.38.32,PhysRevD.21.663,Papavassiliou:2005cs}]. According to the principles of quantum mechanics, with neutrino scattering it is impossible to identify which scatters the neutrino interacts with and thus it is necessary to sum over all possible contributions. In such circumstances, it is appropriate to view the scattering reaction as taking place on the atom as a whole, i.e.,
\begin{align}
\nu+\mathrm{Atom}\longrightarrow\nu+\mathrm{Atom}.
\end{align}

Considering a neutrino elastic scattering off an atom which is composed of $Z$ protons, $N$ neutrons and $Z$ electrons. For the elastic neutrino atom scattering, the low-energy neutrinos scatter off both atomic electrons and nucleus. For nucleus parts, we consider that the neutrinos interact via the $Z^0$ boson with a nucleus as
\begin{align}
\nu+A^{Z}_N\longrightarrow\nu+A^{Z}_N.
\end{align}
In this process a neutrino of any flavor scatters off a nucleus with the same strength. Therefore, the scattering will be insensitive to neutrino flavor. On the other hand, the neutrons can also interact via the $W^\pm$ with nucleus as 
\begin{align}
\nu_l+A^{Z}_N\longrightarrow\,l^-+A^{{Z}+1}_N,
\end{align}
which is a quasi-elastic process for neutrino scattering with the nucleus; we have $A^{Z_e}_N\rightarrow\,A^{{Z_e}+1}_N$. Since this process will change the nucleus state into an excited one, we will not consider its effect here. For detail discussion pf quasi-elastic scattering see ~[\cite{SajjadAthar:2022pjt}].

For atomic electrons, the neutrinos can interact via the $Z^0$ and $W^\pm$ bosons with electrons for different flavors, we have
\begin{align}
&\nu_e+e^-\longrightarrow\nu_e+e^-\,\,\,(\mathrm{Z^0,\,W^\pm\,exchange}),\\
&\nu_{\mu,\tau}+e^-\longrightarrow\nu_{\mu,\tau}+e^-\,\,\,(\mathrm{Z^0\,exchange}).
\end{align}
Because of the fact that the coupling of $\nu_e$ to electrons is quite different from that of $\nu_{\mu,\tau}$, one may expect large differences in the behavior of $\nu_e$ scattering compared to the other neutrino types.

%~~~~~~~~~~~~~~~~~~~~~~~~~~~~~~~~~~~~~~~~~~~~~~~~~~
\subsubsection{Neutrino-atom coherent scattering amplitude/matrix element} 
This section considers how a neutrino scatters from a composite system, assumed to consist of $N$ individual constituents at positions $x_i,\,i=1,2,....N$. Due to the superposition principle, the scattering amplitude $\mathcal{M}_\mathrm{sys}(\bold{p}^\prime,\bold{p})$ for scattering from an incoming momentum $\mathbf{p}$ to an outgoing momentum $\bold{p}^\prime$ is given as the sum of the contributions from each constituent [\cite{Freedman:1977xn,Papavassiliou:2005cs}]:
\begin{align}
\mathcal{M}_\mathrm{sys}(\bold{p}^\prime,\bold{p})=\sum_i^N\,\mathcal{M}_i(\bold{p}^\prime,\bold{p})\,e^{i\bold{q}\cdot\bold{x}_i},
\end{align}
where $\bold{q}=\bold{p}^\prime-\bold{p}$ is the momentum transfer and the individual amplitudes $\mathcal{M}_i(\bold{p}^\prime,\bold{p})$ are added with a relative phase factor determined by the corresponding wave function. %The transition probability is then given by
%\begin{align}
%\mathcal{P}_{\mathrm{sys}}(\bold{p}^\prime,\bold{p})&=|\mathcal{M}_\mathrm{sys}(\bold{p}^\prime,\bold{p})|^2\notag\\
%&=\sum_i|\mathcal{M}_i(\bold{p}^\prime,\bold{p})|^2+\sum_{i,j}^{i\neq\,j}\mathcal{M}_i(\bold{p}^\prime,\bold{p})\mathcal{M}_j^\dagger(\bold{p}^\prime,\bold{p})\,e^{i\bold{q}\cdot\left(\bold{x}_j-\bold{x}_i\right)}.
%\end{align} 
In principle, due to the presence of the phase factors, major cancellation may take place among the terms for the condition $|\bold{q}|R\gg1$, where $R$ is the size of the composite system, and the scattering would be incoherent. However, for the momentum small compared to the inverse target size, i.e., $|\bold{q}|R\ll1$, then all phase factors may be approximated by unity and contributions from individual scatters add coherently.

In the case of neutrino coherent scattering with an atom: If we consider sufficiently small momentum transfer to an atom from a neutrino which satisfies the coherence condition, i.e., $|\bold{q}|R_{\mathrm{atom}}\ll1$, then the relevant phase factors have little effect, allowing us to write the transition amplitude as [\cite{Nicolescu:2013rxa}]
\begin{align}
\label{M_atom}
\mathcal{M}_\mathrm{atom}=\sum_t\frac{G_F}{\sqrt{2}}\left[\overline{u}(p^\prime_\nu)\gamma_\mu\left(1-\gamma_5\right)u(p_\nu)\right]\left[\overline{u}(p^\prime_t)\gamma^\mu\left(c^t_V-c^t_A\gamma^5\right)u(p_t)\right],
\end{align}
where $t$ is all the target constituents (Z protons, N neutrons and Z electrons). The transition amplitude includes contributions from both charged and neutral currents, with
\begin{align}\label{CC_int}
&\mathrm{Charged\,\,Current}: c^t_V=c^t_A=1\\
\label{NC_int}
&\mathrm{Neutral\,\, Current}: c^t_V=I_3-2\mathcal{Q}\sin^2\theta_w,\qquad c^t_A=I_3
\end{align}
where $I_3$ is the weak isospin, $\theta_w$ is the Weinberg angle, and $\mathcal{Q}$ is the particle electric charge. 

Considering the target can be regarded as an equal mixture of spin states $s_z=\pm1/2$, and we can simplify the transition amplitude by summing the coupling constants of the constituents [\cite{PhysRevD.21.663, Sehgal:1986gn}]. We have
\begin{align}
\label{Transition}
\mathcal{M}_\mathrm{atom}=&\frac{G_F}{2\sqrt{2}}\left[\overline{u}(p^\prime_\nu)\gamma_\mu\left(1-\gamma_5\right)u(p_\nu)\right]\notag\\&\bigg[\overline{u}(p^\prime_{a})\sum_t\left(C_L+C_R\right)_t\gamma^\mu\,u(p_{a})-\overline{u}(p^\prime_{a})\sum_t\left(C_L-C_R\right)_t\gamma^\mu\gamma^5u(p_{a})\bigg],
\end{align}
where the $u(p_\nu)$, $u(p^\prime_\nu)$ are the initial and final neutrino states and $u(p_a)$, $u_(p^\prime_a)$ are the initial and final states of the target atom. 
The coupling coefficients $C_L$ and $C_V$ are defined as
\begin{align}
&C_L=c_V+c_A,\,\,\,\,\,C_R=c_V-c_A,
\end{align}
where the coupling constants for neutrino scattering with proton, neutron, and electron are given by Table.~\ref{Table_coupling}. The coupling constants for  $\nu_{\mu,\tau}$ are the same as for the $\nu_e$, excepting  the absence  of a charged current in neutrino-electron scattering.
%%%%%%%%%%%%%%%%%%%%%%%%%%%%%%%%%%%%%%%%%%%%%%%%%%%%%%
\begin{table}[h]
\begin{tabular}[c]{c|c|c|c|c}
\hline\hline
& Electron ($Z^0$ boson) & Electron ($W^\pm$ boson) & Proton (uud) & Neutron (udd)\\
\hline
$C_L$ & $-1+2\sin^2\theta_w$ & $2$ & $1-2\sin^2\theta_w$ & $-1$ \\
\hline
$C_R$ & $2\sin^2\theta_w$ & $0$ &$-2\sin^2\theta_w$ & $0$ \\
\hline\hline
\end{tabular}
\caption{The coupling constants for neutrino scattering with proton, neutron, and electron.}
\label{Table_coupling}  
\end{table}
%%%%%%%%%%%%%%%%%%%%%%%%%%%%%%%%%%%%%%%%%%%%%%%%%%%%%

Given the neutrino-atom coherent scattering amplitude Eq.(\ref{Transition}), the transition matrix element can be written as
\begin{align}
\label{scattering_matrix}
|\mathcal{M}_{\mathrm{atom}}|^2=\frac{G^2_F}{8}L_{\alpha\beta}^{\mathrm{neutrino}}\,\Gamma^{\alpha\beta}_{\mathrm{atom}},
\end{align}
where the neutrino tensor $L_{\alpha\beta}^{\mathrm{neutrino}}$ is given by
\begin{align}
\label{neutrino_tensor}
L_{\alpha\beta}^{\mathrm{neutrino}}
&=\mathrm{Tr}\bigg[\gamma_\alpha\left(1-\gamma_5\right)(\slashed{p}_\nu+m_\nu)\gamma_\beta\left(1-\gamma_5\right)(\slashed{p}^\prime_\nu+m_\nu)\bigg]\notag\\
&=8\bigg[(p_\nu)_\alpha\,(p^\prime_{\nu})_\beta+(p_\nu)^\prime_\alpha\,(p_\nu)_\beta-g_{\alpha\beta}(p_\nu\cdot\,p_\nu^\prime)+i\epsilon_{\alpha\sigma\beta\lambda}(p_\nu)^\sigma(p^\prime_\nu)^\lambda\bigg],
\end{align}
and the atomic tensor $\Gamma^{\alpha\beta}_\mathrm{atom}$ can be written as
\begin{align}
\label{atomic_tensor}
\Gamma^{\alpha\beta}_\mathrm{atom}
&=\mathrm{Tr}\bigg[(C_{LR}\gamma^\alpha-C^\prime_{LR}\gamma^\alpha\gamma^5)(\slashed{p}_a+M_a)(C_{LR}\gamma^\beta-C^\prime_{LR}\gamma^\beta\gamma^5)(\slashed{p}^\prime_a+M_a)\bigg]\notag\\
&=4\bigg\{(C^2_{LR}+C^{\prime2}_{LR})\left[(p_a)^\alpha\,(p^\prime_a)^\beta+(p_a)^{\prime\alpha}\,(p_a)^\beta\right]\notag\\
&\qquad-g^{\alpha\beta}\bigg[(C^2_{LR}-C^{\prime2}_{LR})(p_a\cdot\,p_a^\prime)-(C^2_{LR}-C^{\prime2}_{LR})M^2_a\bigg]\notag\\&\qquad\qquad+2iC_{LR}C^\prime_{LR}\epsilon^{\alpha\sigma^\prime\beta\lambda^\prime}(p_a)_{\sigma^\prime}(p^\prime_a)^{\lambda^\prime}\bigg\},
\end{align}
where $M_a$ is the target atom's mass $(M_a = AM_\mathrm{nucleon}, A=Z+N)$, the coupling constants $C_{LR}$ and $C^\prime_{LR}$ are defined by
\begin{align}
C_{LR}=\sum_t(C_L+C_R)_t,\,\,\,\,\,\,C^\prime_{LR}=\sum_t(C_L-C_R)_t.
\end{align}
Substituting Eq.(\ref{neutrino_tensor}) and Eq.(\ref{atomic_tensor}) into Eq.(\ref{scattering_matrix}), then the transition matrix element for coherent elastic neutrino atom scattering is given by:
\begin{align}
|\mathcal{M}_{\mathrm{atom}}|^2&=\frac{G^2_F}{8}L_{\alpha\beta}^{\mathrm{neutrino}}\,\Gamma^{\alpha\beta}_{\mathrm{atom}}\notag\\
&=8G^2_F\bigg[(C_{LR}+C^\prime_{LR})^2\,(p_\nu\cdot\,p_a)(p^\prime_\nu\cdot\,p^\prime_a)+(C_{LR}-C^\prime_{LR})^2\,(p_\nu\cdot\,p^\prime_a)(p^\prime_\nu\cdot\,p_a)\notag\\&
\,\,\,\,\,\,-(C^2_{LR}-C^{\prime2}_{LR})M^2_a(p_\nu\cdot\,p_\nu^\prime)\bigg].
\end{align}
Taking the atom at rest in the laboratory frame, and considering small momentum transfer to an atom from a neutrino, i.e., $q^2=(p_\nu-p^\prime_\nu)^2=(p_a^\prime-p_a)^2\ll\,M^2_a$, we have
\begin{align}
&p_\nu\cdot\,p_a=E_\nu\,M_a,\\
&p_\nu^\prime\cdot\,p_a=E_\nu^\prime\,M_a\approx\,E_\nu\,M_a,\\
&p^\prime_\nu\cdot\,p^\prime_a=p^\prime_\nu\cdot(p_a+q)=E^\prime_\nu\,M_a\left[\left(1+\frac{q_0}{M_a}\right)-\frac{|p^\prime_\nu||q|}{M_a}\cos\theta\right]\approx\,E_\nu\,M_a,\\
&p_\nu\cdot\,p^\prime_a=p_\nu\cdot(p_a+q)=E_\nu\,M_a\left[\left(1+\frac{q_0}{M_a}\right)-\frac{|p^\prime_\nu||q|}{M_a}\cos\theta\right]\approx\,E_\nu\,M_a.
\end{align}
Then the transition matrix element for neutrino coherent elastic scattering off a rest atom can be written as
\begin{align}\label{M_general}
|\mathcal{M}_{\mathrm{atom}}|^2&=8\,G^2_F\,M_a\,E_\nu^2\left[C^2_{LR}\left(1+\frac{|p_\nu|^2}{E^2_\nu}\cos\theta\right)+3C^{\prime2}_{LR}\left(1-\frac{|p_\nu|^2}{3E_\nu^2}\cos\theta\right)\right],
\end{align}
which is consistent with the results in papers [\cite{PhysRevD.38.32,PhysRevD.21.663,Papavassiliou:2005cs,Smith:1985mta}].
From the above formula we found that the scattering matrix neatly divides into two distinct components: a vector-like component (first term) and an axial-vector like component (second term). They have different angular dependencies: the vector part has a $\left({|p_\nu|^2}/{E^2_\nu}\cos\theta\right)$ dependence, while the axial part has a $\left(-{|p_\nu|^2}/{3E_\nu^2}\cos\theta\right)$ behavior. However, in the case of the nonrelativistic neutrino, both angular dependencies can be neglected because of the limit $p_\nu\ll\,m_\nu$.

Next, we consider the nonrelativistic electron neutrino $\nu_e$ scattering off an general atom with $Z$ protons, $N$ neutrons and $Z$ electrons. Then from Eq.~(\ref{M_general}), the matrix element can be written as
\begin{align}
\label{Probability_e}
|\mathcal{M}_{\mathrm{atom}}|^2&=8\,G^2_F\,M_a\,E_\nu^2\left[\left(3Z-A\right)^2\left(1+\frac{|p_\nu|^2}{E^2_\nu}\cos\theta\right)+3\left(3Z-A\right)^2\left(1-\frac{|p_\nu|^2}{3E_\nu^2}\cos\theta\right)\right]\notag\\
&\approx32\,G^2_F\,M_a\,E_\nu^2\left(3Z-A\right)^2,
\end{align}
where we neglect the angular dependence because of the nonrelativistic limit, and the coefficient $\left(3Z-A\right)^2$ for different target atoms are given in Table.(\ref{Table001}). On the other hand, for nonrelativistic $\nu_{\mu,\tau}$, the scattering matrix is given by
\begin{align}
\label{Probability_mt}
|\mathcal{M}_{\mathrm{atom}}|^2&=8\,G^2_F\,M_a\,E_\nu^2\left[\left(A-Z\right)^2\left(1+\frac{|p_\nu|^2}{E^2_\nu}\cos\theta\right)+3\left(A-Z\right)^2\left(1-\frac{|p_\nu|^2}{3E_\nu^2}\cos\theta\right)\right]\notag\\
&\approx32\,G^2_F\,M_a\,E_\nu^2\left(Z-A\right)^2,
\end{align}
where the coefficient $\left(Z-A\right)^2$ different target atoms are given in Table.(\ref{Table001}). The transition matrix for $\nu_e$ differs from that of $\nu_{\mu,\tau}$; this is due to the charged current reaction with the atomic electrons. Furthermore, the neutral current interaction for the electron and proton will cancel each other because of the opposite weak isospin $I_3$ and charge $\mathcal{Q}$. As a result, the coherent neutrino scattering from an atom is sensitive to the method of the neutrino-electron coupling.
%%%%%%%%%%%%%%%%%%%%%%%%%%%%%%%%%%%%%%%%%%
%%%%%%%%%%%%%%%%%%%%%%%%%%%%%%%%%%%%%%%%%%%%%%%%%%%%%%%%%%%%%%
\begin{table}[h]
\centering
\begin{tabular}{c|c|c}
\hline\hline
 Neutrino Flavor:&$\nu_e$ &$\nu_{\mu,\tau}$\\
\hline\hline
Target Atom & $(3Z-A)^2$  & $(Z-A)^2$\\
\hline
$H_2(A=2, Z=2)$ & $16$ & $0$\\
\hline
${}^{3}H_e(A=3, Z=2)$  & $9$ & $1$\\
\hline
$HD(A=3, Z=2)$ & $9$   & $1$\\
\hline
${}^{4}_2H_e(A=4, Z=2)$  &$4$ & $4$\\
\hline
$DD(A=4, Z=2)$  & $4$ & $4$\\
\hline
${}^{12}_{{}6}C(A=12, Z=6)$  & $36$& $36$\\
\hline\hline
\end{tabular}
\caption{The coefficients for transition amplitude and scattering probability of $\nu_e$ and $\nu_{\mu,\tau}$  coherent elastic scattering off different target atoms. The definition of atomic mass is $A=Z+N$, where $Z$ and $N$ are the number of protons and neutron respectively.}
\label{Table001}  
\end{table}%
%%%%%%%%%%%%%%%%%%%%%%%%%%%%%%%%%%%%%%%%%%%%%%%%%%%%%%%%%%%%%%

%%%%%%%%%%%%%%%%%%%%%%%%%%%%%%%%%%%%%%%%%%%%%%%%%%%%%%%%%%%%%%%%%%%%%%%%
%%%%%%%%%%%%%%%%%%%%%%%%%%%%%%%%%%%%%%%%%%%%%%%%%%%%%%%%%%%%%%%%%%%%%%%%
\subsubsection{Mean field potential for neutrino coherent scattering}
When neutrinos are propagating in matter and interacting with the background particles, they can be described by the picture of free neutrinos traveling in an effective potential~[\cite{PhysRevD.17.2369}]. In the following we describe the effective potential between neutrinos and the target atom, and generalize the potential to the case of neutrino coherent scattering with a multi-atom system.

Let us consider a neutrino elastic scattering off an atom which is composed of Z protons, N neutrons and Z electrons. For the elastic neutrino atom scattering, the low-energy neutrinos are scattering off both atomic electrons and the nucleus. Considering the effective low-energy CC and NC interactions, the effective Hamiltonian in current-current interaction form can be written as 
\begin{align}
\label{H_atom}
\mathcal{H}_I^{\mathrm{atom}}&=\mathcal{H}^\mathrm{electron}_I+\mathcal{H}^\mathrm{nucleon}_I=\frac{G_F}{\sqrt{2}}\,\left(j_\mu\,\mathcal{J}^\mu_{\mathrm{electron}}+j_\mu\,\mathcal{J}^\mu_\mathrm{nucleon}\right),
\end{align}
where $\mathcal{J}^\mu_{\mathrm{nucleon}}$ denote the hadronic current for nucleus, $j^\mu$ and $\mathcal{J}^\mu_{\mathrm{electron}}$ are the lepton currents for neutrino and electron respectively. According to the weak interaction theory, the lepton current for neutrino and  electron can be written as
\begin{align}
&j_\mu=\overline{\psi_{\nu}}\,\gamma_\mu\,\left(1-\gamma_5\right)\,\psi_\nu,\\
\label{Current_e}
&\mathcal{J}^\mu_{\mathrm{electron}}=\overline{\psi_{e}}\,\gamma_\mu\,\left(1-\gamma_5\right)\,\psi_e\,\,\,\,\,(\mathrm{W^\pm\,exchange}),\\
&\mathcal{J}^\mu_{\mathrm{electron}}=\overline{\psi_{e}}\,\gamma_\mu\,\left(c_V^e-c_A^e\gamma_5\right)\,\psi_e\,\,\,\,\,(\mathrm{Z^0\,exchange}),
\end{align}
where  $\psi_\nu$ and $\psi_e$ represent the spinor for the neutrino and electron, respectively. From Eq.~(\ref{NC_int}) the coupling coefficient for electrons are $c^e_V=-1/2+2\sin^2\theta_w$ and $c^e_A=-1/2$. The hadronic current for is given by the expression~[\cite{Giunti:2007ry}]
\begin{align}
\label{Current_h}
\mathcal{J}^\mu_\mathrm{nucleon}\equiv\overline{\psi_t}\,\gamma^\mu\left(c^t_V-c^t_A\gamma^5\right)\psi_t,
\end{align}
where subscript $t$ means the target constituents (protons and neutrons). From Eq.~(\ref{NC_int}) the coupling constants for proton(uud) and neutron(udd) are given by
\begin{align}
&c^p_V=\frac{1}{2}-2\sin^2\theta_w,\,\,\,\,c^p_A=\frac{1}{2},\,\,\,\,\,\mathrm{proton}\\
&c^n_V=-\frac{1}{2}\,\,\,\,c^n_A=-\frac{1}{2},\,\,\,\,\,\mathrm{neutron}.
\end{align}

To obtain the effective potential for atom, we need to average the effective Hamiltonian over the electron and nucleon background. For the neutrino-nucleon (proton,neutron) interaction, we only have the neutral current interaction via $Z^0$ boson. However, for the neutrino-electron interaction, we can have charged-current or neutral current interaction depending on the flavor or neutrino. In following, we consider interaction between $\nu_e$ and electrons first which includes both charged and neutral-currents interaction for general discussion.

Considering atomic electrons as a gas of unpolarized electrons with a statistical distribution function $f(E_e)$, the effective potential for neutrino-electron interaction can be obtained by averaging the effective Hamiltonian over the electron background~[\cite{Giunti:2007ry}]
\begin{align}
\langle{\mathcal{H}^\mathrm{electron}_{I}}\rangle&=\frac{G_F}{\sqrt{2}}\int\,\frac{d^3p_e}{(2\pi)^32E_e}\,f(E_e,T)\left[\overline{\psi_\nu}(x)\,\gamma_\mu\left(1-\gamma_5\right)\,\psi_\nu(x)\right]\notag\\&\times\frac{1}{2}\!\sum_{h_e=\pm1}\!\!\langle\,e^-(p_e,h_e)|\overline{\psi_e}\,\gamma^\mu\big((1+c^e_V)\!-\!(1+c^e_A)\gamma_5\big)\,\psi_e|e^-(p_e,h_e)\rangle,
\end{align}
where $h_e$ denotes the helicity of the electron. The average over helicity of the electron matrix element can be calculated with Dirac spinor and gamma matrix traces ~[\cite{Giunti:2007ry}]. Then the average effective Lagrangian can be written as
\begin{align}
\langle{\mathcal{H}^\mathrm{electron}_{I}}\rangle&=\frac{G_F}{\sqrt{2}}(1+c^e_V)\int\,\frac{d^3p_e}{(2\pi)^3}f(E_e)\left[\overline{\psi_\nu}(x)\,\frac{\gamma^\mu{p_e}_\mu}{E_e}\left(1-\gamma_5\right)\,\psi_\nu(x)\right]\notag\\
&=\frac{G_F}{\sqrt{2}}\,(1+c^e_V)\,\left[\int\,\frac{d^3p_e}{(2\pi)^3}f(E_e)\left(\gamma^0-\frac{\vec{\gamma}\cdot\vec{{p}}_e}{E_e}\right)\right]\overline{\psi_\nu}(x)\left(1-\gamma_5\right)\psi_\nu(x)\notag\\
&=\left[\frac{G_F}{\sqrt{2}}\left(1+c^e_V\right)n_{e}\right]\,\overline{\psi_\nu}(x)\gamma^0\left(1-\gamma_5\right)\psi_\nu(x),
\end{align}
where $n_e$ is the number density of the electron. In this case, the effective potential for neutrino-atomic electron interaction can be written as
\begin{align}
V^{\mathrm{electron}}_{I}=\frac{G_F}{\sqrt{2}}\left(1+c^e_V\right)n_{e}=\frac{G_F}{\sqrt{2}}\left(4\sin^2\theta_w+1\right)n_{e}.
\end{align}
The same method can be applied to the neutrino-nuclear interactions. Following the same approach and averaging the effective neutrino-nuclear Hamiltonian over the nuclear background, the effective potential experienced by a neutrino in a background of neutron/proton is given by~[\cite{Giunti:2007ry}] 
\begin{align}
&V_{I}^{\mathrm{proton}}=\frac{G_F}{\sqrt{2}}\left(1-4\sin^2\theta_w\right)n_{p},\qquad V_{I}^{\mathrm{neutron}}=-\frac{G_F}{\sqrt{2}}\,n_{n},
\end{align}
where $n_p$ and $n_n$ represent the number density of proton and neutron.
Combining the neutron and proton potential together, we define the effective nucleon potential experienced by neutrino as 
\begin{align}
V_I^{\mathrm{nucleon}}\equiv-\frac{G_F}{\sqrt{2}}\bigg[1-\left(1-4\sin^2\theta_w\right)\xi\bigg]n_{n},\qquad\xi=n_{p}/n_{n},
\end{align}
where $\xi$ is the ratio between proton and neutron number density.

In our study, we generalize the effective potential to the case of neutrino coherent scattering with multi-atom system, we consider a neutrino coherent forward scatters from a spherical symmetric system which is composed by atoms. In this case, the neutrino scatters off every atom, and it is impossible to identify which scatterer the neutrino interacts with and thus it is necessary to sum over all possible contributions from each atom. In such circumstances, it is appropriate to assume that the number density of electrons and neutrons can be written as
\begin{align}
&n_e=Z_e\,\left(\frac{N_\mathrm{atom}}{V}\right),\,\,\,\,\mathrm{and}\,\,\,\,n_n=N\,\,\left(\frac{N_\mathrm{atom}}{V}\right),
\end{align}
where $N_\mathrm{atom}$ is the number of atoms inside the system, $V$ is the volume of system, $Z$ is the number of electrons, and $N$ is the number of neutrons.
%%%%%%%%%%%%%%%%%%%%%%%%%%%%%%%%%%%%%%%%%%%%%%%%%%%%%%%%%%%%%%%%%%%%%
Then the effective potential is given by
\begin{align}
\label{Potential}
V_{I}&=V_I^{\mathrm{electron}}+V_I^{\mathrm{nucleon}}\notag\\&=\frac{G_F}{\sqrt{2}}\left(\frac{N_\mathrm{atom}}{V}\right)\bigg\{\left(4\sin^2\theta_w\pm1\right)\,Z_e-\bigg[1-\left(1-4\sin^2\theta_w\right)\xi\bigg]\,N\bigg\},
\end{align}
where the $+$ sign is for electron neutrinos $\nu_e$ and the $-$ sign is for muon(tau) neutrinos $\nu_{\mu,\tau}$, separately. 
From Eq.~(\ref{Potential}), it shows that the effective potential depends on the number density of electrons and nucleons contained within the wavelength. 
Thus by increasing the  atoms contained in the wavelength or selecting different atoms as targets, we can enhance the effective potential and may be able to provide a sensitive way to detect the cosmic neutrino background. Beside the detection of cosmic neutrino background, the effective potential for multi-atom can also provide new approaches for studying other aspects of neutrino physics in the future.

%%%%%%%%%%%%%%%%%%%%
%%%%%%%%%%%%%%%%%%%%%%%%%%%%%%%%%%%%%%%%%%%%%%%%%%%%%%%%%%%%%%%%%%%%%%%%
\subsection{Matrix elements of incoherent neutrino scattering}

To determine the freeze-out temperature (chemical/kinetic freeze-out) for a given flavor of neutrinos, we need to know all the elastic and inelastic interaction processes in the early Universe and compare their interaction rate with Hubble expansion rate. In this section we summarize the matrix elements for the neutrino annihilation/production processes and elastic scattering processes which are relevant for investigating neutrino freezeout. These matrix elements serve as one of the fundamental ingredients for solving the Boltzmann equation ~[\cite{Birrell:2014uka}].

%If we assume the neutrinos are so light that they decoupled from the primordial cosmic plasma at temperature $T>m_\nu$ and 
Considering the Universe with temperature $T\approx\mathcal{O}$(MeV), the   particle species in comisc plasma are given by:
\begin{align}
\mathrm{Particle\,\,species\,\, in \,\,plasma:}
\left\{\gamma,\, l^-,\,l^+,\, \nu_e,\, \nu_\mu,\, \nu_\tau,\, \bar{\nu}_e,\, \bar{\nu}_\mu,\, \bar{\nu}_\tau\right\},
\end{align}
 where $l^\pm$ represents the charged leptons. In this case, neutrinos can interact with all these particles via weak interactions and remain in equilibrium. In Table.~\ref{T005} and Table.~\ref{T006} we present the matrix elements $|M|^2$ for different weak interaction processes in the early Universe.

In the calculation of transition amplitude, we use the low energy approximation for $W^\pm$ and $Z^0$ massive propagators, i.e.
\begin{align}
&\mbox{$Z^0$ boson}:\frac{-i\left[g_{\mu\nu}-\frac{q_\mu q_\nu}{M^2_z}\right]}{q^2-M^2_z}\approx\frac{ig_{\mu\nu}}{M^2_z},\quad
&\mbox{$W^\pm$ boson}:\frac{-i\left[g_{\mu\nu}-\frac{q_\mu q_\nu}{M^2_w}\right]}{q^2-M^2_w}\approx\frac{ig_{\mu\nu}}{M^2_w},
\end{align}
and consider the tree-level Feynman diagram contributions only. Then, following the Feynman rules of weak interaction~[\cite{Griffiths:2008zz}], we obtain the matrix elements $|M|^2$ for different  interaction processes.
%%%%%%%%%%%%%%%%%%%%%%%%%%%%%%%%%%%%%%%%%%%%%%%%%%%%%%%%%%%%%
\begin{table}[h]
\centering
\begin{tabular}{lp{8cm}lp{8cm}l}
\hline\hline
Annihilation/Production \\
\hline\hline
Scattering Process & Transition Amplitude $|M|^2$ \\
\hline
$l^-+l^+\longrightarrow\nu_l+\bar{\nu}_l$ &$ 32G^2_F\bigg[\left(1+2\sin^2\theta_w\right)^2\left(p_1\cdot p_4\right)\left(p_2\cdot p_3\right)$

$+\left(2\sin^2\theta_w\right)^2\left(p_1\cdot p_3\right)\left(p_2\cdot p_4\right)$

$+2\sin^2\theta_w\left(1+2\sin^2\theta_w\right)m^2_l\left(p_3\cdot p_4\right)\bigg]$ \\
\hline
$l^{\prime-}+l^{\prime+}\longrightarrow\nu_l+\bar{\nu}_l$ & $32G^2_F\bigg[\left(1-2\sin^2\theta_w\right)^2\left(p_1\cdot p_4\right)\left(p_2\cdot p_3\right)$

$+\left(2\sin^2\theta_w\right)^2\left(p_1\cdot p_3\right)\left(p_2\cdot p_4\right)$

$-2\sin^2\theta_w\left(1-2\sin^2\theta_w\right)m^2_{l^\prime}\left(p_3\cdot p_4\right)\bigg]$ \\
\hline
$\nu_l+\bar{\nu}_l\longrightarrow\nu_l+\bar{\nu}_l$ &
$32G^2_F\bigg[\left(p_1\cdot p_4\right)\left(p_2\cdot p_3\right)\bigg]$ \\
\hline
$\nu_{l^\prime}+\bar{\nu}_{l^\prime}\longrightarrow\nu_l+\bar{\nu}_l$ &
$32G^2_F\bigg[\left(p_1\cdot p_4\right)\left(p_2\cdot p_3\right)\bigg]$ \\
\hline\hline
\end{tabular}
\caption{The transition amplitude for different annihilation and production processes. The definition of particle number is given by $1+2\leftrightarrow3+4$, where $l,\,l^\prime=e,\,\mu,\,\tau\,(l\neq\,l^\prime)$.}
\label{T005}
\end{table}
%%%%%%%%%%%%%%%%%%%%%%%%%%%%%%%%%%%%%%%%%%%%%%%%%%%
%%%%%%%%%%%%%%%%%%%%%%%%%%%%%%%%%%%%%%%%%%%%%%%%%%%%%%%%%%%%%
\begin{table}[h]
\centering
\begin{tabular}{lp{8cm}lp{8cm}l}
\hline\hline
Elastic Scattering Process ($\nu_e$) \\
\hline\hline
Scattering Process & Transition Amplitude $|M|^2$\\
\hline
$\nu_l+l^-\longrightarrow\nu_l+l^-$ & 
$ 32G^2_F\bigg[
   \left(1+2\sin^2\theta_w\right)^2\left(p_1\cdot p_2\right)\left(p_3\cdot p_4\right)$
   
   $+\left(2\sin^2\theta_w\right)^2\left(p_1\cdot p_4\right)\left(p_2\cdot p_3\right)$
   
   $-2\sin^2\theta_w\left(1+2\sin^2\theta_w\right)m^2_l\left(p_1\cdot p_3\right)\bigg]$ \\
\hline
$\nu_l+l^+\longrightarrow\nu_l+l^+$ &
$ 32G^2_F\bigg[
   \left(1+2\sin^2\theta_w\right)^2\left(p_1\cdot p_4\right)\left(p_2\cdot p_3\right)$
   
   $+\left(2\sin^2\theta_w\right)^2\left(p_1\cdot p_2\right)\left(p_3\cdot p_4\right)$
   
   $-2\sin^2\theta_w\left(1+2\sin^2\theta_w\right)m^2_l\left(p_1\cdot p_3\right)\bigg]$ \\
\hline
$\nu_l+l^{\prime-}\longrightarrow\nu_l+l^{\prime-}$ &
$ 32G^2_F\bigg[
   \left(1-2\sin^2\theta_w\right)^2\left(p_1\cdot p_2\right)\left(p_3\cdot p_4\right)$
   
  $+\left(2\sin^2\theta_w\right)^2\left(p_1\cdot p_4\right)\left(p_2\cdot p_3\right)$
  
  $+2\sin^2\theta_w\left(1-2\sin^2\theta_w\right)m^2_{l^\prime}\left(p_1\cdot p_3\right)\bigg]$ \\
\hline
$\nu_l+l^{\prime+}\longrightarrow\nu_l+l^{\prime+}$ &
$ 32G^2_F\bigg[
   \left(1-2\sin^2\theta_w\right)^2\left(p_1\cdot p_4\right)\left(p_2\cdot p_3\right)$
   
   $+\left(2\sin^2\theta_w\right)^2\left(p_1\cdot p_2\right)\left(p_3\cdot p_4\right)$
   
   $+2\sin^2\theta_w\left(1-2\sin^2\theta_w\right)m^2_{l^\prime}\left(p_1\cdot p_3\right)\bigg]$ \\
\hline
$\nu_l+\nu_l\longrightarrow\nu_l+\nu_l$ &
$\frac{1}{2!}\frac{1}{2!}\times32G^2_F\bigg[4\left(p_1\cdot p_2\right)\left(p_3\cdot p_4\right)\bigg]$ \\
\hline
$\nu_l+\bar{\nu}_l\longrightarrow\nu_l+\bar{\nu}_l$ &
$32G^2_F\bigg[4\left(p_1\cdot p_4\right)\left(p_2\cdot p_3\right)\bigg]$ \\
\hline
$\nu_l+\nu_{l^\prime}\longrightarrow\nu_l+\nu_{l^\prime}$ &
$32G^2_F\bigg[\left(p_1\cdot p_2\right)\left(p_3\cdot p_4\right)\bigg]$ \\
\hline
$\nu_l+\bar{\nu}_{l^\prime}\longrightarrow\nu_l+\bar{\nu}_{l^\prime}$ &
$32G^2_F\bigg[\left(p_1\cdot p_4\right)\left(p_2\cdot p_3\right)\bigg]$ \\
\hline\hline
\end{tabular}
\caption{The transition amplitude for different elastic scattering processes. The definition of particle number is given by $1+2\leftrightarrow3+4$, where $l,\,l^\prime=e,\,\mu,\,\tau\,(l\neq\,l^\prime)$.}
\label{T006}
\end{table}

\clearpage

%~~~~~~~~~~~~~~~~~~~~~~~~~~~~~~~~~~~~~~~~~~~~~~~~~

\section{Neutrinos in the early Universe }
%In this section we will focus on the following:
%\begin{itemize}
%    \item Neutrino decoupling in early Universe: standard scenarios
%    \item Overview of neutrino freeze-out and free streaming
%    \item After freeze-out extra neutrinos from microscope process
%    \item Lepton number and effective number of neutrinos
%\end{itemize}
In the early Universe the neutrinos are kept in equilibrium with cosmic plasma via the weak interaction processes. However, as the Universe expanded, these weak interactions gradually became too slow to maintain equilibrium, then the neutrinos ceased interacting and decoupled from the cosmic background. These relic neutrino background carries great information about our early Universe which can provide the information when our Universe is about $1$ sec old. 

Today the Universe is believed to be filled with relic cosmic neutrinos. Although the cosmic neutrino background is hard to detect, we can use entropy conservation to infer their temperature and find that they should have the temperature $T_\nu^0=1.9\,\mathrm{K}$ in the present epoch for  massless neutrinos. However, from the neutrino oscillation experiment, we know that the the neutrinos are not massless particles. 

The square mass difference $\Delta m^2_{ij}$ has been experimentally measured, from the neutrino oscillation experiment~[\cite{ParticleDataGroup:2022pth}]:
\begin{align}
&\Delta{m}_{21}^2=7.39^{+0.21}_{-0.20}\times10^{-5}\,\mathrm{eV}^2,\\
&\Delta{m}_{32}^2=2.45^{+0.03}_{-0.03}\times10^{-3}\,\mathrm{eV}^2.
\end{align}
and neutrino mass eigenvalue can be ordered in the normal mass hierarchy ($m_1\ll m_2<m_3$) or inverted mass hierarchy ($m_3\ll m_1<m_2$). All three mass states remained relativistic until the temperature dropped below their rest mass. These results allow for the possibility that one mass eigenstate or two mass eigenstates of neutrinos may become non-relativistic today. 

\subsection{Overview of neutrino freeze-out in the early Universe}

The properties of the neutrino background are influenced by the details of the freeze-out or decoupling process at a temperature $T=\mathcal{O}(2\mathrm{MeV})$. In the literature one finds estimates of freeze-out temperatures based on a comparison of Hubble expansion with neutrino scattering length and considering only number changing (i.e. chemical) processes. In the paper~[\cite{Birrell:2014uka}], we employ a similar definition of freeze-out temperature in the context of the Boltzmann equation and refine the results by noting that there are three different freeze-out processes for neutrinos:

%In general, the particle freezeout process include both chemical and kinetic which lead to particle become free-streaming in the early Universe (for detail discussion see Chapter~\ref{Introduction}), we have:

1. Neutrino chemical freeze-out: the temperature at which neutrino number changing processes such as $e^-e^+\to\nu\overline\nu$ effectively cease. After chemical freeze-out, there are no reactions that, in a noteworthy fashion, can
change the neutrino abundance and so particle number is conserved. %Prior to the chemical freezeout temperature, number changing processes are significant and keep the particle in chemical (and thermal) equilibrium, implying the distribution function of neutrino has the Fermi-Dirac form:
%\begin{equation}\label{equilibrium}
%f_{c}(t,E)=\frac{1}{\exp(E/T)+1}, \qquad\text{ for } T> T_{ch}.
%\end{equation}

2. Neutrino kinetic freeze-out: the temperature at which the neutrino momentum exchanging interactions such as $e^\pm\nu\to e^\pm\nu$ are no longer occurring rapidly enough to maintain an equilibrium momentum distribution. %When $T_k<T(t)<T_{ch}$, the number-changing process no longer occurs rapidly enough to keep the distribution in chemical equilibrium but there is still sufficient momentum exchange to keep the distribution in thermal equilibrium. The distribution function is therefore obtained by maximizing entropy, with fixed energy, particle number, and antiparticle number separately. This implies that the distribution function has the form
%\begin{equation}\label{kinetic_equilib}
%f_k(t,E)=\frac{1}{\Upsilon^{-1}\exp(E/T)+1},\qquad \text{ for }T_k< T< T_{ch}.
%\end{equation}
%The time dependent generalized fugacity $\Upsilon(t)$ controls the occupancy of phase space and is necessary once $T(t)<T_{ch}$ in order to conserve particle number.

3. Collisions between neutrinos:
Those collisions $\nu\nu\to\nu\nu$ are capable of reequilibrating energy within and between neutrino flavor families. These processes end at a yet lower temperature and the neutrinos will be truly free-streaming from that point on.
 
%3. Free streaming: for $T<T_k$ there are no longer any significant interactions that couple the particle species of interest and so they begin to free-stream through the Universe. The Einstein-Vlasov equation can be solved~[\cite{choquet2008general,Birrell:2012gg}], to yield the free-streaming momentum distribution
%\begin{equation}\label{free_stream_dist}
%f(t,E)=\frac{1}{\Upsilon^{-1}e^{\sqrt{p^2/T_{fs}^2+m_\nu^2 /T_k^2}}+ 1},\qquad T_{fs}=\frac{T_ka(t_k)}{a(t)}
%\end{equation}
%where the free-streaming effective temperature $T_{fs}$ is obtained by redshifting the temperature at kinetic freeze-out, and $m_\nu$ is the mass of neutrino.

To estimate the freeze-out temperature, we need to solve the Boltzmann equation with different types of collision terms with the transition matrices from Table.~\ref{T005} and Table.~\ref{T006}. The paper~[\cite{Birrell:2014uka} ] developed a new method for analytically simplifying the collision integrals and showing that the neutrino freeze-out temperature is controlled by standard model (SM) parameters. The freeze-out temperature depends only on the magnitude of the Weinberg angle in the form $\sin^2\theta_W$ , and a dimensionless relative interaction strength parameter $\eta$,
\begin{align}
\eta\equiv M_p m_e^3 G_F^2, \qquad M_p^2\equiv \frac{1}{8\pi G_N}, \end{align}
a combination of the electron mass $m_e$, Newton constant $G_N$ (expressed above in terms of Planck mass $M_p$), and the Fermi constant $G_F$. The dimensionless interaction strength parameter $\eta$ in the present-day vacuum has the value
\begin{align}
\eta_0\equiv \left.M_p m_e^3 G_F^2\right|_0 = 0.04421 .
\end{align}

\begin{figure}[ht]
\centerline{\includegraphics[width=0.47\columnwidth]{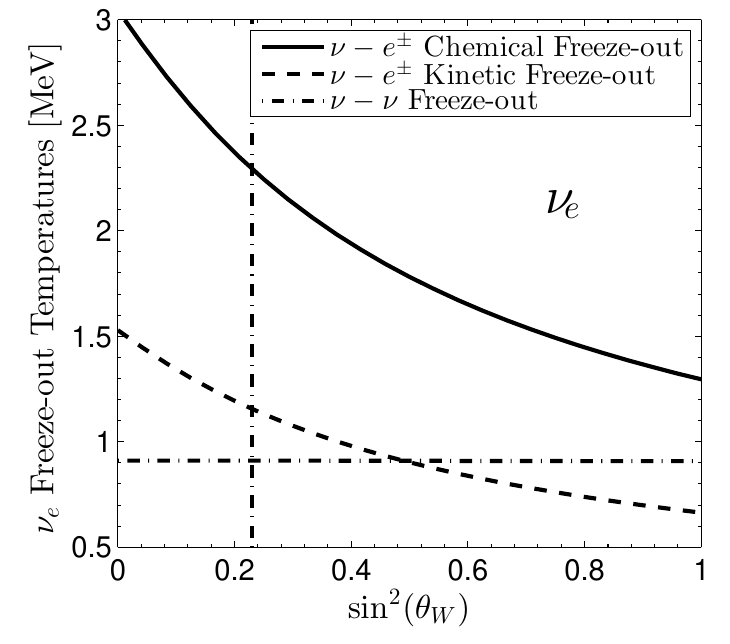}
\hspace{1mm}\includegraphics[width=0.47\columnwidth]{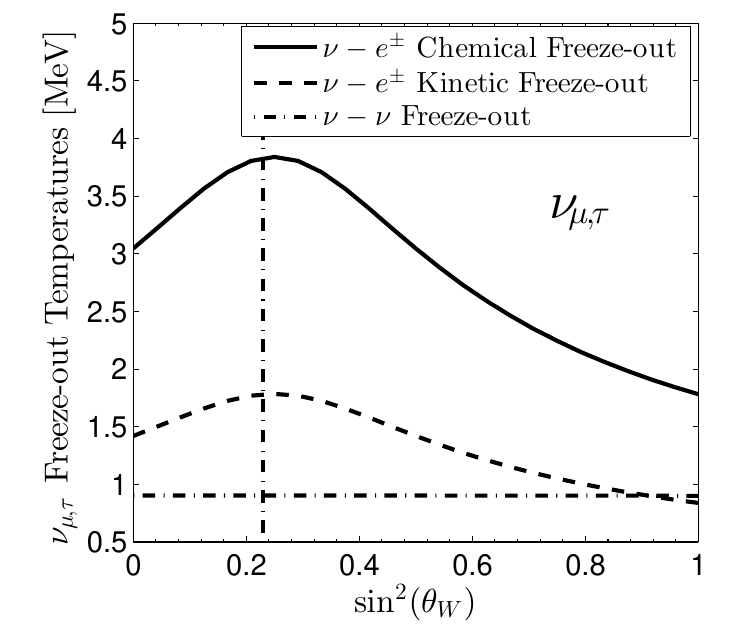}}
\centerline{\includegraphics[width=0.47\columnwidth]{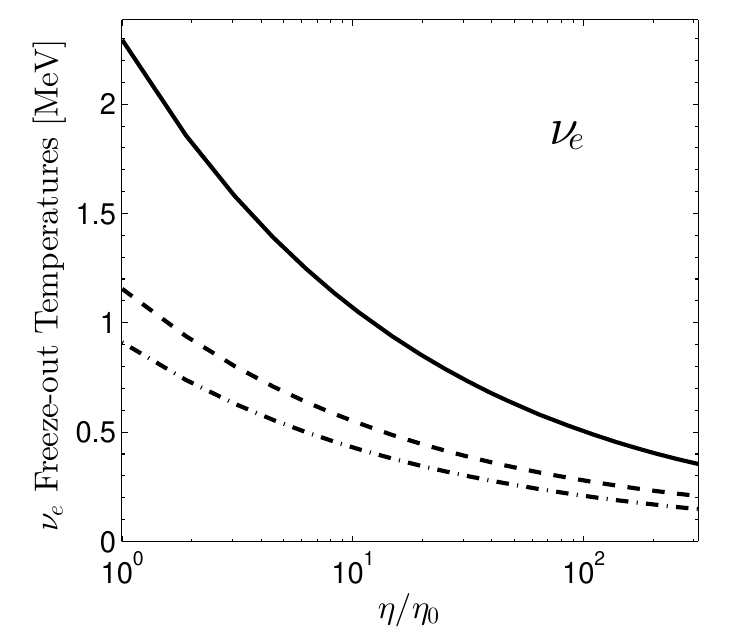}
\hspace{1mm}\includegraphics[width=0.47\columnwidth]{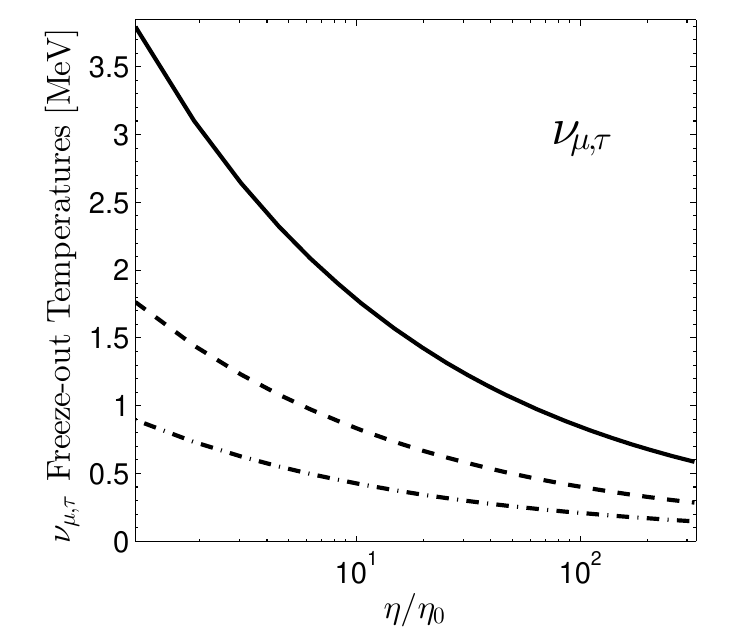}}
\caption{Freeze-out temperatures for electron neutrinos (left) and $\mu$, $\tau$ neutrinos (right) for the three types of freeze-out processes adapted from paper [\cite{Birrell:2014uka}]. Top panels print temperature curves as a function of $\sin^2\theta_W$ for $\eta=\eta_0$, the vertical dashed line is $\sin^2\theta_W=0.23$; bottom panels are printed as a function of relative change in interaction strength $\eta/\eta_0$ obtained for $\sin^2\theta_W=0.23$.}
\label{fig:freezeoutT}
 \end{figure}

The magnitude of $\sin^2\theta_W$ is not fixed within the SM and could be subject to variation as a function of time or temperature. In Fig.~\ref{fig:freezeoutT} we show the dependence of neutrino freeze-out temperatures for $\nu_e$ and $\nu_{\mu,\tau}$ on SM model parameters $\sin^2\theta_W$ and $\eta$ in detail. The impact of SM parameter values on neutrino freeze-out and the discussion of the implications and connections of this work to other areas of physics, namely Big Bang Nucleosynthesis and dark radiation can be found in detail in~[\cite{Dreiner:2011fp,Boehm:2012gr,Blennow:2012de,Birrell:2014uka}]. A comprehensive investigation of neutrino freezeout, and a novel approach to analytically simplify the collision integrals for the Boltzmann equation can be found in Dr. Jeremiah Birrell‘s PhD thesis~[\cite{Birrell:2014ona}].

%We expect that incorporating oscillations into the freeze-out calculation would yield a smaller freeze-out temperature difference between neutrino flavors as oscillation provides a mechanism in which the heavier flavors remain thermally active despite their direct production becoming suppressed. In work by Mangano et. al.~[\cite{Mangano:2005cc}], neutrino freeze-out including flavor oscillations is shown to be a negligible effect.
%\clearpage
%~~~~~~~~~~~~~~~~~~~~~~~~~~~~~~~~~~~~~~~~~~~~~~~~~
%~~~~~~~~~~~~~~~~~~~~~~~~~~~~~~~~~~~~~~~~~~~~~
\clearpage

%%%%%%%%%%%%%%%%%%%
\subsection{Lepton number and effective number of neutrinos}

Neutrinos decoupled from the cosmic plasma in the early Universe at a temperature of $T=\mathcal{O}(2\mathrm{MeV})$ and became free-streaming. However, after freezeout neutrinos still continue to play a significant role in the evolution of the Universe and have a huge impact on cosmological observations such as Big Bang Nucleosynthesis (BBN), the Cosmic Microwave Background (CMB), and the matter spectrum for large scale structure. This is due to the sensitivity of the Hubble parameter to the total energy density in the Universe. Besides photons, neutrinos are the most abundant species and contribute significantly to the relativistic energy density throughout the early Universe, affecting the Hubble expansion rate significantly. 

The contribution of energy density from the neutrino sector can be described by the effective number of neutrinos $N_{\nu}^{\mathrm{eff}}$, which captures the number of relativistic degrees of freedom for neutrinos as well as any reheating that occurred in the sector after freeze-out. The effective number of neutrino is defined as 
\begin{align}\label{Neff}
N_\nu^{\mathrm{eff}}\equiv\frac{\rho^{\mathrm{tot}}_\nu}{\frac{7\pi^2}{120}\left(\frac{4}{11}\right)^{4/3}T_\gamma^4}\;,
\end{align}
where $\rho_\nu^{\mathrm{tot}}$ is the total energy density in neutrinos and $T_\gamma$ is the photon temperature. $N_\nu^{\mathrm{eff}}$ is defined such that three neutrino flavors with zero participation of neutrinos in reheating during $e^\pm$ annihilation results in $N_\nu^{\mathrm{eff}}=3$. The factor of $\left(4/11\right)^{1/3}$ relates the photon temperature to the free-streaming neutrinos temperature, under the assumption of zero neutrino reheating after $e^\pm$ annihilation. The currently accepted theoretical value is $N_\nu^{\mathrm{eff}}=3.046$, after including the slight effect of neutrino reheating [\cite{Mangano:2005cc,Birrell:2014uka}]. The favored value of $N_\nu^{\mathrm{eff}}$ can be found by fitting to CMB data. In 2013 the Planck collaboration found $N_\nu^{\mathrm{eff}}=3.36\pm0.34$ (CMB only) and $N_\nu^{\mathrm{eff}}= 3.62\pm0.25$ (CMB and $H_0$)~[\cite{Planck:2013pxb}].

To explain the experimental value of $N_\nu^{\mathrm{eff}}$, many studies aim to improve the calculation of neutrino decoupling in the early Universe, including exploring the dependence of freeze-out on natural constants~[\cite{Birrell:2014uka}], the entropy transfer from $e^\pm$ annihilation and finite temperature correction~[\cite{Dicus:1982bz,Heckler:1994tv,Fornengo:1997wa}], neutrino decoupling with flavor oscillations~[\cite{Mangano:2001iu,Mangano:2005cc}], and investigating nonstandard neutrino interactions [\cite{Morgan:1981zy,Fukugita:1987uy,Elmfors:1997tt,Vogel:1989iv,Mangano:2006ar,Giunti:2008ve,Mangano:2006ar}].% However, the effective number of neutrino $N_\nu^{\mathrm{eff}}$ can be a consequence of fundamental physics principle.

The standard cosmological model assumes that the lepton asymmetry $L\equiv  [N_\mathrm{L}-N_{\overline{\mathrm{L}}}] /N_\gamma $  (normalized with the photon number) 
between leptons and anti-leptons is small, similar to the baryon asymmetry $B=[N_\mathrm{B}-N_{\overline{\mathrm{B}}}]/N_\gamma $; most often it is assumed $L=B$. Barenboim, Kinney, and Park~[\cite{Barenboim:2016shh,Barenboim:2017dfq}] noted that the lepton asymmetry of the Universe is one of the most weakly constrained parameters is cosmology and they propose that models with leptogenesis are able to accommodate a large lepton number asymmetry surviving up to today.  Moreover, the discrepancy between $H_\mathrm{CMB}$ and $H_0$ has increased~[\cite{riess2018new,Riess:2018byc,Planck:2018vyg}]. The Hubble tension and the possibility that leptogenesis in the early Universe resulted in neutrino asymmetry motivate our study of the dependence of $N_\nu^{\mathrm{eff}}$ on lepton asymmetry, $L$. In our work~[\cite{Yang:2018oqg}] we consider $L\simeq 1$ and explore how this large cosmological lepton yield relates to the effective number of (Dirac) neutrinos $N^{\mathrm{eff}}_\nu$. 

\subsubsection{Relation between $N_\nu^{\mathrm{eff}}$ and neutrino chemical potential}
We consider now neutrinos decouple~[\cite{Birrell:2014gea}] at a temperature of $T_f\simeq 2\,\mathrm{MeV}$ and are subsequently free-streaming. Assuming exact thermal equilibrium at the time of decoupling, the neutrino distribution can be subsequently written as (see~[\cite{Birrell:2012gg}] and references therein)
\begin{align}
\label{fnudef}
&f_\nu=\frac{1}{\exp{\left(\sqrt{\frac{E^2-m_\nu^2}{T_\nu^2}+\frac{m^2_\nu}{T^2_f}}-\sigma\frac{\mu_\nu}{T_f}\right)+1}}\;,\qquad T_\nu\equiv\frac{a(t_f)}{a(t)}T_f,
\end{align}
where $\sigma=+1(-1)$ denotes particles (antiparticles) and we define the effective neutrino temperature $T_\nu$  by the red-shifting of momentum in the comoving volume element of the Universe.

\begin{figure}[t]
\begin{center}
\includegraphics[width=\textwidth]{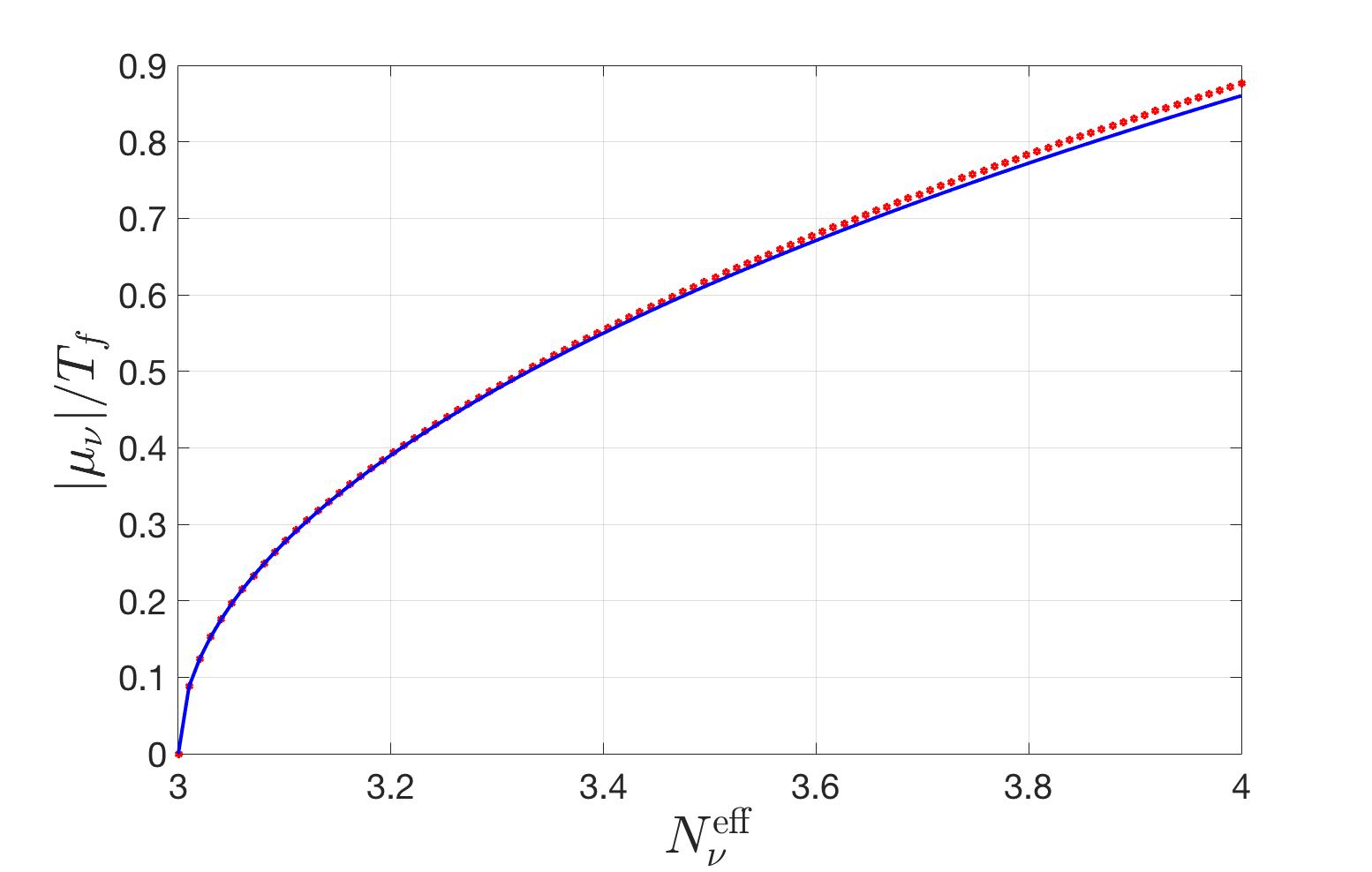}
\caption{The free-streaming neutrino chemical potential $|\mu_\nu|/T_f$ as a function of the effective number of neutrinos $N_\nu^{\mathrm{eff}}$. The solid (blue) line is the exact solution and the (red) dashed line is the approximate solution neglecting the $(\mu_\nu/T_f)^4$ term; the maximum difference in the domain shown is about $2\%$.}
\label{Chemical_Potential_Neff}
\end{center}
\end{figure}

Since the freeze-out temperature $T_f\gg m_\nu$ and also neutrino temperature $T_\nu\gg m_\nu$ in the domain of our analysis, we consider the massless limit in Eq.\;(\ref{fnudef}). Under this approximation, the total neutrino energy density can be written as
\begin{align}
\label{Energy_Density}
\rho_\nu^{\mathrm{tot}}
&=\frac{g_\nu\,T_\nu^4}{2\pi^2}\left[\frac{7\pi^4}{60}+\frac{\pi^2}{2}\left(\frac{\mu_\nu}{T_f}\right)^{\!\!2}+\frac{1}{4}\left(\frac{\mu_\nu}{T_f}\right)^{\!\!4}\right].
\end{align}
Substituting Eq.\;(\ref{Energy_Density}) into the definition of the effective number of neutrinos Eq.~(\ref{Neff}), we obtain 
\begin{align}
\label{Neff_002}
N_\nu^{\mathrm{eff}}\!\!
=\!3\!\left(\frac{11}{4}\right)^{\!\!\frac{4}{3}}\!\!\left(\frac{T_\nu}{T_\gamma}\right)^{\!\!4}\!
\left[1\!+\!\frac{30}{7\pi^2}\!\!\left(\frac{\mu_\nu}{T_f}\right)^{\!\!2} 
\!\!+\frac{15}{7\pi^4}\!\!\left(\frac{\mu_\nu}{T_f}\right)^{\!\!4}\right].
\end{align}
From Eq.\;(\ref{Neff_002}) we have for the standard photon reheating ratio $T_\nu/T_\gamma=(4/11)^{1/3}$ [\cite{Kolb:1990vq}] and degeneracy $g_\nu=3$ (flavor), the relation between the effective number of neutrinos and the chemical potential at freezeout
\begin{align}
\label{Neff_Potential}
N_\nu^{\mathrm{eff}}=3\left[1+\frac{30}{7\pi^2}\left(\frac{\mu_\nu}{T_f}\right)^{\!\!2}+ \frac{15}{7\pi^4} \left(\frac{\mu_\nu}{T_f}\right)^{\!\!4}\right].
\end{align}
To solve the neutrino chemical potential $\mu_\nu/T_f$ as a function of the effective number of neutrinos, we can neglect the $(\mu_\nu/T_f)^4$ term in Eq.\;(\ref{Neff_Potential}) because $m_\nu\ll T_f$ and obtain
\begin{align}\label{Solution}
\frac{\mu_\nu}{T_f}=\pm\sqrt{\frac{7\pi^2}{30}\left(\frac{N_\nu^{\mathrm{eff}}}{3}-1\right)}.
%=&\pm \pi\sqrt{\sqrt{1+\frac{7}{15}\left(\frac{N_\nu^{\mathrm{eff}}}{3}-1\right)}-1}%\approx
\end{align}
In Fig.\;\ref{Chemical_Potential_Neff} we plot the free-streaming neutrino chemical potential $|\mu_\nu|/T_f$ as a function of the effective number of neutrinos $N_\nu^{\mathrm{eff}}$. For comparison, the solid (blue) line is the exact solution of $|\mu_\nu|/T_f$ by solving Eq.~(\ref{{Neff_Potential}}) numerically, and the (red) dashed line is the approximate solution Eq.~(\ref{Solution}) by neglecting the $(\mu_\nu/T_f)^4$ in calculation. In the parameter range of interest, we show that the term $(\mu_\nu/T_f)^4$ only contributes $\approx 2\%$ to the calculation and henceforth we neglect it, and use the approximation Eq.\;(\ref{Solution}). 

The SM value of the effective number of neutrinos, $N_\nu^{\mathrm{eff}}=3$, is obtained under the assumption that the neutrino chemical potentials are not essential, {\it i.e.\/}, $\mu_\nu\ll T_f$. From Fig.\;\ref{Chemical_Potential_Neff}, to interpret the literature values $N_\nu^{\mathrm{eff}}=3.36\pm0.34$ (CMB only) and $N_\nu^{\mathrm{eff}}= 3.62\pm0.25$ (CMB and $H_0$), we require $0.52\leqslant\mu_\nu/T_f\leqslant0.69$. These values suggest  a possible neutrino-antineutrino asymmetry at freezeout, {\it i.e.\/} a difference between the number densities of neutrinos and antineutrinos.
%%%%%%%%%%%%%%%%%%%%%%%%%%%%%%%%%%%%%%%%%%%%%%%%%%%%%%%%%%%%%%%%%%
%%%%%%%%%%%%%%%%%%%%%%%%%%%%%%%%%%%%%%%%%%%%%%%%%%%%%%%%%%%%%%%%%%%

\subsubsection{Dependence of $N_\nu^{\mathrm{eff}}$ on lepton asymmetry}
We now obtain the relation between neutrino chemical potential and the baryon to lepton ratio. Let us consider the neutrino freezeout temperature $T_f\simeq 2.0$ MeV; here we treat neutrino freezeout as occurring instantaneously and prior to $e^\pm$ annihilation (implying zero neutrino reheating). Comoving lepton (and baryon) number is conserved after the epoch of leptogenesis (baryogenesis, respectively) which precedes the epoch  under consideration in this work ($T\lesssim 2$\;MeV). %However, the photon number $N_\gamma$ changes due to reheating, and accommodates the large number of photons originating from $e^+e^-$-annihilation.% We present results in terms of the current epoch photon number. 

The lepton-density asymmetry $\ell $ at neutrino freeze-out can be written as
\begin{align}
\ell_f \equiv\big(n_e-n_{\overline{e}}\big)_f+\sum_{i=e,\mu, \tau}\big(n_{\nu_i}-n_{\overline{\nu}_i}\big)_f,
\end{align}
where we use the subscript $f$ to indicate that the quantities should be evaluated at the neutrino freeze-out temperature. As a first approximation, here we assume that all neutrinos freeze-out at the same temperature and their chemical potentials are the same; {\it i.e.\/},
\begin{align}
\mu_\nu=\mu_{\nu_e}=\mu_{\nu_\mu}=\mu_{\nu_\tau}.
\end{align}
Furthermore, neutrino oscillation implies that neutrino number is freely exchanged between flavors; {\it i.e.\/}, $\nu_e\rightleftharpoons\nu_\mu\rightleftharpoons\nu_\tau$, and we can assume that all neutrino flavors share the same population. Under these assumptions, the lepton-density asymmetry can be written as
\begin{align}
\label{L_asymmetry} 
\ell_f=\big(n_e-n_{\overline{e}}\big)_f+\big(n_{\nu}-n_{\overline{\nu}}\big)_f,
\end{align}
where the three flavors are accounted for by taking the degeneracy $g_\nu=3$ in the last term. The difference in yield of neutrinos and antineutrinos can be written as
\begin{align}
\label{Excess_Neutrino}
\left(n_\nu-n_{\overline{\nu}}\right)_f=\frac{g_\nu}{6\pi^2}T^3_f\bigg[\pi^2\left(\frac{\mu_\nu}{T_f}\right)+\left(\frac{\mu_\nu}{T_f}\right)^{\!\!3}\bigg].
\end{align}

On the other hand, the baryon-density asymmetry $b$ at neutrino freezeout is given by
\begin{align}
\label{B_asymmetry}
b_f \equiv\big(n_p-n_{\overline{p}}\big)_f+\big(n_n-n_{\overline{n}}\big)_f \approx \big(n_p+n_n\big)_f,
\end{align}
where $n_{\overline{n}}$ and $n_{\overline{p}}$ are negligible in the temperature range we consider here. Taking the ratio $\ell_f/b_f$, using charge neutrality, and introducing the entropy density we obtain
\begin{align}\label{Lf_Bf}
\left(\frac{\ell_f}{b_f}\right)  
\approx\left(\frac{n_p}{n_B} \right)_f+\left(n_{\nu}-n_{\overline{\nu}}\right)_f \left(\frac{s}{n_B}\right)_f \frac{1}{s_f},\qquad n_B=(n_p+n_n),
\end{align}
where we introduce the notation $n_B$ for the baryon number density. The proton concentration at neutrino freeze-out is given by
\begin{align}
\label{X_proton}
\left(\frac{n_p}{n_B}\right)_f&=\frac{1}{1+(n_n/n_p)_f}=\frac{1}{1+\exp{\big[-\left(Q+\mu_\nu\right)/T_f\big]}},
\end{align}
with $Q=m_n-m_p=1.293\,\mathrm{MeV}$. We neglect the electron chemical potential in the last step because the $e^\pm$ asymmetry is determined by the proton density, and at energies of order a few MeV, the proton density is small, {\it i.e.\/}, $\mu_e\ll T_f$. 

However, as we will see, for our study of $N_\nu^{\mathrm{eff}}$ we will be interested in the case of a large lepton-to-baryon ratio. From Eq.\;(\ref{X_proton}) it is apparent that this can only be achieved through the second term in Eq.\;(\ref{Lf_Bf}), with the first term then being negligible, as it is smaller than $1$. So we further approximate
\begin{align}\label{L_B_ratio}
\left(\frac{\ell_f}{b_f}\right)  
\approx\left(n_{\nu}-n_{\overline{\nu}}\right)_f \left(\frac{s}{n_B}\right)_f \frac{1}{s_f}.
\end{align}
We retained the full expression Eq.\;(\ref{X_proton}) in our above discussion to show that the presence of a chemical potential $\mu_\nu\simeq 0.2\,Q$ could lead to small, perhaps noticeable, effects on pre-BBN proton and neutron abundance. We defer this unrelated discussion to a separate future work. Note that for large $|\mu_\nu|$, Eq.\;(\ref{L_B_ratio}) implies that the signs of $\mu_\nu$ and $\ell_f$ are the same. However, for very small $\mu_\nu$ the sign of $\ell_f$ is determined by the interplay between (anti)electrons and (anti)neutrinos; {\it i.e.\/}, there is competition between the two terms in Eq.\;(\ref{L_asymmetry}).

In general, the total entropy density at freeze-out can be written
\begin{align}
\label{Entropy_density}
s_f=\frac{2\pi^2}{45}g^s_\ast(T_f)\,T_f^3,
\end{align}
where the $g^s_\ast$ counts the degree of freedom for relativistic particles~[\cite{Kolb:1990vq}]. At $T_f\simeq 2\mathrm{MeV}$, the relativistic species in the early Universe are photons, electron/positrons, and $3$ neutrino species. We have
\begin{align}
g^s_{\ast}&= g_\gamma+\frac{7}{8}\,g_{e^\pm}+\frac{7}{8}\,g_{\nu\bar{\nu}}\left(\frac{T_\nu}{T_\gamma}\right)^{\!\!3}\bigg[1+\frac{15}{7\pi^2}\left(\frac{\mu_\nu}{T_f}\right)^{\!\!2}\bigg]=10.75+\frac{45}{4\pi^2}\left(\frac{\mu_\nu}{T_f}\right)^{\!\!2}\;,
\end{align}
where the degrees of freedom are given by $g_\gamma=2$, $g_{e^\pm}=4$, and $g_{\nu\bar{\nu}}=6$, and we have $T_\nu=T_\gamma=T_f$ at neutrino freeze-out.

%%%%%%%%%%%%%%%%%%%%%%%%%%%%%%%%%
\begin{figure}[h]
\begin{center}
\includegraphics[width=\textwidth]{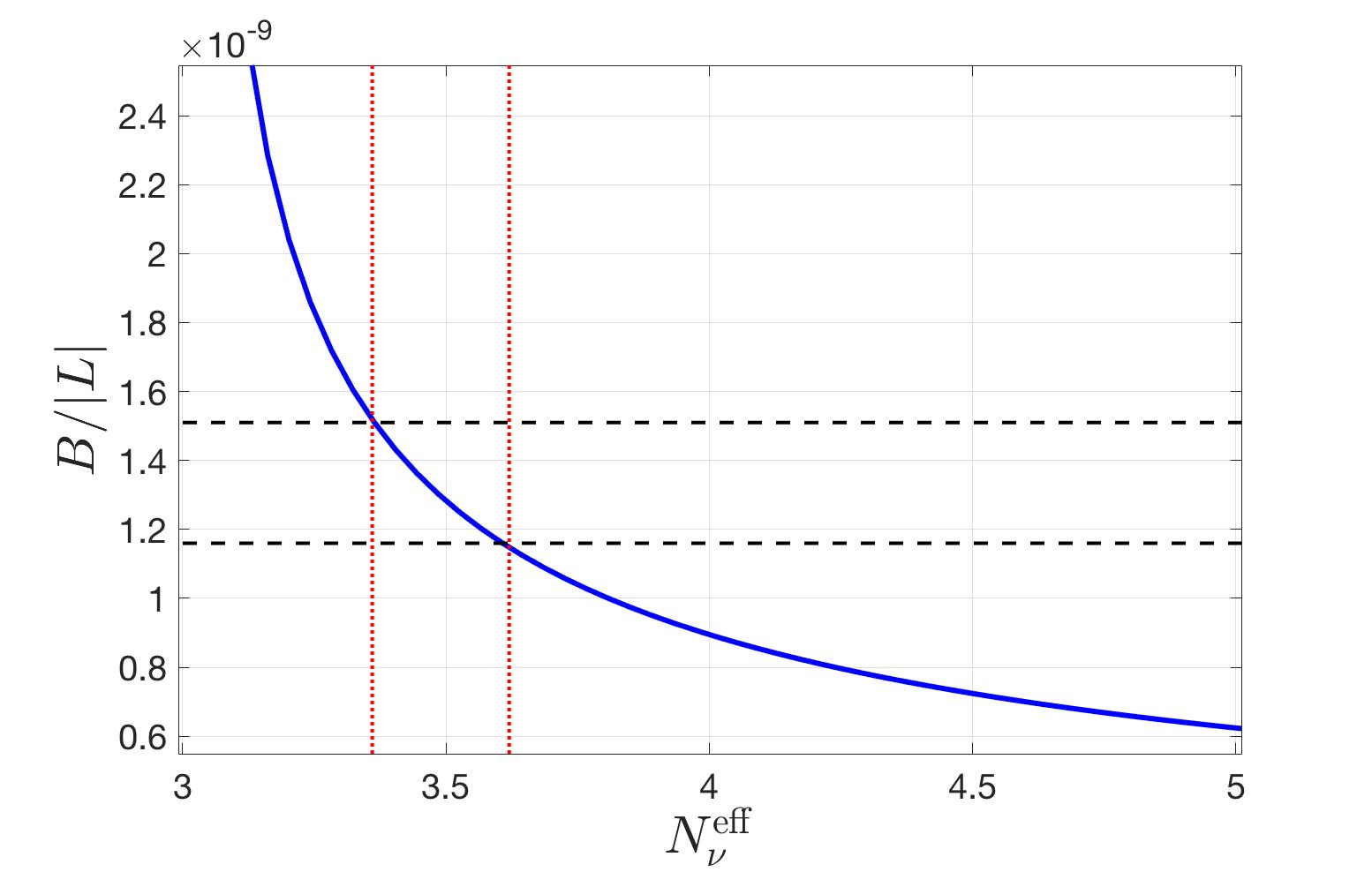}
\caption{The ratio $B/|L|$ between the net baryon number and the net lepton number as a function of $N^{\mathrm{eff}}_\nu$: The solid blue line shows $B/|L|$. The vertical (red) dotted lines represent the values $3.36\leqslant N_\nu^{\mathrm{eff}}\leqslant3.62$, which correspond to $1.16 \times 10^{-9}\leqslant B/|L|\leqslant 1.51 \times 10^{-9}$ (horizontal dashed lines).}
\label{BL_Ratio}
\end{center}
\end{figure}
%%%%%%%%%%%%%%%%%%%%%%%%%%%%%%%%%%%%%%%%%%%%%%%%%%%%%%%%%%%%%%%%%%

Finally, since the entropy-per-baryon from neutrino freeze-out up to the present epoch is constant, we can obtain this value by considering the Universe's entropy content today~[\cite{Fromerth:2012fe}]. For $T\ll1\,\mathrm{MeV}$, the entropy content today is carried by photons and neutrinos, yielding
\begin{align}
\label{Nb_S}
\left(\frac{s}{n_B}\right)_{t_0}&=\frac{\sum_i\,s_i}{n_B}=\frac{n_\gamma}{n_B}\,\bigg(\frac{s_\gamma}{n_\gamma}+\frac{s_\nu}{n_\gamma}+\frac{s_{\bar{\nu}}}{n_\gamma}\bigg)\;\\
&=\left(\frac{1}{B}\right)_{\!\!t_0}\!\!\left[\frac{s_\gamma}{n_\gamma}+\frac{4}{3T_\nu}\frac{\rho_\nu^{\mathrm{tot}}}{n_\gamma}-\frac{\mu_\nu}{T_f}\left(\frac{n_\nu-n_{\bar{\nu}}}{n_\gamma}\right)\right]_{t_0}\;,
\end{align}
where $t_0$ denotes the present day values, we have $B=n_B/n_\gamma= 0.605\times10^{-9}$ (CMB)~[\cite{ParticleDataGroup:2016lqr}] from today's observation. The entropy per particle for a massless boson at zero chemical potential is $(s/n)_{\mathrm{boson}}\approx 3.602$.

Substituting Eq.\;(\ref{Excess_Neutrino}) and Eq.\;(\ref{Entropy_density}) into Eq.\;(\ref{L_B_ratio}) yields the lepton-to-baryon ratio
\begin{align}\label{L_B_ratio_final}
&\frac{L}{B}=\frac{45}{4\pi^4}\frac{\pi^2(\mu_\nu/T_f)+(\mu_\nu/T_f)^3}{10.75+{45}(\mu_\nu/T_f)^2/{4\pi^2}}\left(\frac{s}{n_B}\right)_{\!\!t_0}\;,
\end{align}
in terms of $\mu_\nu/T_f$ which is given by Eq.(\ref{Solution}) and the present day entropy-per-baryon ratio. In Fig.\;\ref{BL_Ratio} we show the ratio between the net baryon number and the net lepton number as a function of the effective number of neutrino species $N^{\mathrm{eff}}_\nu$ with the parameter $ B|_{t_0} =0.605\times 10^{-9}$(CMB). We find that the values $N_\nu^{\mathrm{eff}}=3.36\pm0.34$ and $N_\nu^{\mathrm{eff}}= 3.62\pm0.25$ require the ratio between baryon number and lepton number to be $1.16 \times 10^{-9} \leqslant\, B/|L| \leqslant 1.51\times 10^{-9}$. These values are close to the baryon-to-photon ratio $0.57 \times 10^{-9} \leqslant B  \leqslant 0.67\times 10^{-9}$.

In summary, motivated by the necessity to explain a slightly faster Universe expansion, we believe that there is need for additional unobserved particles,  leading to an increase  in the Universe expansion rate. Considerable effort has been made in this direction, e.g., by introducing exotic and new \lq dark\rq\ particles, see~[\cite{Birrell:2014cja}] and references therein. In this work a similar effect is achieved by introducing lepton asymmetry in the Universe. We connected the lepton asymmetry in the Universe with the chemical neutrino potential $\mu_\nu$,
and further evaluated the consequences for the Universe expansion. We have explored the other natural scenario regarding the baryon number-to-lepton number ratio. Instead of $B\simeq |L|$, we found that $0.4\leqslant|L| \leqslant0.52$ and $B\simeq 1.33\times 10^{-9}|L|$ reconciles the CMB and current epoch results for the Hubble expansion parameter.
%The standard cosmological model assumes (arbitrarily) that the asymmetry between leptons and anti-leptons is small, similar to the baryon asymmetry; most often it is assumed $L=B$. We consider $L\simeq 1$ and explore how this large cosmological lepton yield relates to the effective number of (Dirac) neutrinos $N^{\mathrm{eff}}_\nu$.  we have explored the other natural scenario regarding the baryon number-to-lepton number ratio. Instead of $B\simeq |L|$, we found that $0.4\leqslant|L| \leqslant0.52$ and $B\simeq 1.33\times 10^{-9}|L|$ reconciles the CMB and current epoch results for the Hubble expansion parameter.

The large lepton asymmetry from cosmic neutrino can also affect the neutron lifespan in cosmic plasma which is one of the important parameter controlling BBN element abundances. 
In general the neutron lifespan dependence on temperature of the cosmic medium. When temperature $T=\mathcal{O}(\mathrm{MeV})$, neutron decay occurs in the plasma of electron/positron and 
 neutrino/antineutrino. Electrons and neutrinos in the background plasma can reduce the neutron decay rate by Fermi suppression to the neutron decay rate. Furthermore, the neutrino background can still provide the suppression after electron/positron pair annihilation becomes nearly complete. In this case,the large neutrino chemical potential from lepton asymmetry would play an important role and needs to be accounted for in the precision study of the neutron lifespan in the cosmic plasma.

%% file: Chapter5.tex
\chapter{Charged leptons in cosmic plasma}\label{Electron}

Charged leptons played significant roles in the dynamics and evolution of the early Universe. They were kept in equilibrium via electromagnetic and weak interactions.  In this chapter, I examine a dynamical model of the abundance of charged leptons $\mu^\pm$ and $e^\pm$ in the early Universe obtaining their disappearance temperature, the condition when they disappear from the particle inventory. Of particular interest is the dense electron-positron plasma present during the early Universe evolution. I study the damping rate and the magnetization process in this dense $e^\pm$ plasma in the early Universe.

%{Introduction\daggerfootnote{This chapter has been published previously as \citet{Gottbrath1999}.}}
%~~~~~~~~~~~~~~~~~~~~~~~~~~~~~~~~~~~~~~~~~~~~~~~~~

\section{Overview of charge leptons in early Universe}
%In this section we will focus on the following:
%\begin{itemize}
%    \item Charged leptons in early universe
%    \item Remarks on tau leptons
%\end{itemize}

%In the early universe, charged leptons are kept equilibrium with the cosmic plasma via electromagnetic and weak interactions which played significant roles in the dynamics and evolution of the Universe.  For example, the present $e^\pm$ in early universe can affect  the neutrino decoupling, photon heating, and big bang nucleosynthesis. Although, the massive lepton $\tau^\pm, \mu^\pm$ decay into light leptons ($\nu$, $l^\pm$) and hadrons in their lifespan. These high-energy leptons ($\nu$, $l^\pm$) originating from the decay of $\tau^\pm, \mu^\pm$ continue played a significant role in shaping the particle energy distribution which can affect the property of cosmic plasma.

The $\tau^\pm$ leptons can undergo various decay processes via the weak interaction in the early Universe, and is the only charged lepton that can decay into hadrons because of its heavy mass ($m_\tau=1776.86$ MeV). The principle decay channels of $\tau^\pm$ are given by
\begin{align}
&\tau^-\rightarrow\nu_\tau+e^-+\bar{\nu}_e,\qquad \tau^-\rightarrow\nu_\tau+\mu^-+\bar{\nu}_\mu,\\
&\tau^-\rightarrow\nu_\tau+\pi^-,\qquad\qquad\,\tau^+\rightarrow\bar{\nu}_\tau+\pi^+,
\end{align}
 where the vacuum lifespan for $\tau^\pm$ is given by ~[\cite{ParticleDataGroup:2022pth}]
\begin{align}
&\tau_{\tau}=(290.3\pm0.5)\times10^{-15}\,\mathrm{sec}.
\end{align}

Moreover, following the decay of $\tau^\pm$ into pions, these pions subsequently decay into a muon and a neutrino through the reaction
\begin{align}
\pi^-\rightarrow\nu_\mu+\mu^-,\qquad\qquad\,\pi^+\rightarrow\bar{\nu}_\mu+\mu^+,
\end{align}
with pion vacuum lifespan $\tau_\pi=2.6033\times10^{-8}$ sec~[\cite{ParticleDataGroup:2022pth}].
In this scenario, $\tau^\pm$ disappears from the Universe via multiparticle decay processes.
These decay processes can contribute as one of the sources for the production of neutrinos and muons in the early Universe.

The $\mu^\pm$ lepton abundance is an important quantity required for the understanding of several fundamental questions regarding properties of the primordial Universe,  particularly in relation to the freeze-out of strangeness flavor in the early Universe. We recall that the strangeness decay often proceeds into muons, energy thresholds permitting, as the charged kaons K$^\pm$ have a 63\% branching into $\mu+\bar \nu_\mu$. Should muons fall out of thermal abundance equilibrium this would directly impact the detailed balance back-reaction processes. Another, indirect influence on strangeness in early Universe arises through the nearly exclusive decay of charged pions into $\mu+\bar \nu_\mu$. Without chemical abundance equilibrium this back reaction stops too impacting pions and thus all other hadronic particles in the Universe. 

On the other hand, we will show that the lightest charged leptons $e^\pm$ can persist via the reaction $\gamma\gamma\to e^-e^+$ until the temperature $T=20$ keV in the early Universe.  After $T=20$ keV, the positron rapidly disappears through annihilation, leaving only residual electrons to maintain the Universe's charge neutrality. The existence of an electron-positron plasma plays a pivotal role in several aspects of the early Universe as follows: 

1. The role of electron-positron plasma has not received the appropriate attention in the days of precision Big-Bang nucleosynthesis studies. The standard BBN model indicates that the synthesis of light elements typically takes place at temperatures around  $86\,\mathrm{keV}>T_{BBN}>50\,\mathrm{keV}$~[\cite{Pitrou:2018cgg}]. Within this temperature range there are millions of electron-positron pairs per charged nucleon, providing an electron-positron-rich plasma environment for nucleosynthesis which leads to modifications in the Coulomb potential due to the screening effect. Furthermore, the electron-positron densities can reach millions of times normal atomic densities. The presence of  these $e\bar e$-pairs before and during BBN has been acknowledged by Wang, Bertulani and Balantekin~[\cite{Wang:2010px}] nearly a decade ago.

2. The Universe today is filled with magnetic fields at various scales and strengths both within galaxies and in deep extra-galactic space. The origin of these magnetic fields is currently unknown. In the early Universe, when temperature $T>20$ keV, we have dense $e^\pm$ plasma. The significant magnetic moments of electrons and positrons also provide opportunities to investigate spin magnetization process.

Understanding the abundances of muons and electrons/positrons provides essential insights into the evolution of the primordial Universe.  In the following we discuss the muon density at persistence temperature in section \ref{section_muon}, and explore the electron/positron plasma properties, including the damped rate and magnetization in section \ref{section_electron}.

%~~~~~~~~~~~~~~~~~~~~~~~~~~~~~~~~~~~~~~~~~~~~~~~~~

\section{Muon–antimuon in the early Universe}\label{section_muon}
%In this section we will focus on the following:
%\begin{itemize}
%    \item Vanishing of muon in early Universe
%    \item Muon density at persistence temperature
%\end{itemize}

%\subsection{Vanishing of muon in early Universe}
Our interest in strangeness flavor freeze-out in the early Universe requires the understanding of the abundance of muons in the early Universe. The specific question needing an answer is at which temperature muons remain in abundance (chemical) equilibrium established predominantly by electromagnetic and weak interaction processes, allowing detailed-balance back-reactions to influence strangeness abundance.

In the early Universe in the the cosmic plasma muons of mass $m_\mu=105.66$\,MeV can be produced by the following interaction processes
\begin{align} 
&\gamma+\gamma\longrightarrow\mu^++\mu^-,\qquad & e^++e^-\longrightarrow \mu^++\mu^-\;,\\
&\pi^-\longrightarrow\mu^-+\bar{\nu}_\mu,\qquad & \pi^+\longrightarrow\mu^++\nu_\mu\;.
\end{align}
The back reactions for all above processes are in detailed balance, provided all particles shown on the right hand side (RHS) exist in chemical abundance equilibrium in the Universe. We recall the empty space (no plasma) at rest lifetime of pions $\tau_\pi=2.6033\times10^{-8}$ sec. 

However, all produced muons can also decay via the reactions
\begin{equation}
\mu^-\rightarrow\nu_\mu+e^-+\bar{\nu}_e,\qquad \mu^+\rightarrow\bar{\nu}_\mu+e^++\nu_e\,,
\end{equation} 
with the empty space (no plasma) at rest lifetime $\tau_{\mu}=2.197 \times 10^{-6}\,\mathrm{sec}$. We thus must establish the range of temperature in which production processes exceed in speed the decay process.
 
 The temperature range of our interests is the Universe when $m_\mu\gg T$. In this case the the Boltzmann approximation is appropriate for studying massive particles muons and pions. The thermal decay rate per volume and time  for muons $\mu^\pm$ (and pions $\pi^\pm$) in the Boltzmann limit  are given by~[\cite{PhysRevC.82.035203}]:
\begin{align}
&R_\mu=\frac{g_\mu}{2\pi^2}\left(\frac{T^3}{\tau_\mu}\right)\left(\frac{m_\mu}{T}\right)^2K_1(m_\mu/T)\;,\\
&R_\pi=\frac{g_\pi}{2\pi^2}\left(\frac{T^3}{\tau_\pi}\right)\left(\frac{m_\pi}{T}\right)^2K_1(m_\pi/T)\;, 
\end{align}
where the lifespan of $\mu^\pm$ and $\pi^\pm$ in the vacuum were given above. This rate accounts for both the density of particles in chemical abundance equilibrium and the effect of time dilation present when particles are in thermal motion with respect to observer at rest in the local reference frame. The effects of Fermi blocking or boson stimulated emission have been neglected.

The thermal averaged reaction rate per volume for the reaction $a\overline{a}\rightarrow b\overline{b}$ in Boltzmann approximation is given by [\cite{Letessier:2002ony}]
\begin{align}\label{pairR}
R_{a\overline{a}\rightarrow b\overline{b}}=\frac{g_ag_{\overline{a}}}{1+I}\frac{T}{32\pi^4}\int_{s_{th}}^\infty ds\frac{s(s-4m^2_a)}{\sqrt{s}}\sigma_{a\overline{a}\rightarrow b\overline{b}}~K_1(\sqrt{s}/T),
\end{align}
where $s_{th}$ is the threshold energy for the reaction, $\sigma_{a\overline{a}\rightarrow b\overline{b}}$ is the cross section for the given reaction, and $K_1$ is the modified
Bessel function of integer order $”1”$. We introduce the factor $1/1+I$ to avoid the double counting of indistinguishable pairs of particles; we have $I=1$ for an identical pair and $I=0$ for a distinguishable pair.

The leading order invariant matrix elements for the reactions $e^++e^-\to\mu^++\mu^-$ and $\gamma+\gamma\to\mu^++\mu^-$, are introduced in this work by [\cite{Kuznetsova:2008jt}]
\begin{align}\label{Mee}
|M_{e\bar e\to\mu\bar\mu}|^2=&32\pi^2\alpha^2\frac{(m_\mu^2-t)^2+(m_\mu^2-u)^2+2m_\mu^2s}{s^2},\quad m_\mu\gg m_e\;,\\[0.2cm]
\label{Mgg}
|M_{\gamma\gamma\to\mu\bar\mu}|^2=&32\pi^2\alpha^2\bigg[\left(\frac{m_\mu^2-u}{m_\mu^2-t}+\frac{m_\mu^2-t}{m_\mu^2-u}\right)+4\left(\frac{m_\mu^2}{m_\mu^2-t}+\frac{m_\mu^2}{m^2_\mu-u}\right)\\[0.1cm]  \nonumber
&\hspace{1cm}-4\left(\frac{m_\mu^2}{m^2_\mu-t}+\frac{m^2_\mu}{m^2_\mu-u}\right)^2\bigg]\;,
\end{align}
 where $s, t, u$ are the Mandelstam variables. The cross section required in Eq.\,(\ref{pairR}) can be obtained by integrating the matrix elements Eq.\,(\ref{Mee}) and Eq.\,(\ref{Mgg}) over the Mandelstam variable $t$ ~[\cite{PhysRevC.82.035203}]. We have
\begin{align}
&\sigma_{e\bar e\to\mu\bar\mu} 
=\frac{64\pi\alpha^2}{48\pi}\left(\frac{1+2m^2_\mu/s}{s-4m_e^2}\right)\sqrt{1-\frac{4m^2_\mu}{s}},\\
&\sigma_{\gamma\gamma\to\mu\bar\mu}=\frac{\pi}{2}\left(\frac{\alpha}{m_\mu}\right)^2(1-\beta^2)\left[(3-\beta^4)\ln\frac{1+\beta}{1-\beta}-2\beta(2-\beta^2)\right],\\
&\beta=\sqrt{1-4m^2_\mu/s}
\end{align}
Substituting the cross sections into Eq.\,(\ref{pairR}) we obtain the production rates for $e\bar e\to\mu\bar\mu$ and $\gamma\gamma\to\mu\bar\mu$ respectively.

\begin{figure}[ht]
\begin{center}
\includegraphics[width=5.0in]{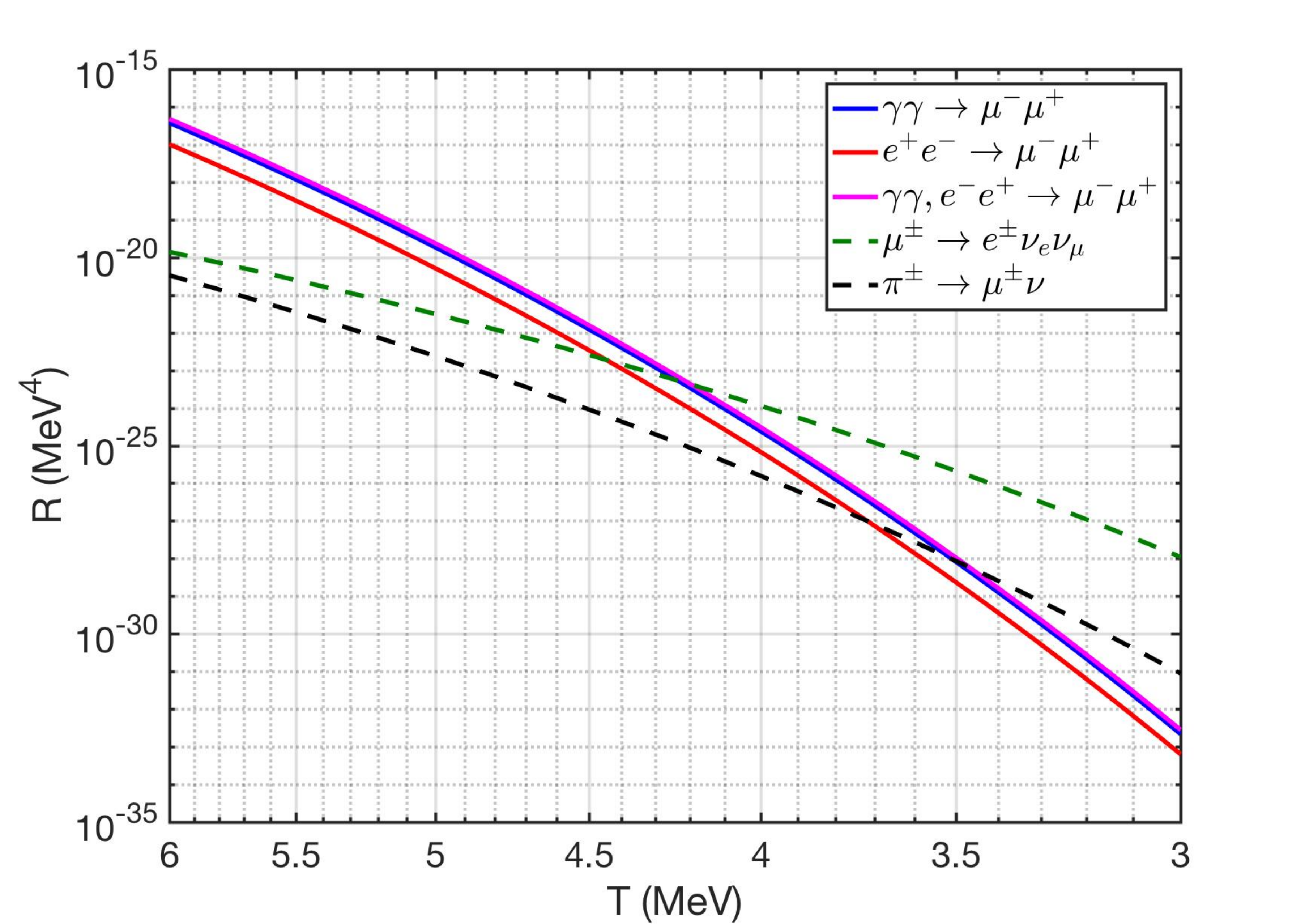}
\caption{We plot the thermal reaction rate per volume for different reactions as a function of temperature. We found that dominant reactions for $\mu^\pm$ production are ${\gamma+\gamma\to\mu^++\mu^-}$ and $e^++e^-\to\mu^++\mu^-$, and the total production rate crosses the decay rate of $\mu^\pm$ at temperature $T_{dissapear}\approx 4.195$ MeV.}
\label{MuonRatenew_fig}
\end{center}
\end{figure}

In Fig.~\ref{MuonRatenew_fig} we show the invariant thermal reaction rates per volume and time for rates of relevance, as a function of temperature $T$.
As the temperature decreases in the expanding Universe, the initially dominant production rates ($e\bar e,\gamma\gamma\to\mu\bar\mu$) decrease with decreasing temperature, and eventually cross the $\mu^\pm$ decay rates. 
Muon abundance disappears as soon as any decay rate is faster than the fastest production rate. Specifically after the Universe cools below the temperature $T_\mathrm{disappear}=4.195$ MeV, the dominant reaction is the muon decay. Due to the relatively slow expansion of the Universe, the disappearance of muons is sudden, and the abundance of muons vanishes as soon as a decay rate surpasses the dominant production rate.

%~~~~~~~Figure~~~~~~ ~~~~~~~~~~~~~~~~~~~~~~~~~~~~~~~~~~~~~~~~~~~~~~~~~~~~~~~~~~~~~~~~~~~
%~~~~~~~~~~~~~~~~~~~~~~~~~~~~~~~~~~~~~~~~~~~~~~~~~~~~~~~~~~~~~~~~~~~~~~~~~~~~~~~~~~~~~~~~~~~~~~~~

On the other hand, considering the number density for nonrelativistic $\mu^\pm$ in the Boltzmann approximation, we have
\begin{align}\label{nmupm}
n_{\mu^\pm}=\frac{g_{\mu^\pm}}{2\pi^2}T^3\left(\frac{m_\mu}{T}\right)^2 K_2(m_\mu/T)=g_{\mu^\pm}\left(\frac{m_\mu T}{2\pi}\right)^{3/2}e^{-{m_\mu}/{T}}\;. 
\end{align}
then the number density between $n_{\mu^\pm}$ and baryon $n_B$ can be written as
\begin{align}
\frac{n_{\mu^\pm}}{n_\mathrm{B}}=\frac{n_{\mu^\pm}}{s}\frac{s}{n_\mathrm{B}}=
\frac{n_{\mu^\pm}}{s}\left(\frac{s}{n_\mathrm{B}}\right)_{\!t_0},
\end{align}
where we used that $s/n_\mathrm{B}$ remains constant and $t_0$ represent present day value. The present value is given by $(n_B/s)_{t_0}\approx8.69\times10^{-11}$ (detail please see Chapter~\ref{Introduction}). The entropy density $s$ can be characterized introducing $g^s_\ast$, the total number of \lq entropic\rq\ degrees of freedom
\begin{align}\label{entrop}
s=\frac{2\pi^2}{45}g^s_\ast T^3\;.
\end{align}
For temperature $10\,\mathrm{MeV} >T>3 $\,MeV, the massless photons, nearly relativistic electron/positrons, and practically massless neutrinos contribute to the degree of freedom $g^s_\ast$.  In this case, the number density between $n_{\mu^\pm}$ and baryon $n_B$ in the temperature interval we consider $10\,\mathrm{MeV} >T>3 $\,MeV is given by
\begin{align}\label{nmuperbF} 
\frac{n_{\mu^\pm}}{n_\mathrm{B}}=\frac{45}{2\pi^2}\frac{g_{\mu^\pm}}{g^s_\ast}\left(\frac{m_\mu}{2\pi T}\right)^{3/2}e^{-{m_\mu}/{T}}\;\left(\frac{s}{n_\mathrm{B}}\right)_{\!t_0}.
\end{align}

%Figure~~~~~~~~~~~~~~~~~~~~~~~~~~~~~~~~~~~~~~~~~~~~~~~~~~~~~~~~~~~~~~~~~~~~~~~~~
\begin{figure}[t]
\begin{center}
\includegraphics[width=\linewidth]{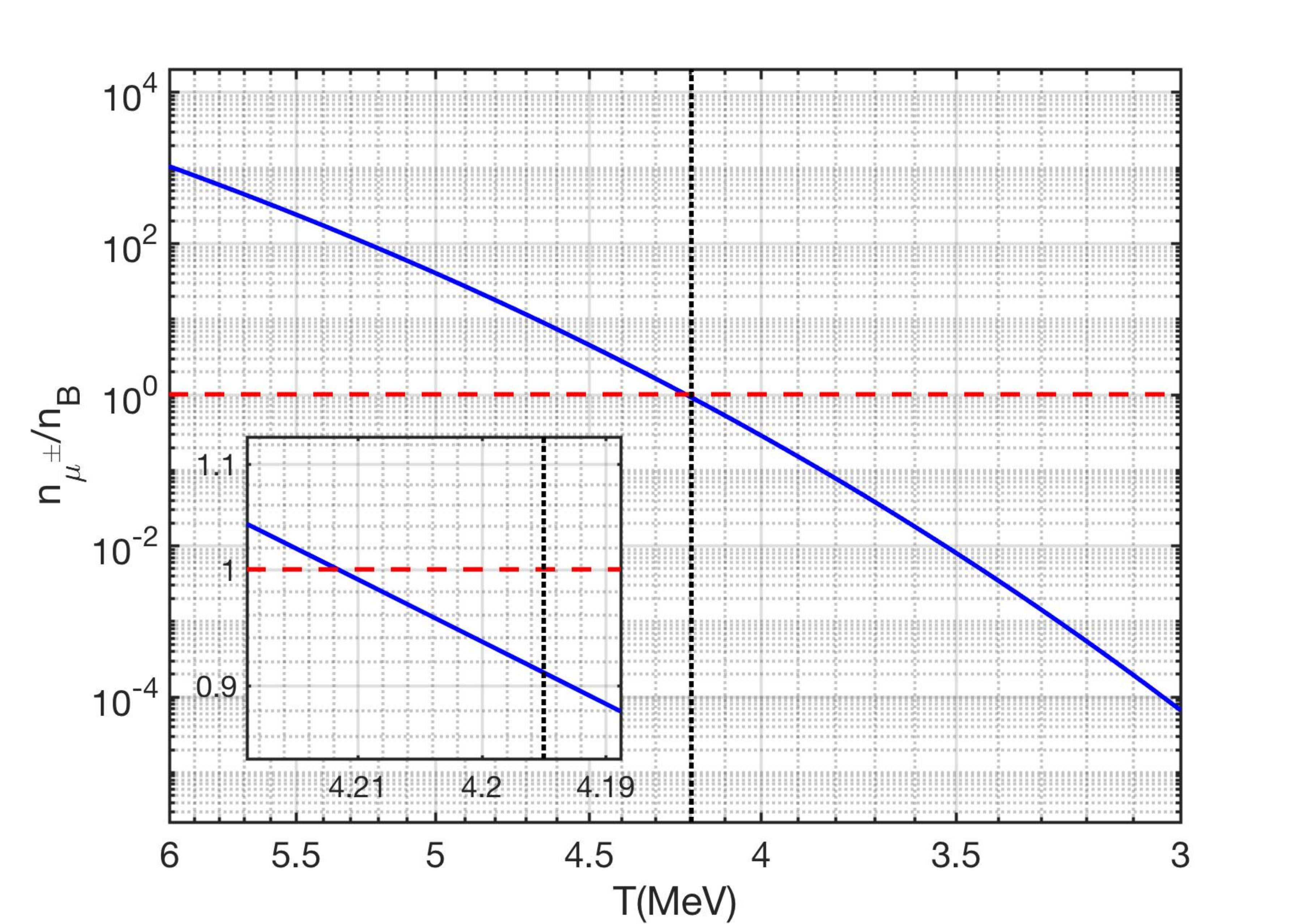}
\caption{
The density ratio between $\mu^\pm$ and baryons as a function of temperature. The density ratio at muon disappearance temperature is about $n_{\mu^\pm}/n_\mathrm{B}(T_\mathrm{disappear})\approx0.911$, and around the temperature $T\approx4.212$ MeV the density ratio $n_{\mu^\pm}/n_\mathrm{B}\approx1$.}
\label{DensityRatio_fig}
\end{center}
\end{figure}
%~~~~~~~~~~~~~~~~~~~~~~~~~~~~~~~~~~~~~~~~~~~~~~~~~~~~~~~~~~~~~~~~~~~~~~~~~~~~~

In Fig.\,\ref{DensityRatio_fig} we show the muon to baryon density ratio Eq.\,(\ref{nmuperbF}) as a function of $T$. We see that the muon abundance $T=10$\,MeV exceeds that of baryons by a factor 500,000 while at muon disappearance temperature $n_{\mu^\pm}/n_\mathrm{B}(T_\mathrm{disappear})\approx0.911$. The number density $n_{\mu^\pm}$ and $n_\mathrm{B}$  abundances are equal at around the temperature $T_\mathrm{equal}\approx4.212\,\mathrm{MeV} >  T_\mathrm{disappear}$.  This means that the muon abundance may still be able to influence baryon evolution because their number density is comparable to the baryon density.% However, we also find that at the temperature $T_\mathrm{equal}\approx4.212$\,MeV the density ratio is unity $n_{\mu^\pm}/n_\mathrm{B}\approx1$.

The primary insight of this work is that aside of protons, neutrons and other nonrelativistic particles, both positively and negatively charged muons $\mu^\pm$ are present in thermal equilibrium and in non-negligible abundance for $T>T_\mathrm{dissapear}\approx 4.195$\,MeV. This offers a new and tantalizing model building opportunity for anyone interested in baryon-antibaryon separation in the primordial Universe, strangelet formation, and perhaps other exotic primordial structure formation mechanisms.

%~~~~~~~~~~~~~~~~~~~~~~~~~~~~~~~~~~~~~~~~~~~~~~~~~

\section{ Electron-positron plasma in the early Universe}\label{section_electron}
%In this section we will focus on the following:
%\begin{itemize}
%    \item Chemical potential of electron in early universe
%    \item Electron-positron plasma in BBN (Damped sccreening)
%    \item Electron-positron magnetization
%%    \item Neutron Lifespan in magnetized electron/positron plasma.
%\end{itemize}

In the early Universe, after the neutrino freeze-out at $T\approx 2$\,MeV, the Universe is controlled by the electron-positron-photon plasma. In this section, we demonstrate the rich electron-positron plasma in the early Universe by examining the chemical potential $\mu_e$ in the charge-neutral and entropy-conserving Universe. We study the  microscope collision property of electron-positron plasma and explore the spin response of the electron-positron plasma to external and self-magnetization fields, thus developing methods for future detailed study.

%In this section, we will quantify the dynamical picture of $e^\pm$ plasma and show that the $e^+$ abundance can persist in early universe at relatively low temperature $T = 20$ keV which provide the dense $e^\pm$ plasma environment for the big-bang nucleosynthesis (BBN) in the early universe. 

%The role of electron-positron plasma has not received the appropriate attention in the days of precision big bang nucleosynthesis studies. The standard BBN model indicates that the synthesis of light elements typically takes place at temperatures around  $86\,\mathrm{keV}>T_{BBN}>50\,\mathrm{keV}$~[\cite{Pitrou:2018cgg}]. Within this temperature range there are millions of electron-positron pairs per charged nucleon, providing an electron-positron-rich plasma environment for nucleosynthesis. Furthermore, the electron-positron densities can reach millions of times normal atomic densities. The presence of  these $e\bar e$-pairs before and during BBN has been acknowledged by Wang, Bertulani and Balantekin~[\cite{Wang:2010px}] nearly a decade ago.

%On the other hand, the Universe today filled with magnetic fields at various scales and strengths both within galaxies and in deep extra-galactic space.It is currently unknown the origin for these magnetic fields today. In early Universe when temperature $T>20$ keV , we have dense $e^\pm$ plasma. The significant magnetic moments of electrons and positrons also provide opportunities to investigate spin magnetization process.

%~~~~~~~~~~~~~~~~~~~~~~~~~~~~~~~~~~~~~~~~~~~~~~~~~~~~~~~~~~~~~~~~~~~~~~~~~

\subsection{Electron chemical potential in the early Universe}
In this section, we derive the dependence of electron chemical potential, and hence $e^\pm$ density, on the photon background temperature by employing the following physical principles:
\begin{enumerate}
\item Charge neutrality of the Universe:
\begin{align}\label{neutrality}
n_e-n_{\overline{e}}=n_p-n_{\overline{p}}\approx\,n_p,
\end{align}
where $n_e$ and $n_{\overline{e}}$ denotes the number density of electron and positron.
\item Neutrinos decouple (freeze-out) at a temperature $T_f\simeq 2$ MeV, after which they free stream through the Universe with an effective temperature~[\cite{Birrell:2012gg}]
\begin{align}
T_\nu(t)=T_f a(t_f)/a(t),
\end{align}
 where $a(t)$ is the FLRW Universe scale factor.
\item Total comoving entropy is conserved. At $T\leq T_f$ the dominant contributors to entropy are photons, $e^\pm$, and neutrinos.
In addition, after neutrino freeze-out, neutrino comoving entropy is independently conserved ~[\cite{Birrell:2012gg}]. This  implies that the combined comoving entropy in $\gamma$, $e^\pm$ is also conserved for $T_\gamma\leq T_f$.
\end{enumerate}

Motivated by the fact that comoving entropy in $\gamma$, $e^\pm$ is conserved after neutrino freeze-out, we rewrite the charge neutrality condition, Eq.(\ref{neutrality}) in the form
\begin{align}\label{charge_neutral_cond2}
n_e-n_{\overline{e}}=X_p\frac{n_B}{s_{\gamma,e,\overline{e}}} s_{\gamma,e,\overline{e}},\qquad X_p\equiv\frac{n_p}{n_B},
\end{align}
where $n_B$ is the number density of baryons, and $s_{\gamma,e,\overline{e}}$ is the combined entropy density in photons, electrons, and positrons. During the Universe expansion, the comoving entropy and baryon number are conserved quantities, hence the ratio $n_B/s_{\gamma,e,\overline{e}}$ is conserved. We have
\begin{align}
\frac{n_B}{s_{\gamma,e,\overline{e}}}=\left(\frac{n_B}{s_{\gamma,e,\overline{e}}}\right)_{t_0}\!\!\!\!=\left(\frac{n_B}{s_{\gamma}}\right)_{t_0}\!\!\!\!=\left(\frac{n_B}{n_\gamma}\right)_{t_0}\left(\frac{n_\gamma}{s_{\gamma}}\right)_{t_0},
\end{align}
where the subscript $t_0$ denotes the present day value, and the second equality is obtained by observing that the present day $e^\pm$-entropy density is negligible compared to the photon entropy density. We can evaluate the ratio by giving the present day baryon-to-photon ratio: $n_B/n_\gamma= 6.05\times10^{-10}$(CMB) ~[\cite{ParticleDataGroup:2022pth}] and the entropy per particle for a massless boson:$(s/n)_{\mathrm{boson}}\approx 3.602$~[\cite{Letessier:2002ony}].

The total entropy density of photons and electron/positron can be written as
\begin{align}\label{entropy_per_baryon}
s_{\gamma,e,\overline{e}}=\frac{2\pi^2}{45}g_\gamma\,T_\gamma^3+\frac{\rho_{e,\overline{e}}+P_{e,\overline{e}}}{T_\gamma}-\frac{\mu_e}{T_\gamma}(n_e-n_{\overline{e}}),
\end{align}
where $ \rho_{e,\overline{e}}=\rho_{e}+\rho_{\overline{e}}$ and $P_{e,\overline{e}}=P_{e}+P_{\overline{e}}$ are the total energy density and pressure of electrons/positron respectively.
The energy density and pressure in electrons and positrons are given by
\begin{align}\label{rho_e}
\frac{\rho_{e,\overline{e}}}{T_\gamma^4}=\frac{g_e}{2\pi^2}M_e^4 \bigg[&\int_{1}^\infty \frac{ u^2\sqrt{ u^2-1} du}{\exp(M_e u-b_e)+1}+\int_{1}^\infty \frac{ u^2\sqrt{ u^2-1} du}{\exp(M_e u+b_e)+1}\bigg]\,,
\end{align}
and
\begin{align}\label{P_e}
\frac{P_{e,\overline{e}}}{T_\gamma^4}=\frac{g_e}{6\pi^2}M_e^4\bigg[&\int_{1}^\infty   \frac{(u^2-1)^{3/2} du}{\exp(M_e u-b_e)+1}+\int_{1}^\infty   \frac{(u^2-1)^{3/2} du}{\exp(M_e u+b_e)+1}\bigg],
\end{align}
where we introduce the dimensionless variables as follows: 
\begin{align}\label{Variables}
u=\frac{E}{m_e},\qquad M_e=\frac{m_e}{T_\gamma},\qquad b_e=\frac{\mu_e}{T_\gamma}.
\end{align}

By incorporating Eq.(\ref{charge_neutral_cond2}) and Eq.(\ref{entropy_per_baryon}), the charge neutrality condition can be expressed as
\begin{align}\label{charge_neutral_cond3}
&\left[1+X_p\left(\frac{n_B}{n_\gamma}\right)_{t_0}\left(\frac{n_\gamma}{s_{\gamma}}\right)_{t_0}\frac{\mu_e}{T_\gamma}\right]\frac{n_e-n_{\overline{e}}}{T_\gamma^3}\notag\\
&\qquad\qquad\qquad=X_p\left(\frac{n_B}{n_\gamma}\right)_{t_0}\left(\frac{n_\gamma}{s_{\gamma}}\right)_{t_0} \left(\frac{2\pi^2}{45}g_\gamma+\frac{\rho_{e,\overline{e}}+P_{e,\overline{e}}}{T_\gamma^4}\right).
\end{align}
Using the Fermi distribution, the number density of electrons over positrons in the early Universe is given by
\begin{align}\label{ee_density}
n_e-n_{\overline{e}}&=\frac{g_e}{2\pi^2}\left[\int_0^\infty\frac{p^2dp}{\exp{\left((E-\mu_e)\right)/T_\gamma}+1}\right.\left.-\int_0^\infty\frac{p^2dp}{\exp{\left((E+\mu_e)/T_\gamma\right)}+1}\right]\notag\\
&=\frac{g_e}{2\pi^2}{T_\gamma^3}\tanh(b_e)M_e^3\int_{1}^\infty \!\!\!\!\frac{  u \sqrt{u^2-1} du}{1+\cosh(M_eu)/\cosh(b_e)}.
\end{align}
Substituting Eq.(\ref{ee_density}) into Eq.(\ref{charge_neutral_cond3}) and giving the value of $X_p$, the charge neutrality condition can be solved to determine $\mu_e/T_\gamma$ as a function of $M_e$ and $T_\gamma$. 
%Fig~~~~~~~~~~~~~~~~~~~~~~~~~~~~~~~~~~~~~~~~~~~~~~~~~~~~~
\begin{figure}[ht]
\begin{center}
\includegraphics[width=\linewidth]{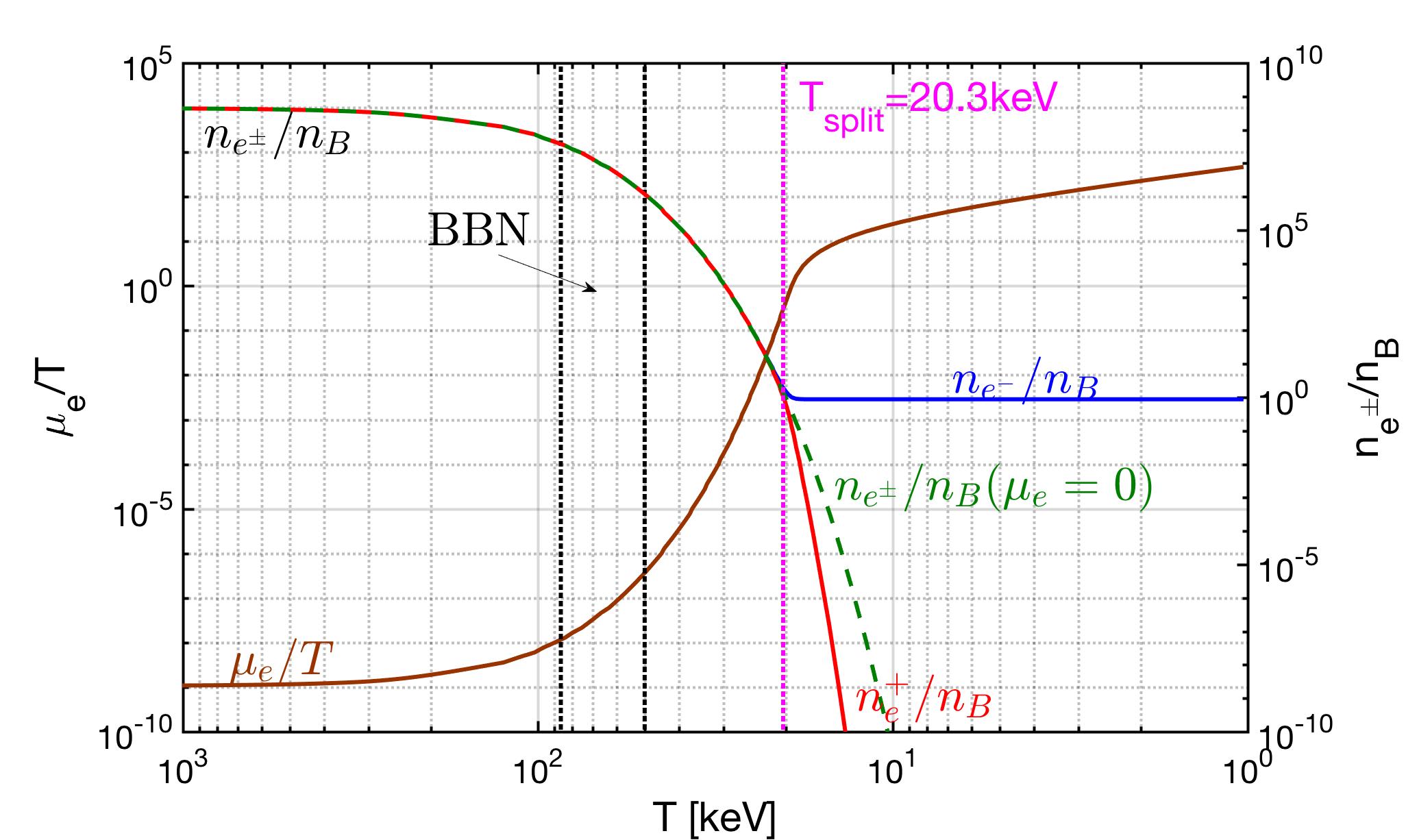}
\caption{Left axis: The chemical potential of an electron as a function of photon temperature $T=T_\gamma$ with $X_p=0.878$ and $n_B/n_\gamma=6.05\times10^{-10}$. Right axis: the ratio of electron(positron) number density to baryon density as a function of temperature. The blue solid line is the electron density, the red dashed line is the positron density, and the green dotted line is the number density with $\mu_e=0$. We found that when electron chemical potential $\mu_e\approx T=0.02\,\mathrm{MeV}$ the positron density decreases because of the annihilation.}
\label{BBN_Electron}
\end{center}
\end{figure}
%~~~~~~~~~~~~~~~~~~~~~~~~~~~~~~~~~~~~~~~~~~~~~~~~~~~~~

In Fig.~\ref{BBN_Electron} (left axis) we solve Eq.(\ref{charge_neutral_cond3}) numerically and plot the electron chemical potential as a function of temperature with the following parameters: proton concentration $X_p=0.878$ from 
 observation~[\cite{ParticleDataGroup:2022pth}] and  $n_B/n_\gamma=6.05\times10^{-10}$ from CMB. We can see the value of chemical potential is comparatively small $\mu_e/T\approx10^{-6}\sim10^{-7}$ during the BBN temperature range, implying an equal number of electrons and positrons in plasma. From the ratio of electron (positron) number density to baryon density in Fig.~\ref{BBN_Electron} (right axis) we can see that during the accepted BBN temperature range the Universe was filled with an electron-positron rich plasma.
For example when the temperature is around $T=70\,\mathrm{keV}$ the density of electrons and positrons is comparatively large in the early Universe $n_{e^\pm}\approx10^7\,n_B$. Later when the temperature is around $T=20.3\,\mathrm{keV}$, the positron density decreases, leading to the transformation of the pair plasma to an electron-proton plasma.
%~~~~~~~~~~~~~~~~~~~~~~~~~~~~~~~~~~~~~~~~~~~~~~~~~
\subsection{Microscope damping rate of electron-positron plasma}\label{relax}
In electron-positron plasma, the major reactions between photons and $e^+e^-$ pairs are inverse Compton scattering, M{\o}ller scattering, and Bhabha scattering:
\begin{align}
&e^\pm+\gamma\longrightarrow e^\pm+\gamma,\qquad e^\pm+e^\pm\longrightarrow e^\pm+e^\pm,\qquad e^\pm+e^\mp\longrightarrow e^\pm+e^\mp.
\end{align}
The general formula for thermal reaction rate per volume is discussed in~[\cite{Letessier:2002ony}] (Eq.(17.16), Chapter 17). For inverse Compton scattering we have
\begin{align}
R_{e^{\pm}\gamma}=\frac{g_eg_\gamma}{16\left(2\pi\right)^5}T\int_{m_e^2}^\infty\!\!\!\!ds\frac{K_1(\sqrt{s}/T)}{\sqrt{s}}\int^0_{-(s-m_e^2)^2/s}\!\!\!\!\!\!\!\!\!\!\!\!\!\!\!\!dt\, |M_{e^{\pm}\gamma}|^2,
\end{align} 
and for M{\o}ller and Bhabha reactions we have
\begin{align}
&R_{e^\pm e^\pm}=\frac{g_eg_e}{16\left(2\pi\right)^5}T\!\!\int_{4m_e^2}^\infty\!\!\!\!ds\frac{K_1(\sqrt{s}/T)}{\sqrt{s}}\int^0_{-(s-4m_e^2)}\!\!\!\!\!\!\!\!\!\!\!\!\!\!\!\!dt\,|M_{e^\pm e^\pm}|^2,\\
&R_{e^\pm e^\mp}=\frac{g_eg_e}{16\left(2\pi\right)^5}T\!\!\int_{4m_e^2}^\infty\!\!\!\!ds\frac{K_1(\sqrt{s}/T)}{\sqrt{s}}\int^0_{-(s-4m_e^2)}\!\!\!\!\!\!\!\!\!\!\!\!\!\!\!\!dt\,|M_{e^\pm e^\mp}|^2,
\end{align}
where $g_i$ is the degeneracy of particle $i$, $|M|^2$ is the matrix element for a given reaction, $K_1$ is the Bessel function of order $1$, and $s,t,u$ are Mandelstam variables. The leading order matrix element associated with inverse Compton scattering can be expressed in the Mandelstam variables~[\cite{Kuznetsova:2011wt, Kuznetsova:2009bq}] we have
\begin{align}
|M_{e^\pm\gamma}|^2\!=32 \pi^2\alpha^2\bigg[&4\left(\frac{m_e^2}{m_e^2-s}+\frac{m_e^2}{m_e^2-u}\right)^2\notag\\
&\qquad\qquad-\frac{4m_e^2}{m_e^2-s}-\frac{4m_e^2}{m_e^2-u} -
 \frac{m_e^2-u}{m_e^2-s} -\frac{m_e^2-s}{m_e^2-u}\bigg],
\end{align}
and for M{\o}ller and Bhabha scattering we have 
\begin{align}
|M_{e^{\pm}e^{\pm}}|^{2}\!=64\pi^{2}\alpha^{2}\bigg[&
\frac{s^{2}+u^{2}+8m_e^{2}(t-m_e^{2})}{2(t-m^2_{\gamma})^{2}}\notag\\
&\quad+\frac{{s^{2}+t^{2}}+8m_e^{2}
(u-m_e^{2})}{2(u-m_{\gamma}^2)^{2}} + \frac{\left( {s}-2m_e^{2}\right)\left({s}-6m_e^{2}\right)}
{(t-m_{\gamma}^2)(u-m_{\gamma}^2)} \bigg],
\end{align}
and
\begin{align}
|M_{e^\pm e^\mp}|^{2}=64\pi^{2}\alpha^{2}
\bigg[&\frac{s^{2}+u^{2}+8m_e^{2}(t-m_e^{2})}{2(t-m^2_{\gamma})^{2}}\notag\\
&\quad+\frac{u^{2}+t^{2}+8m_e^{2}
(s-m_e^{2})}{2(s-m^2_{\gamma})^{2}}  +   \frac{\left({u}-2m_e^{2}\right)\left({u}-6m_e^{2}\right)}
   {(t-m^2_{\gamma})(s-m^2_{\gamma})} \bigg],
\label{M_fi_b}
\end{align}
where we introduce the photon mass $m_\gamma$ to account the plasma effect and avoid singularity in reaction matrix elements. 

The photon mass $m_\gamma$ in plasma is equal to the plasma frequency $\omega_p$, where we have~[\cite{Kislinger:1975uy}]
\begin{align}
m^2_\gamma=\omega^2_{p}=8\pi\alpha\int\frac{d^3p_e}{(2\pi)^3}\left(1-\frac{p_e^2}{3E_e^2}\right)\frac{f_e+f_{\bar e}}{E_e},
\end{align}
where $E_e=\sqrt{p_e^2+m^2_e}$. In the BBN temperature range $86\,\mathrm{keV}>T_{BBN}>50\,\mathrm{keV}$ we have $m_e\gg T$ and considering the nonrelativistic limit for electron-positron plasma, we obtain
\begin{align}
m^2_\gamma=\frac{4\pi\alpha}{2m_e}\left(\frac{2m_eT}{\pi}\right)^{3/2}e^{-m_e/T}\cosh\left(\frac{\mu_e}{T}\right).
\end{align}
In the BBN temperature range, we have $\mu_e/T\ll1$, which implies the equal number of electrons and positrons in plasma.

\begin{figure}[h]
\begin{center}
\includegraphics[width=\linewidth]{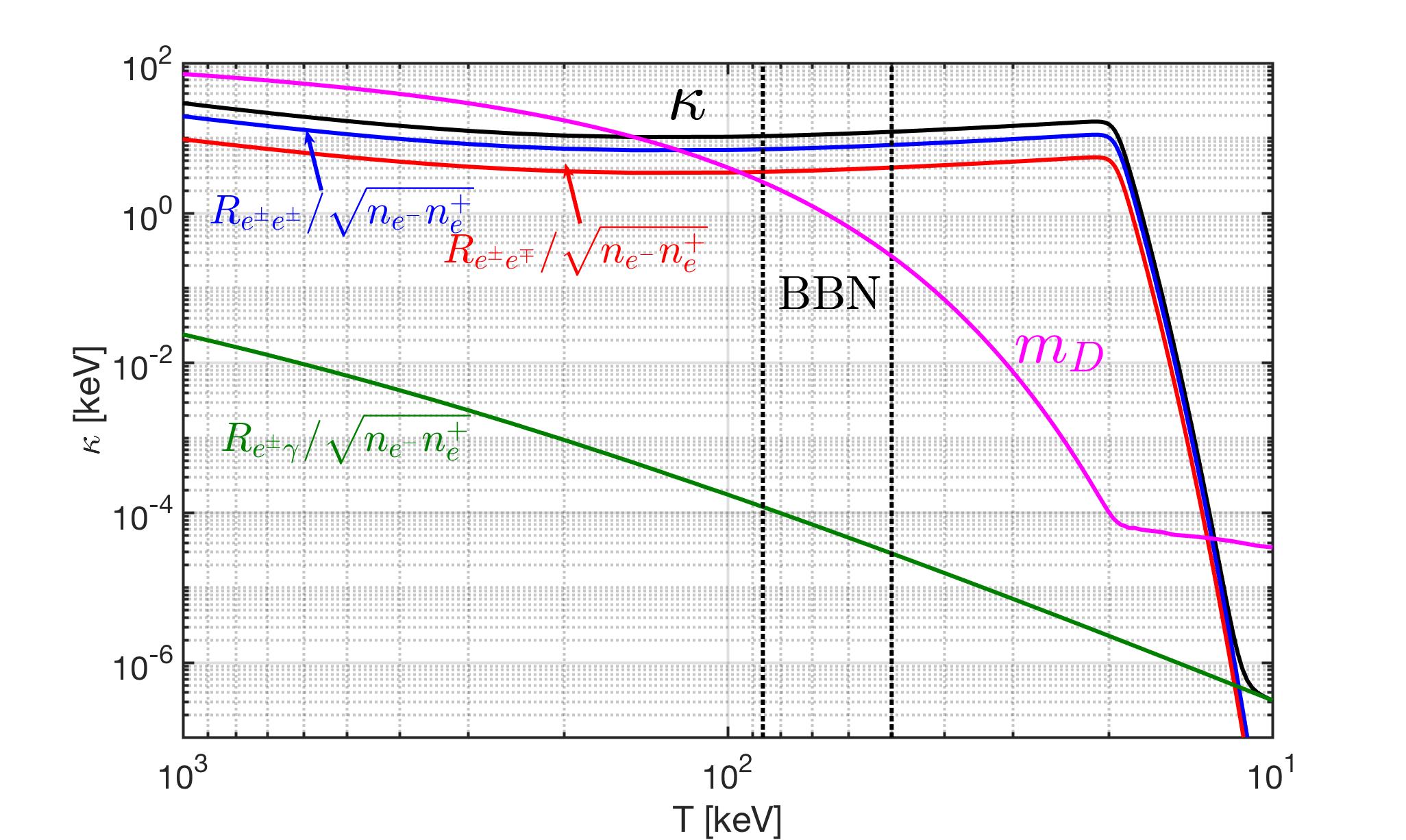}
\caption{The relaxation rate $\kappa$ as a function of temperature in nonrelativistic electron-positron plasma. For comparison, we show  reaction rates  for M{\o}ller reaction $e^-+e^-\to e^-+e^-$ (blue line), Bhabha reaction $e^-+e^+\to e^-+e^+$ (red line), and inverse Compton scattering $e^-+\gamma\to e^-+\gamma$ (green line) respectively. It shows that the dominant reactions during BBN are the M{\o}ller and Bhabha scatterings between electrons and positrons. The total relaxation rate Eq.(\ref{Kappa}) is shown in the black line. It shows that we have $\kappa=10\sim12$ keV during the BBN temperature range. For comparison, the Debye mass $m_D=\omega_{p}\sqrt{m_e/T}$(purple line) is shown as a function of temperature.
}
\label{RelaxationRate_fig}
\end{center}
\end{figure}

To discuss the collisional plasma by the linear response theory, it is convenient to define the average relaxation rate for the electron-positron plasma as follows:
\begin{align}\label{Kappa}
\kappa=\frac{R_{e^\pm e^\pm}+R_{e^\pm e^\mp}+R_{e^\pm\gamma}}{\sqrt{n_{e^-}n_{e^+}}}\approx\frac{R_{e^\pm e^\pm}+R_{e^\pm e^\mp}}{\sqrt{n_{e^-}n_{e^+}}},
\end{align}
where the density function ${\sqrt{n_{e^-}n_{e^+}}}$ in the Boltzmann limit is given by
\begin{align}
{\sqrt{n_{e^-}n_{e^+}}}=\frac{g_e}{2\pi^3}T^3\left(\frac{m_e}{T}\right)^2K_2(m_e/T).
\end{align}
In Fig.~\ref{RelaxationRate_fig}, we show the reaction rates for M{\o}ller reaction, Bhabha reaction, and inverse Compton scattering as a function of temperature. For temperatures $T>12.0$ keV, the dominant reactions in plasma are M{\o}ller and Bhabha scatterings between electrons and positrons. Thus in the BBN temperature range, we can neglect the inverse Compton scattering. The total relaxation rate is approximately constant $\kappa=10\sim12$ keV during the BBN. For $T<20.3$ keV the relaxation rate $\kappa$ decreases rapidly because of positron annihilation. At this temperature, the composition of plasma begins to change from an electron-positron plasma to an electron-baryon plasma.
%~~~~Figure~~~~~~~~~~~~~~~~~~~~~~~~~~~
%~~~~Figure~~~~~~~~~~~~~~~~~~~~~~~~~~~

\subsubsection{From static to damped dynamic screening}

At present, the observation of light element (e.g. D, $^3$He, $^4$He, and $^7$Li) abundances produced in Big-Bang nucleosynthesis (BBN) offers a reliable probe of the early Universe before the recombination. Much effort of the BBN study is currently being made to reconcile the discrepancies and tensions between theoretical predictions and observations of light element abundances, e.g. $^7$Li problem ~[\cite{Pitrou:2018cgg, Fields:2011zzb}].
Current models assume that the Universe was essentially void of anything but reacting light nucleons and electrons needed to keep the local baryon density charge-neutral, a situation similar to the experimental environment where empirical nuclear reaction rates are obtained.

The electron-positron plasma influences light element abundances through electromagnetic screening of the nuclear potential. The electron cloud surrounding the charge of an ion screens other nuclear charges far from its own radius and reduces the Coulomb barrier. In nuclear reactions, the reduction of Coulomb barrier makes the penetration probability easier and enhance the thermonuclear reaction rates. In this case, the modification of the nuclei interaction due to the plasma screening effect may plays a key role in the formation of light element in the BBN. 

The enhancement factor of thermonuclear reaction rates and screening potential are calculated by Salpeter in 1954~[\cite{Salpeter:1954nc}], which describes the static screening effects for the thermonuclear reactions. In an isotropic and homogeneous plasma the Coulomb potential of a point-like particle with charge $Ze$ at rest is modified into~[\cite{Salpeter:1954nc}]
\begin{align}
\phi_\text{stat}(r)=\frac{Ze}{4\pi\epsilon_0 r}e^{-m_Dr},
\end{align}
where $m_D$ is the Debye mass. After that it has been exploited widely in BBN for static screening ~[\cite{1969ApJ...155..183S,Famiano:2016hhs}]. 

Subsequently, the study of dynamical screening for moving ions has been taken into account~[\cite{1988ApJ...331..565C,Gruzinov:1997as,Hwang:2021kno}]. When a test charge moves with a velocity that is enough to react with the background charge in plasma, the Coulomb potential is modified by the dynamical effect. However, the applications focus on the weakly interacting electron-positron plasma only. 

In our separate work~[\cite{Grayson:2023flr}] we use the linear response theory adapted by C.Grayson to to describe the inter nuclear potential in electron-positron plasma during BBN. We improve the prior efforts by evaluation and inclusion of the collision damping rate due to scattering in the dense plasma medium and provide an approximate analytic formula that can be readily used to estimate the effect of screening on internuclear potential. For comprehensive discussion and the application of the damped dynamic screening see~[\cite{Grayson:2023flr}].

%~~~~~~~~~~~~~~~~~~~~~~~~~~~~~~~~~~~~~~~~~~~~~~~~~~~~
\subsection{Magnetization of the electron-positron plasma}

In the present-day Universe, we have magnetic fields~[\cite{giovannini2003magnetized, Kronberg:1993vk,kronberg1994extragalactic}] at various scales and strengths both within galaxies and in deep extra-galactic space far away from matter sources. Current observations suggest the upper and lower bounds for the Extra-Galactic Magnetic Field (EGMF) are given by~[\cite{neronov2010evidence,taylor2011extragalactic,pshirkov2015new,jedamzik2019stringent,vernstrom2021discovery}]
\begin{align}
    \label{egmf}
    10^{-8}{\mathrm G}>B_{\mathrm{EGFM}}>10^{-16}{\mathrm G}\,.
\end{align}
The origin for EGMF today is currently unknown; different models are considered in lectures~[\cite{Widrow:2011hs,Vazza:2021vwy}]. In our work~[\cite{Rafelski:2023emw}], we investigate the hypothesis that the observed EGMF are primordial in nature, predating even the recombination
epoch. Under this hypothesis, the first best candidate is the electron-positron plasma. This is because for the temperature range $ 200\,\mathrm{keV} > T > 20$ keV, we still have relatively large quantity of both $e^\pm$ in the the early Universe plasma. In addition, electrons and positrons have the largest magnetic moments in nature, are likely to have been magnetized in the early Universe due to spin orientation. These  provide the possibility origins for a primordial magnetic field.

As the Universe undergoes the isentropic expansion,  the temperature gradually decreases as $T\propto1/a(t)$, where $a(t)$ represents the scale factor. The assumption is made that the magnetic flux is conserved over comoving surfaces, implying that the primordial relic field is expected to dilute as $B\propto1/a(t)^{2}$~[\cite{Rafelski:2023emw}]. Combining these cosmological redshift relations, we can introduce a dimensionless cosmic magnetic scale that remains unchanged during the evolution of the Universe 
\begin{align}
    \label{tbscale}
    b \equiv\frac{e{B}}{T^{2}}=\left(\frac{e{B}}{T^{2}}\right)_{t_0}=b_0={\rm\ const.}\qquad10^{-3}>b_{0}>10^{-11}\,.
\end{align}
The upper and lower bounds for $b_0$ are estimated by using the present day EGMF observations Eq.~(\ref{egmf}) and the present CMB temperature $T_{0}=2.7\,\mathrm{K}\approx2.3\times10^{-4}$ eV~[\cite{aghanim2018planck}].
As $b_0$ is a constant of expansion, this means the contemporary small bounded values of may have once represented large magnetic fields in the early Universe and require detailed study in a different epoch of the Universe. Therefore, correctly describing the dynamics of this $e^{\pm}$ plasma is of interest when considering the origin of extra-galactic magnetic fields (EGMF). 

In the following,  we will demonstrate that fundamental quantum statistical analysis can lead to further insights on the behavior of magnetized plasma, and show that the $e^\pm$ plasma is overall paramagnetic and yields a positive overall magnetization, which
is contrary to the traditional assumption that matter-antimatter plasma lack significant magnetic responses. For more detailed discussion  of electron-positron plasma magnetization, please see~[\cite{Andrew:2023abc}].

\subsubsection{Electron-positron partition function}
To study the statistical behavior of the $e^\pm$ system in a magnetic field, we utilize the general Fermion partition function~[\cite{Elze:1980er}]
\begin{align}
 \label{PartFunc} \ln\mathcal{Z}=\sum_{\alpha}\ln\left(1+e^{-\beta(E-\eta)}\right)\,,
\end{align}
where $\beta=1/T$, $\alpha$ is the set of all quantum numbers in the system, and $\eta$ is the generalized chemical potential. In the case of a magnetized $e^{\pm}$ system, we consider it as a system of four quantum species: Particles and antiparticles, and spin aligned and anti-aligned. Taken together, we consider a system where electrons and positrons can be spin aligned or anti-aligned with the magnetic field $B$ and the partition function of the system can be written as
\begin{align}\label{PartFuncB}
%&\ln\mathcal{Z}_{tot}=&\frac{2eBV}{(2\pi)^2}\sum_{\sigma}^{\pm1}\sum_{s}^{\pm1/2}\sum_{n=0}^\infty\int^\infty_{0}dp_z\left[\ln\left(1+\Upsilon_{\sigma}^{s}(x)e^{-\beta E_{n}^{s}}\right)\right]\,\\
\ln\mathcal{Z}_{tot}=\frac{2eBV}{(2\pi)^2}\sum_{\sigma}^{\pm1}\sum_{s}^{\pm1/2}\sum_{n=0}^\infty\int^\infty_{0}dp_z\left[\ln\left(1+\Upsilon(x)e^{(\sigma\eta_{e}+s\eta_s)/T}e^{-\beta E_{n}^{s}}\right)\right]\,,
\end{align}
where $n$ is the principle quantum number for the Landau levels. The parameter $\eta_{e}$ is the electron chemical potential and $\eta_s$ is the spin chemical potential~[\cite{Andrew:2023abc}]. The parameter $\Upsilon(x)$ is the fugacity of the Fermi gas. In this thesis we will focus on the case $\Upsilon(x)=1$ and $\eta_s=0$ , 
we leave the general case $\Upsilon(x)\neq1$ and $\eta_s\neq0$ for future work.

%In general, $\Upsilon=1$ represents the maximum entropy and corresponds to the normal Fermi distribution. The deviation of $\Upsilon\neq1$ represents the configurations of reduced entropy without pulling the system off a thermal temperature. This scenario is well studied for quarks in QGP. The situation for $e^\pm$ plasma is similar to the case of the quarks during QGP, but instead here the deviation is spatial rather than temporal. Inhomogeneity can arise from the influence of other forces on the gas such as gravitational forces. This is precisely the kind of behavior that may arise in the $e^{\pm}$ epoch as the dominant photon thermal bath keeps the Fermi gas in thermal equilibrium while spatial inequilibrium could spontaneously develop. 

In the following, we will retain $\Upsilon(x)=1$ and consider the case $\eta_s/T\ll1$ for the first approximation. Then the partition function becomes
\begin{align}
\ln\mathcal{Z}_{tot}=\frac{2eBV}{(2\pi)^2}\sum_{s}^{\pm1/2}\sum_{n=0}^\infty\int^\infty_{0} \!\!dp_z\bigg[\ln\left(1+e^{-\beta(E_{n}^s-\eta_e)}\right)+\ln\left(1+e^{-\beta(E_{n}^s+\eta_e)}\right)\bigg].
\end{align}
Considering the $e^\pm$ plasma in a uniform magnetic field $B$ pointing along the $z$-axis, the energy $E_{n}^\pm$ of electron/positron system can be written as~[\cite{Rafelski:2023emw}]
\begin{align}
&E_{n}^\pm=\sqrt{p^2_z+\tilde m^2_\pm+2eBn},\qquad\tilde{m}^2_\pm=m^2_e+eB\left(1\mp\frac{g}{2}\right)\,,
\end{align}
where the $\pm$ script refers to spin aligned and anti-aligned eigenvalues. The parameter $g$ is the gyro-magnetic ($g$-factor) of the particle. 

To simplify the partition function, we consider the expansion of the logarithmic function as follows:
\begin{align}
\ln\left(1+x\right)=\sum^{\infty}_{k=1}\frac{(-1)^{k+1}}{k}x^k, \,\,\,\,\,\,\,\mathrm{for}\,|x|<1.
\end{align}
Then the partition function of electron/positron system can be written as
\begin{align}
\ln\mathcal{Z}_{tot}=&\frac{2eBV}{(2\pi)^2}\sum_{n=0}^\infty\int^\infty_{0} \!\!dp_z\sum^{\infty}_{k=1}\frac{(-1)^{k+1}}{k}\bigg[e^{k\beta\mu_e}+e^{-k\beta\mu_e}\bigg]e^{-k\beta E_n^\pm}\notag\\
&=\frac{2eBV}{(2\pi)^2}\sum_{n=0}^\infty\sum^{\infty}_{k=1}\frac{(-1)^{k+1}}{k}\bigg[2\cosh{(k\beta\mu_e)}\bigg]\int_0^\infty dp_z\,e^{-k\beta E_n^\pm}.
\end{align}
Using the general definition of Bessel function:
\begin{align}
K_\nu(\beta m)=\frac{\sqrt{\pi}}{\Gamma({\nu-1/2})}\frac{1}{m}\left(\frac{\beta}{2m}\right)^{\nu-1}\int_0^\infty\,dp\,p^{2\nu-2}e^{-\beta E} \,\,\,\,\,\,\,\mathrm{for}\,\nu>1/2,
\end{align}
the integral over $dp_z$ can be written as
\begin{align}
\int_0^\infty dp_z\,e^{-k\beta E_n^\pm}&=\frac{\Gamma{(1/2)}}{\sqrt{\pi}}\sqrt{\tilde{m}^2_\pm+2eBn}\,\,K_1\!\!\left({k\sqrt{\tilde{m}^2_\pm+2eBn}}/{T}\right)\notag\\&=\sqrt{\tilde{m}^2_\pm+2eBn}\,\,K_1\!\!\left({k\sqrt{\tilde{m}^2_\pm+2eBn}}/{T}\right).
\end{align}
In this case, the partition function becomes
\begin{align}
\ln\mathcal{Z}_{tot}&=\frac{2eBV}{(2\pi)^2}\sum_{n=0}^\infty\sum^{\infty}_{k=1}\frac{(-1)^{k+1}}{k}\bigg[2\cosh{(k\beta\mu_e)}\bigg]\sqrt{\tilde{m}^2_\pm+2eBn}\,\,K_1({k\sqrt{\tilde{m}^2_\pm+2eBn}}/{T})\notag\\
&=\frac{2eBTV}{(2\pi)^2}\sum^{\infty}_{k=1}\frac{(-1)^{k+1}}{k^2}\bigg[2\cosh{(k\beta\mu_e)}\bigg]\sum_{n=0}^\infty W^\pm_1(n),
\end{align}
where we introduce the function $W^\pm_1(n)$ as follows
\begin{align}
W^\pm_1(n)\equiv\frac{k\sqrt{\tilde{m}^2_\pm+2eBn}}{T}\,\,K_1\!\!\left({k\sqrt{\tilde{m}^2_\pm+2eBn}}/{T}\right).
\end{align}

Considering the Euler-Maclaurin formula to replace the sum over Landau levels, we have
\begin{align}
\sum^{\infty}_{n=0}W^\pm_1(n)=\int^\infty_0\!\!dn\,W^\pm_1(n)&+\frac{1}{2}\bigg[W^\pm_1(\infty)+W^\pm_1(0)\bigg]\notag\\
&\qquad+\frac{1}{12}\bigg[\left.\frac{\partial W^\pm_1}{\partial n}\right|_{\infty}-\left.\frac{\partial W^\pm_1}{\partial n}\right|_{0}\bigg]+R,
\end{align}
where $R$ is the error remainder which is defined by integrals over Bernoulli polynomials which is small and can be neglected~[\cite{Elze:1980er}]. Using the properties of Bessel function we have
\begin{align}
&\frac{\partial W^\pm_1}{\partial n}=-\frac{k^2eB}{T^2}K_0\left({k\sqrt{\tilde{m}^2_\pm+2eBn}}/{T}\right),\qquad W^\pm_1(\infty)=0,\\
&\int^\infty_a\!\!dx\,x^2K_1(x)=a^2K_2(a),
\end{align}
then we obtain
\begin{align}
\sum^{\infty}_{n=0}W^\pm_1(n)
&=\left(\frac{T^2}{k^2eB}\right)\left[\left(\frac{k\tilde{m}_\pm}{T}\right)^2K_2(k\tilde m_\pm/T)\right]+\frac{1}{2}\left[\left(\frac{k\tilde{m}_\pm}{T}\right)K_1(k\tilde m_\pm/T)\right]\notag\\
&\qquad+\frac{1}{12}\left[\left(\frac{k^2eB}{T^2}\right)K_0(k\tilde m_\pm/T)\right].
\end{align}
Replacing the sum over Landau levels by the integral, the partition function becomes
\begin{align}
\ln\mathcal{Z}_{tot}=\ln\mathcal{Z}_{free}+\ln\mathcal{Z}_B\,,
\end{align}
where we define the partition functions as  
\begin{align}
 \label{FreePart}&\ln\mathcal{Z}_{free}=\frac{T^3V}{2\pi^2}\left[2\cosh{\left(\frac{\eta_{e}}{T}\right)}\right]\sum_{i=\pm}x_i^2K_2\left(x_i\right)\,,\qquad x_i=\frac{\tilde{m}_i}{T}\\
 \label{MagPart}&\ln\mathcal{Z}_B=\frac{eBTV}{2\pi^2}\left[2\cosh{\left(\frac{\eta_{e}}{T}\right)}\right]\sum_{i=\pm}\bigg[\frac{x_i}{2}K_1\left(x_i\right)+\frac{b_0}{12}K_0\left(x_i\right)\bigg]\,.
\end{align}
The partition function $\ln(\mathcal{Z}_{free})$ in Eq.~(\ref{FreePart}) represents the general form of the Fermi partition function for $e^\pm$ with "effective mass" $\tilde{m}_\pm$ in our system. When the magnetic field $B=0$ the function $\ln(\mathcal{Z}_{free})$ will go back to the general form of the Fermi partition function without the external field. The partition function $\ln\mathcal{Z}_B$ gives us the partition with magnetic field effect to the order  $\mathcal{O}(eB)$ and  $\mathcal{O}(eB)^2$.

In the temperature domain $ 200\,\mathrm{keV} > T > 20$ keV, we have $m_e\gg T$, and it suffices to consider the Boltzmann limit of the quantum distributions. Considering the Boltzmann approximation for non-relativistic electrons and positrons we can rewrite Eq.~(\ref{FreePart}) - Eq.~(\ref{MagPart}) and obtain
\begin{align}
 \label{lnZ}
&\ln\mathcal{Z}_{tot}\!=\!\frac{T^3V}{2\pi^2}\left[2\cosh\left(\frac{\eta_{e}}{T}\right)\right]\sum_{i=\pm}\left\{x_i^{2} K_2\left(x_i\right)+\frac{b_0}{2}x_iK_1\left(x_i\right)+\frac{b^2_0}{12}K_0\left(x_i\right)\right\}.
\end{align}
Given the partition function Eq.~(\ref{lnZ}), we can explore the chemical potential and magnetization of $e^\pm$ plasma in the early Universe
under the hypothesis of charge neutrality and entropy conservation.

\subsubsection{Electron chemical potential under magnetic field}
\begin{figure}[ht]
\begin{center}
\includegraphics[width=\linewidth]{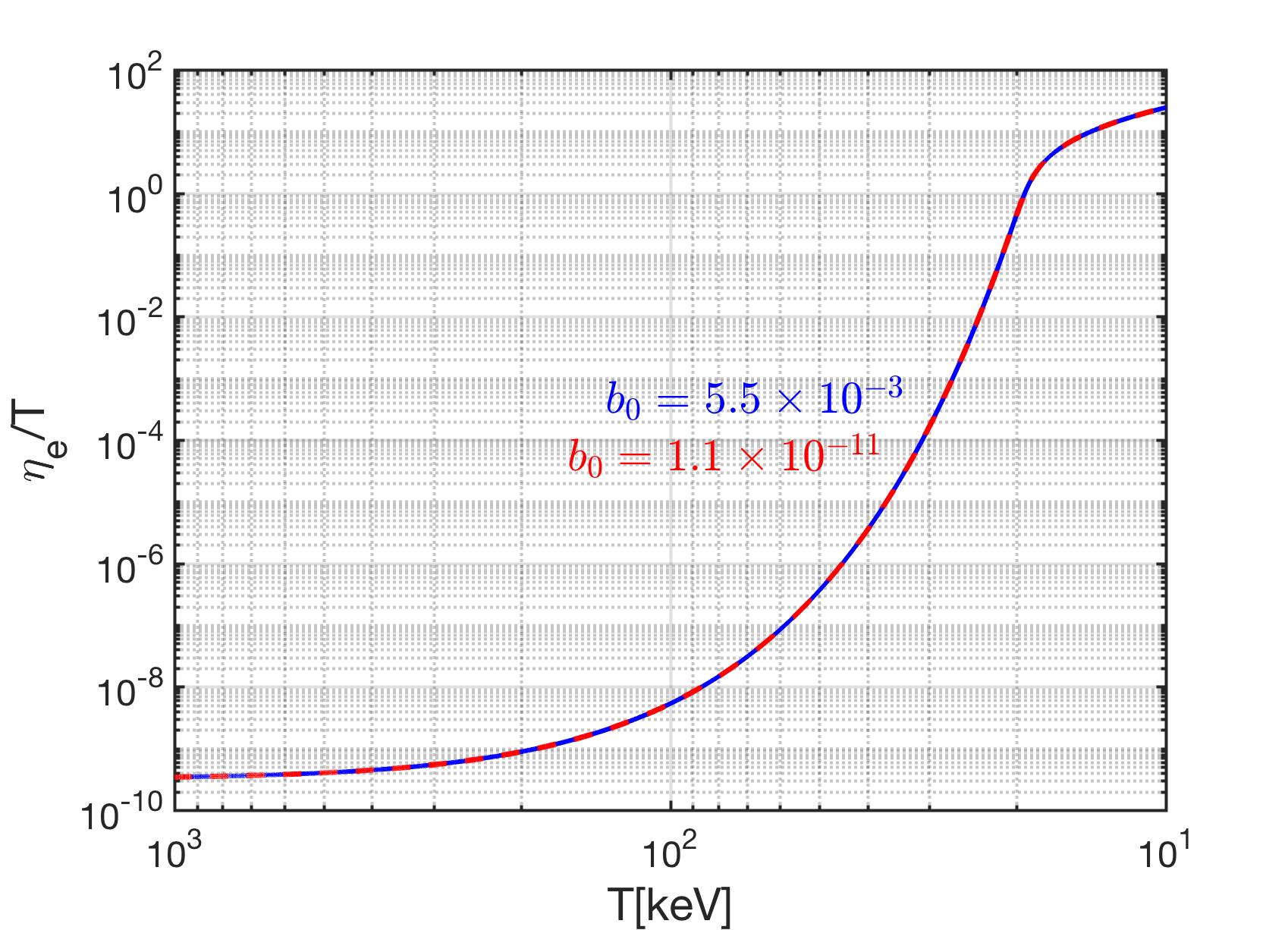}
\caption{The chemical potential of electron as a function of temperature in the magnetic field $b_0$ with $X_p=0.878$ and $n_B/n_\gamma=6.05\times10^{-10}$. The red dashed line represents the magnetic field $b_0=1.1\times10^{-11}$ and blue line labels the magnetic field $b_0=5.5\times10^{-3}$}
\label{ChemicalPotential_B}
\end{center}
\end{figure}

We explore the chemical potential of electron-positron plasma in a uniform magnetic field $B$ in the early Universe under the hypothesis of charge neutrality and entropy conservation. Considering the temperature after neutrino freeze-out, the charge neutrality condition can be written as
\begin{align}
 \label{density_proton}
 \left(n_{e}-n_{\bar{e}}\right)=n_{p}=X_p\,\left(\frac{n_{B}}{s_{\gamma,e}}\right)\,s_{\gamma,e},\qquad X_p\equiv\frac{n_p}{n_B}\,,
\end{align}
where $n_{p}$ and $n_B$ is the number density of protons and baryons respectively. Using the partition function Eq.~(\ref{lnZ}), the net number density of electrons in Boltzmann approximation can be written as
\begin{align}\label{NetElectron}
\left(n_e-n_{\bar e}\right)&=\frac{T}{V}\frac{\partial}{\partial \eta_{e}}\ln\mathcal{Z}_{tot}\notag\\
&=\frac{T^3}{2\pi^2}\left[2\sinh{(\eta_{e}/T)}\right]\sum_{i=\pm}\left[x_i^2K_2(x_i)+\frac{b_0}{2}x_i K_1(x_i)+\frac{b^2_0}{12}K_0(x_i)\right]\,.
\end{align}
Substituting Eq.~(\ref{NetElectron}) into the charge neutrality condition Eq.~(\ref{density_proton}), we can solve the chemical potential of electron $\eta_e/T$ numerically. We have
\begin{align}\label{ChemicalPotential}
\sinh{(\eta_{e}/T)}&=\frac{2\pi^2}{2T^3}\,\frac{X_p(n_B/s_{\gamma,e})s_{\gamma,e}}{\sum_{i=\pm}\left[x_i^2K_2(x_i)+\frac{b_0}{2}x_i K_1(x_i)+\frac{b^2_0}{12}K_0(x_i)\right]}\,,\\
&\longrightarrow\frac{2\pi^2n_p}{2T^3}\,\frac{X_p(n_B/s_{\gamma,e})s_{\gamma,e}}{2x^2K_2(x)},\qquad x=m_e/T,\qquad \mathrm{for}\,\,b_0=0\label{ChemiticalPotential_000}.
\end{align}
We see in Eq.~(\ref{ChemiticalPotential_000}) that for the case $b_0=0$, the chemical potential agrees with the free particle result in~[\cite{Grayson:2023flr}]. In Fig.~\ref{ChemicalPotential_B} we plot the chemical potential of electron as a function of temperature with different value of $b_0$. It shows that the chemical potential is not sensitive to the magnetic field because the small value of $10^{-3}>b_0>10^{-11}$ can be neglected in Eq.~(\ref{ChemicalPotential}).
%~~figure~~~~~~~~~~~~~~~~~~~~~~~~~~~~~~
%~~~~~~~~~~~~~~~~~~~~~~~~~~~~~~~~~~~~~~~~

\subsubsection{Electron-positron magnetization}
%We consider the electron-positron plasma in the mean field approximation where the external field is representative of the \lq\lq bulk\rq\rq\ internal magnetization of the gas. Each particle is therefore responding to the averaged magnetic flux generated by its neighbors as well as any global external field contribution. 

Considering the magnetized electron-positron partition function Eq.~(\ref{lnZ}), it is convenient to introduce the dimensionless magnetization $\overline{\mathcal{M}}$ and the critical field $B_c$ as follows
\begin{align}
\label{Mdef}
\overline{\mathcal{M}}\equiv\frac{M}{\mathcal{B}_{c}}=\frac{1}{\mathcal{B}_{c}}\left(\frac{T}{V}\frac{\partial \ln\mathcal{Z}_{tot}}{\partial B}\right)\,\qquad \mathcal{B}_{c}=\frac{m_{e}^{2}}{e}\,.
\end{align}
Substituting the partition function Eq.~(\ref{lnZ}) into Eq.~(\ref{Mdef}), the total magnetization ${\overline{\mathcal M}}$ can be broken into the sum of spin parallel $\overline{\mathcal M}_{+}$ and spin anti-parallel $\overline{\mathcal M}_{-}$ magnetization. We have
\begin{align}\label{Magnetization}
&{\overline{\mathcal M}}={\overline{\mathcal M}_+}+{\overline{\mathcal M}_-},\\
&\overline{\mathcal M}_{\pm}=\frac{e^2T^{2}}{2\pi^2m_e^2}\left[2\cosh\left(\frac{\eta_{e}}{T}\right)\right]\left\{c_{1}(x_{\pm})K_1(x_i)+c_{0}K_0(x_\pm)\right\}\,,
\end{align}
where the coefficients are given by
\begin{align}
    c_{1}(x_{\pm}) &= \left[\frac{1}{2}-\left(\frac{1}{2}\pm\frac{g}{4}\right)\left(1+\frac{b^2_0}{12x^2_\pm}\right)\right]x_\pm\,,\qquad c_{0} = \left[\frac{1}{6}-\left(\frac{1}{4}\pm\frac{g}{8}\right)\right]b_0\,.
\end{align}
Substituting the chemical potential Eq.~(\ref{ChemicalPotential}) into Eq.~(\ref{Magnetization}), we can solve the magnetization numerically.
%~Figure~~~~~~~~~~~~~~~~~~~~~~~~~~~~~
\begin{figure}[ht]
    \centering
    \includegraphics[width=\textwidth]{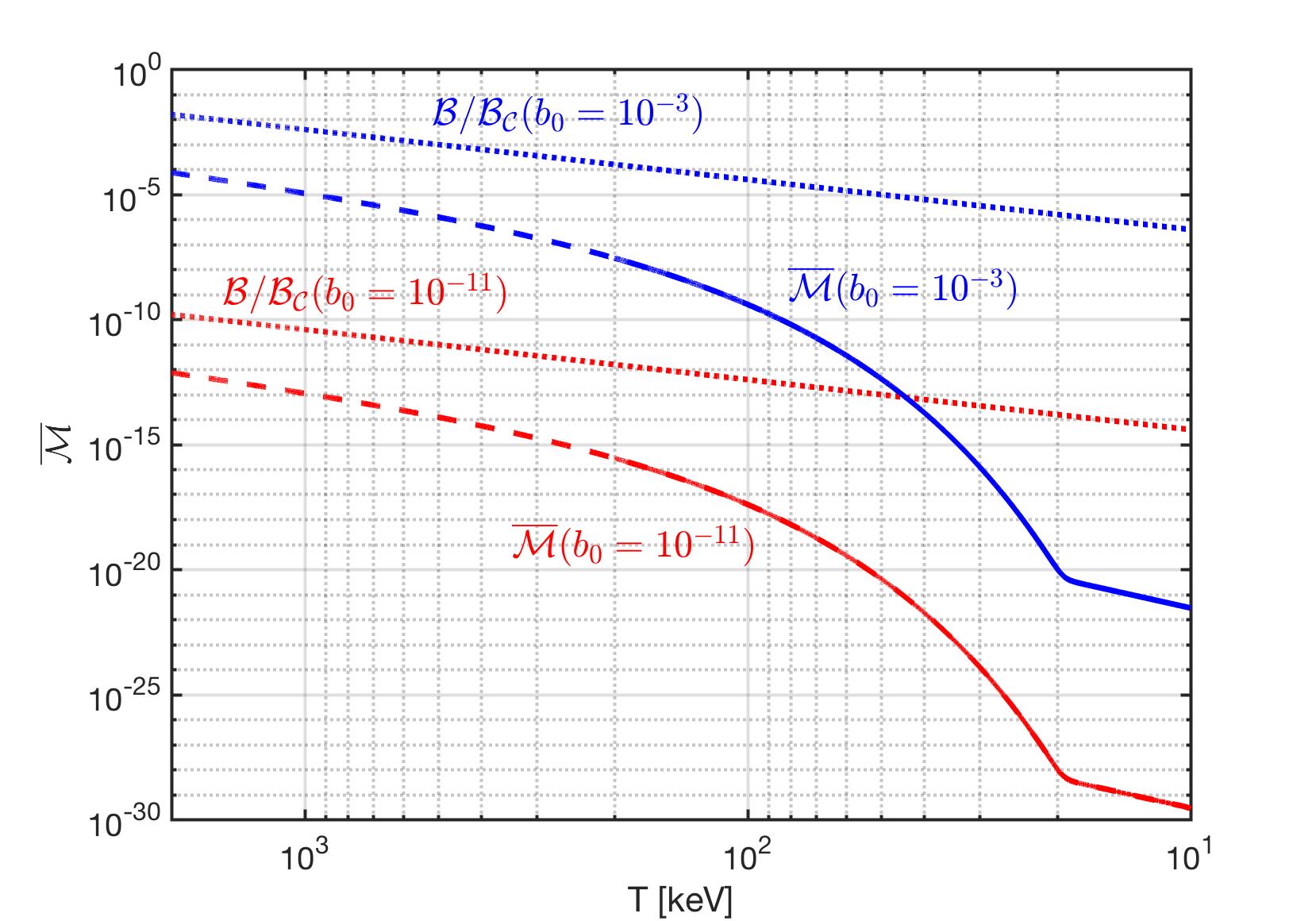}
    \caption{The magnetization $\overline{\cal M}=\mathcal{M}/\mathcal{B}_C$, with $g=2$, of the primordial $e^{+}e^{-}$ plasma is plotted as a function of temperature. The lower (solid red) and upper (solid blue) bounds for cosmic magnetic scale $b_{0}$ are included. The external magnetic field strength ${\cal B}/{\cal B}_{C}$ is also plotted in for lower (dashed red) and upper (dashed blue) bounds. The spin fugacity is set to unity.}
    \label{fig:magnet} 
\end{figure}
%~Figure~~~~~~~~~~~~~~~~~~~~~~~~~~~~~

In this thesis we focus on considering the case for $g=2$. In this case, the electron-positron magnetization can be written as 
\begin{align}\label{Magnetization_g2}
&{\overline{\mathcal M}}={\overline{\mathcal M}_+}+{\overline{\mathcal M}_-}\\
&{\overline{\cal M}}_{+}=\frac{e^{2}}{\pi^{2}}\frac{T^{2}}{m_{e}^{2}}\cosh{\frac{\eta_e}{T}}\left[\frac{1}{2}x_{+}K_{1}(x_{+})+\frac{b_{0}}{6}K_{0}(x_{+})\right]\,,\\
&{\overline{\cal M}}_{-}=-\frac{e^{2}}{\pi^{2}}\frac{T^{2}}{m_{e}^{2}}\cosh{\frac{\eta_e}{T}}\left[\left(\frac{1}{2}+\frac{b_{0}^{2}}{12x_{-}^{2}}\right)x_{-}K_{1}(x_{-})+\frac{b_{0}}{3}K_{0}(x_{-})\right]\,,
\end{align}
where $x_\pm$ are given by
\begin{align}
x_{+}=\frac{m_{e}}{T},\qquad   x_{-}=\sqrt{\frac{m_{e}^{2}}{T^{2}}+2b_{0}}
\end{align}
The discussion for the case $g\neq2$ can be found in~~[\cite{Andrew:2023abc}].

In Fig.~\ref{fig:magnet}, we present the magnetization Eq.~(\ref{Magnetization_g2}) for the case $g=2$ as a function of temperature. It shows that the magnetization depends on the magnetic scale $b_0$ and the $e^{+}e^{-}$ plasma possesses an overall paramagnetic property, resulting in a positive magnetization $\overline{\mathcal{M}}$. This paramagnetic property is contrary to the conventional assumption that matter-antimatter plasmas lack significant inherent magnetic responses. However, the magnetization never exceeds the external field under the parameters considered, which shows a lack of ferromagnetic behavior. As the Universe cooled, the dropping magnetization slowed at $T_{\mathrm{split}}=20.3$ keV, where positrons vanished. Thereafter the remaining electron density diluted with cosmic expansion.

In this section, we have explored the electron-positron plasma considering  external and self-magnetization fields  without spin potential $\eta_s/T\ll1$. However the nonzero spin potential $\eta_s\neq0$  would have an impact on the primordial $e^{+}e^{-}$ plasma. In general, the  magnetization is also a function of the spin potential $\eta_s$, and would be one important parameter that control the spin direction of primordial gas which allows for magnetization even in the absence of external magnetic fields. For further discussion see ~[\cite{Andrew:2023abc}].

%% file: Chapter6.tex
\chapter{{Outlook: research publications underway}}\label{Outlook}
In this chapter, I address the ongoing research projects I hope to complete by the end of year $2023$, which include: The low temperature baryongenesis originating in nonequilibrium of bottom flavor; Population of Higgs in the early Universe; Extra neutrino from microscopic processes after freeze-out; and Self-consistent relaxation rate for electron-positron plasma in the early Universe.%; Effective screening potential in QGP/hadronic plasma to study heavy particle condensation.
.

\section{{Possibility of bottom-catalyzed low temperature baryongenesis}}
From the PDG, the observed baryon-to-photon density ratio $\eta$ today is given by  $5.8\times10^{-10} \leqslant\eta\leqslant6.5\times10^{-10}$ [\cite{ParticleDataGroup:2018ovx}] where the $\eta=(6.12\pm0.04)\times10^{-10}$~[\cite{ParticleDataGroup:2022pth}] is used in our calculation. This observed value is the evidence of baryon asymmetry and quantifies the matter-antimatter asymmetry in the Universe. In the past, the small value of the baryon asymmetry could be interpreted as simply due to the initial conditions in the Universe. However, in the current standard cosmological model, the inflation can erase any pre-existing asymmetry between baryons and anti-baryons. In this case, we need baryogenesis to generate of excess of baryon number compared to anti-baryon number in order to create the observed baryon number today.

The precise epoch responsible for the observed matter genesis $\eta$  in the early Universe has not been established yet. 
Several mechanisms have been proposed to explain baryogenesis with investigations typically focusing on the temperature range between GUT phase transition $T_\mathrm{G}\simeq10^{16}\,\mathrm{GeV}$ and the electroweak phase transition near $T_\mathrm{W}\simeq130\,\mathrm{GeV}$~[\cite{Kuzmin:1985mm,Kuzmin:1987wn,Arnold:1987mh,Kolb:1996jt,Riotto:1999yt,Nielsen:2001fy,Giudice:2003jh,Davidson:2008bu,Morrissey:2012db}].

In following section we present arguments that the Sakharov conditions~[\cite{Sakharov:1967dj}] for matter asymmetry to form also appear during QGP hadronization era near to $T_\mathrm{H}\simeq150\,\mathrm{MeV}$, and show the possibility that bottom catalyzed the baryongenesis.

\subsubsection{Overview of Sakhraov conditions}

In 1967, the Soviet physicist Sakharov first formulated the three conditions necessary to permit baryongenesis in the early Universe~[\cite{Sakharov:1967dj}] and in 1991 he refined the three conditions as follows~[\cite{Sakharov:1988vdp}]:
\begin{itemize}
  \item Absence of baryonic charge conservation 
  \item Violation of CP-invariance
  \item Non-stationary conditions in absence of local thermodynamic equilibrium
\end{itemize}

%The first condition, a violation of baryon $B$, lepton $L$ number conservation coincident around hadronization era, needs future theoretical and experimental consideration, constrained by the experimental limit on  proton life span $\mathcal{O}(10^{32}\mathrm{y}$). The relevant thermal environment can be explored in the laboratory since $T_\mathrm{H}=150$\,MeV is readily available in relativistic heavy ion (RHI) collision experiments~[\cite{Rafelski:2019twp}]. However, $T_\mathrm{H}$  may not suffice for catalysis of baryogenesis~[\cite{Kuzmin:1985mm,Kuzmin:1987wn,Arnold:1987mh}]. The second Sakharov condition requiring $CP$ assures us that we can recognize a universal difference between matter and antimatter, thus one abundance can be enhanced compared to the other.

\noindent Since the inflation erases any initial asymmetry between baryons and anti-baryons, the first condition, a violation of baryon number $B$, must exist to have the present baryon asymmetry. The second Sakharov condition requiring $CP$ violation assures us that we can recognize a universal difference between matter and antimatter, thus one abundance can be enhanced compared to the other.

The third condition, departure from thermal equilibrium, is one of the important prerequisites for matter genesis, this is because it creates the arrow in time for the Universe. In general, the thermal equilibrium implies both chemical (abundance) equilibrium and kinetic (equipartition of energy) equilibrium. The observed baryon and lepton numbers cannot be generated in a full thermal (chemical and kinetic) equilibrium, because even if the required processes are occurring, the net effect is cancelled out by the equal number of back-reactions. We believe that the presence of chemical non-equilibrium is more relevant -- kinetic equilibrium is usually established much more quickly and has less impact on the actual particle abundances~[\cite{Koch:1986ud,Birrell:2014gea}]. 

Chemical non-equilibrium can be achieved by breaking the detailed balance between particle production reaction and annihilation/decay. 
When the Universe expands and temperature cools down, the production process slows down and is not able to keep up with decay reactions. Then the detailed balance is broken and creates the arrow in time for the Universe. In this case, the third condition of Sakharov can be interpreted as:
\begin{itemize}
  \item Non-stationary conditions in absence of detailed balance,
\end{itemize}
where the ideal thermodynamic equilibrium is generalized to the concept of detailed balance.

%~~~~~~~~~~~~~~~~~~~~~~~~~~~~~~~~~~~~~~~~~~~~~~~~~

%\subsection{Possibility of bottom-catalyzed matter genesis}
%Given that the non-equilibrium of bottom flavor arises at relatively low QGP temperature allows the Sakharov conditions around QGP hadronization. In this section, our interest is to study what conditions in the primordial QGP-HG phase transition can yield the observed baryon asymmetry today and provide the values for experiment investigation. 

 \subsubsection{Is there enough bottom flavor to matter?} Considering that FLRW-Universe evolves conserving entropy, and that baryon and lepton number following on the era of matter genesis is conserved, the current day baryon $B$ to entropy $S$, $B/S$-ratio must be achieved during matter genesis. The estimates of present day baryon-to-photon density ratio $\eta$ allows the determination of the present value of baryon per entropy ratio [\cite{Rafelski:2019twp,Letessier:2002ony,Fromerth:2002wb,Fromerth:2012fe}]:
\begin{align}
\left(\frac{B}{S}\right)_{t_0}\!\!\!\!=\eta\left(\frac{n_\gamma}{\sigma_\gamma+\sigma_\nu}\right)_{\!t_0}\!\!\!\!=(8.69\pm0.05)\!\!\times\!\!10^{-11},
\end{align}
where the subscript $t_0$ denotes the present day value, where $\eta=(6.12\pm0.04)\times10^{-10}$~[\cite{ParticleDataGroup:2018ovx}] is used in calculation. Here we consider that the Universe today is dominated by photons and free-streaming low mass neutrinos~[\cite{Birrell:2012gg}], and $\sigma_\gamma$ and $\sigma_\nu$ are the entropy density for photons and neutrinos, respectively. 
 
In chemical equilibrium the ratio of bottom quark (pair) density $n_b^{th}$ to entropy density $\sigma=S/V$ just above quark-gluon hadronization temperature $T_\mathrm{H}=150\sim160\,\mathrm{MeV}$ is $n_b^{th}/\sigma=10^{-10}\sim10^{-13}$ (see Fig.~\ref{number_entropy_b002}). By studying the bottom density per entropy near to the hadronization temperature and comparing it to the baryon-per-entropy ratio $B/S$  we found there is sufficient abundance of bottom quarks for the proposed matter genesis mechanism to be relevant.

\subsubsection{Non-stationary conditions in absence of detailed balance}
We have demonstrated that the bottom quark nonequliibrium occurs near to QGP phase transition around the temperature $T=0.3\sim0.15$ GeV in Fig.~\ref{fugacity_bc} and Fig.~\ref{NonFugacity}. We demonstrate that the competition between weak interaction decay and the strong interaction fusion processes is responsible for driving the bottom quark departure from the equilibrium in the early Universe. In all cases we see prolonged non-equilibrium which provides the arrow of time for baryongenesis.

%After formation, the heavy $b, \bar b$ quark can bind with any of the available lighter quarks. %Here, we assume that due to the enhanced binding effect of $\mathrm{B}_c^\pm$ all bottom $b$, $\bar b$ quarks are found in $\mathrm{B}_c^\pm$ eventually.
%Formation of a matter excess over antimatter at a relatively late Universe evolution period has the additional merit of depending on an experimentally accessible Universe evolution stage: Further insight can be derived from  experimental study of bottom flavor in RHI collisions, and  more generally, exploration of the  the properties of bottom flavored particles, including but not limited to the search for a specific baryon non-conservation mechanism.

\subsubsection{Violation of $CP$-invariance}
In general, violation of $CP$ asymmetry can occur in the amplitudes of hadron decay. The weak interaction $CP$ violation arises from the components of Cabibbo-Kobayashi-Maskawa (CKM) matrix associated with quark-level transition amplitude and $CP$-violating phase.

Given that the non-equilibrium of bottom flavor arises at relatively low QGP temperature, the bottom quark decay occurs from preformed $\mathrm{B}_x$ meson states, $x=u,d,s,c$~[\cite{Karsch:1987pv,Brambilla:2010vq,Aarts:2011sm,Brambilla:2017zei,Bazavov:2018wmo,Offler:2019eij}]. These decays violate aside of the $CP$ symmetry, see for example~[\cite{LHCb:2019jta,LHCb:2020vut}]. The exploration of the here interesting $CP$ symmetry breaking in B$_c(b\bar c)$ decay is in progress~[\cite{Tully:2019ltb,HFLAV:2019otj,ParticleDataGroup:2018ovx}]. 
 Present measurements of $CP$-violation suggest that the CP asymmetry parameter is around $\delta_{CP}\approx10^{-3}$ ~[\cite{ParticleDataGroup:2018ovx}].

\subsubsection{Nonequilibrium bottom and baryon asymmetry}
 The off equilibrium phenomenon of bottom quark around the temperature range $T=0.3\sim0.15$ GeV can provide the non-chemical equilibrium condition for baryogenesis to occur in the primordial-QGP hadronization era. Furthermore, let us consider the scenario where all bottom quarks are confined within $B_c^\pm$ meson. In this case, the decay of charged mesons in the primordial-QGP can be a source of $CP$ violation. However, it remains uncertain whether the decay of $B_c^\pm$ mesons contributes to baryon violation. Our postulation is as follows: the baryon asymmetry is produced by the bottom quark disappearance via the irreversible decay of $B^\pm_c$ meson during the off-equilibrium process. Once a baryon symmetry exists in universe, it will also produce the asymmetry between leptons and anti-leptons which is similar to the baryon asymmetry by the $L=B$.

The heavy $B_c^\pm$ meson decay into multi-particles in plasma is associated with the irreversible process. This is because after decay the daughter particles can interact with plasma and distribute their energy to other particles and reach equilibrium with the plasma quickly. In this case the  energy required for the inverse reaction to produce $B_c^\pm$ meson is difficult to overcome and therefore we have an irreversible process for multi-particle decay in plasma.

The rapid $B_c^\pm$ decay and bottom reformation speed at picosecond scale assures that there are millions of individual microscopic processes involving bottom quark production and decay before and during the hadronization epoch of QGP. In this case, we have an Urca process for the bottom quark, i.e. a cycling reaction that produces the bottom quark which subsequently  disappears via the $B_c^\pm$ meson decay. The Urca process is a fundamental physical process  and has been stuying  the realms of in astrophysics and nuclear physics. In our case, for bottom quark as a example: at low temperature, the number of bottom quark cycling can be estimated as
\begin{align}
\left.\mathrm{C_{cycle}}\right|_{T=0.2\mathrm{GeV}}=\frac{\tau_H}{\tau_{B_c}}\approx2\times10^7,
\end{align}
where the lifespan of $B_c^\pm$ is  $\tau_{\mathrm{B}_c}\approx0.51\times10^{-12}\,\mathrm{sec}$ and at temperature $T=0.2$ GeV the Hubble time is $\tau_H=1/H=1.272\times10^{-5}$ sec. The Urca process plays a significant role by potentially amplifying any small and currently unobserved violation of baryon number associated with the bottom quark. The small baryon asymmetry is enhanced by the Urca-like process with cycling ${\tau^\ast_H}/{\tau_\ast}$ in the early Universe.
This amplification would be crucial for achieving the required strength for today's observation.

Full understanding of the possibility for baryogenesis to occur in primordial-QGP hadronization still requires a detailed study of the relation between baryon asymmetry and irreversible process; we are working on it now. Here, our interest is to show that our results provide a strong  motivation to explore the physics of baryon nonconservation involving the bottomnium mesons or/and bottom quarks in thermal environment.

%%%%%%%%%%%%%%%%%%%%%%%%%%%%%%%%%%%%%%%%%%%%%%%
\section{{Population of Higgs in the early Universe}}
In our earlier study, the bottom $(b)$ quark abundance depends on the competition between the strong interaction fusion and weak interaction decay rates. This lead to the off equilibrium phenomenon of the bottom quark near the hadronization temperature. The same idea can be applied to other heavy particles in QGP. In following we focus on the Higgs abundance first and develop methods for future detailed study.

Considering the temperature range $10\,\mathrm{GeV}>T>1\,\mathrm{GeV}$ in the early Universe, the number density of the Higgs can be written as
\begin{align}
n_{H}=\frac{\Upsilon_H}{2\pi^2}T^3\left(\frac{m_H}{T}\right)^2 K_2(m_H/T),
\end{align}
where $\Upsilon_H$ is the Higgs fugacity parameter, and $m_H=125$ GeV is the mass of Higgs. In the temperature range we consider here, we have $m_H\gg T$ which allows us to consider the Boltzmann limit for the calculation of Higgs number density.
%%%%%%%%%%%%%%%%%%%%%%%%%%%%%%%%%%%%%%%%%%%%%%%%%%%%%%%%%%%%%%%%%%
\begin{figure}[ht]
\begin{center}
\includegraphics[width=\textwidth]{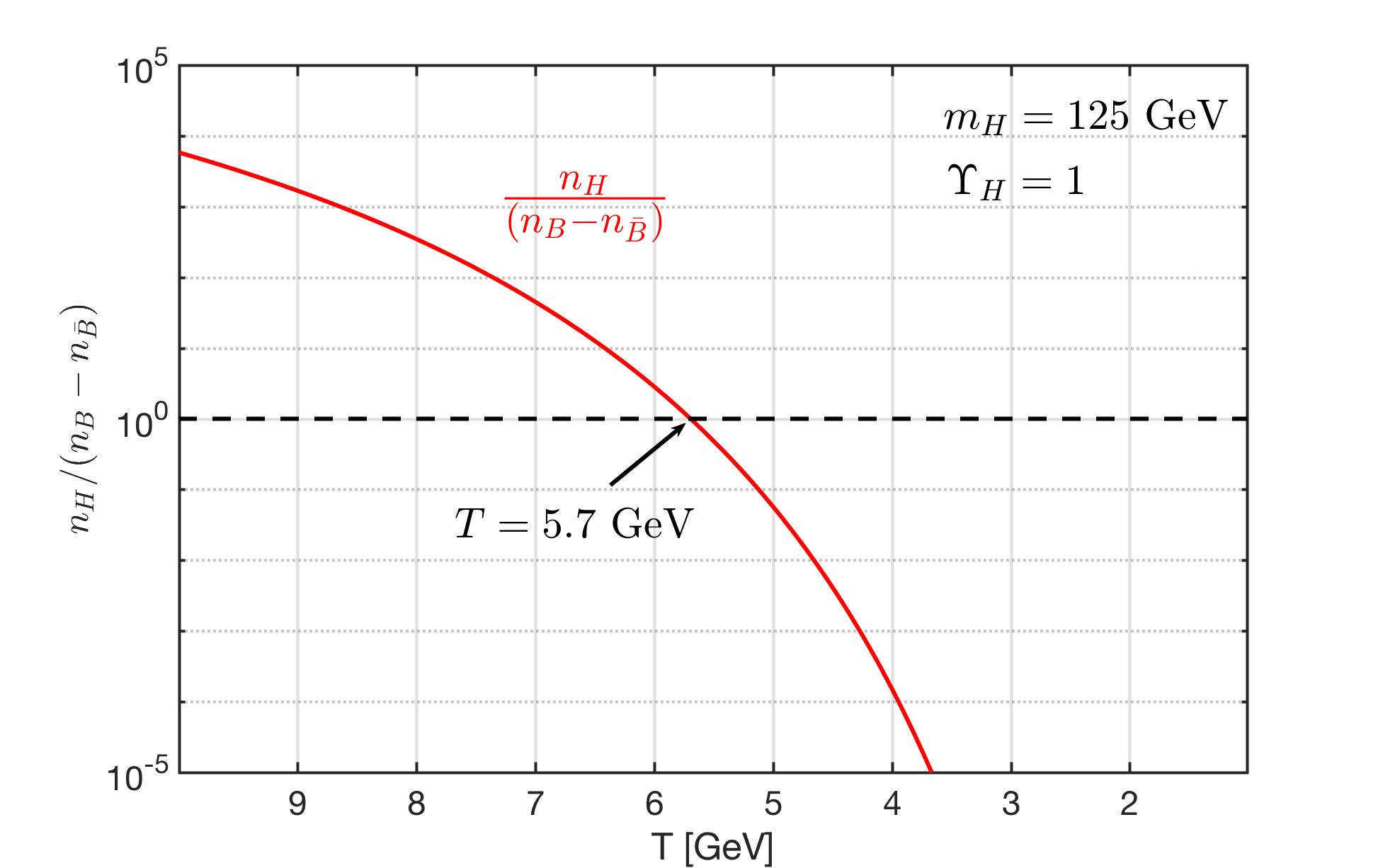}
\caption{The density ratio between Higgs and baryon asymmetry as a function of temperature with condition $\Upsilon_H=1$. It shows that the $n_H=(n_B-n_{\bar{B}})$ at temperature $T=5.7$ GeV.}
\label{HiggsDensity_fig}
\end{center}
\end{figure}
%%%%%%%%%%%%%%%%%%%%%%%%%%%%%%%%%%%%%%%%%%%%%%%%%%%%%%%%%%%%%%%%%%%
Using constant baryon-per-entropy ratio, the density between Higgs and baryon asymmetry ($u,d$ quark-antiquark asymmetry) can be written as
\begin{align}
\frac{n_H}{(n_B-n_{\bar{B}})}=\frac{n_{H}}{s_{tot}}\,\left(\frac{s_{tot}}{n_B-n_{\bar{B}}}\right)=
\frac{n_{H}}{s_{tot}}\left(\frac{s_{\gamma,\nu}}{n_B-n_{\bar{B}}}\right)_{\!t_0},
\end{align}
where the present day value of baryon per entropy ratio is given by Eq.~(\ref{BaryonEntropyRatio}). The entropy density in QGP can be written as
\begin{align}
    &s_{tot}=\frac{2\pi^2}{45}g^s_\ast T_\gamma^3,\qquad g^s_\ast=\sum_{i=\mathrm{g},\gamma}g_i\left({\frac{T_i}{T_\gamma}}\right)^3+\frac{7}{8}\sum_{i=l^\pm,\nu,u,d}g_i\left({\frac{T_i}{T_\gamma}}\right)^3,
\end{align}
where we consider the massless particles in QGP only as a first estimation. In Fig.~\ref{HiggsDensity_fig}, we plot the density ratio between Higgs and baryon asymmetry for the case $\Upsilon_H=1$. It shows that at temperature $T=5.7$ GeV, the ratio is equal to one, i.e., $n_H=(n_B-n_{\bar{B}})$ which means that the Higgs abundance could influence baryon evolution because the Higgs number density is comparable to the baryon number density. This result motivated us to examine the Higgs boson and to study its dynamic abundance in detail during the QGP epoch.

In the QGP epoch, the dominant production of the Higgs boson is the bottom fusion reaction: 
\begin{align}
b+\overline{b}\longrightarrow H,
\end{align}
which is the inverse decay process of $H\to b+\overline{b}$. On the other hand, Higgs abundance disappears via the $W,Z$ decay channel as follows:
\begin{align}
H\longrightarrow WW^\ast, ZZ^\ast\longrightarrow\mathrm{anything}.
\end{align}
where $W^\ast,Z^\ast$ represent the virtual bosons. Once Higgs decays into the $W$ and $Z$ bosons, the short lifespan of $W,Z$ mean they decay into multi-particles and reequilibrate with the plasma quickly. In this case
the energy required for the inverse decay reaction to produce the Higgs boson is difficult to overcome in the temperature we are interested in. Similar to the case of bottom quark we studied, the competition between the production and decay of Higgs require the dynamic study of the particle abundance in the early Universe. We aim to apply the knowledge from our study of bottom quark to the Higgs boson, and to examine all other possible sources for Higgs production in QGP and developing methods for future study before the end of year 2023.

\section{{After neutrino freeze-out: Extra neutrinos from microscope processes}}

After neutrinos chemical freeze-out, the number of neutrinos is independently conserved. However, the presence of electron-positron rich plasma until $T=20$ keV permits the reaction $\gamma\gamma\to e^-e^+\to\nu\bar{\nu}$ to occur even after neutrinos decouple from the cosmic plasma. This suggests the small amount of extra neutrinos can be produced until temperature $T=20$ keV and can modify the free streaming distribution and the effective number of neutrinos. In this section, we examine the possible source of extra neutrino from electron-positron plasma and develop methods for
future detailed study.

\begin{figure}[ht]
\begin{center}
\includegraphics[width=\textwidth]{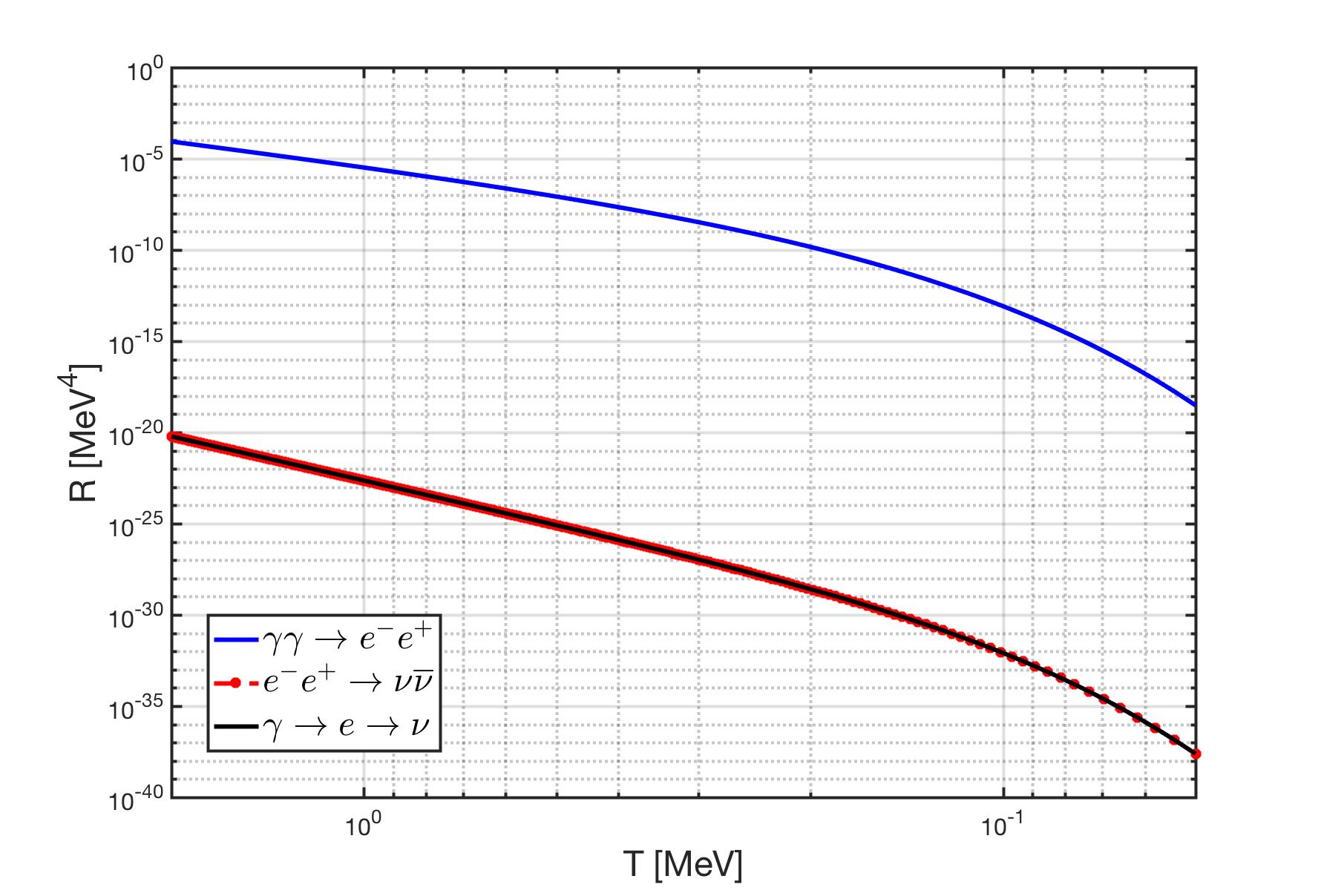}
\caption{The thermal reaction rate per volume as a function of temperature $2\,\mathrm{MeV}>T>0.05\,\mathrm{MeV}$. The dominant reaction for the process $\gamma\gamma\to e^-e^+\to\nu\bar{\nu}$ is the $e\overline{e}\to\nu\overline{\nu}$ and we have $R_{\gamma\to e\to\nu}=R_{e\overline{e}\to\nu\overline{\nu}}$.}
\label{ExtraNeutrinoRate}
\end{center}
\end{figure}

Considering that neutrinos decouple at $T_f=2$ MeV and become free streaming after freeze-out. The presence of electron-positron plasma environment from $2\,\mathrm{MeV}>T>0.02$ MeV can allow the following weak reaction to occur:
\begin{align}
\gamma+\gamma\longrightarrow e^-+e^+\longrightarrow \nu+\bar{\nu}.
\end{align}
Given the  thermal reaction rate per volume $R_{\gamma\gamma\to e\overline{e}}$ for reaction $\gamma\gamma\to e\overline{e}$ and $R_{e\overline{e}\to\nu\overline{\nu}}$ for reaction $e\overline{e}\to\nu\overline{\nu}$, then the thermal reaction rate per volume for $\gamma\gamma\to e^-e^+\to\nu\bar{\nu}$ can be written as
\begin{align}
R_{\gamma\to e\to\nu}=R_{\gamma\gamma\to e\overline{e}}\left(\frac{R_{e\overline{e}\to\nu\overline{\nu}}}{R_{\gamma\gamma\to e\overline{e}}+R_{e\overline{e}\to\nu\overline{\nu}}}\right)\approx R_{e\overline{e}\to\nu\overline{\nu}}
\end{align}
In Fig.~\ref{ExtraNeutrinoRate} we plot the thermal reaction rate per volume for relevant reactions as a function of temperature $2\,\mathrm{MeV}>T>0.05\,\mathrm{MeV}$. It shows that the dominant reaction for the process $\gamma\gamma\to e^-e^+\to\nu\bar{\nu}$ is the $e\overline{e}\to\nu\overline{\nu}$ and can be approximated $R_{\gamma\to e\to\nu}=R_{e\overline{e}\to\nu\overline{\nu}}$ in the temperature we are interested in.
%%%%%%%%%%%%%%%%%%%%%%%%%%%%%%%%%%%%%%%%%%%%%%%%%%%%%%%%%%%%%%%%%%
%%%%%%%%%%%%%%%%%%%%%%%%%%%%%%%%%%%%%%%%%%%%%%%%%%%%%%%%%%%%%%%%%%%

Given the thermal reaction rate, the dynamic equation describing the relic neutrino abundance after freeze-out can be expressed as:
\begin{align}\label{ExtraNeutrio_eq}
\frac{dn_\nu}{dt}+3Hn_\nu=R_{e\overline{e}\to\nu\overline{\nu}}(T_{\gamma,e^\pm})-R_{\nu\overline{\nu}\to e\overline{e}}(T_\nu),
\end{align}
where $n_\nu$ is the number density of neutrinos and $H$ is the Hubble parameter. The parameter $T_{\gamma,e^\pm}$ is the equilibrium temperature between photons and $e^\pm$ and $T_\nu$ is the temperature for free-streaming neutrinos: 
\begin{align}
T_\nu=\frac{a(t_f)}{a(t)}T_f,
\end{align}
where $T_f$ is the neutrino freezeout temperature. After neutrinos decoupled from the cosmic plasma, we have $T_\nu\neq T_{\gamma,e^\pm}$. This is because
the conservation of entropy,  after freezeout, the relic neutrino entropy is conserved independently and the entropy from $e^\pm$ annihilation flows solely into photons and reheats the photons' temperature. However, after neutrino freezeout, extra entropy from electron-positron plasma can still flow into the free-streaming neutrino sector via the reaction $\gamma\gamma\to e^-e^+\to\nu\bar{\nu}$. To describe this novel situation, kinetic theory for entropy production needs to be adapted, a topic we will address in the future. Here we neglect this extra entropy and consider the standard scenario for first approximation.

\begin{figure}[ht]
\begin{center}
\includegraphics[width=\textwidth]{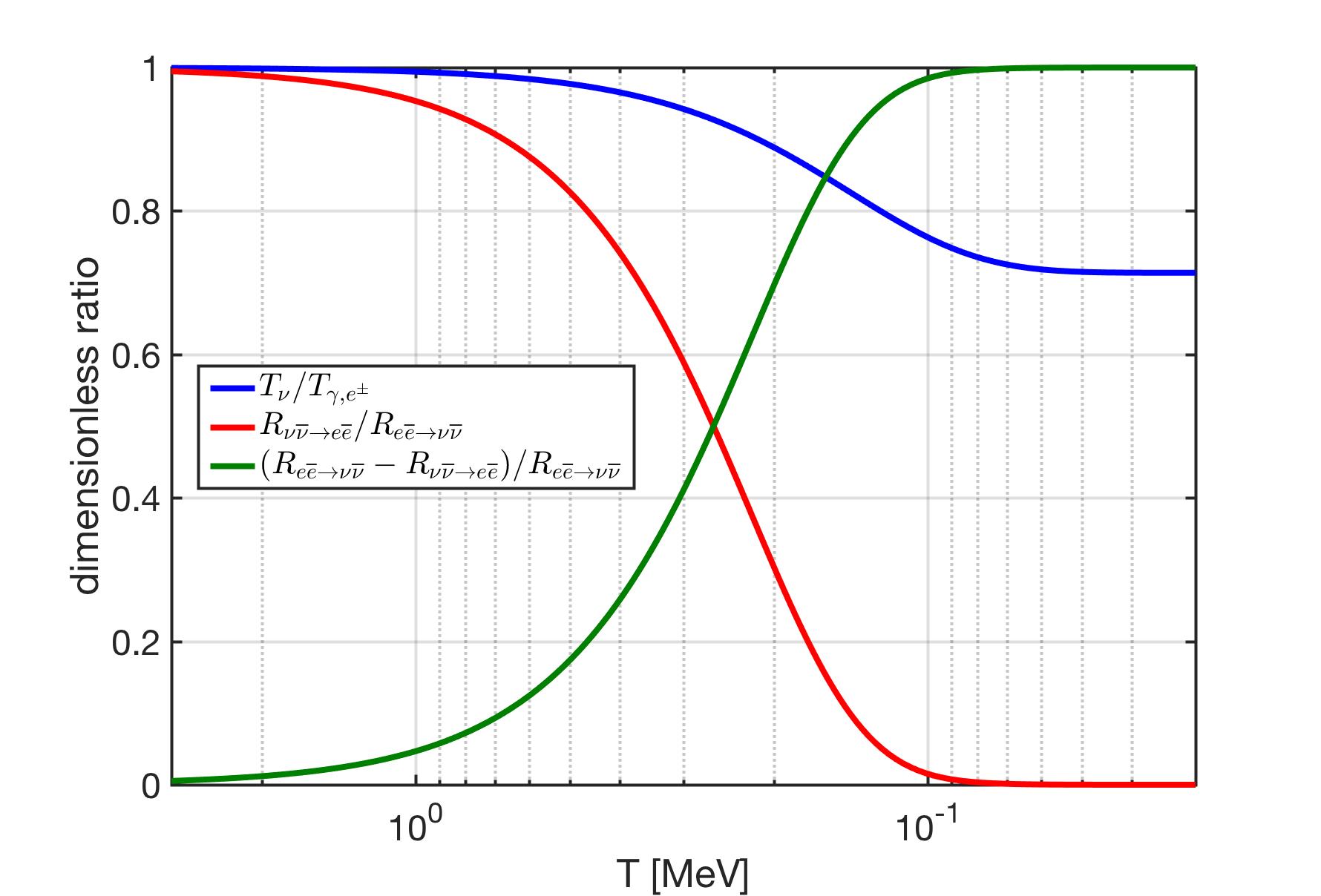}
\caption{The temperature ratio $T_\nu/T_{\gamma,e^\pm}$ (blue line), the rate ratio $R_{\nu\overline{\nu}\rightarrow e\overline{e}}/R_{e\overline{e}\rightarrow\nu\overline{\nu}}$ (red line) and $(R_{e\overline{e}\rightarrow\nu\overline{\nu}}-R_{\nu\overline{\nu}\rightarrow e\overline{e}})/R_{e\overline{e}\rightarrow\nu\overline{\nu}}$ (green line) as a function of temperature. It shows that the reaction $\nu\overline{\nu}\rightarrow e\overline{e}$ is small compare to the reaction $e\overline{e}\rightarrow\nu\overline{\nu}$ as temperature cooling down. }
\label{DimensionlessRatio}
\end{center}
\end{figure}

In Fig.~\ref{DimensionlessRatio} we plot the temperature ratio $T_\nu/T_{\gamma,e^\pm}$, the rate ratio $R_{\nu\overline{\nu}\rightarrow e\overline{e}}/R_{e\overline{e}\rightarrow\nu\overline{\nu}}$ and $(R_{e\overline{e}\rightarrow\nu\overline{\nu}}-R_{\nu\overline{\nu}\rightarrow e\overline{e}})/R_{e\overline{e}\rightarrow\nu\overline{\nu}}$ as a function of temperature. It shows that after neutrino freezeout, the back reaction $\nu\overline{\nu}\rightarrow e\overline{e}$ becomes smaller compared to the reaction $e\overline{e}\rightarrow\nu\overline{\nu}$ as the temperature cools down. This is because as $T_\nu$ cools down, the density of relic neutrinos becomes so low and their energy becomes too small to interact. However, the hot and rich electron-positron plasma can still annihilate into neutrino pairs without any difficulties.
%%%%%%%%%%%%%%%%%%%%%%%%%%%%%%%%%%%%%%%%%%%%%%%%%%%%%%%%%%%%%%%%%
%%%%%%%%%%%%%%%%%%%%%%%%%%%%%
%%%%%%%%%%%%%%%%%%%%%%%%%%%%%%%%%%%%%%%%%%%%%%%%%%%

Solving the dynamic equation of neutrino abundance Eq.(\ref{ExtraNeutrio_eq}), the general solution can be written as
\begin{align}
n_\nu(T)=n_\mathrm{relic}(T)+n_\mathrm{extra}(T),\qquad T=T_{\gamma,e^\pm},
\end{align}
where $n_\mathrm{relic}$ represents the relic neutrino number density and $n_\mathrm{extra}$ is the extra number density from the $e^\pm$ annihilation. The relic neutrino density is given by
\begin{align}  &n_\mathrm{relic}=n_\nu^0\exp\left(-3\int_{t_i}^t{dt^\prime}H(t^\prime)\right)=n_\nu^0\exp\left(3\int_{T_i}^T\frac{dT^\prime}{T^\prime}(1+\mathcal{F})\right),\\
&n^0_\nu=g_\nu\frac{3\zeta(3)}{4\pi^2}T^3_i,\qquad F=\frac{T}{3g^\ast_s}\frac{dg^\ast_s}{dT},
\end{align}
where $T_i$ is the initial temperature and $g^\ast_s$ is the entropy degrees of freedom. The extra neutrino density can be written as
\begin{align}
n_\mathrm{extra}=-&\exp\left(3\int_{T_i}^T\frac{dT^\prime}{T^\prime}(1+\mathcal{F})\right)\notag\\
\times&\int_{T_i}^T\frac{dT^\prime}{T^\prime}\frac{R_{e\overline{e}}(T^\prime)-R_{\nu\overline{\nu}}(T^\prime_\nu)}{H(T^\prime)}\left(1+\mathcal{F}\right)\exp\left(-3\int_{T_i}^{T^{\prime}}\frac{dT^{\prime\prime}}{T^{\prime\prime}}(1+\mathcal{F})\right).
\end{align}
%%%%%%%%%%%%%%%%%%%%%%%%%%%%%%%%%%%%%%%%%%%%%%%%%%%%%%%%%%%%%%%%%%
\begin{figure}[ht]
\begin{center}
\includegraphics[width=\textwidth]{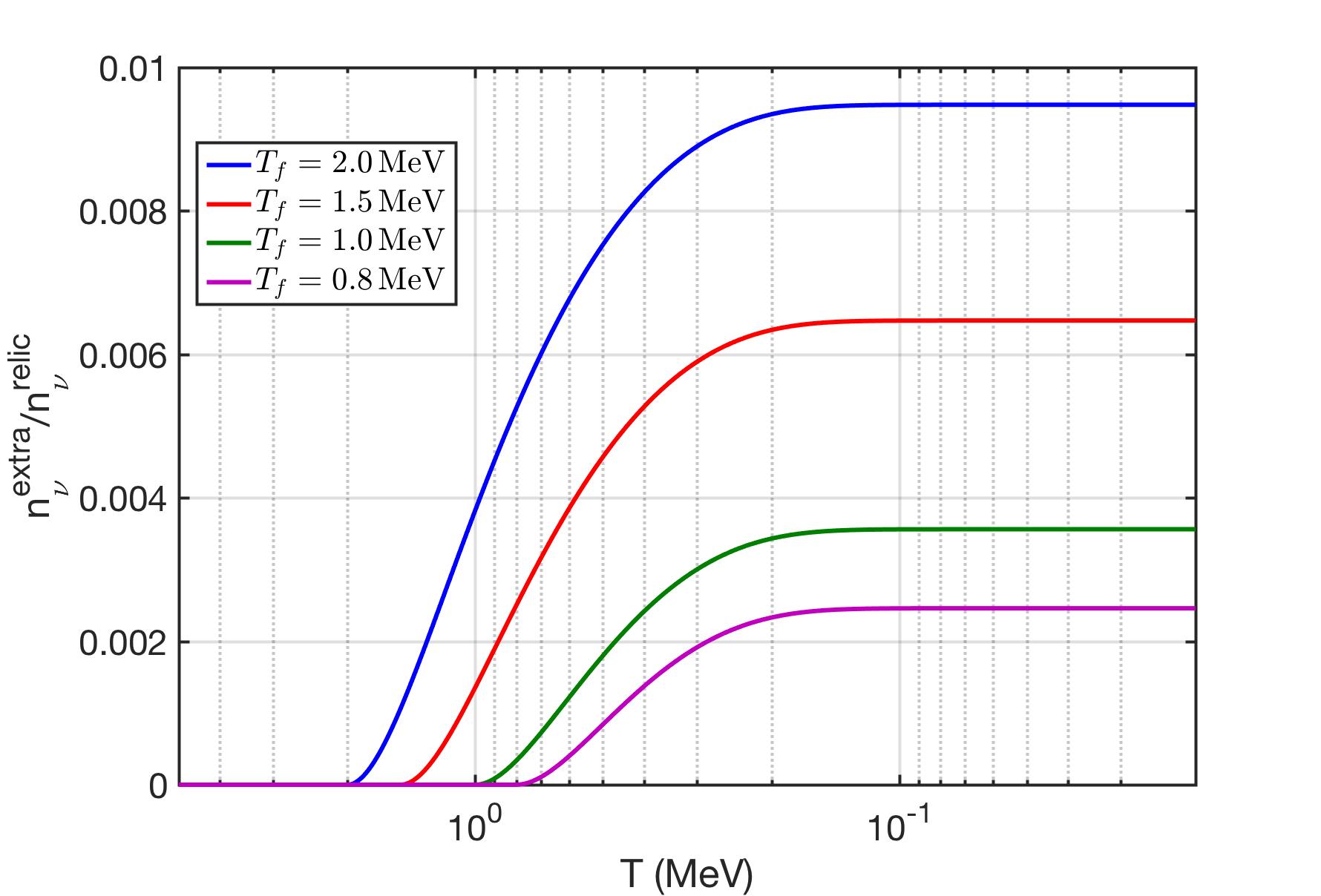}
\caption{the ratio between $n_{extra}/n_{relic}$ as a function of temperature with different neutrino freezeout temperature $T_f$. It shows that the higher freezeout temperature $T_f$ the higher number of extra neutrinos can be produced.}
\label{ExtraNeutrinoRatio}
\end{center}
\end{figure}
%%%%%%%%%%%%%%%%%%%%%%%%%%%%%%%%%%%%%%%%%%%%%%%%%%%%%%%%%%%%%%%%%%%

In Fig.~\ref{ExtraNeutrinoRatio} we plot the ratio between $n_\mathrm{extra}/n_\mathrm{relic}$ as a function of temperature with different neutrino freeze-out temperature $T_f$. It shows that the number of extra neutrinos depends strongly on the parameter $T_f$. This is because the freeze-out temperature determines the timing of the entropy transfer between $e^\pm$ and photon, which subsequently affects the evolution of temperature ratio between neutrinos and photons in the early Universe. The temperature ratio affects the rate ratio between $\nu\overline{\nu}\to e\overline{e}$ and $ e\overline{e}\to\nu\overline{\nu}$, because once the neutrino is too cold and the back reaction $\nu\overline{\nu}\to e\overline{e}$ can not maintain the balance, the $e^\pm$ annihilation starts to feed the extra neutrinos to the relic neutrino background. 

In addition to the annihilation of electron-positron pairs, there are other sources that can contribute to the presence of extra neutrinos in the early Universe. These additional sources include particle physics phenomena and plasma effects: neutrinos from charged leptons $\mu^\pm,\tau^\pm$ decay, neutrinos from the $\pi^\pm$ decay, and neutrino radiation from massive photon decay in electron-positron rich plasma. All of these potential sources of extra neutrinos can impact the distribution of freely streaming neutrinos and the effective number of neutrinos. Understanding these effects is crucial to comprehending how the neutrino component influences the expansion of the Universe, as well as the potential implications for large-scale structure formation and the spectrum of relic neutrinos.

\section{The $e^\pm$ plasma relaxation rate: Self-consistence approach}
In electron-positron plasma, the photon mass appears as $m_\gamma^2$ in the transition matrices for M{\o}ller and Bhabha reactions, which is one of important parameters in the calculation of the relaxation rate in $e^\pm$ plasma. When evaluating M{\o}ller and Bhabha scattering, we include the temperature-dependent mass of the photon obtained in plasma theory without damping. In general, the effective mass of the photon depends on the property of the plasma. Considering the linear response theory, the dispersion relation for the photon in nonrelativistic $e^\pm$ plasma is given by~[\cite{Formanek:2021blc}]
\begin{align}\label{dispersion_damping}
w^2=|k|^2+\frac{w}{w+i\kappa}w_{pl}^2,
\end{align}
where $w_{pl}$ is the plasma frequency and $\kappa$ is the average collision rate of $e^\pm$ plasma. The effective plasma frequency in damped plasma can be solved by considering the case $|k|^2=0$~[\cite{Formanek:2021blc}]
\begin{align}\label{plasmafrequency_damped}
w_{\pm}=-i\frac{\kappa}{2}\pm\sqrt{w^2_{pl}-\frac{\kappa^2}{4}}.
\end{align}
It shows that the plasma frequency in damped plasma $w_\pm$ is a function of $\kappa.$  In this case, the effective photon mass in damped plasma is also a function of the scattering rate. We have
\begin{align}\label{PhotonMass_self}
m_\gamma=w_\pm(w_{pl},\kappa)=m_\gamma(w_{pl},\kappa),
\end{align}
where the photon mass $m_\gamma=w_+$ for the underdamped plasma $w_{pl}>\kappa/2$, and $m_\gamma=w_-$ for overdamped plasma $w_{pl}<\kappa/2$. Eq.~(\ref{PhotonMass_self}) shows that computed damping strength $\kappa$ is the dominant scale for collisional plasma and it is also the main parameter determining the photon mass in plasma. 

Substituting the effective photon mass Eq.~(\ref{PhotonMass_self}) into the definition of the average relaxation rate Eq.~(\ref{Kappa}), we obtain the self-consistent equation for damping rate $\kappa$ as 
\begin{align}\label{RealaxtionSelf}
\kappa\,&\left[\frac{g_e}{2\pi^3}T^3\left(\frac{m_e}{T}\right)^2K_2(m_e/T)\right]\notag\\&=\frac{g_eg_e}{32\pi^4}T\!\!\int_{4m_e^2}^\infty\!\!\!\!ds\frac{s(s-4m^2_e)}{\sqrt{s}}K_1(\sqrt{s}/T)
\bigg[\sigma_{e^\pm e^\pm}(s,w_{pl},\kappa)+\sigma_{e^\pm e^\mp}(s,w_{pl},\kappa)\bigg],
\end{align}
where the cross sections depend on the parameter $w_{pl}$ and $\kappa$, and the variable $\kappa$ appears on both sides of the equation so we need solve the equation numerically to determine the $\kappa$ value that satisfies this condition.

\begin{figure}[h]
\begin{center}
\includegraphics[width=\linewidth]{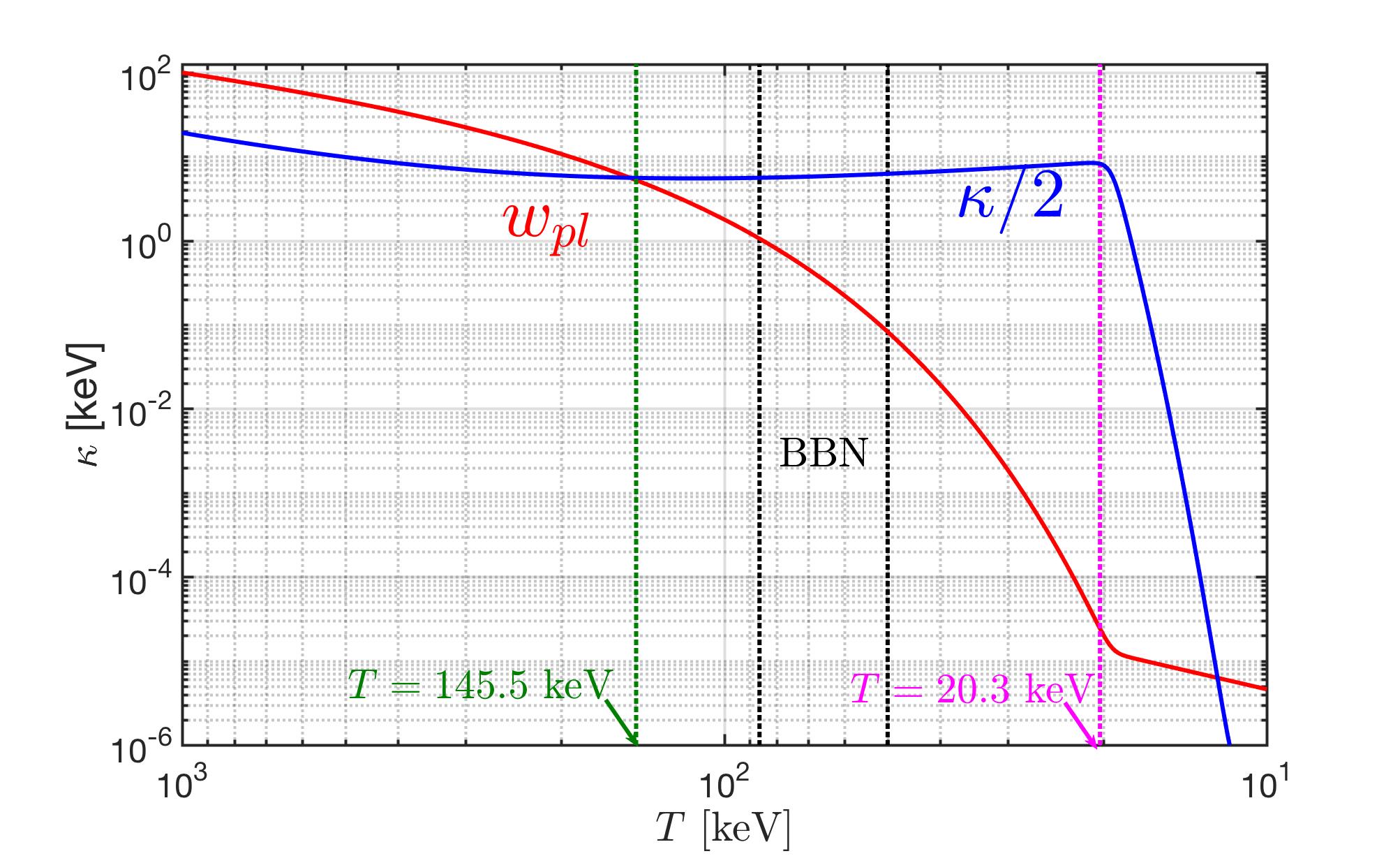}
\caption{The relaxation rate $\kappa/2$ (blue line) as a function of temperature in nonrelativistic electron-positron plasma. For comparison, we show the plasma frequency $\omega_{pl}$ in the red line. It shows that for $T>145.5$ MeV, the plasma frequency is larger than the collision rate $w_{pl}>\kappa/2$; for temperature $T<145.5$ MeV, we have $\kappa/2>w_{pl}$. For temperature $T<20.3$ keV, the composition of plasma is changed to electron and proton, which is beyond our current study because of unequal numbers of electrons and positrons.}
\label{RelaxationRate002_fig}
\end{center}
\end{figure}

Depending on the nature of the plasma (overdamped or underdamped plasma), we can establish the photon mass in collision plasma based on two distinct conditions as follows:
\begin{itemize}
\item Case 1. The plasma frequency is larger than the collision rate $w_{pl}>\kappa/2$, we have
\begin{align}
m_\gamma=w_+=-i\frac{\kappa}{2}+\sqrt{w^2_{pl}-\frac{\kappa^2}{4}}.
\end{align}
\item Case 2. The plasma frequency is smaller than the collision rate $w_{pl}<\kappa/2$, we have
\begin{align}\label{PhotonMassPlasma}
m_\gamma=w_-=-i\left(\frac{\kappa}{2}+\sqrt{\frac{\kappa^2}{4}-w^2_{pl}}\right).
\end{align}
\end{itemize}
In Fig.~\ref{RelaxationRate002_fig} it shows that during the BBN temperature range $50\leqslant T\leqslant 86$ keV, the plasma frequency is smaller than the collision rate $w_{pl}<\kappa/2$.  In this case, the effective photon mass in collision plasma during BBN epoch is given by Eq.(\ref{PhotonMassPlasma}).

%~~~~Figure~~~~~~~~~~~~~~~~~~~~~~~~~~~
%~~~~Figure~~~~~~~~~~~~~~~~~~~~~~~~~~~

To calculate the cross section for  M{\o}ller and Bhabha scattering we need to include the imaginary photon mass in the calculation of transition matrix elements. In general, the real part of photon mass in the calculation includes the effective photon-electron/positron scattering in plasma, and the imaginary part of photon mass contributes to the decay width of massive photon in plasma. To estimate the effect of photon mass on the damping rate $\kappa$, we first consider the effective mass corresponding to photon-electron/positron scattering in plasma, and leave the photon decay for future study.

For BBN temperature $50\leqslant T\leqslant 86$ keV,
we have $w_{pl}<\kappa$ and the effective photon mass can be approximated as
\begin{align}
m^2_\gamma=w_-w_-^\ast&=\left(\frac{\kappa}{2}+\sqrt{\frac{\kappa^2}{4}-w^2_{pl}}\right)^2
=\frac{\kappa^2}{2}\left[\left(1-\frac{2w^2_{pl}}{\kappa^2}\right)+\sqrt{1-\frac{4w^2_{pl}}{\kappa^2}}\right]\notag\\
&=\frac{\kappa^2}{2}\left[\left(1-\frac{2w^2_{pl}}{\kappa^2}\right)+\left(1-\frac{2w^2_{pl}}{\kappa^2}+\cdots\right)\right]\approx\kappa^2.
\label{PhotonMassPlasma002}
\end{align}
where we consider the limit $w^2_{pl}/\kappa^2\ll1$ and effective photon mass is equal to the average collision rate in plasma $m^2_\gamma\approx\kappa$.

Substituting the photon mass $m^2_\gamma=\kappa^2$ for overdamping plasma into the relaxation rate of electron-positron Eq.~(\ref{RealaxtionSelf}), and introducing the following dimensionless variables
\begin{align}
x=\sqrt{s}/T,\qquad a=m_\gamma/T=\kappa/T,\qquad b=m_e/T,
\end{align}
%~~~~~~~~~~~~~~~~~~~~~~~~~~~~~~~~~~~~~~~~~~~~~~~~~~~~~~~~~~~~~~~~~~~~~~~~~~~~~~~~
\begin{figure}[ht]
%\begin{center}
\centering
\includegraphics[width=\linewidth]{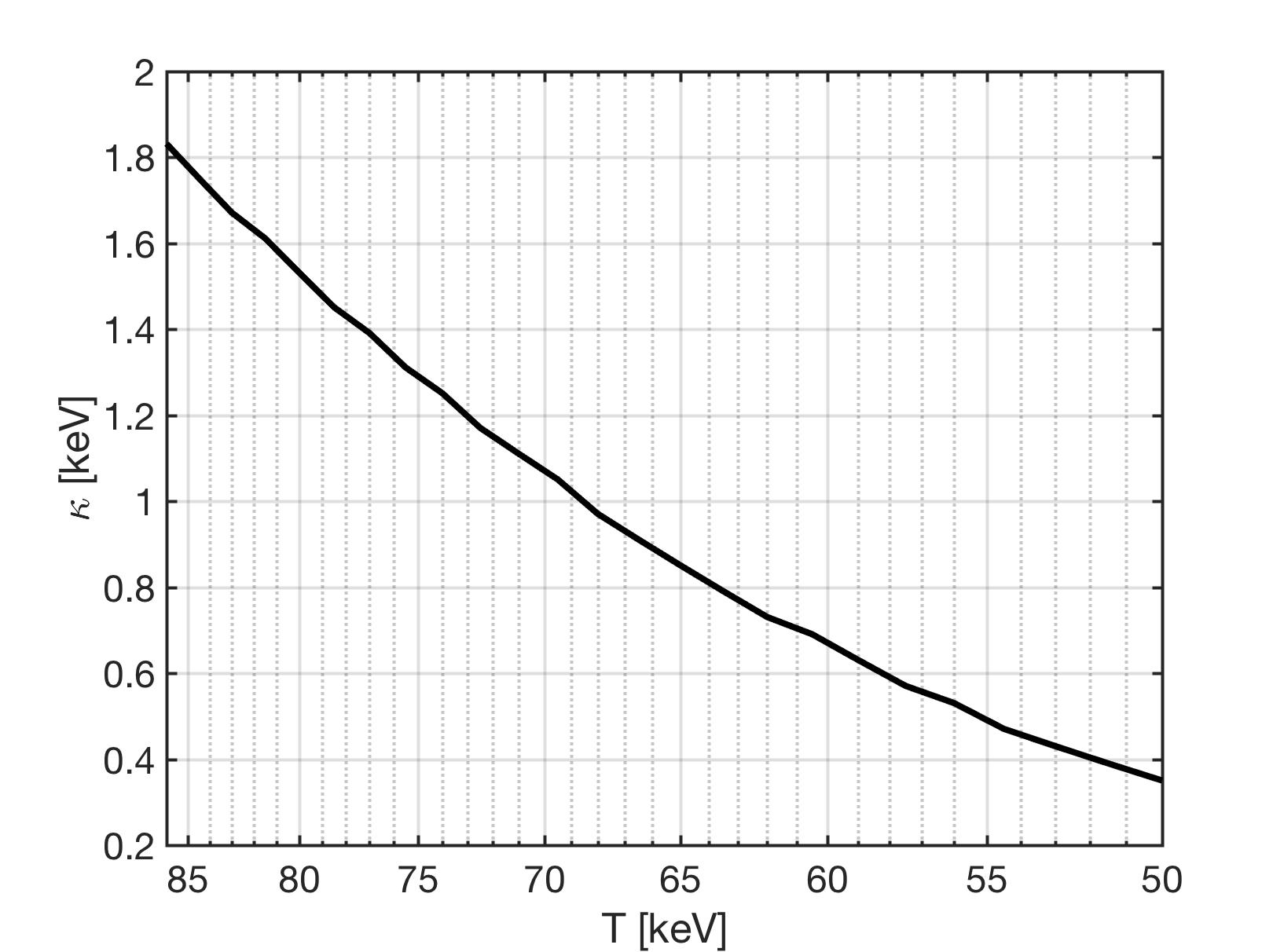}
\caption{The relaxation rate $\kappa$ that satisfies Eq.(\ref{Numerical_eq}) as a function of temperature $50\leqslant T\leqslant 86$\,keV. It shows that for overdamping plasma, we have $m^2_\gamma=\kappa^2$, and $\kappa=1.832$\,keV when $T=86$\,keV and $\kappa=0.350$\,keV when $T=50$\,keV. The minor fluctuations are a result of the restricted numerical precision.}
\label{KappaSol.fig} 
\end{figure}
%~~~~~~~~~~~~~~~~~~~~~~~~~~~~~~~~~~~~~~~~~~~~~~~~~~~~~~~~~~~~~~~~~~~~~~~~~~~~~~~~
then the relaxation rate of electron-positron can be written as
\begin{align}\label{Numerical_eq}
&\left[\frac{g_e}{2\pi^2}T^4\left(\frac{m_e}{T}\right)^{\!2}\!K_2(m_e/T)\right]\,\left(\frac{\kappa}{T}\right)\notag\\
&\qquad\qquad\qquad=\frac{g^2_e\alpha^2}{8\pi^3}T^4\!\!\int_{2b}^\infty\!dxK_1(x)\left[\mathcal{F}_{e^\pm e^\pm}(x,\kappa/T)+\mathcal{F}_{e^\pm e^\mp}(x,\kappa/T)\right],
\end{align}
where the functions $\mathcal{F}_{e^\pm e^\pm}$ and $\mathcal{F}_{e^\pm e^\mp}$ are given by
\begin{align}
\mathcal{F}_{e^\pm e^\pm}(x,a=\kappa/T)&=\left\{2\left[3a^2+4b^2+\frac{4(b^4-a^4)}{x^2-4b^2+2a^2}\right]\ln\left(\frac{a^2}{x^2-4b^2+a^2}\right)\right.\notag\\
&\left.+\frac{(x^2-4b^2)(8b^4+2a^4+3a^2x^2+2x^4-4b^2(2x^2+a^2))}{a^2(x^2-4b^2+a^2)}\right\}
\end{align}
and 
\begin{align}
\mathcal{F}_{e^\pm e^\mp}(x,a=\kappa/T)&=\left\{\frac{2x^2(a^2+x^2)-4b^4}{x^2-a^2}\ln\left(\frac{a^2}{x^2-4b^2+a^2}\right)\right.\notag\\
&+\frac{(x^2-4b^2)(3x^2+4b^2+2a^2)}{(x^2-a^2)}+\frac{x^6-12b^4x^2-16b^6}{3(x^2-a^2)^2}\notag\\
&\left.+\frac{(x^2-4b^2)(8b^4+2a^4+3a^2x^2+2x^4-4b^2(2x^2+a^2))}{a^2(x^2-4b^2+a^2)}\!\right\}.
\end{align}

To determine the $\kappa$ that satisfies Eq.~(\ref{Numerical_eq}), we can solve it numerically. 
In Fig.~\ref{KappaSol.fig}, we plot the relaxation rate $\kappa$ that satisfies Eq.~(\ref{Numerical_eq}) as a function of temperature $50\,\mathrm{keV} \leqslant T\leqslant 86$ keV. It shows that during the BBN temperature range, we have overdamping electron/positron plasma $w_{pl}<\kappa$, and the effective photon mass $m^2_\gamma=\kappa^2$. The relaxation rate $\kappa=1.832\sim0.350$ keV during the BBN temperature range, which is smaller than the relaxation rate without damping photon mass (See Fig.~\ref{RelaxationRate_fig}, where the relaxation rate $\kappa=10\sim12$ keV during the BBN temperature).

Our first estimation implies that the relaxation rate is sensitive to the photon mass in damped plasma. To address the self-consistent evaluation of damping  rate in plasma requires the development of a well-defined, self-consistent approach, where both damping and photon properties in plasma are determined in a mutually consistent manner. We aim to include the full photon mass effect (both real and imaginary parts) in the study and improve our calculation for the next step to complete the project.

%% file: Chapter7.tex
\chapter{Summary and conclusion}\label{Summary}
%{Introduction\daggerfootnote{This chapter has been published previously as \citet{Gottbrath1999}.}}
We have studied the evolution of the early Universe from the QGP epoch to the $e^\pm$ plasma $300\,\mathrm{MeV}>T>0.02\,\mathrm{MeV}$. Our effort focuses on understanding how the plasma state impacts and modifies elementary interactions between particles and thus the properties of the early Universe.
In the QGP epoch, 
we investigate the heavy-quark bottom and charm near the QGP hadronization temperature $300\,\mathrm{MeV}>T>150$ MeV. We show that the  faster quark-gluon pair fusion keeps the charm quarks in chemical equilibrium up until hadronization. After hadronization, charm quarks form heavy mesons that decay quickly into multi-particles and causes charm quarks to vanish from inventory of particles in the Universe. For the bottom quarks, the
quark production rate competes with bottom decay rate as a function of temperature. When the Universe's temperature is near to the QGP phase transition $300\,\mathrm{MeV}>T>150$ MeV, the bottom quark breaks the detail balance and disappearance from particle inventory provides the arrow in time and which can be the ‘sweet-spot’ for Sakharov conditions.

After hadronization, the free quarks and gluons become confined within baryon/mesons and the Universe becomes hadronic-matter dominated. In the temperature range $ 150\,\mathrm{MeV}>T>20\,\mathrm{MeV}$, the Universe is rich in physics phenomena involving strange mesons and (anti)baryons including (anti)hyperon abundances. 
Considering the inventory of the Universe  with strange mesons and baryons, the antibaryons disappear from the Universe at temperature $T=38.2$ MeV. Strangeness can be produced by the inverse decay reactions via weak, electromagnetic, and strong interactions in the early Universe until $T\approx13$ MeV. For $T>20$ MeV, the strangeness is dominantly present in the meson. For $20 >T > 13$ MeV, the strangeness can present in the hyperons and anti-strangeness in kaon keeping symmetry $s=\overline{s}$. Below the temperature $T<13$ MeV a new regime opens up in which the tiny residual strangeness abundance in hyperons and is governed by weak decays with no reequilibration with mesons.

When the temperature in the Universe is about $T\approx10$ MeV, the main ingredients that control the Universe evolution are: photons, neutrinos, electrons and positrons. The massive meson, baryon, and $\mu^\pm$ and $\tau^\pm$ are also present, but their small number density can be neglected when considering the energy density of the Universe expansion. For temperature $10\,\mathrm{MeV}>T>2$ MeV the Universe is controlled by the weak interaction between neutrinos and matter. We explore the neutrino coherent and incoherent scatterings with matter and apply them to study the neutrino freeze-out (incoherent scattering) in the early Universe and  long wavelength cosmic neutrino-atom scattering (coherent scattering). After neutrino freeze-out $T_f\approx2\,\mathrm{MeV}$, it becomes free-streaming and the number of neutrinos is independently conserved. We investigate the relation between the lepton number $L$ and the effective number of neutrinos $N^{\mathrm{eff}}_\nu$, and explore the impact of a large cosmological lepton asymmetry on the Universe evolution. Instead of $B\simeq |L|$, we found that $0.4\leqslant|L| \leqslant0.52$ and $B\simeq 1.33\times 10^{-9}|L|$ reconciles the CMB and current epoch results for the Hubble expansion parameter.

Considering the temperature after neutrino freeze-out: $2\,\mathrm{MeV}>T>0.02$ MeV, the cosmic plasma is dominated by the photon, electrons and positrons. The massive $\mu^\pm$ abundance disappears at $T_\mathrm{disappear}=4.195$ MeV as soon as the muon decay rate becomes faster than muon production rate. The muon-baryon density ratio at  muon persistence temperature is equal to $n_\mu^\pm/n_B(T_\mathrm{disappear})\approx0.911$, which implies that the muon abundance could influence baryon evolution because muon number density is comparable to the baryon number density. 

We demonstrate that the presence of rich electron-positron plasma can last until temperature $T=20.3$ keV by calculating the chemical potential of electrons in the universe that maintains charge neutrality and entropy conservation.
We evaluate the microscope damping rate with temperature dependent photon mass in electron-positron plasma which is one important variable to study the inter-nuclear electromagnetic potentials  in damped plasma. Finally  we examine the magnetization process within dense electron-positron plasma and show that it has paramagnetic properties when subjected to an external field. Our study of magnetization can provide insights to understand the magnetized plasma and the possible origin of primordial magnetic field and  developing methods for
future detailed study.

Looking forward, we will refine our understanding of the evolution of the early Universe by studying the topics that follow. For electron-positron plasma: we will improve our calculation on the damping rate $\kappa$ and develop a self-consistent approach where both damping and photon properties in plasma are determined in a mutually consistent manner. For the neutrino sector: we will examine the sources of extra neutrino and entropy transfer from microscope process after neutrino freezeout in details. 
%Applying our understanding of neutrino-matter coherent scattering and effective potential to further explore the relic neutrino in early Universe. 
By studying these phenomena, we aim to enhance our understanding of the role of neutrinos in the evolution of universe. For the heavy particle in QGP: we will focus on understanding the behavior of heavy quarks in extreme environments, and refine our study of nonequilibrium bottom and its potential applications on baryongenesis at low temperature. We will also study the dynamic Higgs abundance under the competition between production and decay to understand the departure from equilibrium of Higgs in the early Universe.

In summary, our research offers an initial glimpse into the first hour of the Universe's history, exploring the intricate relationship between fundamental particles and plasma in the early Universe. We hope that the output of our study will be beneficial to all parties concerned while at the same time contribute to the knowledge enhancement in the understanding of the early Universe.

%% file: appendix.tex
\chapter{{Decomposition of Fermi gas into zero and finite temperature components}}

The most interesting physics of finite temperature Fermi gasses occurs around the Fermi surface, which needs a mathematical tool that can capture the finite temperature behavior of the Fermi-Dirac distribution in an analytic fashion. We provide a novel form of the Fermi distribution that can separate the Fermi gas into zero and finite temperature components analytically which is useful for addressing physics beyond the zero temperature approximation. This is an ongoing research project and will be submitted to \textit{International Journal of Theoretical Physics}.

\subsubsection{A novel form of Fermi distribution}
We have empirically identified the following novel way to state the form of the Fermi distribution. We obtained this form while seeking to carry out cosmological computations involving shifts in behavior from high to very low-temperature physics. 
\begin{align}\label{NFF1}
f&\equiv \frac{1}{e^{ (E-\widetilde\mu)/T} +1}\notag\\
&=\Theta\left(\frac{\widetilde\mu - E}{T}\right) +  e^{ - |E-\widetilde\mu|/T }
\; \left[\frac{1}{2}\mathrm{sgn}\left(\frac{E-\widetilde\mu}{T}\right) 
 +\frac{1}{2}\tanh\left(\frac{E-\widetilde\mu}{2T}\right)\right], 
\end{align}
 where the Heaviside step function $\Theta(x)$ and sign function $\mathrm{sgn}(x)$ are given by
\begin{align}
\Theta(x)=\left\{
\begin{array}{r}
1,\quad\mathrm{for}\quad{x}>0\\
1/2,\quad\mathrm{for}\quad{x}=0\\
0,\quad\mathrm{for}\quad{x}<0
\end{array}\right.\,,\qquad
\mathrm{sgn}(x)=\left\{
\begin{array}{r}
+1,\quad\mathrm{for}\quad{x}>0\\
0,\quad\mathrm{for}\quad{x}=0\\
-1,\quad\mathrm{for}\quad{x}<0\\
\end{array}\right.\,.
\end{align}
The first term in Eq.~(\ref{NFF1}) represents the zero temperature limit of the Fermi distribution, and  the finite temperature contributions in square brackets are weighted by a decaying exponential function. One of the benefits of weighting the finite temperature contributions with decaying exponential function is that it can provide a good asymptotic limit numerically and the numerical evaluations will naturally center around the Fermi surface of the system.

\subsubsection{Mathematical proof}
 We note  that the right hand side (RHS) of Eq.\,(\ref{NFF1}) comprises several non-analytical functions, also called distributions. On first sight it is hard to believe that these cancel to create the analytical Fermi function format seen on the left hand side (LHS). In the following, we will demonstrate that the two sides equal each other. To demonstrate the novel form of the Fermi distribution, it is convenient to introduce the dimensionless variable as follows:
\begin{align}
x=\frac{E-\widetilde\mu}{T}.
\end{align}
With this dimensionless variable, the Fermi function can be written as
\begin{align}
f(x)=\Theta(-x)+e^{-|x|}\left[\frac{1}{2}\mathrm{sgn}\left(x\right) 
 +\frac{1}{2}\tanh\left(\frac{x}{2}\right)\right].
\end{align}

We use following properties of the sign function to help simplify the expression: 
\begin{align}\label{NFF2a}
&\mathrm{sgn}(x)%\equiv \frac{d|x|}{dx}
\equiv  \frac{|x|}{x}\equiv \frac{x}{|x|}\;,
   \quad \mathrm{sgn}(0)=0,\qquad
 \mathrm{sgn}(x)=2\Theta(x)-1\;,\\
 \label{NFFa1}
&\mathrm{sgn}^{2}(x)\sinh(x)=\sinh(x)\;.
 \end{align}
Since $\sinh(x)$  vanishes  at $x=0$,  we don't need to worry about what value to assign to $\mathrm{sgn}^{2}(x)$ at $x=0$. Using the properties above, we can replace the step function and exponential function as follows:
 \begin{align}\label{NFF4}
&\Theta(-x)=\frac 1 2 (1-\mathrm{sgn}(x)\;,\\ 
&e^{-|x|}=\cosh|x|-\sinh|x|=\cosh x- \mathrm{sgn}(x)\sinh x\;.
\end{align}
Then the Fermi distribution function can be written as
\begin{align}
f=\frac{1}{2} &+\left[\cosh {x}- \mathrm{sgn}(x)\sinh{x} -1\right]\frac{1}{2} \mathrm{sgn}(x)\notag\\
 &\qquad\qquad\qquad+\left[\cosh{x}- \mathrm{sgn}(x)\sinh{x} \right]\frac{1}{2}\tanh{(x/2)}\;.
\end{align}
Using the properties of the singular function  Eq.\,(\ref{NFFa1}), the distribution function can be written as
\begin{align}
&f=f_R+f_I\\
\label{NFF5a}
 &f_R= \frac{1}{2}\left(1-\sinh x +\cosh x \tanh (x/2)\right) \\
 &f_I= \mathrm{sgn}(x) \frac{1}{2} \left(\cosh x-1 - \sinh x \tanh (x/2)\right),
 \label{NFF5b}
\end{align}
where $f_R$ and $f_I$ represent the regular and irregular part of the distribution, respectively.

For the irregular part $f_I$, we use the properties of hyperbolic functions
\begin{align}
\cosh x-1= 2\sinh^2(x/2), \qquad\sinh x=2 \sinh(x/2) \cosh(x/2).
\end{align}
Then it can be written as
\begin{align}
f_I&=\mathrm{sgn}(x) \frac{1}{2} \left[ 2\sinh^2(x/2) - 2 \sinh(x/2) \cosh(x/2) \tanh (x/2)\right]\\
&=\mathrm{sgn}(x) \frac{1}{2}\left[2\sinh^2(x/2)-2\sinh^2(x/2)\right]\\
&=0.
\end{align}
We show that the irregular part $f_I$ vanishes as an identity. On the other hand, the regular part $f_R$ can be simplified, as shown below:
\begin{align}
f_R&=\frac{1}{2}\left[1-\frac{1}{2}\left(e^x-e^{-x}\right)+\frac{1}{2}\left(e^x+e^{-x}\right)\frac{e^x-1}{e^x+1}\right]\\
&=\frac{1}{2(e^x+1)}\left[\left(e^x+1\right)-\frac{1}{2}\left[\left(e^x+1\right)\left(e^x-e^{-x}\right)-\left(e^x-1\right)\left(e^x+e^{-x}\right)\right]\right]\\
&=\frac{1}{2(e^x+1)}\left[\left(e^x+1\right)-\left(e^x-1\right)\right]\\
&=f.
\end{align}

Finally, we consider the LHS and RHS of Eq.\,(\ref{NFF1}) at $x=0$. With the singular function properties as given we see that at $x=0$ both LHS and RHS of Eq.\,(\ref{NFF1}) are equal to $1/2$, and in the first derivative of the RHS at $x=0$ the two $\delta(x)$-terms 
\begin{align}\label{NFF1b}
&\frac{d\Theta((\widetilde\mu-E)/T)}{dE}=-\frac{1}{T}\delta\left(\frac{\widetilde\mu-E}{T}\right)\;,\,\,\frac{d\,\mathrm{sgn}((E -\widetilde\mu)/T)}{dE}=\frac{2}{T}\delta\left(\frac{E-\widetilde\mu}{T}\right)\;, 
 \end{align}
cancel exactly as required, since there is no $\delta(x)$ on LHS. This encourages us to believe that all of the singular expressions cancel. This completes the demonstration of the exact validity of  Eq.\,(\ref{NFF1}). 

\subsubsection{Numerical illustration}

\begin{figure}[ht]
\begin{center}
\includegraphics[width=0.5\textwidth]{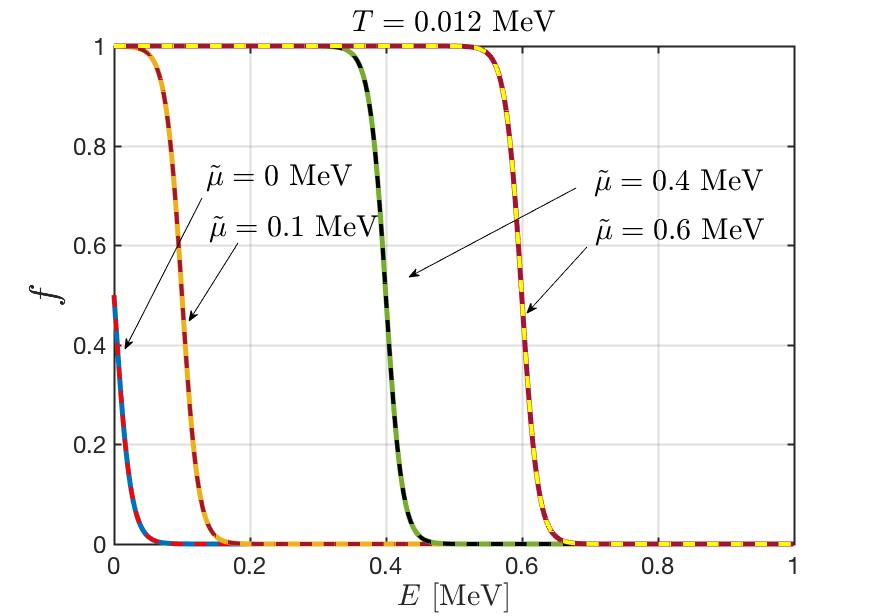}\includegraphics[width=0.5\textwidth]{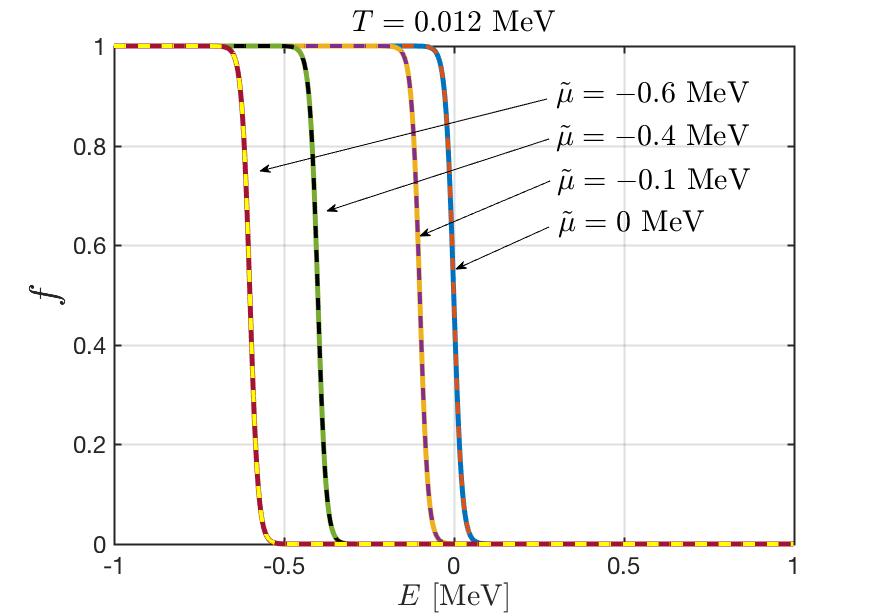}
\includegraphics[width=0.5\textwidth]{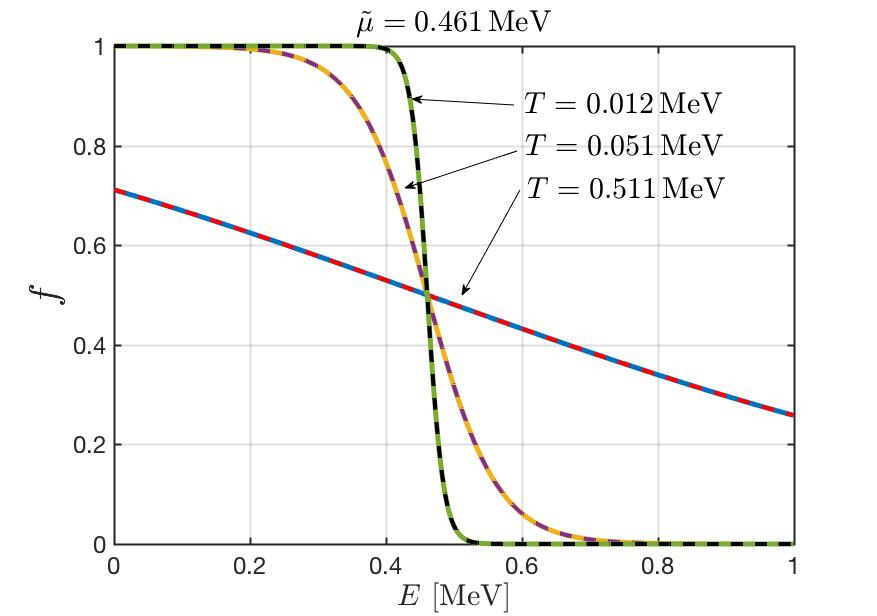}\includegraphics[width=0.5\textwidth]{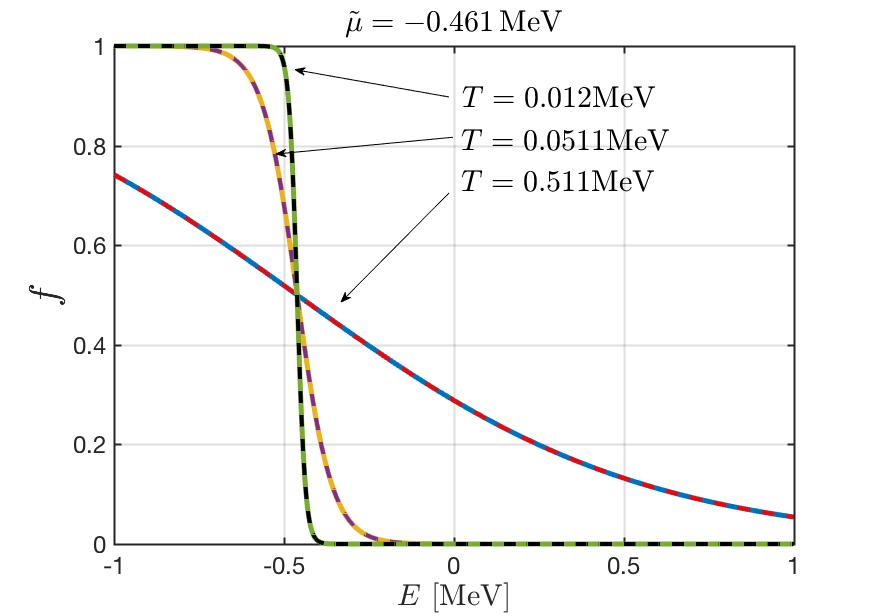}
\caption{The exact Fermi-distribution (LHS of Eq.~(\ref{NFF1}) with solid lines and novel form of Fermi-distribution (RHS of Eq.~(\ref{NFF1})) with dashed lines as a function of energy with different parameters. Top: we compare the RHS and LHS of Eq.~(\ref{NFF1}) with different chemical potential: $\widetilde\mu=0, \pm0.1, \pm0.4\,\pm0.6\, \mathrm{MeV}$ at temperature $T=0.012\,\mathrm{MeV}$. Bottom: the Fermi distribution with different temperatures $T=0.511, 0.0511, 0.012\,\mathrm{MeV}$ for chemical potential $\widetilde\mu=\pm0.461\,\mathrm{MeV}$.}
\label{Fermi_Checking}
\end{center}
\end{figure}

In Fig.~\ref{Fermi_Checking} we plot the exact Fermi-distribution (LHS of Eq.~(\ref{NFF1}) with solid lines and novel form of Fermi-distribution (RHS of Eq.~(\ref{NFF1})) with dashed lines as a function of energy with different parameters. It demonstrates that the 
LHS and RHS of Eq.~(\ref{NFF1}) are equivalent to each other numerically.

%~~~~~~~Figure~~~~~~~~~~~~~~~~~~~~~~~~~~~~~~~~~~~~~~~~~~~~~~~~~~~~~~~~~~~~~~~~~~~~~~~~~~~~~~~~~~~~~
\begin{figure}[ht]
\begin{center}
\includegraphics[width=0.5\textwidth]{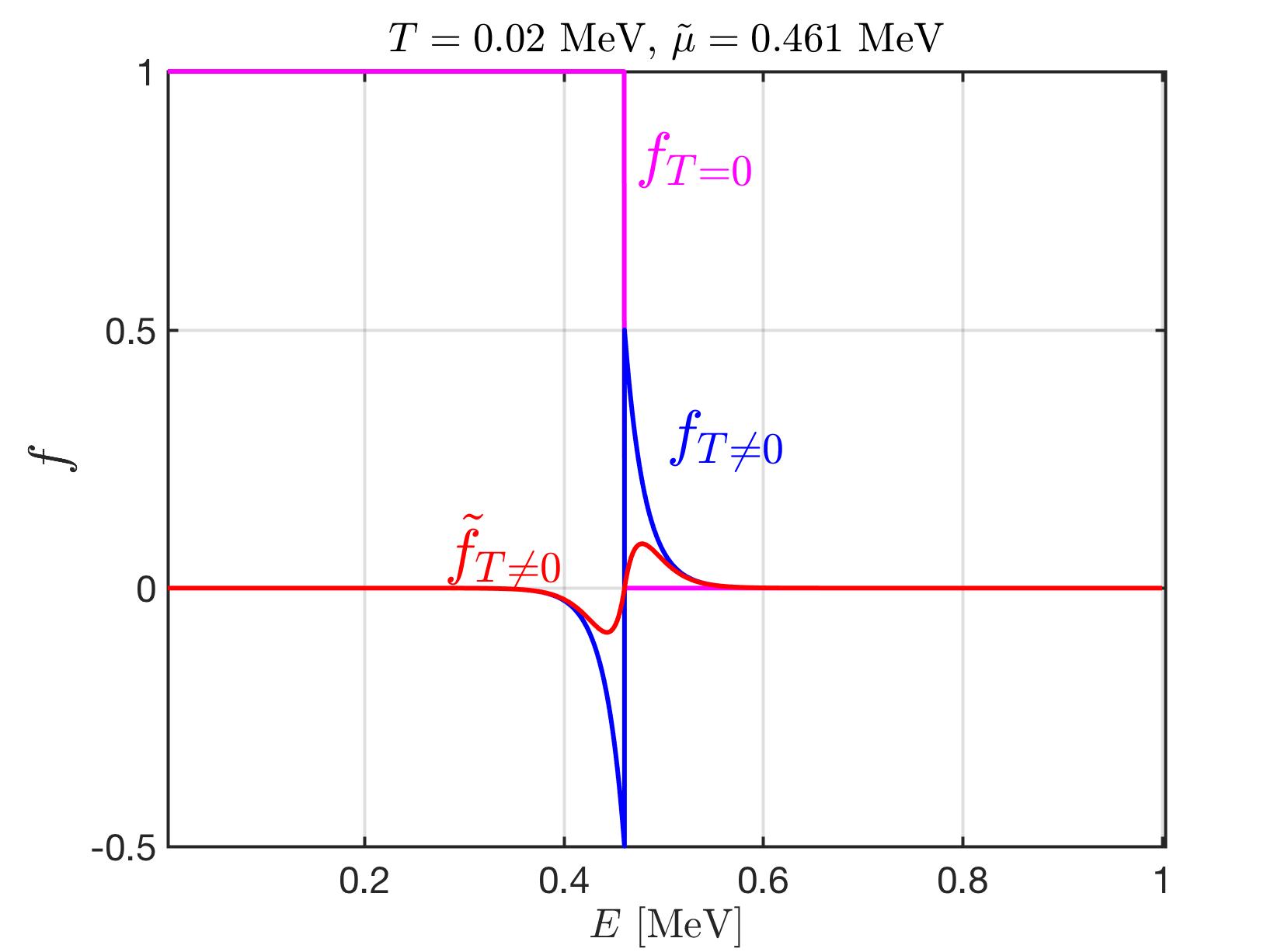}\includegraphics[width=0.5\textwidth]{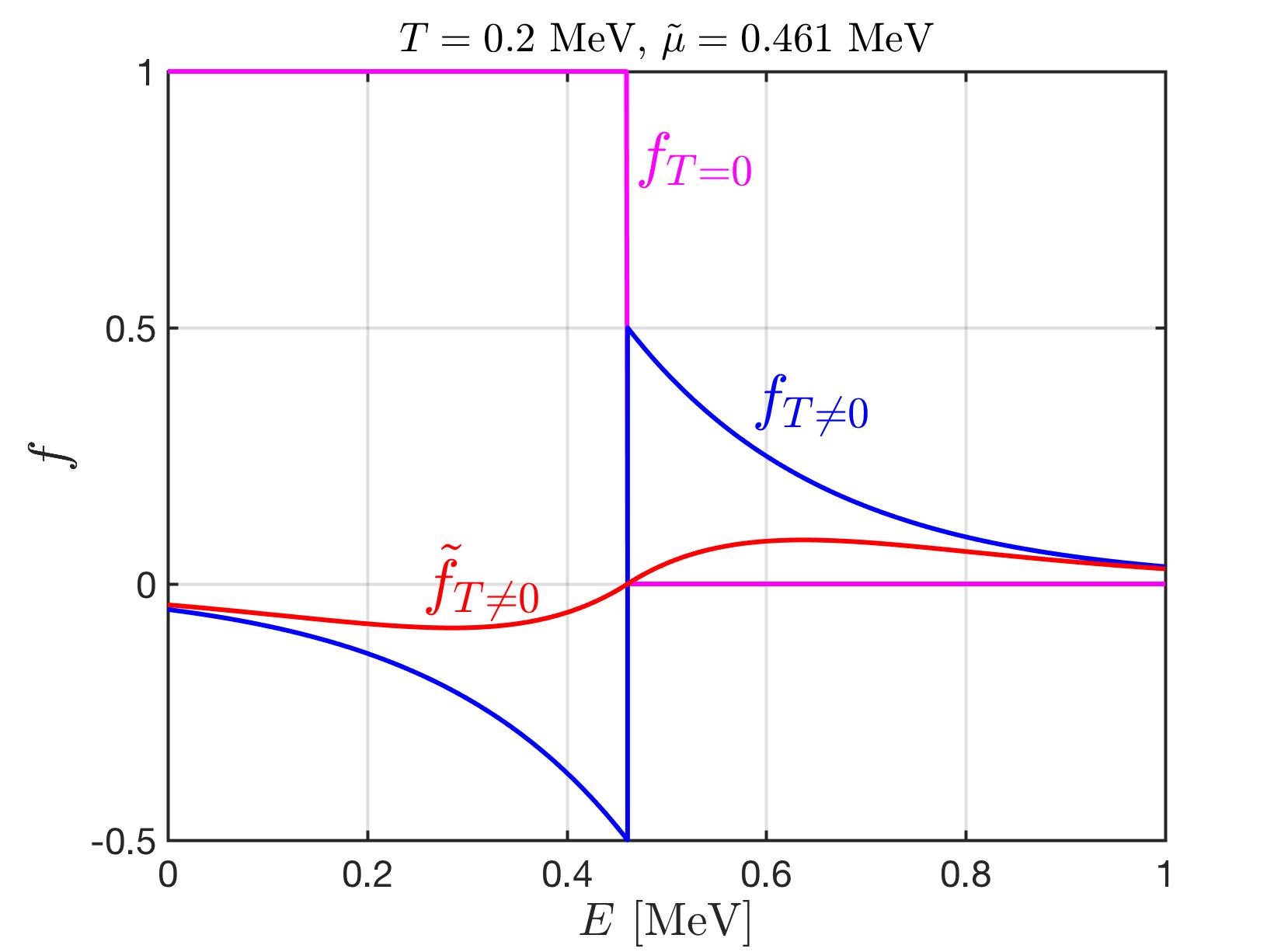}
\caption{The zero and finite temperature components of the decomposition here considered for Fermi distribution as a function of energy with chemical potential $\widetilde\mu=0.461$ MeV at temperature $T=0.02$ MeV and $T=0.2$ MeV. The purple line represents the zero temperature component $f_{\mathrm{T}=0}$, and blue and red lines represent the finite temperature components $f_\mathrm{T\neq0}$ and $\tilde f_\mathrm{T\neq0}$ respectively. }
\label{Fermi_Component}
\end{center}
\end{figure}

%~~~~~~~~~~~~~~~~~~~~~~~~~~~~~~~~~~~~~~~~~~~~~~~~~~~~~~~~~~~~~~
To illustrate the zero and finite temperature contribution to the Fermi distribution function, it is convenient to rewrite Eq.~(\ref{NFF1}) into the following form
\begin{align}
&f=f_{\mathrm{T=0}}+f_\mathrm{T\neq0}+\tilde f_\mathrm{T\neq0}
\end{align}
where the function are defined as
\begin{align}
&f_{\mathrm{T=0}}=\Theta\left(\frac{\widetilde\mu - E}{T}\right),\\ &f_\mathrm{T\neq0}=\frac{1}{2}e^{ - |E-\widetilde\mu|/T }\mathrm{sgn}\left(\frac{E-\widetilde\mu}{T}\right),\quad\tilde f_\mathrm{T\neq0}=\frac{1}{2}e^{ - |E-\widetilde\mu|/T }\tanh\left(\frac{E-\widetilde\mu}{2T}\right)
\end{align}
In Fig.~(\ref{Fermi_Component}) we plot the zero (purple lines, $f_{T=0}$) and finite temperature components of the Fermi distribution as a function of energy choosing in this example the chemical potential $\widetilde\mu=0.461$ MeV at temperature $T=0.02$ MeV and at $T=0.2$ MeV.
%~~~~~~~Figure~~~~~~~~~~~~~~~~~~~~~~~~~~~~~~~~~~~~~~~~~~~~~~~~~~~~~~~~~~~~~~~~~~~~~~~~~~~~~~~~~~~~~

This Fig.~(\ref{Fermi_Component}) shows  that the two finite temperature components $f_\mathrm{T\neq0}$ and $\tilde f_\mathrm{T\neq0}$ are naturally centered around the Fermi surface of the system $E=\tilde\mu$. Both finite $T$ always have the same sign, so there is no opportunity generate numerical noise; this can be the case when considering brute force removal of $T=0$ limit from the low temperature distribution. The dominant contribution (blue lines, $f_{T\ne 0}$) are containing the  decaying exponential function of the Fermi distribution at large argument. The remainder (red lines, $\tilde f_{T\ne 0}$) has the shape of a derivative of a finite support representation of delta-function. 
\vfill\eject
%~~~~~~~~~~~~~~~~~~~~~~~~~~~~~~~~~~~~~~~~~~~~~~~~~~~~~~~~~~~~~~

%% file: dissertation.bbl
\begin{thebibliography}{119}
\addcontentsline{toc}{chapter}{\bibname}
\providecommand{\natexlab}[1]{#1}
\expandafter\ifx\csname urlstyle\endcsname\relax
  \providecommand{\doi}[1]{doi:\discretionary{}{}{}#1}\else
  \providecommand{\doi}{doi:\discretionary{}{}{}\begingroup
  \urlstyle{rm}\Url}\fi

\bibitem[{Aaij et~al.(2020{\natexlab{a}})}]{LHCb:2020vut}
Aaij, R. et~al. (2020{\natexlab{a}}).
\newblock {Measurement of CP observables in B$^{\pm}$ $\rightarrow$ DK$^{\pm}$
  and B$^{\pm}$ $\rightarrow$ D\ensuremath{\pi}$^{\pm}$ with D $\rightarrow$ $
  {K}_{\mathrm{S}}^0{K}^{\pm }{\pi}^{\mp } $ decays}.
\newblock \emph{JHEP}, \textbf{06}, p. 058.
\newblock \doi{10.1007/JHEP06(2020)058}.

\bibitem[{Aaij et~al.(2020{\natexlab{b}})}]{LHCb:2019jta}
Aaij, R. et~al. (2020{\natexlab{b}}).
\newblock {Observation of Several Sources of $CP$ Violation in $B^+ \to \pi^+
  \pi^+ \pi^-$ Decays}.
\newblock \emph{Phys. Rev. Lett.}, \textbf{124}(3), p. 031801.
\newblock \doi{10.1103/PhysRevLett.124.031801}.

\bibitem[{Aarts et~al.(2011)Aarts, Allton, Kim, Lombardo, Oktay, Ryan,
  Sinclair, and Skullerud}]{Aarts:2011sm}
Aarts, G., C.~Allton, S.~Kim, M.~P. Lombardo, M.~B. Oktay, S.~M. Ryan, D.~K.
  Sinclair, and J.~I. Skullerud (2011).
\newblock {What happens to the $\Upsilon$ and $\eta_b$ in the quark-gluon
  plasma? Bottomonium spectral functions from lattice QCD}.
\newblock \emph{JHEP}, \textbf{11}, p. 103.
\newblock \doi{10.1007/JHEP11(2011)103}.

\bibitem[{Ade et~al.(2014)}]{Planck:2013pxb}
Ade, P. A.~R. et~al. (2014).
\newblock {Planck 2013 results. XVI. Cosmological parameters}.
\newblock \emph{Astron. Astrophys.}, \textbf{571}, p. A16.
\newblock \doi{10.1051/0004-6361/201321591}.

\bibitem[{Ade et~al.(2016)}]{Planck:2015fie}
Ade, P. A.~R. et~al. (2016).
\newblock {Planck 2015 results. XIII. Cosmological parameters}.
\newblock \emph{Astron. Astrophys.}, \textbf{594}, p. A13.
\newblock \doi{10.1051/0004-6361/201525830}.

\bibitem[{Aghanim et~al.(2020{\natexlab{a}})}]{Planck:2018vyg}
Aghanim, N. et~al. (2020{\natexlab{a}}).
\newblock {Planck 2018 results. VI. Cosmological parameters}.
\newblock \emph{Astron. Astrophys.}, \textbf{641}, p.~A6.
\newblock \doi{10.1051/0004-6361/201833910}.
\newblock [Erratum: Astron.Astrophys. 652, C4 (2021)].

\bibitem[{Aghanim et~al.(2020{\natexlab{b}})}]{aghanim2018planck}
Aghanim, N. et~al. (2020{\natexlab{b}}).
\newblock {Planck 2018 results. VI. Cosmological parameters}.
\newblock \emph{Astron. Astrophys.}, \textbf{641}, p.~A6.
\newblock \doi{10.1051/0004-6361/201833910}.
\newblock [Erratum: Astron.Astrophys. 652, C4 (2021)].

\bibitem[{Amhis et~al.(2021)}]{HFLAV:2019otj}
Amhis, Y.~S. et~al. (2021).
\newblock {Averages of b-hadron, c-hadron, and $\tau $-lepton properties as of
  2018}.
\newblock \emph{Eur. Phys. J. C}, \textbf{81}(3), p. 226.
\newblock \doi{10.1140/epjc/s10052-020-8156-7}.

\bibitem[{Anderson and Witting(1974)}]{ANDERSON1974466}
Anderson, J. and H.~Witting (1974).
\newblock A relativistic relaxation-time model for the Boltzmann equation.
\newblock \emph{Physica}, \textbf{74}(3), pp. 466--488.
\newblock ISSN 0031-8914.
\newblock \doi{https://doi.org/10.1016/0031-8914(74)90355-3}.

\bibitem[{Arnold and McLerran(1987)}]{Arnold:1987mh}
Arnold, P.~B. and L.~D. McLerran (1987).
\newblock {Sphalerons, Small Fluctuations and Baryon Number Violation in
  Electroweak Theory}.
\newblock \emph{Phys. Rev. D}, \textbf{36}, p. 581.
\newblock \doi{10.1103/PhysRevD.36.581}.

\bibitem[{Barenboim et~al.(2017)Barenboim, Kinney, and
  Park}]{Barenboim:2016shh}
Barenboim, G., W.~H. Kinney, and W.-I. Park (2017).
\newblock {Resurrection of large lepton number asymmetries from neutrino flavor
  oscillations}.
\newblock \emph{Phys. Rev. D}, \textbf{95}(4), p. 043506.
\newblock \doi{10.1103/PhysRevD.95.043506}.

\bibitem[{Barenboim and Park(2017)}]{Barenboim:2017dfq}
Barenboim, G. and W.-I. Park (2017).
\newblock {A full picture of large lepton number asymmetries of the Universe}.
\newblock \emph{JCAP}, \textbf{04}, p. 048.
\newblock \doi{10.1088/1475-7516/2017/04/048}.

\bibitem[{Bazavov et~al.(2018)Bazavov, Brambilla, Petreczky, Vairo, and
  Weber}]{Bazavov:2018wmo}
Bazavov, A., N.~Brambilla, P.~Petreczky, A.~Vairo, and J.~H. Weber (2018).
\newblock {Color screening in (2+1)-flavor QCD}.
\newblock \emph{Phys. Rev. D}, \textbf{98}(5), p. 054511.
\newblock \doi{10.1103/PhysRevD.98.054511}.

\bibitem[{Birrell(2014)}]{Birrell:2014ona}
Birrell, J. (2014).
\newblock \emph{{Non-Equilibrium Aspects of Relic Neutrinos: From Freeze-out to
  the Present Day}}.
\newblock Ph.D. thesis, Arizona U.

\bibitem[{Birrell and Rafelski(2015)}]{Birrell:2014cja}
Birrell, J. and J.~Rafelski (2015).
\newblock {Quark\textendash{}gluon plasma as the possible source of
  cosmological dark radiation}.
\newblock \emph{Phys. Lett. B}, \textbf{741}, pp. 77--81.
\newblock \doi{10.1016/j.physletb.2014.12.033}.

\bibitem[{Birrell et~al.(2015)Birrell, Wilkening, and
  Rafelski}]{Birrell:2014gea}
Birrell, J., J.~Wilkening, and J.~Rafelski (2015).
\newblock {Boltzmann Equation Solver Adapted to Emergent Chemical
  Non-equilibrium}.
\newblock \emph{J. Comput. Phys.}, \textbf{281}, pp. 896--916.
\newblock \doi{10.1016/j.jcp.2014.10.056}.

\bibitem[{Birrell et~al.(2013)Birrell, Yang, Chen, and
  Rafelski}]{Birrell:2013gpa}
Birrell, J., C.-T. Yang, P.~Chen, and J.~Rafelski (2013).
\newblock {Fugacity and Reheating of Primordial Neutrinos}.
\newblock \emph{Mod. Phys. Lett. A}, \textbf{28}, p. 1350188.
\newblock \doi{10.1142/S0217732313501885}.

\bibitem[{Birrell et~al.(2014{\natexlab{a}})Birrell, Yang, Chen, and
  Rafelski}]{Birrell:2012gg}
Birrell, J., C.-T. Yang, P.~Chen, and J.~Rafelski (2014{\natexlab{a}}).
\newblock {Relic neutrinos: Physically consistent treatment of effective number
  of neutrinos and neutrino mass}.
\newblock \emph{Phys. Rev. D}, \textbf{89}, p. 023008.
\newblock \doi{10.1103/PhysRevD.89.023008}.

\bibitem[{Birrell et~al.(2014{\natexlab{b}})Birrell, Yang, and
  Rafelski}]{Birrell:2014uka}
Birrell, J., C.-T. Yang, and J.~Rafelski (2014{\natexlab{b}}).
\newblock {Relic Neutrino Freeze-out: Dependence on Natural Constants}.
\newblock \emph{Nucl. Phys. B}, \textbf{890}, pp. 481--517.
\newblock \doi{10.1016/j.nuclphysb.2014.11.020}.

\bibitem[{Blennow et~al.(2012)Blennow, Fernandez-Martinez, Mena, Redondo, and
  Serra}]{Blennow:2012de}
Blennow, M., E.~Fernandez-Martinez, O.~Mena, J.~Redondo, and P.~Serra (2012).
\newblock {Asymmetric Dark Matter and Dark Radiation}.
\newblock \emph{JCAP}, \textbf{07}, p. 022.
\newblock \doi{10.1088/1475-7516/2012/07/022}.

\bibitem[{Boehm et~al.(2012)Boehm, Dolan, and McCabe}]{Boehm:2012gr}
Boehm, C., M.~J. Dolan, and C.~McCabe (2012).
\newblock {Increasing Neff with particles in thermal equilibrium with
  neutrinos}.
\newblock \emph{JCAP}, \textbf{12}, p. 027.
\newblock \doi{10.1088/1475-7516/2012/12/027}.

\bibitem[{Brambilla et~al.(2010)Brambilla, Escobedo, Ghiglieri, Soto, and
  Vairo}]{Brambilla:2010vq}
Brambilla, N., M.~A. Escobedo, J.~Ghiglieri, J.~Soto, and A.~Vairo (2010).
\newblock {Heavy Quarkonium in a weakly-coupled quark-gluon plasma below the
  melting temperature}.
\newblock \emph{JHEP}, \textbf{09}, p. 038.
\newblock \doi{10.1007/JHEP09(2010)038}.

\bibitem[{Brambilla et~al.(2018)Brambilla, Escobedo, Soto, and
  Vairo}]{Brambilla:2017zei}
Brambilla, N., M.~A. Escobedo, J.~Soto, and A.~Vairo (2018).
\newblock {Heavy quarkonium suppression in a fireball}.
\newblock \emph{Phys. Rev. D}, \textbf{97}(7), p. 074009.
\newblock \doi{10.1103/PhysRevD.97.074009}.

\bibitem[{{Carraro} et~al.(1988){Carraro}, {Schafer}, and
  {Koonin}}]{1988ApJ...331..565C}
{Carraro}, C., A.~{Schafer}, and S.~E. {Koonin} (1988).
\newblock {Dynamic Screening of Thermonuclear Reactions}.
\newblock \emph{\apj}, \textbf{331}, p. 565.
\newblock \doi{10.1086/166582}.

\bibitem[{Choquet-Bruhat(2008)}]{choquet2008general}
Choquet-Bruhat, Y. (2008).
\newblock \emph{General relativity and the Einstein equations}.
\newblock OUP Oxford.

\bibitem[{Coc et~al.(2006)Coc, Olive, Uzan, and Vangioni}]{Coc:2006rt}
Coc, A., K.~A. Olive, J.-P. Uzan, and E.~Vangioni (2006).
\newblock {Big bang nucleosynthesis constraints on scalar-tensor theories of
  gravity}.
\newblock \emph{Phys. Rev. D}, \textbf{73}, p. 083525.
\newblock \doi{10.1103/PhysRevD.73.083525}.

\bibitem[{Cugnon and Lombard(1984)}]{Cugnon:1984pm}
Cugnon, J. and R.~M. Lombard (1984).
\newblock {K+ PRODUCTION IN A CASCADE MODEL FOR HIGH-ENERGY NUCLEUS NUCLEUS
  COLLISIONS}.
\newblock \emph{Nucl. Phys. A}, \textbf{422}, pp. 635--653.
\newblock \doi{10.1016/0375-9474(84)90369-5}.

\bibitem[{Czarnecki et~al.(2004)Czarnecki, Marciano, and
  Sirlin}]{Czarnecki:2004cw}
Czarnecki, A., W.~J. Marciano, and A.~Sirlin (2004).
\newblock {Precision measurements and CKM unitarity}.
\newblock \emph{Phys. Rev. D}, \textbf{70}, p. 093006.
\newblock \doi{10.1103/PhysRevD.70.093006}.

\bibitem[{Davidson et~al.(2008)Davidson, Nardi, and Nir}]{Davidson:2008bu}
Davidson, S., E.~Nardi, and Y.~Nir (2008).
\newblock {Leptogenesis}.
\newblock \emph{Phys. Rept.}, \textbf{466}, pp. 105--177.
\newblock \doi{10.1016/j.physrep.2008.06.002}.

\bibitem[{Dicus et~al.(1982)Dicus, Kolb, Gleeson, Sudarshan, Teplitz, and
  Turner}]{Dicus:1982bz}
Dicus, D.~A., E.~W. Kolb, A.~M. Gleeson, E.~C.~G. Sudarshan, V.~L. Teplitz, and
  M.~S. Turner (1982).
\newblock {Primordial Nucleosynthesis Including Radiative, Coulomb, and Finite
  Temperature Corrections to Weak Rates}.
\newblock \emph{Phys. Rev. D}, \textbf{26}, p. 2694.
\newblock \doi{10.1103/PhysRevD.26.2694}.

\bibitem[{Dodelson(2003)}]{Dodelson:2003ft}
Dodelson, S. (2003).
\newblock \emph{{Modern Cosmology}}.
\newblock Academic Press, Amsterdam.
\newblock ISBN 978-0-12-219141-1.

\bibitem[{Dreiner et~al.(2012)Dreiner, Hanussek, Kim, and
  Sarkar}]{Dreiner:2011fp}
Dreiner, H.~K., M.~Hanussek, J.~S. Kim, and S.~Sarkar (2012).
\newblock {Gravitino cosmology with a very light neutralino}.
\newblock \emph{Phys. Rev. D}, \textbf{85}, p. 065027.
\newblock \doi{10.1103/PhysRevD.85.065027}.

\bibitem[{Elmfors et~al.(1997)Elmfors, Enqvist, Raffelt, and
  Sigl}]{Elmfors:1997tt}
Elmfors, P., K.~Enqvist, G.~Raffelt, and G.~Sigl (1997).
\newblock {Neutrinos with magnetic moment: Depolarization rate in plasma}.
\newblock \emph{Nucl. Phys. B}, \textbf{503}, pp. 3--23.
\newblock \doi{10.1016/S0550-3213(97)00382-9}.

\bibitem[{Elze et~al.(1980)Elze, Greiner, and Rafelski}]{Elze:1980er}
Elze, H.~T., W.~Greiner, and J.~Rafelski (1980).
\newblock {The relativistic ideal Fermi gas revisited}.
\newblock \emph{J. Phys. G}, \textbf{6}, pp. L149--L153.
\newblock \doi{10.1088/0305-4616/6/9/003}.

\bibitem[{Famiano et~al.(2016)Famiano, Balantekin, and
  Kajino}]{Famiano:2016hhs}
Famiano, M.~A., A.~B. Balantekin, and T.~Kajino (2016).
\newblock {Low-lying Resonances and Relativistic Screening in Big Bang
  Nucleosynthesis}.
\newblock \emph{Phys. Rev. C}, \textbf{93}(4), p. 045804.
\newblock \doi{10.1103/PhysRevC.93.045804}.

\bibitem[{Fields(2011)}]{Fields:2011zzb}
Fields, B.~D. (2011).
\newblock {The primordial lithium problem}.
\newblock \emph{Ann. Rev. Nucl. Part. Sci.}, \textbf{61}, pp. 47--68.
\newblock \doi{10.1146/annurev-nucl-102010-130445}.

\bibitem[{Formanek et~al.(2018)Formanek, Evans, Rafelski, Steinmetz, and
  Yang}]{Formanek:2017mbv}
Formanek, M., S.~Evans, J.~Rafelski, A.~Steinmetz, and C.-T. Yang (2018).
\newblock {Strong fields and neutral particle magnetic moment dynamics}.
\newblock \emph{Plasma Phys. Control. Fusion}, \textbf{60}, p. 074006.
\newblock \doi{10.1088/1361-6587/aac06a}.

\bibitem[{Formanek et~al.(2021)Formanek, Grayson, Rafelski, and
  M\"uller}]{Formanek:2021blc}
Formanek, M., C.~Grayson, J.~Rafelski, and B.~M\"uller (2021).
\newblock {Current-conserving relativistic linear response for collisional
  plasmas}.
\newblock \emph{Annals Phys.}, \textbf{434}, p. 168605.
\newblock \doi{10.1016/j.aop.2021.168605}.

\bibitem[{Fornengo et~al.(1997)Fornengo, Kim, and Song}]{Fornengo:1997wa}
Fornengo, N., C.~W. Kim, and J.~Song (1997).
\newblock {Finite temperature effects on the neutrino decoupling in the early
  universe}.
\newblock \emph{Phys. Rev. D}, \textbf{56}, pp. 5123--5134.
\newblock \doi{10.1103/PhysRevD.56.5123}.

\bibitem[{Freedman et~al.(1977)Freedman, Schramm, and Tubbs}]{Freedman:1977xn}
Freedman, D.~Z., D.~N. Schramm, and D.~L. Tubbs (1977).
\newblock {The Weak Neutral Current and Its Effects in Stellar Collapse}.
\newblock \emph{Ann. Rev. Nucl. Part. Sci.}, \textbf{27}, pp. 167--207.
\newblock \doi{10.1146/annurev.ns.27.120177.001123}.

\bibitem[{Fromerth et~al.(2012)Fromerth, Kuznetsova, Labun, Letessier, and
  Rafelski}]{Fromerth:2012fe}
Fromerth, M.~J., I.~Kuznetsova, L.~Labun, J.~Letessier, and J.~Rafelski (2012).
\newblock {From Quark-Gluon Universe to Neutrino Decoupling: 200 \ensuremath{<}
  T \ensuremath{<} 2MeV}.
\newblock \emph{Acta Phys. Polon. B}, \textbf{43}(12), pp. 2261--2284.
\newblock \doi{10.5506/APhysPolB.43.2261}.

\bibitem[{Fromerth and Rafelski(2002)}]{Fromerth:2002wb}
Fromerth, M.~J. and J.~Rafelski (2002).
\newblock {Hadronization of the quark Universe}.
\newblock \emph{arXiv astro-ph/0211346}.

\bibitem[{Fukugita and Yazaki(1987)}]{Fukugita:1987uy}
Fukugita, M. and S.~Yazaki (1987).
\newblock {Reexamination of Astrophysical and Cosmological Constraints on the
  Magnetic Moment of Neutrinos}.
\newblock \emph{Phys. Rev. D}, \textbf{36}, p. 3817.
\newblock \doi{10.1103/PhysRevD.36.3817}.

\bibitem[{Giovannini(2004)}]{giovannini2003magnetized}
Giovannini, M. (2004).
\newblock {The Magnetized universe}.
\newblock \emph{Int. J. Mod. Phys. D}, \textbf{13}, pp. 391--502.
\newblock \doi{10.1142/S0218271804004530}.

\bibitem[{Giudice et~al.(2004)Giudice, Notari, Raidal, Riotto, and
  Strumia}]{Giudice:2003jh}
Giudice, G.~F., A.~Notari, M.~Raidal, A.~Riotto, and A.~Strumia (2004).
\newblock {Towards a complete theory of thermal leptogenesis in the SM and
  MSSM}.
\newblock \emph{Nucl. Phys. B}, \textbf{685}, pp. 89--149.
\newblock \doi{10.1016/j.nuclphysb.2004.02.019}.

\bibitem[{Giunti and Kim(2007)}]{Giunti:2007ry}
Giunti, C. and C.~W. Kim (2007).
\newblock \emph{{Fundamentals of Neutrino Physics and Astrophysics}}.
\newblock Oxford University Press.
\newblock ISBN 978-0-19-850871-7.

\bibitem[{Giunti and Studenikin(2009)}]{Giunti:2008ve}
Giunti, C. and A.~Studenikin (2009).
\newblock {Neutrino electromagnetic properties}.
\newblock \emph{Phys. Atom. Nucl.}, \textbf{72}, pp. 2089--2125.
\newblock \doi{10.1134/S1063778809120126}.

\bibitem[{Grayson et~al.(2023)Grayson, Yang, Formanek, and
  Rafelski}]{Grayson:2023flr}
Grayson, C., C.~T. Yang, M.~Formanek, and J.~Rafelski (2023).
\newblock {Electron\textendash{}positron plasma in BBN: Damped-dynamic
  screening}.
\newblock \emph{Annals Phys.}, \textbf{458}, p. 169453.
\newblock \doi{10.1016/j.aop.2023.169453}.

\bibitem[{Griffiths(2008)}]{Griffiths:2008zz}
Griffiths, D. (2008).
\newblock \emph{{Introduction to elementary particles}}.
\newblock ISBN 978-3-527-40601-2.

\bibitem[{Gruzinov(1998)}]{Gruzinov:1997as}
Gruzinov, A.~V. (1998).
\newblock {Dynamic screening and thermonuclear reaction rates}.
\newblock \emph{Astrophys. J.}, \textbf{496}, p. 503.
\newblock \doi{10.1086/305349}.

\bibitem[{Heckler(1994)}]{Heckler:1994tv}
Heckler, A.~F. (1994).
\newblock {Astrophysical applications of quantum corrections to the equation of
  state of a plasma}.
\newblock \emph{Phys. Rev. D}, \textbf{49}, pp. 611--617.
\newblock \doi{10.1103/PhysRevD.49.611}.

\bibitem[{Hwang et~al.(2021)Hwang, Jang, Park, Kusakabe, Kajino, Balantekin,
  Maruyama, Ryu, and Cheoun}]{Hwang:2021kno}
Hwang, E., D.~Jang, K.~Park, M.~Kusakabe, T.~Kajino, A.~B. Balantekin,
  T.~Maruyama, C.-M. Ryu, and M.-K. Cheoun (2021).
\newblock {Dynamical screening effects on big bang nucleosynthesis}.
\newblock \emph{JCAP}, \textbf{11}, p. 017.
\newblock \doi{10.1088/1475-7516/2021/11/017}.

\bibitem[{Jedamzik and Saveliev(2019)}]{jedamzik2019stringent}
Jedamzik, K. and A.~Saveliev (2019).
\newblock Stringent limit on primordial magnetic fields from the cosmic
  microwave background radiation.
\newblock \emph{Physical review letters}, \textbf{123}(2), p. 021301.
\newblock \doi{10.1103/PhysRevLett.123.021301}.

\bibitem[{Karsch et~al.(1988)Karsch, Mehr, and Satz}]{Karsch:1987pv}
Karsch, F., M.~T. Mehr, and H.~Satz (1988).
\newblock {Color Screening and Deconfinement for Bound States of Heavy Quarks}.
\newblock \emph{Z. Phys. C}, \textbf{37}, p. 617.
\newblock \doi{10.1007/BF01549722}.

\bibitem[{Kislinger and Morley(1976)}]{Kislinger:1975uy}
Kislinger, M.~B. and P.~D. Morley (1976).
\newblock {Collective Phenomena in Gauge Theories. 1. The Plasmon Effect for
  Yang-Mills Fields}.
\newblock \emph{Phys. Rev. D}, \textbf{13}, p. 2765.
\newblock \doi{10.1103/PhysRevD.13.2765}.

\bibitem[{Koch et~al.(1986)Koch, Muller, and Rafelski}]{Koch:1986ud}
Koch, P., B.~Muller, and J.~Rafelski (1986).
\newblock {Strangeness in Relativistic Heavy Ion Collisions}.
\newblock \emph{Phys. Rept.}, \textbf{142}, pp. 167--262.
\newblock \doi{10.1016/0370-1573(86)90096-7}.

\bibitem[{Kolb et~al.(1996)Kolb, Linde, and Riotto}]{Kolb:1996jt}
Kolb, E.~W., A.~D. Linde, and A.~Riotto (1996).
\newblock {GUT baryogenesis after preheating}.
\newblock \emph{Phys. Rev. Lett.}, \textbf{77}, pp. 4290--4293.
\newblock \doi{10.1103/PhysRevLett.77.4290}.

\bibitem[{Kolb and Turner(1990{\natexlab{a}})}]{kolb1990early}
Kolb, E.~W. and M.~S. Turner (1990{\natexlab{a}}).
\newblock \emph{The early universe}.
\newblock CRC Press.

\bibitem[{Kolb and Turner(1990{\natexlab{b}})}]{Kolb:1990vq}
Kolb, E.~W. and M.~S. Turner (1990{\natexlab{b}}).
\newblock \emph{{The Early Universe}}, volume~69.
\newblock ISBN 978-0-201-62674-2.
\newblock \doi{10.1201/9780429492860}.

\bibitem[{Kronberg(1994{\natexlab{a}})}]{Kronberg:1993vk}
Kronberg, P.~P. (1994{\natexlab{a}}).
\newblock {Extragalactic magnetic fields}.
\newblock \emph{Rept. Prog. Phys.}, \textbf{57}, pp. 325--382.
\newblock \doi{10.1088/0034-4885/57/4/001}.

\bibitem[{Kronberg(1994{\natexlab{b}})}]{kronberg1994extragalactic}
Kronberg, P.~P. (1994{\natexlab{b}}).
\newblock Extragalactic magnetic fields.
\newblock \emph{Reports on Progress in Physics}, \textbf{57}(4), p. 325.
\newblock \doi{10.1088/0034-4885/57/4/001}.

\bibitem[{Kuzmin et~al.(1985)Kuzmin, Rubakov, and Shaposhnikov}]{Kuzmin:1985mm}
Kuzmin, V.~A., V.~A. Rubakov, and M.~E. Shaposhnikov (1985).
\newblock {On the Anomalous Electroweak Baryon Number Nonconservation in the
  Early Universe}.
\newblock \emph{Phys. Lett. B}, \textbf{155}, p.~36.
\newblock \doi{10.1016/0370-2693(85)91028-7}.

\bibitem[{Kuzmin et~al.(1987)Kuzmin, Rubakov, and Shaposhnikov}]{Kuzmin:1987wn}
Kuzmin, V.~A., V.~A. Rubakov, and M.~E. Shaposhnikov (1987).
\newblock {Anomalous Electroweak Baryon Number Nonconservation and GUT
  Mechanism for Baryogenesis}.
\newblock \emph{Phys. Lett. B}, \textbf{191}, pp. 171--173.
\newblock \doi{10.1016/0370-2693(87)91340-2}.

\bibitem[{Kuznetsova et~al.(2008)Kuznetsova, Habs, and
  Rafelski}]{Kuznetsova:2008jt}
Kuznetsova, I., D.~Habs, and J.~Rafelski (2008).
\newblock {Pion and muon production in e-, e+, gamma plasma}.
\newblock \emph{Phys. Rev. D}, \textbf{78}, p. 014027.
\newblock \doi{10.1103/PhysRevD.78.014027}.

\bibitem[{Kuznetsova et~al.(2010)Kuznetsova, Habs, and
  Rafelski}]{Kuznetsova:2009bq}
Kuznetsova, I., D.~Habs, and J.~Rafelski (2010).
\newblock {Thermal reaction processes in a relativistic QED plasma drop}.
\newblock \emph{Phys. Rev. D}, \textbf{81}, p. 053007.
\newblock \doi{10.1103/PhysRevD.81.053007}.

\bibitem[{Kuznetsova and Rafelski(2010{\natexlab{a}})}]{Kuznetsova:2010pi}
Kuznetsova, I. and J.~Rafelski (2010{\natexlab{a}}).
\newblock {Unstable Hadrons in Hot Hadron Gas in Laboratory and in the Early
  Universe}.
\newblock \emph{Phys. Rev. C}, \textbf{82}, p. 035203.
\newblock \doi{10.1103/PhysRevC.82.035203}.

\bibitem[{Kuznetsova and Rafelski(2010{\natexlab{b}})}]{PhysRevC.82.035203}
Kuznetsova, I. and J.~Rafelski (2010{\natexlab{b}}).
\newblock Unstable hadrons in hot hadron gas: In the laboratory and in the
  early Universe.
\newblock \emph{Phys. Rev. C}, \textbf{82}, p. 035203.
\newblock \doi{10.1103/PhysRevC.82.035203}.

\bibitem[{Kuznetsova and Rafelski(2012)}]{Kuznetsova:2011wt}
Kuznetsova, I. and J.~Rafelski (2012).
\newblock {Electron-Positron Plasma Drop Formed by Ultra-Intense Laser Pulses}.
\newblock \emph{Phys. Rev. D}, \textbf{85}, p. 085014.
\newblock \doi{10.1103/PhysRevD.85.085014}.

\bibitem[{Letessier and Rafelski(2002)}]{Letessier:2002ony}
Letessier, J. and J.~Rafelski (2002).
\newblock \emph{{Hadrons and Quark\textendash{}Gluon Plasma}}.
\newblock Oxford University Press.
\newblock ISBN 978-1-00-929075-3, 978-0-511-53499-7, 978-1-00-929070-8,
  978-1-00-929073-9.
\newblock \doi{10.1017/9781009290753}.

\bibitem[{Letessier and Rafelski(2008)}]{Letessier:2005qe}
Letessier, J. and J.~Rafelski (2008).
\newblock {Hadron production and phase changes in relativistic heavy ion
  collisions}.
\newblock \emph{Eur. Phys. J. A}, \textbf{35}, pp. 221--242.
\newblock \doi{10.1140/epja/i2007-10546-7}.

\bibitem[{Lewis(1980)}]{PhysRevD.21.663}
Lewis, R.~R. (1980).
\newblock Coherent detector for low-energy neutrinos.
\newblock \emph{Phys. Rev. D}, \textbf{21}, pp. 663--668.
\newblock \doi{10.1103/PhysRevD.21.663}.

\bibitem[{Mangano et~al.(2002)Mangano, Miele, Pastor, and
  Peloso}]{Mangano:2001iu}
Mangano, G., G.~Miele, S.~Pastor, and M.~Peloso (2002).
\newblock {A Precision calculation of the effective number of cosmological
  neutrinos}.
\newblock \emph{Phys. Lett. B}, \textbf{534}, pp. 8--16.
\newblock \doi{10.1016/S0370-2693(02)01622-2}.

\bibitem[{Mangano et~al.(2005)Mangano, Miele, Pastor, Pinto, Pisanti, and
  Serpico}]{Mangano:2005cc}
Mangano, G., G.~Miele, S.~Pastor, T.~Pinto, O.~Pisanti, and P.~D. Serpico
  (2005).
\newblock {Relic neutrino decoupling including flavor oscillations}.
\newblock \emph{Nucl. Phys. B}, \textbf{729}, pp. 221--234.
\newblock \doi{10.1016/j.nuclphysb.2005.09.041}.

\bibitem[{Mangano et~al.(2006)Mangano, Miele, Pastor, Pinto, Pisanti, and
  Serpico}]{Mangano:2006ar}
Mangano, G., G.~Miele, S.~Pastor, T.~Pinto, O.~Pisanti, and P.~D. Serpico
  (2006).
\newblock {Effects of non-standard neutrino-electron interactions on relic
  neutrino decoupling}.
\newblock \emph{Nucl. Phys. B}, \textbf{756}, pp. 100--116.
\newblock \doi{10.1016/j.nuclphysb.2006.09.002}.

\bibitem[{Morgan(1981)}]{Morgan:1981zy}
Morgan, J.~A. (1981).
\newblock {Cosmological upper limit to neutrino magnetic moments}.
\newblock \emph{Phys. Lett. B}, \textbf{102}, pp. 247--250.
\newblock \doi{10.1016/0370-2693(81)90868-6}.

\bibitem[{Morrissey and Ramsey-Musolf(2012)}]{Morrissey:2012db}
Morrissey, D.~E. and M.~J. Ramsey-Musolf (2012).
\newblock {Electroweak baryogenesis}.
\newblock \emph{New J. Phys.}, \textbf{14}, p. 125003.
\newblock \doi{10.1088/1367-2630/14/12/125003}.

\bibitem[{Mukhanov(2005)}]{Mukhanov:2005sc}
Mukhanov, V. (2005).
\newblock \emph{{Physical Foundations of Cosmology}}.
\newblock Cambridge University Press, Oxford.
\newblock ISBN 978-0-521-56398-7.
\newblock \doi{10.1017/CBO9780511790553}.

\bibitem[{Neronov and Vovk(2010)}]{neronov2010evidence}
Neronov, A. and I.~Vovk (2010).
\newblock Evidence for strong extragalactic magnetic fields from Fermi
  observations of TeV blazars.
\newblock \emph{Science}, \textbf{328}(5974), pp. 73--75.
\newblock \doi{10.1126/science.1184192}.

\bibitem[{Nicolescu(2013)}]{Nicolescu:2013rxa}
Nicolescu, G. (2013).
\newblock A heuristic approach to the detection of solar neutrinos.
\newblock \emph{Journal of Physics G: Nuclear and Particle Physics},
  \textbf{40}, p. 055201.
\newblock \doi{10.1088/0954-3899/40/5/055201}.

\bibitem[{Nielsen and Takanishi(2001)}]{Nielsen:2001fy}
Nielsen, H.~B. and Y.~Takanishi (2001).
\newblock {Baryogenesis via lepton number violation in anti-GUT model}.
\newblock \emph{Phys. Lett. B}, \textbf{507}, pp. 241--251.
\newblock \doi{10.1016/S0370-2693(01)00357-4}.

\bibitem[{Offler et~al.(2019)Offler, Aarts, Allton, Glesaaen, J\"ager, Kim,
  Lombardo, Ryan, and Skullerud}]{Offler:2019eij}
Offler, S., G.~Aarts, C.~Allton, J.~Glesaaen, B.~J\"ager, S.~Kim, M.~P.
  Lombardo, S.~M. Ryan, and J.-I. Skullerud (2019).
\newblock {News from bottomonium spectral functions in thermal QCD}.
\newblock \emph{PoS}, \textbf{LATTICE2019}, p. 076.
\newblock \doi{10.22323/1.363.0076}.

\bibitem[{Papavassiliou et~al.(2006)Papavassiliou, Bernabeu, and
  Passera}]{Papavassiliou:2005cs}
Papavassiliou, J., J.~Bernabeu, and M.~Passera (2006).
\newblock {Neutrino-nuclear coherent scattering and the effective neutrino
  charge radius}.
\newblock \emph{PoS}, \textbf{HEP2005}, p. 192.
\newblock \doi{10.22323/1.021.0192}.

\bibitem[{Patrignani et~al.(2016)}]{ParticleDataGroup:2016lqr}
Patrignani, C. et~al. (2016).
\newblock {Review of Particle Physics}.
\newblock \emph{Chin. Phys. C}, \textbf{40}(10), p. 100001.
\newblock \doi{10.1088/1674-1137/40/10/100001}.

\bibitem[{Pitrou et~al.(2018)Pitrou, Coc, Uzan, and Vangioni}]{Pitrou:2018cgg}
Pitrou, C., A.~Coc, J.-P. Uzan, and E.~Vangioni (2018).
\newblock {Precision big bang nucleosynthesis with improved Helium-4
  predictions}.
\newblock \emph{Phys. Rept.}, \textbf{754}, pp. 1--66.
\newblock \doi{10.1016/j.physrep.2018.04.005}.

\bibitem[{Pshirkov et~al.(2016)Pshirkov, Tinyakov, and Urban}]{pshirkov2015new}
Pshirkov, M.~S., P.~G. Tinyakov, and F.~R. Urban (2016).
\newblock {New limits on extragalactic magnetic fields from rotation measures}.
\newblock \emph{Phys. Rev. Lett.}, \textbf{116}(19), p. 191302.
\newblock \doi{10.1103/PhysRevLett.116.191302}.

\bibitem[{Rafelski(2020)}]{Rafelski:2019twp}
Rafelski, J. (2020).
\newblock {Discovery of Quark-Gluon-Plasma: Strangeness Diaries}.
\newblock \emph{Eur. Phys. J. ST}, \textbf{229}(1), pp. 1--140.
\newblock \doi{10.1140/epjst/e2019-900263-x}.

\bibitem[{Rafelski and Birrell(2014)}]{Rafelski:2013yka}
Rafelski, J. and J.~Birrell (2014).
\newblock {Traveling Through the Universe: Back in Time to the Quark-Gluon
  Plasma Era}.
\newblock \emph{J. Phys. Conf. Ser.}, \textbf{509}, p. 012014.
\newblock \doi{10.1088/1742-6596/509/1/012014}.

\bibitem[{Rafelski et~al.(2023)Rafelski, Birrell, Steinmetz, and
  Yang}]{Rafelski:2023emw}
Rafelski, J., J.~Birrell, A.~Steinmetz, and C.~T. Yang (2023).
\newblock {A short survey of matter-antimatter evolution in the primordial
  universe}.

\bibitem[{Rafelski et~al.(2001)Rafelski, Letessier, and
  Torrieri}]{Rafelski:2001hp}
Rafelski, J., J.~Letessier, and G.~Torrieri (2001).
\newblock {Strange hadrons and their resonances: A Diagnostic tool of QGP
  freezeout dynamics}.
\newblock \emph{Phys. Rev. C}, \textbf{64}, p. 054907.
\newblock \doi{10.1103/PhysRevC.64.054907}.
\newblock [Erratum: Phys.Rev.C 65, 069902 (2002)].

\bibitem[{Rafelski and Yang(2021)}]{Rafelski:2021aey}
Rafelski, J. and C.~T. Yang (2021).
\newblock {The muon abundance in the primordial Universe}.
\newblock \emph{Acta Phys. Polon. B}, \textbf{52}, p. 277.
\newblock \doi{10.5506/APhysPolB.52.277}.

\bibitem[{Rafelski and Yang(2022)}]{Rafelski:2020ajx}
Rafelski, J. and C.~T. Yang (2022).
\newblock {Reactions Governing Strangeness Abundance in Primordial Universe}.
\newblock \emph{EPJ Web Conf.}, \textbf{259}, p. 13001.
\newblock \doi{10.1051/epjconf/202225913001}.

\bibitem[{Riess et~al.(2018{\natexlab{a}})Riess, Casertano, Yuan, Macri,
  Anderson, MacKenty, Bowers, Clubb, Filippenko, Jones et~al.}]{riess2018new}
Riess, A.~G., S.~Casertano, W.~Yuan, L.~Macri, J.~Anderson, J.~W. MacKenty,
  J.~B. Bowers, K.~I. Clubb, A.~V. Filippenko, D.~O. Jones, et~al.
  (2018{\natexlab{a}}).
\newblock New parallaxes of galactic cepheids from spatially scanning the
  hubble space telescope: Implications for the hubble constant.
\newblock \emph{The Astrophysical Journal}, \textbf{855}(2), p. 136.

\bibitem[{Riess et~al.(2018{\natexlab{b}})}]{Riess:2018byc}
Riess, A.~G. et~al. (2018{\natexlab{b}}).
\newblock {Milky Way Cepheid Standards for Measuring Cosmic Distances and
  Application to Gaia DR2: Implications for the Hubble Constant}.
\newblock \emph{Astrophys. J.}, \textbf{861}(2), p. 126.
\newblock \doi{10.3847/1538-4357/aac82e}.

\bibitem[{Riotto and Trodden(1999)}]{Riotto:1999yt}
Riotto, A. and M.~Trodden (1999).
\newblock {Recent progress in baryogenesis}.
\newblock \emph{Ann. Rev. Nucl. Part. Sci.}, \textbf{49}, pp. 35--75.
\newblock \doi{10.1146/annurev.nucl.49.1.35}.

\bibitem[{Sajjad~Athar et~al.(2023)Sajjad~Athar, Fatima, and
  Singh}]{SajjadAthar:2022pjt}
Sajjad~Athar, M., A.~Fatima, and S.~K. Singh (2023).
\newblock {Neutrinos and their interactions with matter}.
\newblock \emph{Prog. Part. Nucl. Phys.}, \textbf{129}, p. 104019.
\newblock \doi{10.1016/j.ppnp.2022.104019}.

\bibitem[{Sakharov(1967)}]{Sakharov:1967dj}
Sakharov, A.~D. (1967).
\newblock {Violation of CP Invariance, C asymmetry, and baryon asymmetry of the
  universe}.
\newblock \emph{Pisma Zh. Eksp. Teor. Fiz.}, \textbf{5}, pp. 32--35.
\newblock \doi{10.1070/PU1991v034n05ABEH002497}.

\bibitem[{Sakharov(1990)}]{Sakharov:1988vdp}
Sakharov, A.~D. (1990).
\newblock {Baryon asymmetry of the universe}.
\newblock pp. 65--80.
\newblock \doi{10.1070/PU1991v034n05ABEH002504}.

\bibitem[{Salpeter(1954)}]{Salpeter:1954nc}
Salpeter, E.~E. (1954).
\newblock {Electron screening and thermonuclear reactions}.
\newblock \emph{Austral. J. Phys.}, \textbf{7}, pp. 373--388.
\newblock \doi{10.1071/PH540373}.

\bibitem[{{Salpeter} and {van Horn}(1969)}]{1969ApJ...155..183S}
{Salpeter}, E.~E. and H.~M. {van Horn} (1969).
\newblock {Nuclear Reaction Rates at High Densities}.
\newblock \emph{\apj}, \textbf{155}, p. 183.
\newblock \doi{10.1086/149858}.

\bibitem[{Schroedter et~al.(2000)Schroedter, Thews, and
  Rafelski}]{Schroedter:2000ek}
Schroedter, M., R.~L. Thews, and J.~Rafelski (2000).
\newblock {$B_c$ meson production in nuclear collisions at RHIC}.
\newblock \emph{Phys. Rev. C}, \textbf{62}, p. 024905.
\newblock \doi{10.1103/PhysRevC.62.024905}.

\bibitem[{Sehgal and Wanninger(1986)}]{Sehgal:1986gn}
Sehgal, L.~M. and M.~Wanninger (1986).
\newblock {Atomic Effects in Coherent Neutrino Scattering}.
\newblock \emph{Phys. Lett. B}, \textbf{171}, pp. 107--112.
\newblock \doi{10.1016/0370-2693(86)91008-7}.

\bibitem[{Smith(1984)}]{Smith:1985mta}
Smith, P.~F. (1984).
\newblock Coherent neutrino scattering—Relativistic and nonrelativistic.
\newblock \emph{II Nuovo Cimento A (1965-1970)}, \textbf{83}, pp. 263--274.
\newblock \doi{10.1007/BF02902601}.

\bibitem[{Steinmetz et~al.(2023{\natexlab{a}})Steinmetz, Yang, and
  Rafelski}]{Andrew:2023abc}
Steinmetz, A., C.~T. Yang, and J.~Rafelski (2023{\natexlab{a}}).
\newblock {Antimatter origin of cosmic magnetism: A first look}.
\newblock \emph{(in preparation)}.

\bibitem[{Steinmetz et~al.(2023{\natexlab{b}})Steinmetz, Yang, and
  Rafelski}]{Steinmetz:2023nsc}
Steinmetz, A., C.~T. Yang, and J.~Rafelski (2023{\natexlab{b}}).
\newblock {Matter-antimatter origin of cosmic magnetism}.

\bibitem[{Tanabashi et~al.(2018)}]{ParticleDataGroup:2018ovx}
Tanabashi, M. et~al. (2018).
\newblock {Review of Particle Physics}.
\newblock \emph{Phys. Rev. D}, \textbf{98}(3), p. 030001.
\newblock \doi{10.1103/PhysRevD.98.030001}.

\bibitem[{Taylor et~al.(2011)Taylor, Vovk, and
  Neronov}]{taylor2011extragalactic}
Taylor, A.~M., I.~Vovk, and A.~Neronov (2011).
\newblock Extragalactic magnetic fields constraints from simultaneous GeV--TeV
  observations of blazars.
\newblock \emph{Astronomy \& Astrophysics}, \textbf{529}, p. A144.
\newblock \doi{10.1051/0004-6361/201116441}.

\bibitem[{Tully(2019)}]{Tully:2019ltb}
Tully, A.~M. (2019).
\newblock \emph{{Doubly charmed $B$ decays with the LHCb experiment}}.
\newblock Ph.D. thesis, Cambridge U.
\newblock \doi{10.17863/CAM.44796}.

\bibitem[{Vazza et~al.(2021)}]{Vazza:2021vwy}
Vazza, F. et~al. (2021).
\newblock {Magnetogenesis and the Cosmic Web: A Joint Challenge for Radio
  Observations and Numerical Simulations}.
\newblock \emph{Galaxies}, \textbf{9}(4), p. 109.
\newblock \doi{10.3390/galaxies9040109}.

\bibitem[{Vernstrom et~al.(2021)Vernstrom, Heald, Vazza, Galvin, West,
  Locatelli, Fornengo, and Pinetti}]{vernstrom2021discovery}
Vernstrom, T., G.~Heald, F.~Vazza, T.~J. Galvin, J.~L. West, N.~Locatelli,
  N.~Fornengo, and E.~Pinetti (2021).
\newblock {Discovery of magnetic fields along stacked cosmic filaments as
  revealed by radio and X-ray emission}.
\newblock \emph{Monthly Notices of the Royal Astronomical Society},
  \textbf{505}(3), pp. 4178--4196.
\newblock \doi{10.1093/mnras/stab1301}.

\bibitem[{Vogel and Engel(1989)}]{Vogel:1989iv}
Vogel, P. and J.~Engel (1989).
\newblock {Neutrino Electromagnetic Form-Factors}.
\newblock \emph{Phys. Rev. D}, \textbf{39}, p. 3378.
\newblock \doi{10.1103/PhysRevD.39.3378}.

\bibitem[{Wang et~al.(2011)Wang, Bertulani, and Balantekin}]{Wang:2010px}
Wang, B., C.~A. Bertulani, and A.~B. Balantekin (2011).
\newblock {Electron screening and its effects on Big-Bang nucleosynthesis}.
\newblock \emph{Phys. Rev. C}, \textbf{83}, p. 018801.
\newblock \doi{10.1103/PhysRevC.83.018801}.

\bibitem[{Weber(1988)}]{PhysRevD.38.32}
Weber, J. (1988).
\newblock Apparent observation of abnormally large coherent scattering cross
  sections using keV and MeV range antineutrinos, and solar neutrinos.
\newblock \emph{Phys. Rev. D}, \textbf{38}, pp. 32--39.
\newblock \doi{10.1103/PhysRevD.38.32}.

\bibitem[{Widrow et~al.(2012)Widrow, Ryu, Schleicher, Subramanian, Tsagas, and
  Treumann}]{Widrow:2011hs}
Widrow, L.~M., D.~Ryu, D.~R.~G. Schleicher, K.~Subramanian, C.~G. Tsagas, and
  R.~A. Treumann (2012).
\newblock {The First Magnetic Fields}.
\newblock \emph{Space Sci. Rev.}, \textbf{166}, pp. 37--70.
\newblock \doi{10.1007/s11214-011-9833-5}.

\bibitem[{Wolfenstein(1978)}]{PhysRevD.17.2369}
Wolfenstein, L. (1978).
\newblock Neutrino oscillations in matter.
\newblock \emph{Phys. Rev. D}, \textbf{17}, pp. 2369--2374.
\newblock \doi{10.1103/PhysRevD.17.2369}.

\bibitem[{Workman et~al.(2022)}]{ParticleDataGroup:2022pth}
Workman, R.~L. et~al. (2022).
\newblock {Review of Particle Physics}.
\newblock \emph{PTEP}, \textbf{2022}, p. 083C01.
\newblock \doi{10.1093/ptep/ptac097}.

\bibitem[{Yang et~al.(2018{\natexlab{a}})Yang, Birrell, and
  Rafelski}]{Yang:2018oqg}
Yang, C.~T., J.~Birrell, and J.~Rafelski (2018{\natexlab{a}}).
\newblock {Lepton Number and Expansion of the Universe}.

\bibitem[{Yang et~al.(2018{\natexlab{b}})Yang, Birrell, and
  Rafelski}]{Yang:2018qrr}
Yang, C.~T., J.~Birrell, and J.~Rafelski (2018{\natexlab{b}}).
\newblock {Temperature Dependence of the Neutron Lifespan}.

\bibitem[{Yang and Rafelski(2020)}]{Yang:2020nne}
Yang, C.~T. and J.~Rafelski (2020).
\newblock {Possibility of bottom-catalyzed matter genesis near to primordial
  QGP hadronization}.

\bibitem[{Yang and Rafelski(2022)}]{Yang:2021bko}
Yang, C.~T. and J.~Rafelski (2022).
\newblock {Cosmological strangeness abundance}.
\newblock \emph{Phys. Lett. B}, \textbf{827}, p. 136944.
\newblock \doi{10.1016/j.physletb.2022.136944}.

\end{thebibliography}
